\documentclass[a4paper,11pt]{article}
\usepackage{jheppub} %
\usepackage[utf8]{inputenc}
\graphicspath{ {./figures/}}

\usepackage{multirow}

\usepackage{amssymb}
\usepackage{amsmath}
\usepackage{cancel}
\usepackage{tabu,booktabs}
\usepackage{color}
\usepackage{braket}
\usepackage{xcolor}
\usepackage{graphicx}
\usepackage{multirow}
\usepackage{verbatim}
\usepackage{amsthm}
\usepackage{slashed}
\usepackage{wasysym}
\usepackage{simplewick}
\usepackage{mathtools}
\usepackage{soul}
\usepackage{xspace}
\usepackage{mdframed}
\usepackage{appendix}
\usepackage{subcaption}
\usepackage[utf8]{inputenc}
\usepackage[T1]{fontenc}
\usepackage{overpic}
\usepackage{hyperref}

\definecolor{dancomment}{RGB}{0,159,0}

\def\cA{\mathcal{A}}

\def\cD{\mathcal{D}}

\def\cL{\mathcal{L}}

\def\cN{\mathcal{N}}
\def\cO{\mathcal{O}}

\def\cP{\mathcal{P}}

\def\nn{{\nonumber}}

\def\be#1\ee{\begin{align}#1\end{align}}

\def\({\left(}
\def\){\right)}
\def\[{\left[}
\def\]{\right]}

\allowdisplaybreaks[3]

\newcommand{\namedref}[2]{\hyperref[#2]{#1~\ref*{#2}}}
\newcommand{\secref}[1]{\namedref{Section}{#1}}
\newcommand{\appref}[1]{\namedref{Appendix}{#1}}
\newcommand{\tabref}[1]{\namedref{Table}{#1}}
\newcommand{\figref}[1]{\namedref{Figure}{#1}}

\newcommand{\avgBS}[1]{\left< #1 \right>_{SO(4)}}
\newcommand{\avgCut}[1]{\left< #1 \right>_{SO(1,3)}}
\newcommand{\avgSTD}[1]{\left< #1 \right>_{SO(3)}}

\newcommand{\msubJ}[1]{m_{1, #1}}

\newcommand{\Ga}{\Gamma}

\newcommand{\eps}{\epsilon}

\usepackage[font={small}]{caption}

\newcommand{\eeq}{\end{equation}}

\newcommand{\eeqq}{\end{equation*}}

\newcommand\eeqaa{\end{eqnarray*}}

\newcommand\eeqa{\end{array}}

\newcommand{\eea}{\end{eqnarray}}

\renewcommand{\Im}{\operatorname{Im}}

\newcommand\mgap{{m_{\text{gap}}^2}}

\def\PJfive{\tilde{P}_J}

\def\ODisp{\mathcal{A}}

\title{\boldmath

From Partons to Strings: Scattering on the Coulomb Branch of ${\cal N}=4$ SYM 
}

\author[a]{Luis F. Alday,}
\author[b,c]{Elisabetta Armanini,}
\author[d]{Kelian H\"{a}ring,}
\author[b]{Alexander Zhiboedov}
\affiliation[a]{Mathematical Institute, University of Oxford, Andrew Wiles Building, Radcliffe Observatory Quarter, Woodstock Road, Oxford, OX2 6GG, U.K.}
\affiliation[b]{CERN, Theoretical Physics Department, Geneva, Switzerland}
\affiliation[c]{Fields and Strings Laboratory, Institute of Physics,
\'Ecole Polytechnique F\'ed\'eral de Lausanne (EPFL), Route de la Sorge, CH-1015 Lausanne, Switzerland}
\affiliation[d]{Institute for Theoretical Physics, University of Amsterdam, 1090 GL Amsterdam, The Netherlands}

\abstract{We study scattering on the Coulomb branch of planar ${\mathcal{N}}=4$ SYM at finite 't Hooft coupling. 
This setup defines a family of classical open‑string S‑matrices that smoothly interpolates  between perturbative parton scattering at weak coupling and flat‑space string scattering at strong coupling. We focus on the four-point amplitude, which exhibits a remarkably rich structure: nonlinear Regge trajectories, dual conformal invariance, an intricate spectrum of bound states with an accumulation point, and a two-particle cut.
Dual conformal invariance relates the spectrum of Regge trajectories to the energy spectrum of the Maldacena-Wilson cusp Hamiltonian.
This connection allows us to use integrability to compute the leading and subleading Regge trajectories at finite coupling, which we then input into the bootstrap analysis.
At strong coupling, we use the worldsheet bootstrap to construct the first $AdS$-curvature correction to the Veneziano amplitude. 
We apply dispersion relations and S‑matrix bootstrap techniques to derive bounds on Wilson coefficients, couplings to bound states, and the overall shape of the amplitude. We find that the $\mathcal{N}=4$ amplitude saturates the bootstrap bounds at weak coupling and nearly saturates them at strong coupling.
At intermediate coupling, the amplitude traces a nontrivial path through the allowed space of observables. To characterize this path, we combine the weak- and strong-coupling information about the amplitude to construct a finite-coupling model for Wilson coefficients using a Pad\'{e} approximation. The resulting model satisfies bootstrap constraints and yields sharp predictions for the finite-coupling behavior of the amplitude. We provide evidence that complete monotonicity of the scattering amplitude, previously observed perturbatively, persists at finite coupling.
}

\begin{document}
\begin{flushright}
    {\ttfamily CERN-TH-2025-202}
\end{flushright}

\maketitle
\flushbottom

\newpage

\section{Introduction}
\label{sec:intro}

String theory appears to be the only framework that allows relativistic weakly-coupled massive higher-spin particles to interact consistently. There is a tension between Lorentz invariance, unitarity, and causality when massive higher-spin particles are exchanged \cite{Adams:2006sv,Camanho:2014apa}. 
Restoring causality requires a physical mechanism, and the only mechanism known to achieve this is the reorganization of higher-spin states into excitations of an extended object, a string.\footnote{See, for instance, Appendix G in \cite{Maldacena:2012sf} for the precise formulation of this conjecture. For graviton scattering, a much stronger hypothesis was put forward in \cite{Chowdhury:2019kaq}.}
Two canonical realizations of this idea are perturbative string theory and confining flux tubes in large-$N_c$ gauge theories. Although they might appear quite different at first glance, the two are deeply connected via the gauge/string duality, which equates gauge theory to string theory in a curved background \cite{Maldacena:1997re}.

Constructing S-matrices with massive higher spin particles is a hard task, even if one assumes that each such S-matrix arises from a worldsheet conformal field theory (CFT). The only known explicit examples up to date are just the Virasoro-Shapiro amplitude for closed strings, the Veneziano amplitude for open strings, and their higher-point cousins.\footnote{By the existence of an S-matrix, we mean that all $m \to n$ scattering amplitudes are well-defined, unitary, and causal.} These examples are very special: they exhibit an equidistant spectrum of particles and exactly linear Regge trajectories. Both properties are highly non-generic, and we expect them to be lifted if we deform the string background. It is therefore of great interest to construct a stringy S-matrix that ``steps away'' from these familiar but
highly non-generic examples.

\begin{figure}[h!]
    \centering
    \includegraphics[width=0.6\linewidth]{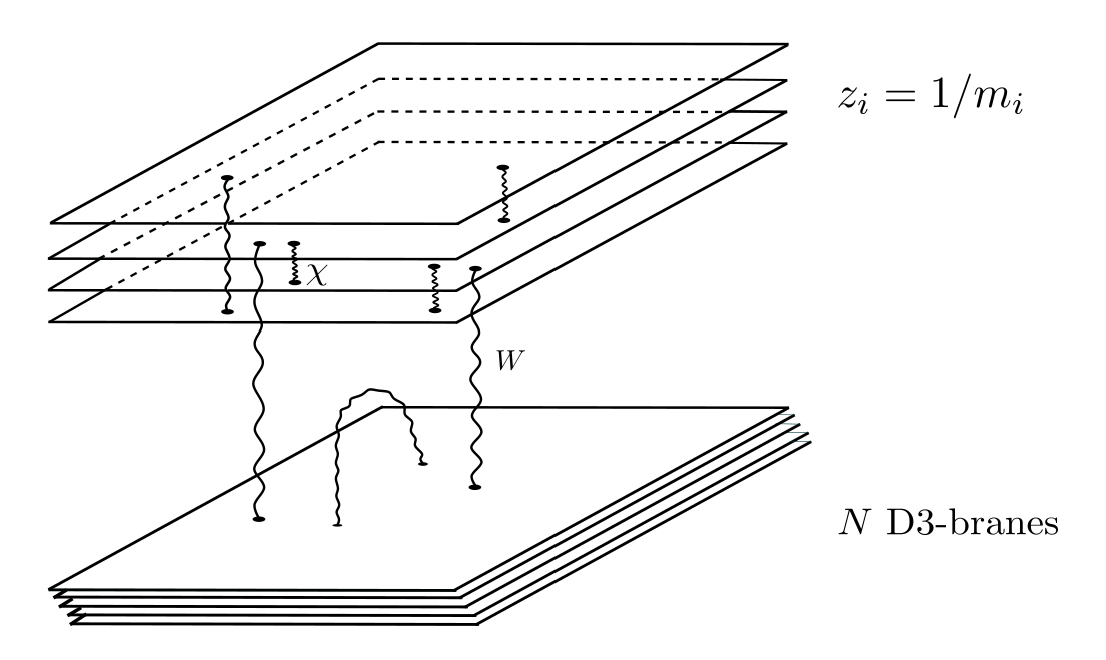}
    \caption{
    Scattering of open strings $\chi$ attached to distinct probe D3-branes at positions $z_i = 1/m_i$.
    The strings stretched between the probe branes and the stack of $N$ coincident D3-branes correspond to the massive W-bosons.
    }
    \label{fig:D3_branes}
\end{figure}

In this paper, we study scattering in ${\cal N}=4$ SYM on the Coulomb branch in the planar limit. Via the $AdS$/CFT correspondence it describes scattering of open strings on the probe D3-branes at fixed radial position in the dual $AdS_5 \times S^5$ geometry \cite{Maldacena:1997re}, see \figref{fig:D3_branes}. At weak 't Hooft coupling, the relevant degrees of freedom are partons (quarks, gluons, and W-bosons) of the boundary gauge theory, and the amplitude is given by a sum of planar Feynman diagrams \cite{Alday:2009zm}. Massive W-bosons created in the collision event interact via the attractive Coulomb potential,\footnote{This statement is true at leading order. At higher orders, the potential acquires nontrivial corrections, see \cite{Beneke:2013jia}.} generating a spectrum of bound states identical to that of the hydrogen atom \cite{Caron-Huot:2014gia}. At strong coupling, the spectrum of bound states becomes stringy, and remarkably \emph{low-energy} scattering is captured by the flat space super-Veneziano amplitude \cite{Maldacena:2022ckr}.\footnote{This is to be contrasted with the flat space limit of closed string scattering in $AdS$. There,  the flat space limit is encoded in the \emph{high-energy} part of the correlator and is related to probing the bulk-point features \cite{Gary:2009ae,Penedones:2010ue,Maldacena:2015iua,Caron-Huot:2025hmk}. This difference between the flat space limit for closed strings and open strings attached to a probe D3-brane has a clear geometrical explanation that we discuss in the bulk of the paper.} 
As the coupling changes, with the dual string becoming stiffer at larger coupling \cite{Klebanov:2006jj}, the amplitude smoothly interpolates between these two regimes.\footnote{In more complicated examples, such as large-$N_c$ QCD, both regimes are realized simultaneously at high and low energies due to the running of the coupling.} %
Therefore, scattering on the Coulomb branch defines a family of consistent tree-level open string S-matrices smoothly connected to the flat space string scattering.\footnote{At the level of the four-point amplitude many deformations of the Veneziano amplitude were investigated in the past %
\cite{Coon:1969yw,khuri1969derivation,Matsuda:1969zz,Mandelstam:1968czc, Cheung:2023adk,Cheung:2023uwn,Eckner:2024ggx,Haring:2023zwu}, however, it is not known if any of them can be uplifted to a fully consistent S-matrix, see e.g. \cite{Gross:1969db,Arkani-Hamed:2023jwn,Geiser:2023qqq,Cheung:2025krg}. The present paper describes one such family.}

The goal of the present work is to exploit integrability, perturbative results at weak coupling, the worldsheet bootstrap at strong coupling, and the S-matrix bootstrap to study this four-point amplitude at finite coupling. Let us introduce our setup. It is convenient to break the gauge group of ${\cal N}=4$ SYM spontaneously $U(N+4) \to U(N) \times U(1)^4$, which allows us to define the scattering amplitudes. We consider the scattering of `meson' fields $\chi_{i,j}$, with $1\leq i < j \leq 4$, see \figref{fig:D3_branes}, of mass $|m_i - m_j|$, where $m_i$'s are related to the vacuum expectation values of the scalar fields in ${\cal N}=4$ SYM and the radial position of the probe D3-branes in $AdS$ correspondingly. Explicitly, in Poincaré coordinates the positions of the probes are $z_i \sim 1/m_i$.
In the planar limit, the leading contribution to the meson scattering amplitude comes from creating a pair of off-shell W-bosons, which transform in the fundamental representation of the unbroken $U(N)$-theory. In other words, the meson scattering amplitude provides an interesting dynamical probe of the conformal gauge theory flux tube \cite{Klebanov:2006jj}.\footnote{Recall that conformal symmetry fixes the quark–antiquark potential, defined through the vacuum expectation value of the Wilson loop, to $V(r) = - {\alpha \over r}$. For dynamical quarks of finite mass, however, the potential receives corrections \cite{Beneke:2013jia}, and we will see that the corresponding scattering amplitude encodes a rich spectrum of bound states.}

\begin{figure}[h!]
    \centering
    \includegraphics[width=0.9\linewidth]{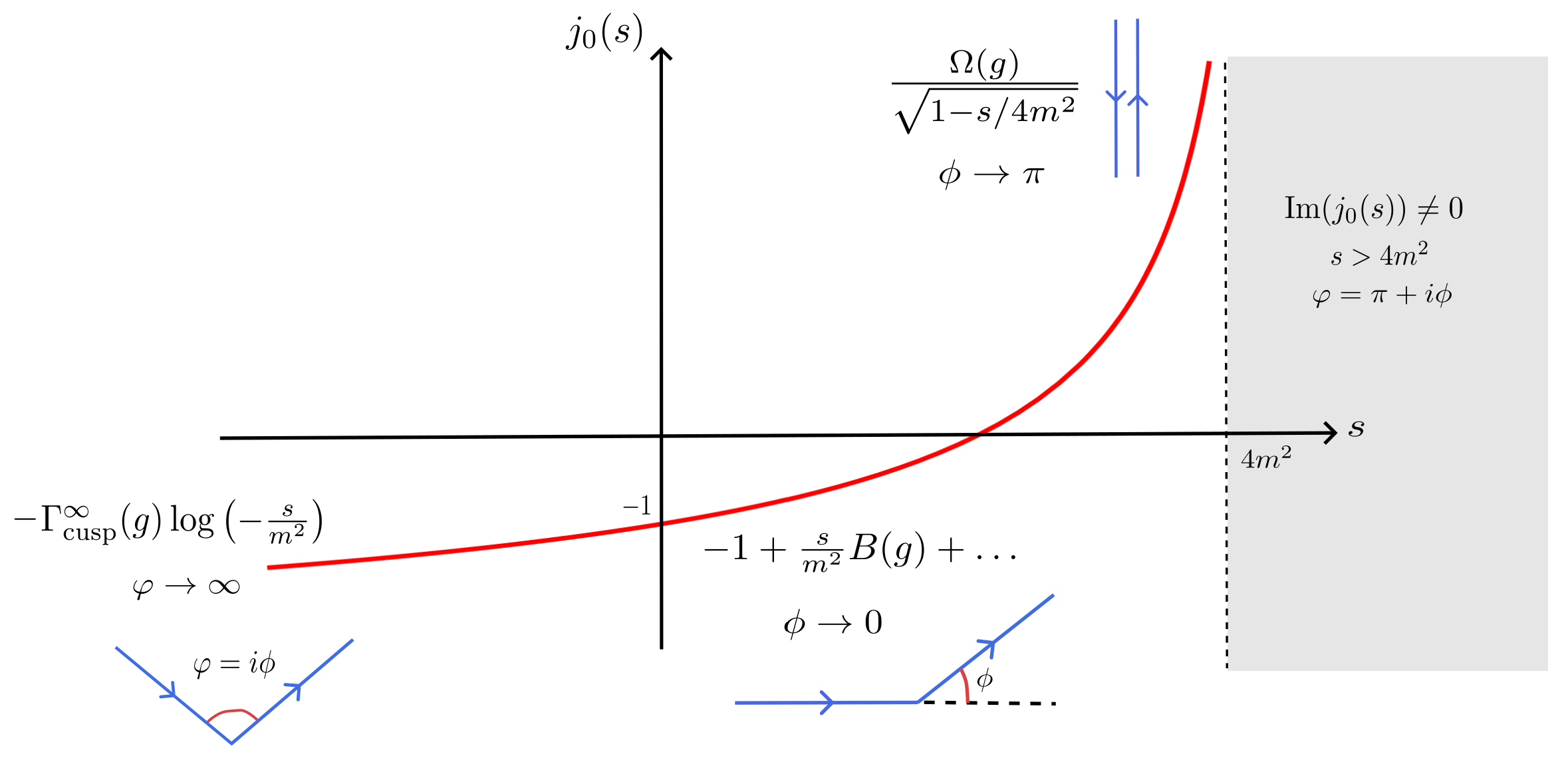}
    \caption{A characteristic shape of the leading Regge trajectory at finite coupling. We also depict the relevant limits of the Wilson line with a cusp. The Regge trajectory is curved, it has a logarithmic asymptotic at large $s$, and it exhibits an accumulation point at $s=4m^2$. For $s>4 m^2$ it develops a non-zero imaginary part.}
    \label{fig:lead_regge}
\end{figure}

At the heart of our work lies dual conformal invariance \cite{Drummond:2008vq,Berkovits:2008ic,Alday:2009zm,Alday:2007hr}. It reflects the underlying symmetries of string dynamics in the $AdS_5 \times S^5$ background and is already apparent in the perturbation theory. It has several important consequences. First, it implies that the scattering amplitude is only a function of two cross-ratios
\be
\label{eq:dualconf}
u = {4 m_1 m_3 \over -s + (m_1 - m_3)^2}, ~~~ v = {4 m_2 m_4 \over -t + (m_2 - m_4)^2} \ . 
\ee
Second, the fact that the amplitudes realize an action of the dual conformal invariance group $SO(4,2)$ organizes exchanged states into \emph{larger} multiplets compared to the naive $SO(3)$ partial waves familiar from four-dimensional scattering \cite{Caron-Huot:2014gia,Bruser:2018jnc}.\footnote{We will see that bound states are organized into $SO(4)$ multiplets, and scattering states into the $SO(1,3)$ multiplets. For the hydrogen atom, this was demonstrated in a beautiful paper by Fock \cite{Fock:1935vv}, where he analyzed the Schrodinger equation in momentum space.} This fact leads to the strictly stronger unitarity conditions on the four-point scattering amplitude compared to the standard situation. The fact that the amplitude only depends on the cross-ratios \eqref{eq:dualconf} leads to an interesting UV/IR connection, where the Regge limit $t \to \infty$ is equivalent to the soft limit $m_2 \to 0$ \cite{Henn:2010bk,Correa:2012nk}. In the former limit, the amplitude is organized in terms of the Regge trajectories. In the latter limit, the amplitude becomes IR divergent, and the dependence on the regulator $m_2$ is controlled by the anomalous dimension of the Wilson line with a cusp \cite{Polyakov:1980ca,Korchemsky:1987wg}. Equating the two limits naturally leads to the correspondence between the leading Regge trajectory and the Maldacena-Wilson line with a cusp
\be
\label{eq:ReggeCusp}
j_0(s) = - \Gamma_{\text{cusp}}(g,\phi) - 1, ~~~ s=4m^2 \sin^2 \phi/2 ,
\ee
where, for simplicity, we set $m_i=m$, and this relationship was explored in \cite{Henn:2010bk}.\footnote{In \eqref{eq:ReggeCusp} $\phi$ is the Euclidean angle, and $\Gamma_{\text{cusp}}(g,\phi)$ is the Euclidean cusp anomalous dimension.} 
The characteristic shape of the leading trajectory is shown in \figref{fig:lead_regge}. One peculiar property of the scattering amplitudes on the Coulomb branch in ${\cal N}=4$ SYM is that they exhibit an accumulation point of high-spin resonances in their spectrum at the two-particle threshold. This is nothing but an ionization threshold of an attractive Coulomb potential \cite{Klebanov:2006jj,Caron-Huot:2014gia,Maldacena:2022ckr}. The existence of the continuum of states for $s \geq 4 m^2$ is mapped to the existence of the $AdS$ horizon in the dual geometry.
For the first subleading Regge trajectory, where we put all D3-branes on the same point on $S^5$ \cite{Alday:2009zm}, the relationship \eqref{eq:ReggeCusp} generalizes to the subleading trajectory
\be
\label{eq:ReggeCuspPhi}
j_1(s) = - \Gamma_{\text{cusp},\Phi}(g,\phi) - 1, ~~~ s=4m^2 \sin^2 \phi/2 ,
\ee
where $\Phi(x)$ is an insertion of the so-called \emph{parallel} scalar at the cusp. It was put forward in \cite{Caron-Huot:2014gia} and further explored in \cite{Bruser:2018jnc}, see also \cite{Flieger:2025ekn}. The connection between the Regge trajectories $j_n(s)$ and the energy spectrum of the cusp Hamiltonian $H_{\text{cusp}}(\phi)$ has not been explored beyond the first two Regge trajectories. We propose that \emph{the Regge/cusp correspondence} extends to higher trajectories, and our results provide nontrivial evidence in support of this hypothesis. The precise proposal and its tests are presented in the main text.

The relations \eqref{eq:ReggeCusp} and \eqref{eq:ReggeCuspPhi} allow us to harness the power of integrability methods. Indeed, the Wilson line with insertions of local operators at the cusp has been recently computed using the Quantum Spectral Curve (QSC) \cite{Gromov:2015dfa, Grabner:2020nis}. In particular, we will use this technique to compute the leading $j_0(s)$, and the first subleading $j_1(s)$ Regge trajectories. 
For $s<0$ the Regge trajectories control the high-energy behavior of the amplitude, and for $0\leq s<4m^2$ they encode the spectrum of the bound states. We will therefore use integrability to compute the spectrum of the lightest states at finite 't Hooft coupling $g$.

As already mentioned, at strong coupling the low-energy scattering is captured by the flat space super-Veneziano amplitude. The first subleading corrections around strong coupling can be computed following the ideas of \cite{Alday:2023jdk,Alday:2023mvu,Alday:2024yax,Alday:2024ksp}. In short, a worldsheet ansatz of a specific form is supplemented with the expected structure of poles for the amplitude, feeding in particular the known correction to the leading Regge trajectory. In this way we are able to fix completely the answer, which passes several nontrivial tests. 

Next, we turn to S-matrix bootstrap techniques to derive constraints on the amplitude at finite coupling. We write down dispersion relations and impose crossing symmetry and unitarity to derive bounds on various observables.  Crucially, dual conformal invariance leads to an \emph{enhanced} form of the unitarity constraint.
As input, we use the spectrum of the leading and subleading Regge trajectories computed from integrability. Combining these ingredients yields novel bounds on the Wilson coefficients at finite coupling. Remarkably, in the weak-coupling limit $g\ll 1$, we find the $\cN=4$ amplitude to be at a \emph{kink} in the space of Wilson coefficients. 

Finally, we combine the weak-coupling expansion with our strong-coupling results to construct a \emph{Pad\'{e} model} for the Wilson coefficients at finite coupling. Imposing this model for the Wilson coefficients, we derive two-sided bounds on the three-point coupling of the first massive state, as well as on the amplitude itself. Curiously, the bounds obtained at finite coupling are consistent with the property of \emph{complete monotonicity}, recently discussed in a perturbative context in \cite{Henn:2024qwe}.

\subsection{Plan of the paper}

\begin{figure}[h!]
  \centering
  \begin{overpic}[width=\textwidth]{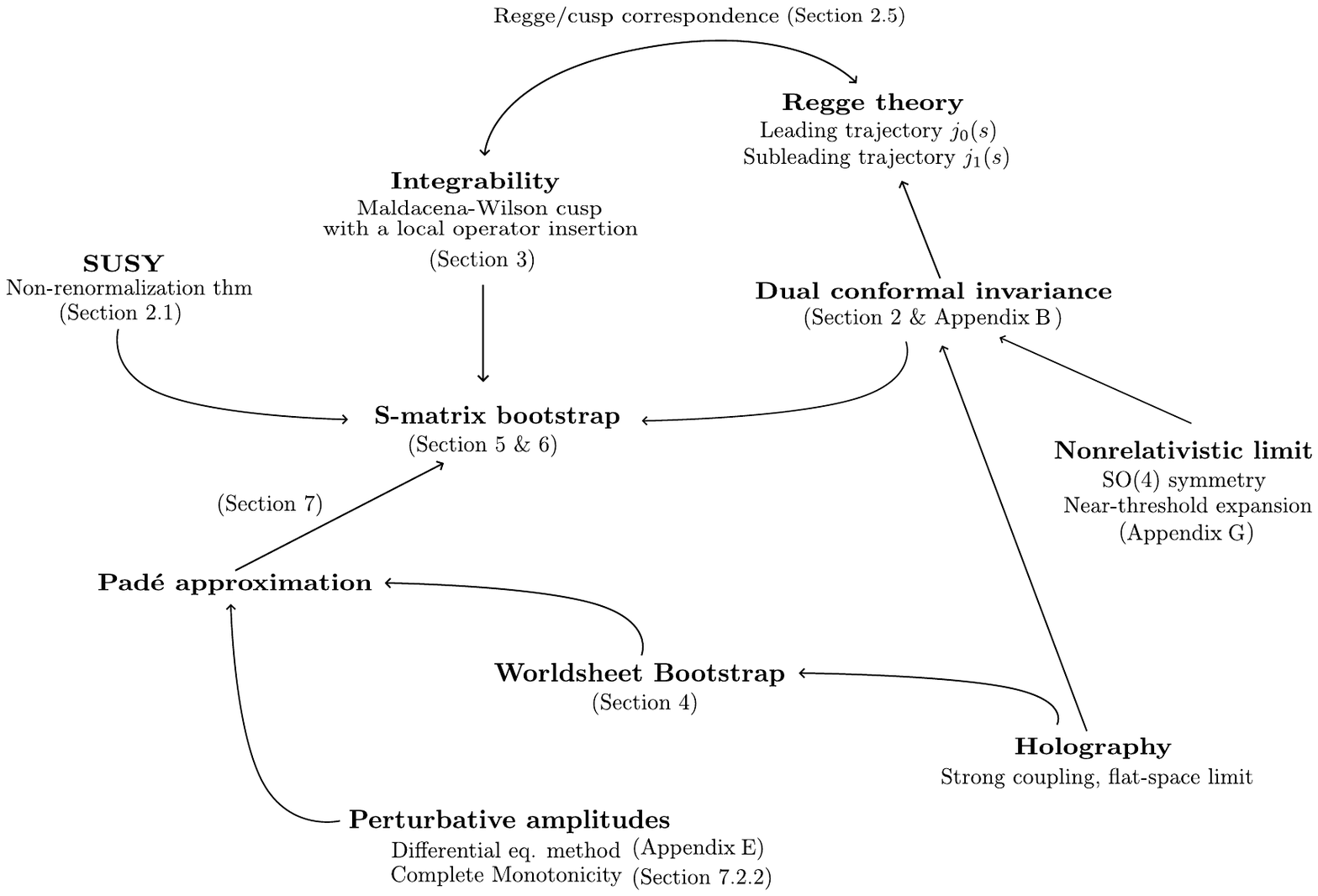}
\put(60.5,44.4){%
\hyperref[sec:setup]{%
    \fboxrule=0.4pt\fboxsep=0pt%
    \fbox{\phantom{\makebox[2.8em][c]{\rule{0pt}{0.8em}}}}%
  }%
}
\put(4.7,44.8){%
\hyperref[sec:Perturbative results]{%
    \fboxrule=0.4pt\fboxsep=0pt%
    \fbox{\phantom{\makebox[3.7em][c]{\rule{0pt}{0.8em}}}}%
  }%
}
\put(58.9,67.0){%
\hyperref[sec:sublRegge]{%
    \fboxrule=0.4pt\fboxsep=0pt%
    \fbox{\phantom{\makebox[3.5em][c]{\rule{0pt}{0.8em}}}}%
  }%
}
    \put(32.4,48.8){%
  \hyperref[sec:integrability]{%
    \fboxrule=0.4pt\fboxsep=0pt%
    \fbox{\phantom{\makebox[3.1em][c]{\rule{0pt}{0.8em}}}}%
  }%
}
\put(44.4,16){%
  \hyperref[sec:Worldsheet bootstrap]{%
    \fboxrule=0.4pt\fboxsep=0pt%
    \fbox{\phantom{\makebox[3.1em][c]{\rule{0pt}{0.8em}}}}%
  }%
}
\put(31.4,35.1){%
  \hyperref[sec:bootstrapping_setup]{%
    \fboxrule=0.4pt\fboxsep=0pt%
    \fbox{\phantom{\makebox[2.7em][c]{\rule{0pt}{0.8em}}}}%
  }%
}
\put(40.3,35.1){%
  \hyperref[sec:Universal_bounds]{%
    \fboxrule=0.4pt\fboxsep=0pt%
    \fbox{\phantom{\makebox[0.5em][c]{\rule{0pt}{0.8em}}}}%
  }%
}
\put(16.6,30.7){%
  \hyperref[sec:bounds_using_pade_model]{%
    \fboxrule=0.4pt\fboxsep=0pt%
    \fbox{\phantom{\makebox[3.1em][c]{\rule{0pt}{0.8em}}}}%
  }%
}
\put(47.5,2.9){%
  \hyperref[sec:Bounds on the amplitude itself]{%
    \fboxrule=0.4pt\fboxsep=0pt%
    \fbox{\phantom{\makebox[4.1em][c]{\rule{0pt}{0.8em}}}}%
  }%
}
\put(69.8,44.4){%
\hyperref[app:extunit]{%
    \fboxrule=0.4pt\fboxsep=0pt%
    \fbox{\phantom{\makebox[3.6em][c]{\rule{0pt}{0.8em}}}}%
  }%
}

\put(83.5,28.4){%
\hyperref[app:residue]{%
    \fboxrule=0.4pt\fboxsep=0pt%
    \fbox{\phantom{\makebox[4.0em][c]{\rule{0pt}{0.8em}}}}%
  }%
}
\put(47.5,5.1){%
  \hyperref[app:weak_and_strongCoupling]{%
    \fboxrule=0.4pt\fboxsep=0pt%
    \fbox{\phantom{\makebox[4em][c]{\rule{0pt}{0.8em}}}}%
  }%
}
  \end{overpic}
  \caption{Roadmap illustrating the topics discussed in the paper and the connections among them.}
  \label{fig:roadmap}
\end{figure}

The plan of the paper is as follows
\begin{itemize}
    \item In \secref{sec:setup}, we  introduce the setup and summarize the main properties of the four-point scattering amplitude at finite coupling.
    \item In \secref{sec:integrability}, we compute the leading and the subleading Regge trajectory at finite coupling using integrability.
    \item In \secref{sec:Worldsheet bootstrap}, we derive the leading $1/g$ correction to the super-Veneziano amplitude starting from the worldsheet representation of the scattering amplitude. 
    \item In \secref{sec:bootstrapping_setup}, we introduce the S-matrix bootstrap setup used to derive bounds on various observables.
    \item In \secref{sec:Universal_bounds}, we derive the S-matrix bootstrap bounds on various observables.
    \item In \secref{sec:bounds_using_pade_model}, we combine our weak- and strong-coupling results to construct a finite-coupling Pad\'{e} model for the Wilson coefficients. We use this model together with the S-matrix bootstrap to characterize the finite-coupling amplitude.
    \item In \secref{sec:conclusions} we conclude and discuss future directions.
\end{itemize}
Many technical aspects of the work are collected in the appendices. 
To facilitate the reader’s navigation through the paper, we provide a visual overview of the paper in \figref{fig:roadmap}.

\section{Scattering on the Coulomb branch of ${\cal N}=4$ SYM}\label{sec:setup}

We consider planar ${\cal N}=4$ SYM on the Coulomb branch, where the conformal symmetry is spontaneously broken. Let us, for example, consider the gauge group $U(N+4)$, which we spontaneously break to $U(N) \times U(1)^4$. It is conveniently described by writing the fundamental fields in the Lagrangian in the block-diagonal form
\be \label{eq:funda_fields}
{\cal O} = \begin{pmatrix}
{\cal O}_{a,b} & {\cal O}_{a,N+j} \\
{\cal O}_{N+i,b} & {\cal O}_{N+i,N+j}
\end{pmatrix} ,
\ee
where ${\cal O} = \left\{ A_{\mu}, \Psi, \Phi \right\}$, $a,b = 1, ... , N$
and $i,j=1,...,4$. We consider the nontrivial vacuum expectation values (VEVs) for the scalars $\langle \Phi^I_{N+i, N+j} \rangle = {\sqrt{2} \over g_{YM}} M^I_{N+i, N+j}$, where $I=4,...,9$ and the matrix of VEVs takes the form
\be
\overrightarrow{M}_{N+i,N+j} =
\begin{pmatrix}
\vec n_1 m_1 & 0 & 0 & 0 \\
0 & \vec n_2 m_2 & 0 & 0 \\
0 & 0 & \vec n_3 m_3 & 0 \\
0 & 0 & 0 & \vec n_3 m_4
\end{pmatrix} \ , 
\ee   
with $n_i^{I}$, such that $\vec n_i^2=1$, labelling a point on $S^5$. The complete description of the gauge-fixed action and the Feynman rules in this setup can be found, for example, in \cite{Alday:2009zm,Ivanovskiy:2024vel}.
We will be interested in the planar limit $N \to \infty$ with the 't Hooft coupling
\be
g^2 = {g_{YM}^2 N \over (4\pi)^2}\equiv \frac{\lambda}{(4\pi)^2} \ , 
\ee
kept fixed. 
In addition to the massless fields associated with the $U(N)$ gauge factor ${\cal O}_{a,b}$, the theory contains `heavy' massive W-bosons ${\cal O}_{a, N+i}$ of mass $m_i$, and `light' meson fields ${\cal O}_{N+i ,N+j}$ of mass 
\be
m_{N+i, N+j}=|m_i \vec n_i - m_j \vec n_j|,
\label{eq:mesonmass}
\ee
see Figure \ref{fig:D3_branes}. 

\begin{figure}[h!]
    \centering
    \includegraphics[width=0.7\linewidth]{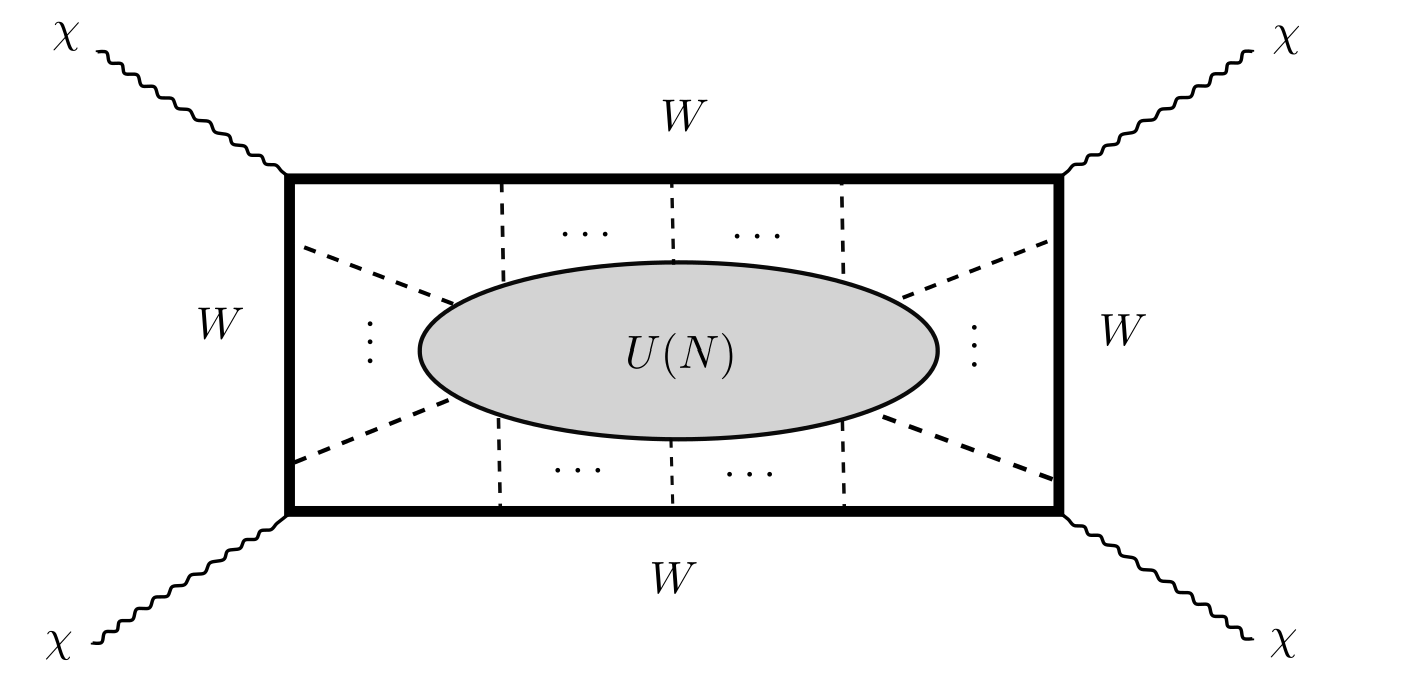}
    \caption{Planar diagrams with external $\chi$ fields (wavy lines) attached to a frame made of W-bosons (thick lines), filled with massless fields from the unbroken gauge group $U(N)$ (dashed lines).}
    \label{fig:Feynman_diagram_1}
\end{figure}

Let us discuss the properties of the scattering amplitudes of scalar meson fields $\chi_{i,j}\equiv \Phi_{N+i,N+j}$. We will focus on the simplest case considered in \cite{Alday:2009zm}, where all VEVs are aligned along $S^5$, 
\be
n_i^{I} = \delta^{I9} \ .
\label{eq:n_all_aligned}
\ee
To start, we can consider the following color-ordered four-point amplitude
\be
\label{eq:defampl}
i A(s,t,m_i) = \langle (\Phi_I)_{N+1,N+2}(p_1) (\Phi_J)_{N+2,N+3}(p_2) (\Phi_I)_{N+3,N+4}(p_3) (\Phi_J)_{N+4,N+1}(p_4) \rangle , 
~~~ I,J \neq 9 \ ,
\ee
where $s=-(p_1+p_2)^2$, $t=-(p_2+p_3)^2$, and we use the all-ingoing convention for the scattering momenta and mostly-plus signature for the metric. Thanks to \eqref{eq:n_all_aligned}, the on-shell condition becomes simply $-p_i^2 =(m_i-m_{i+1})^2$. 
In the planar limit, the leading diagrams consist of planar graphs, where the external particles are attached to a rectangular frame made out of W-bosons, filled with the massless fields belonging to the unbroken gauge group \cite{Alday:2009zm}, see \figref{fig:Feynman_diagram_1}. 

The scattering amplitude $A$ is IR finite in the planar theory, and it can be conveniently written as
\be
A(s,t,m_i) &=g_{YM}^2 M , \\
M &= 1 + g^2 M^{(1)} + g^4 M^{(2)} + \cdots .
\ee
As already emphasized in the introduction, it was found in  \cite{Alday:2009zm}  that the amplitude $A(s,t,m_i)$ is in fact only a function 
of \emph{two} cross-ratios
\be
\label{eq:crossratios}
u = {4 m_1 m_3 \over -s + (m_1 - m_3)^2}, ~~~ v = {4 m_2 m_4 \over -t + (m_2 - m_4)^2} \ . 
\ee
This is a consequence of dual conformal invariance \cite{Drummond:2008vq,Berkovits:2008ic}, 
and it is particularly clear in the bulk, where dual conformal invariance is associated with the isometries of the $AdS_5$ factor, see \appref{app:extunit} for details. We will therefore consider $M(u,v)$ with \eqref{eq:crossratios} in mind, which hopefully will not cause any confusion. As a consequence of the cyclic symmetry of the definition \eqref{eq:defampl}, $M(u,v)$ satisfies crossing
\be
M(u,v) = M(v,u) \ . 
\ee

As a next step, it is convenient to set $m_i=m$ for which external particles are massless. Finally, it will be useful to introduce a slightly different function $f(s,t)$ defined as follows
\be
f(s,t) \equiv -{M(-4m^2/s,-4m^2/t) \over s t} .
\label{eq:f_and_M_rel}
\ee
It satisfies crossing symmetry $f(s,t) = f(t,s)$, 
and is simply related to the \emph{elastic} amplitude of charged mesons \cite{Berman:2023jys},\footnote{In the all-ingoing notation of \cite{Berman:2023jys}, this amplitude was denoted by $A[zz\bar z \bar z]$.}
\be
\label{eq:elscat}
A_{z,z \to z,z} = s^2 f(s,t) \ ,
\ee
where $z = \Phi_I + i \Phi_J$.
This fact will be important when discussing unitarity below.

The main goal of the present paper is to explore $f(s,t)$ at finite 't Hooft coupling $g$.

\subsection{Perturbative results}
\label{sec:Perturbative results}

\begin{figure}[h!]
\centering
\includegraphics[width=6cm]{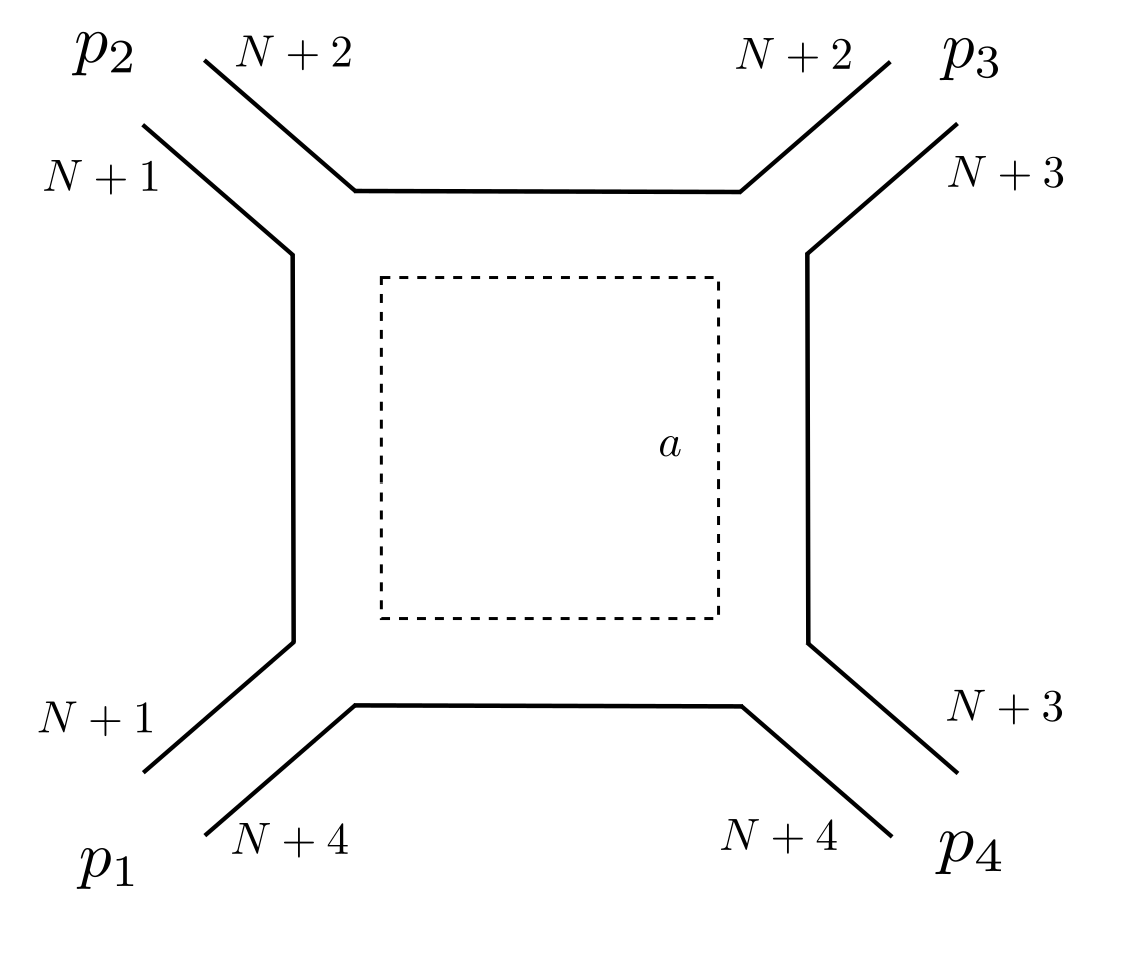}
\caption{One-loop box diagram in double-line notation.}
\label{fig:1_loop}
\end{figure}

Let us quote the leading order results for $f(s,t)$ both at weak and strong coupling. At weak coupling, the leading correction is a one-loop box diagram, see \figref{fig:1_loop}, and  takes the form \cite{Davydychev:1993ut} 
\be
\label{eq:weak1loop}
f(s,t) 
&=- {1 \over s t} + \frac{g^2}{6 m^4} F_{3}\left(1,1,1,1;5/2 \Big{|} \frac{s}{4 m^2}, \frac{t}{4 m^2} \right) + \mathcal{O}(g^4) \ , ~~~ g \ll 1 \ . 
\ee
where $F_3$ is the third Appell hypergeometric function. 
This form makes the low-energy expansion of the one-loop result particularly simple.
It can also be written in terms of logarithms and dilogarithms. 

\begin{figure}[h!]
\centering
\includegraphics[scale=0.5]{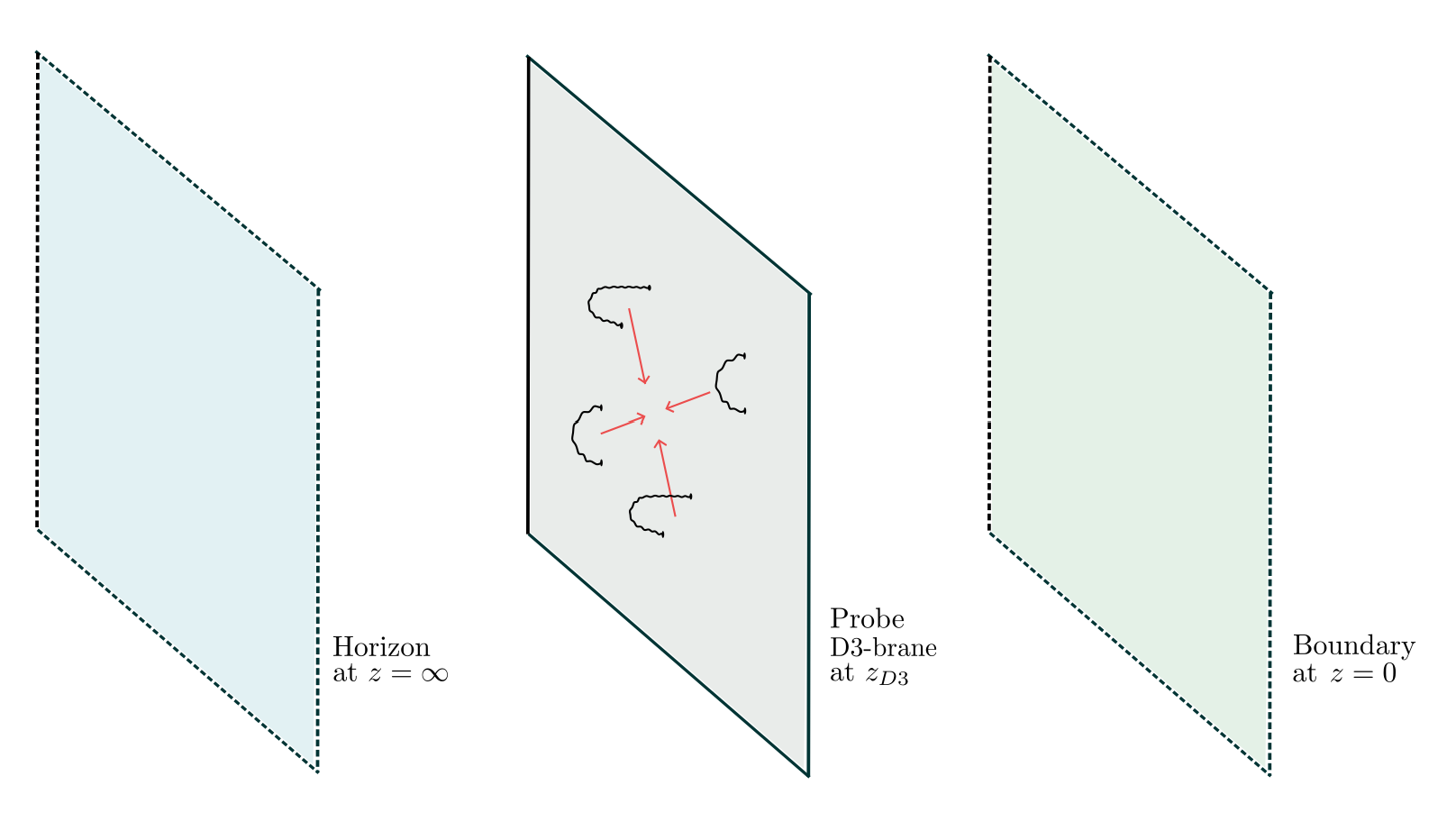}
\caption{Scattering of open strings on a probe D3-brane at fixed radial position $z_{D3}$ in $AdS$. }
\label{fig:D3_brane_1}
\end{figure}
Similarly, the two-loop result can be written explicitly, and three-loop corrections are also known \cite{Caron-Huot:2014lda,Bruser:2018jnc}, and we will utilize these results in the paper. At the level of the amplitude integrand, there exists an all-loop proposal \cite{Bourjaily:2011hi,Caron-Huot:2021usw}.

At strong coupling, the leading answer at low energies is controlled by the flat space tree-level string scattering \cite{Maldacena:2022ckr}, and is given by the super-Veneziano amplitude which captures the scattering of strings on the D3-brane, see \figref{fig:D3_brane_1}. The precise expression takes the form
\be
f(s,t) &={g \over \pi m^2}{1 \over s+t} \int_0^1 dx \ x^{-1-{g s \over m^2 \pi}} (1-x)^{-1-{g t \over m^2 \pi}}  + ... \nonumber \\
&=-{g^2 \over \pi^2 m^4} \frac{\Gamma \left(-\frac{g s}{m^2 \pi }\right) \Gamma \left(-\frac{g t}{m^2 \pi}\right)}{\Gamma \left(1-\frac{g s}{m^2 \pi }-\frac{g t}{m^2 \pi }\right)} + \dots \ , ~~~ g \gg 1 \ , 
\label{eq:super_Veneziano}
\ee
where in taking the large $g$ limit we kept $(\frac{g s}{m^2 \pi },\frac{g t}{m^2 \pi })$  fixed. Let us comment on a slightly unusual property of the expression above. As we increase the coupling $g \to \infty$ the stringy states appear at \emph{lower} energies. This contrasts with the fact that in the bulk the stringy states become \emph{heavier} as we increase the 't Hooft coupling. The resolution of this apparent puzzle, which is of course instrumental for the existence of the $AdS$/CFT correspondence, lies in the $AdS$ redshift factor. In the amplitude above, we keep the energy of the two-particle threshold fixed to be $4m^2$. In the bulk, it corresponds to the mass of a pair of strings that stretch from the probe brane to the $AdS$ horizon, and to keep its mass fixed as we increase the tension of the string ${1 \over 2 \pi \alpha'}=2g$ in units $R_{AdS}=1$, we have to set the location of the D3-brane to be $z_{D3} = {2 g \over m}$, see \appref{app:stringmass}. Therefore, we move the probe D3-brane closer and closer to the $AdS$ horizon as we send $g \to \infty$. This introduces an additional redshift factor between the energies measured in the field theory and locally on the D3-brane
\be
s_{{\cal N}=4} = {s_{\text{D3}} \over z_{\text{D3}}^2 } , 
\ee
where $z$ is the radial position in $AdS$, $d s_{AdS}^2 = {d z^2 + dx_\mu d x^\mu \over z^2}$, and the field theory energy $p^\mu$ conjugate to ${\partial \over \partial x^\mu}$ is measured at $z=1$ in this convention. Switching in \eqref{eq:super_Veneziano} to the local energy on the D3-brane $(s_{{\cal N}=4}, t_{{\cal N}=4}) = {m^2 \over 4 g^2}( s_{\text{D3}},t_{\text{D3}})$ and remembering that $g = {1 \over 4 \pi \alpha'}$ recovers the usual flat space formula $-{\Gamma(-\alpha' s_{\text{D3}}) \Gamma(-\alpha' t_{\text{D3}}) \over \Gamma(1-\alpha' s_{\text{D3}}-\alpha' t_{\text{D3}})}$ and the standard intuition that stringy states get heavier as we increase the coupling.

The amplitude above is a particular component of the corresponding super-amplitude, which can also be conveniently written in the on-shell superspace, see e.g. \cite{Berman:2023jys}.\footnote{For the four-dilaton scattering amplitude we get $A(\phi\phi\phi\phi) = 2 g_{YM}^2 (s^2 + t^2 + u^2)(f(s,t) + f(s,u) + f(t,u))$.} For us, however, another implication of supersymmetry will be much more important. The non-renormalization theorem proven in \cite{Dine:1997nq} states that the low-energy expansion at any coupling takes the form
\be
\label{eq:protectedopen}
f(s,t) = - {1 \over s t} + {g^2 \over 6 m^4} + {\cal O} \left( s/m^2,t/m^2 \right) \ .
\ee
This is indeed correctly reproduced by the weak and strong coupling perturbative expressions above.

\subsection{Analyticity}

Let us next discuss the analytic structure of the amplitude. The weak and strong expressions above have notably different analytic properties. 

The weak coupling result \eqref{eq:weak1loop} exhibits a two-particle cut $s,t=4m^2$ and obeys maximal analyticity\footnote{By this we mean analyticity in the cut complex plane $(s,t) \in (\mathbb{C} \setminus [4m^2,\infty) ])^2$.} which is neatly encapsulated by the Mandelstam representation
\be
\label{eq:oneloopf}
f(s,t) =- {1 \over s t} + 2 g^2 \int_{4m^2}^\infty {d s' d t' \over (s'-s)(t'-t)} {\theta \Big(s' t' - 4 m^2 (s'+t')\Big) \over \sqrt{s' t' (s' t' - 4 m^2 (s'+t'))}} + \cO(g^4),
\ee
where the Heaviside theta-function $\theta \Big(s t - 4 m^2 (s+t)\Big)$ imposes that the double discontinuity of the amplitude is only non-zero above the Landau curve of the one-loop box diagram, see \figref{fig:1_loop}. Since all the higher-order graphs are obtained by adding massless particles inside the one-loop frame created by massive W-bosons, we expect that the \emph{support} of the finite coupling double spectral density is given by the one-loop Landau curve as well. The Mandelstam representation was also explicitly checked for the two-loop result in \cite{Bruser:2018jnc}, and maximal analyticity was argued to hold up to three loops. 

The strong-coupling result \eqref{eq:protectedopen} is, on the other hand, meromorphic. As $g \to \infty$, the poles of the amplitude $\frac{g s}{m^2 \pi } = \mathbb{Z}_+$ are located in the region $0<s<4m^2$, therefore they are nothing but the bound states. Physically, as we increase the coupling, the gauge theory flux tube becomes stiffer and supports more  excitations \cite{Klebanov:2006jj}. 

\begin{figure}[h!]
\centering\includegraphics[width=0.7\linewidth]{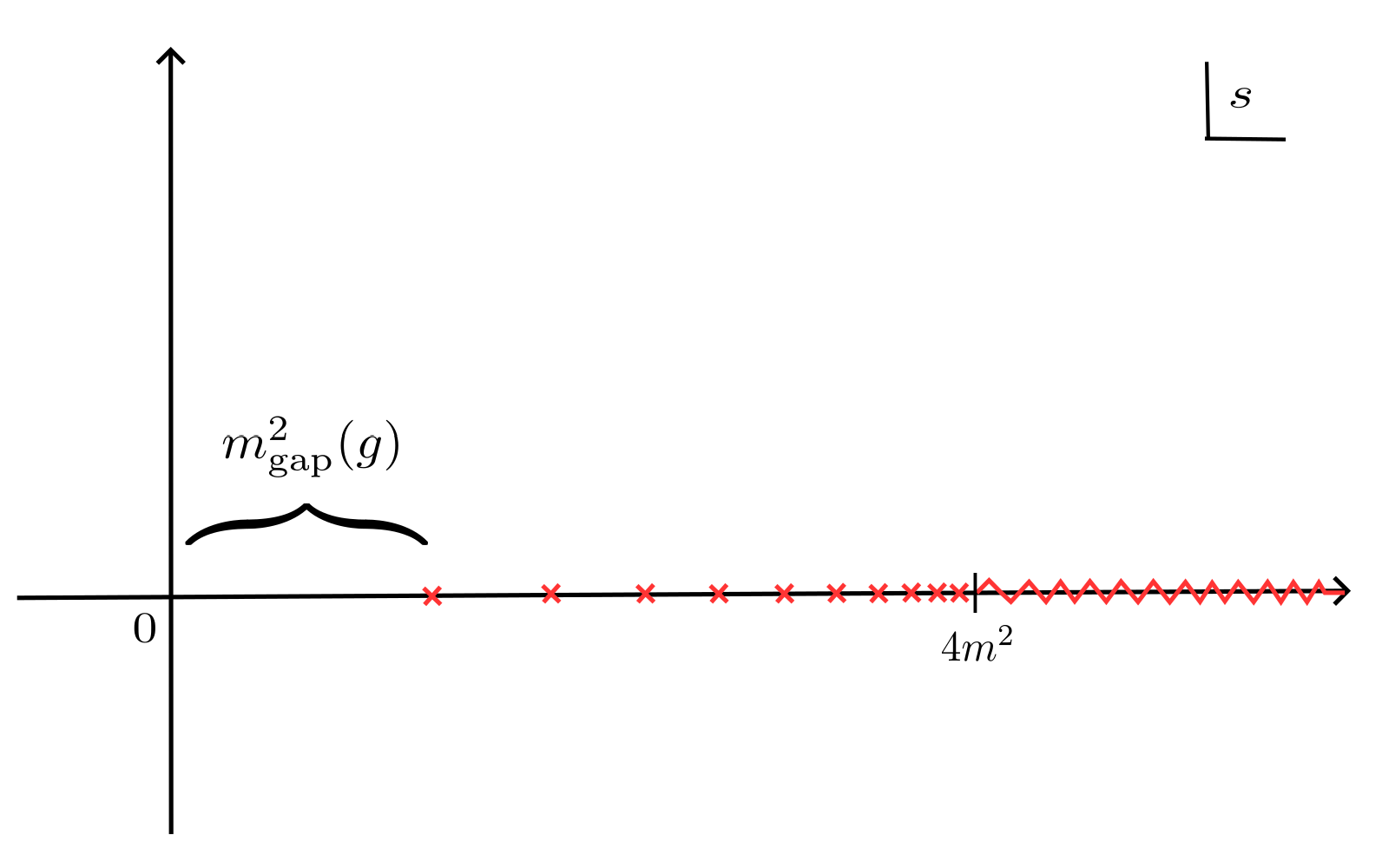}
    \caption{Analytic structure of $f(s,t)$ at finite coupling $g$ for fixed $t$. There is an infinite set of bound states for $s < 4m^2$ and a two-particle cut from $s = 4m^2$.}
    \label{fig:spectrum}
\end{figure}

At finite coupling $g$, the expected analytic structure takes the form depicted in \figref{fig:spectrum}. We have infinitely many bound states for $s<4m^2$, and we have a two-particle cut starting from $s=4m^2$. %
We denote by $m_{\text{gap}}$ the mass of the lightest bound state, which depends nontrivially on the coupling $g$.
We will discuss the detailed structure of the bound states and the associated Regge trajectories below. 

For $t< \mgap$, we know from the form of the leading Regge trajectory that $\lim_{|s| \to \infty}f(s,t)=0$, and therefore the amplitude admits the unsubtracted dispersion relation
\be\label{eq:Disprel}
        f(s,t) =- \frac{1}{st} + \frac{1}{\pi}\int_{m_{\rm gap}^2(g)}^{\infty} ds' \frac{f_s(s',t)}{s'-s}\,, 
\ee
where $f_s(s,t) \equiv {f(s+i \eps,t) - f(s-i \eps,t) \over 2 i}$ is the discontinuity of the amplitude. The amplitude also admits an anti-subtracted dispersion relation
\be \label{eq:unsubtractedDR}
-{1 \over t} = \oint {ds' \over 2 \pi i} f(s',t) =  \frac{1}{\pi} \int_{m_{\rm gap}^2(g)}^{\infty} ds' f_s(s',t) \ , 
\ee
where we used the fact that the leading Regge trajectory satisfies $j_0(t)<-1$ for $t<0$, as shown in \figref{fig:lead_regge}.

Based on the perturbative discussion, one might hope that at finite coupling, the amplitude satisfies maximal analyticity with the double discontinuity of the amplitude being non-zero above the leading Landau curve $s t - 4 m^2 (s+t) \geq 0$. It is an interesting possibility that the Mandelstam representation holds at finite coupling
\be\label{eq:mandelstamRepFiniteg}
f(s,t) \overset{?}{=} - \frac{1}{st} + \frac{1}{\pi}\int_{m_{\rm gap}^2(g)}^{4 m^2} ds' \frac{f_s(s',t)}{s'-s} +  \int_{4m^2}^\infty {d s' d t' \  \theta \Big(s' t' - 4 m^2 (s'+t')\Big)\over (s'-s)(t'-t)}  \rho(s',t') \ , 
\ee
where the double spectral density satisfies crossing symmetry $\rho(s,t) = \rho(t,s)$. We will show in \appref{sec:nomandelstam} that this ansatz is \emph{not} consistent with crossing at finite coupling $g$, therefore the Mandelstam representation as expressed by the formula above is only a perturbative property of the amplitude in this theory.

Finally, let us mention that if we sent $m\to0$, or, equivalently, consider the high-energy fixed angle scattering $s/t$ fixed and $s,t \to \infty$, the amplitude takes the following universal form \cite{Alday:2009zm,Bruser:2018jnc} 
\be
\label{eq:GMADS}
\log f(s,t) &= -{\Gamma_{\text{cusp}}^{\infty}(g) \over 4} \left(2 \log({m^2 \over -s})\log({m^2 \over -t}) - \pi^2\right)+ \Gamma_{\text{coll}}(g) \left( \log({m^2 \over -s}) + \log({m^2 \over -t}) \right) \nonumber \\
&+ c(g) + \mathcal{O}(m^2) \ ,
\ee
where $\Gamma_{\text{cusp}}^{\infty}(g)$ is the light-like cusp anomalous dimension and $\Gamma_{\text{coll}}(g)$ is a collinear anomalous dimension.\footnote{For future reference, at weak coupling we have $\Gamma_{\text{cusp}}^{\infty}(g) = 4 g^2 -8 \zeta_2 g^4+88 \zeta_4 g^6 + ...$, and at strong coupling $\Gamma_{\text{cusp}}^{\infty}(g) = 2 g - {3 \log 2 \over 2 \pi} + ...$, see e.g. \cite{Basso:2007wd}.} 
We see that the amplitude at high energies and fixed angles decays faster than a polynomial (a polynomial decay is expected in QCD), but much slower than the flat-space string amplitude \cite{Gross:1987kza}. %
At strong coupling $g \gg 1$, the amplitude develops an intermediate Gross-Mende high-energy regime $1 \gg ({-s \over 4 m^2},{-t \over 4 m^2}) \gg {1 \over g}$
\be
\label{eq:GMregime}
\log f(s,t) &={g \over m^2 \pi}\Big[(s+t) \log (-s-t) - s \log(-s) - t \log (-t) \Big] + \dots  \ ,
\ee
which describes high-energy fixed-angle scattering in flat space.

\subsection{Enhanced unitarity}
\label{sec:enhunit}

Unitarity of the elastic amplitude \eqref{eq:elscat} in the planar limit leads to the positivity property of its imaginary part (or its discontinuity). To formulate it, we expand the discontinuity of the amplitude $f_s(s,t) \equiv {f(s+i \epsilon,t)-f(s-i \epsilon,t) \over 2 i}$ in the $SO(3)$ partial waves
\be
\label{eq:usualPW}
\textbf{SO(3) unitarity:}~~~ f_s(s,t) = s^{-2} \sum_{J=0}^\infty \rho_J(s) P_J (\cos \theta_s)\,, ~~~ s\geq 0\,, 
\ee
where $\cos\theta_s = 1+ 2t/s$ is the cosine of the scattering angle and $P_J(x)$ are the Legendre polynomials. We also included the prefactor $s^{-2}$ to make the spectral density dimensionless. Unitarity states that
\begin{equation}\label{eq:SO3Unitarity}
    \rho_J(s) \geq 0 \,,~~~ s> 0\,.
\end{equation}
Physically, $\rho_J(s)$ has an interpretation of the probability of producing states with energy $s$ and the angular momentum $J$ and hence it has to be non-negative.

A stronger version of unitarity exists due to dual conformal invariance \cite{Bruser:2018jnc}. The basic idea is that there are extra degeneracies that appear in the expansion \eqref{eq:SO3Unitarity}, see also \cite{Caron-Huot:2014gia}, that can be understood by thinking about how dual conformal $SO(4,2)$ acts on the intermediate states that appear in the scattering process. We review this in \appref{app:extunit}.

As a result, the relevant little group depends on the energy of the state. For $4m^2>s>0$, intermediate states appear in the  $SO(4)$ multiplets, and the relevant scattering angle is
\be
\cos\tilde\theta_s &= 1+ \frac{2t}{s} - \frac{t}{2m^2}\,,~~~ 0<s<4m^2 \ . \label{eq:SO4angle}
\ee
We can then expand the amplitude discontinuity in effectively five-dimensional partial waves
\be
\textbf{SO(4) unitarity:}~~~f_{s}(s,t) &= s^{-2}\sum_{J=0}^\infty c_J(s)\PJfive(\cos\tilde\theta_s),~c_J(s)\geq 0,~~0\leq \,s<4m^2 \label{eq:fsInSO4PW}   \ , 
\ee
where the $SO(4)$ Gegenbauer polynomials are given by
\be \label{eq:PJ_s04}
\PJfive(x) &=\,_2F_1\left(-J,J+2;\frac{3}{2};\frac{1-x}{2}\right) = {\sin (J+1) \tilde \theta_s \over (J+1) \sin  \tilde \theta_s} \ , ~~~ x =\cos \tilde \theta_s \ . 
\ee
The extended $SO(4)$ unitarity condition \eqref{eq:fsInSO4PW} immediately implies the standard $SO(3)$ unitarity \eqref{eq:usualPW} due to the fact that
\begin{align}
    \PJfive(\cos\tilde\theta_s) &= \sum_{j=0}^J \alpha_{J,j}(s) P_j(\cos\theta_s) \,, ~~ \alpha_{J,j}(s) \geq 0\,, ~~~ 0<s<4m^2\, ,
\end{align}
which describes the decomposition of a single $SO(4)$ multiplet into the standard $SO(3)$ partial waves. Therefore dual conformal invariance leads to a stronger unitarity constraint on the amplitude. If we formally write the scattering angle as
\be\label{eq:cosWith5dEnergy}
\cos\tilde\theta_s &= 1+ \frac{2t}{s_{5d}}, ~~~ s_{5d}={s \over 1 - {s \over 4 m^2}} \ . 
\ee
We see that \emph{effective} five-dimensional scattering energy changes from $0 < s_{5d} < \infty$ as we change the scattering energy $0<s<4m^2$. We will see that this point of view on the $SO(4)$ unitarity will be convenient when performing the bootstrap analysis of the amplitude.

For $s \geq 4m^2$ the relevant little group is $SO(1,3)$. The scattering (hyperbolic) angle in this case takes the form
\be
  \cosh\tilde\theta_s&= 1+ \frac{2t}{s} - \frac{t}{2m^2}\,,~~~ s\geq4m^2\,\label{eq:SO31angle}  . 
\ee
We can then expand the amplitude as
\be
\textbf{SO(1,3) unitarity:}~~~f_{s}(s,t) &= s^{-2}\int_0^\infty d\nu c_\nu(s) \Omega_\nu(\cosh\tilde\theta_s),~c_\nu(s)\geq 0,~s\geq4m^2\,,\label{eq:fsInSO31PW}
\ee
where the $SO(1,3)$ harmonic functions are given by
\be
 \Omega_\nu(x) &= \, _2F_1\left(1-i \nu ,1+i \nu;\frac{3}{2};\frac{1-x}{2}\right) = {\sin \nu \theta \over \nu \sinh \theta}, ~~~ x = \cosh \theta \ .
\ee
As above, the $SO(1,3)$ unitarity \eqref{eq:fsInSO31PW} automatically implies the $SO(3)$ unitarity \eqref{eq:usualPW} as we decompose the $SO(1,3)$ multiplets into the $SO(3)$ partial waves
\begin{align}
    \Omega_\nu(\cosh\tilde\theta_s) &= \sum_{j=0}^\infty \tilde \alpha_{\nu,j}(s) P_j(\cos\theta_s) \,, ~~ \tilde \alpha_{\nu,j}(s) \geq 0\,, ~~~ s \geq 4m^2\, \ . 
\end{align}
As above, we can introduce $\infty>\tilde s_{5d}={s \over {s \over 4 m^2}-1} >4 m^2$ in terms of which the bootstrap analysis of unitarity becomes simpler.

\subsection{Spectrum of Regge trajectories}
\label{sec:spectrum_bound_states}

Let us next discuss the spectrum of bound states below the two-particle threshold $s<4 m^2$ in more detail. It is convenient to organize this discussion by considering the Regge expansion of $f(s,t)$ in the $t \to \infty$ limit. A starting point for this discussion is the $SO(4)$ partial wave expansion \eqref{eq:fsInSO4PW} applied to the full amplitude. The standard Sommerfeld-Watson manipulations then lead to the following Regge expansion of the elastic amplitude $A_{z,z \to z,z}$, see \cite{Bruser:2018jnc}, %
\be
f(s,t) = - {1 \over s \sqrt{t (t+{4 m^2 s \over 4m^2-s})}} \sum_{i} r_i(s) Y^{-j_i(s)-1}, ~~~ Y = e^{i \tilde \theta_s} \ . 
\label{eq:Regge_limit}
\ee
At large $t$, we have ${1 \over Y} \simeq {4 m^2 - s \over s} {t \over m^2} + ...$ \ . In the formula above, $j_i(s)$ are \emph{the Regge trajectories}, which correspond to poles in the $J$-plane of the analytically continued $SO(4)$ partial wave harmonics defined above. For $s<0$ they control the high-energy behavior of the amplitude, for $0<s<4m^2$, they encode the spectrum of bound states $m_{i,J}$, which is given by the following condition
\be
\label{eq:boundstates}
j_i(m^2_{i,J}) = J, ~~~ J \in \mathbb{Z}_{\geq 0} \ , 
\ee
This equation states that a spin $J$ bound state of mass $m_{i,J}$ belongs to the $i$'th Regge trajectory $j_i(s)$.
At any coupling $g$ we have infinitely many Regge trajectories present in the high-energy expansion of the amplitude in agreement with \cite{Eckner:2024pqt}, however, as we will see below, not all trajectories admit bound states, or, equivalently, nontrivial solutions to \eqref{eq:boundstates}. 

Let us start our discussion with the leading Regge trajectory $j_0(s)$. The Regge limit of interest is described by taking $t \to \infty$ while holding $s$ and $m$ fixed. In this limit, the cross-ratio $v \to 0$, while $u$ is kept fixed. Using dual conformal invariance the same limit can be achieved by taking $m_2 \to 0$ while keeping $s$, $t$, and other masses fixed. In this limit, the amplitude is expected to develop an IR divergence which is controlled by the cusp anomalous dimension of the Maldacena-Wilson line \cite{Polyakov:1980ca,Korchemsky:1987wg,Maldacena:1998im}, defined by
\begin{equation}
    \langle W(C) \rangle \sim \left( \frac{\Lambda_{\text{UV}}}{\Lambda_{\text{IR}}} \right)^{-\Gamma_{\text{cusp}}(g,\phi)}\,.
\end{equation}
Equating the two effects leads to the following relation \cite{Henn:2010bk,Correa:2012nk,Caron-Huot:2014gia}
\be
j_0(s) = - 1 - \Gamma_{\text{cusp}}(g,\phi), ~~~ s = 4m^2 \sin^2 \phi/2 \ . 
\label{eq:j0_and_gamma_cusp}
\ee
Notice that $\Gamma_{\text{cusp}}(g,\phi)$ can be efficiently computed using the integrability methods \cite{Drukker:2012de, Correa:2012hh, Gromov:2015dfa}. Using the relationship \eqref{eq:j0_and_gamma_cusp}, we can immediately infer the shape of the leading Regge trajectory. 

Its key property follows the relationship between $\Gamma_{\text{cusp}}(g,\phi)$ and the quark-antiquark potential $V(r) = - {\Omega(g) \over r}$, see e.g. \cite{Erickson:1999qv, Gromov:2016rrp},
\begin{equation}
    \lim_{\phi \to \pi} \Gamma_{\text{cusp}}(g,\phi) = - {\Omega(g) \over \pi - \phi} \, .
\label{eq:Gamma_cusp_pi}
\end{equation}
This immediately implies that the amplitude of interest exhibits an accumulation point of bound states at the two-particle threshold
\begin{equation}
    j_0(s) \simeq \frac{\Omega(g)}{\sqrt{1-s/4m^2}}, \quad s \to 4m^2 .
\end{equation}
On the other hand, we have 
\be
j_0(s) \simeq -1 + B(g) {s \over m^2} + \mathcal{O}(s^2) , ~~~ s \to 0 ,
\label{eq:bremss_j0}
\ee
where $B(g) = \frac{g}{\pi} \frac{ \hspace{1mm} I_2(4 \pi g)}{I_1(4 \pi g)}$ is the so-called bremsstrahlung function \cite{Correa:2012at, Fiol:2012sg, Correa:2012hh}. We can  use the leading Regge trajectory to define \emph{the mass gap} (an energy scale where the amplitude develops discontinuity) as follows
\be
\label{eq:gapdef}
j_0(m_{\text{gap}}^2) = 0 \ .
\ee
We will discuss its detailed behavior as a function of the coupling below. 

At weak coupling the spectrum of the bound states is localized very closed to $s=4m^2$ and is effectively described by non-relativistic quantum mechanics of two particles interacting via Coulomb potential. It is well-approximated by the formula $j_0(s) = -1 +  \frac{\Omega(g)}{\sqrt{1-s/4m^2}}$, where $\Omega \ll 1$. Inverting this formula we get a formula for the spectrum of bound states which takes a familiar form of the hydrogen atom
\be
m_{J} = 2m - {m \Omega(g)^2 \over (J+1)^2} ,
\ee
where $J$ is the $SO(4)$ spin and each bound state $m_{J}$ is degenerate when decomposed in terms of the more familiar $SO(3)$ harmonics, see e.g. \cite{Sakurai:2011zz}.\footnote{Traditionally, the hydrogen spectrum is written as $E_n \sim -{1 \over n^2}$, so that $n=J_{SO(4)}+1$.} 

\begin{figure}[h!]
    \centering
    \includegraphics[width=0.9\linewidth]{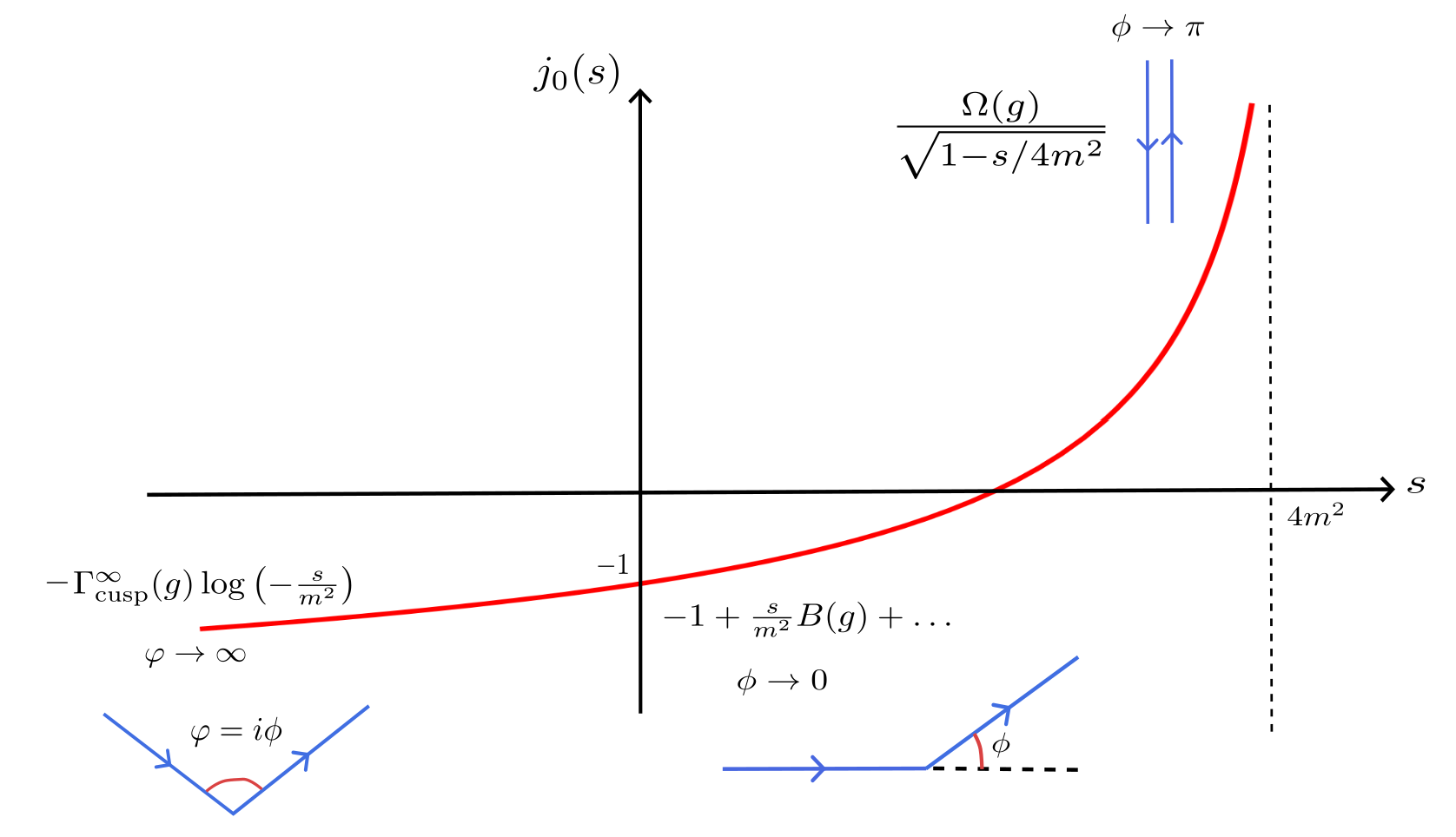}
    \caption{Summary of the different limits captured by $j_0(s) = -1-\Gamma_{\text{cusp}}(g,\phi)$, with $s = 4m^2 \sin^2(\phi/2)$.  Close to $s = 0$ (i.e. $\phi = 0$), $j_0(s)$ is related to the Bremsstrahlung function $B(g)$, which is exactly known from supersymmetric localization \cite{Correa:2012at} and integrability \cite{Correa:2012hh}. When $s \to 4m^2$ (i.e. $\phi \to \pi$), the Wilson lines become anti-parallel and the Regge trajectory diverges. This limit captures the quark-antiquark potential \eqref{eq:Gamma_cusp_pi}. The Euclidean angle $\phi$ can be analytically continued to its Lorentzian version $\varphi = i \phi$: the behavior as $\varphi \to \infty$ of $\Gamma_{\text{cusp}}$ captures the limit $|s| \to \infty$ of $j_0(s)$ \eqref{eq:light_like_cusp}. In this regime, the trajectory scales as a logarithm and is controlled by $\Gamma_{\text{cusp}}^{\infty}(g)$, which is the cusp anomalous dimension of a light-like Wilson line \cite{Korchemsky:1987wg}. The region $s>4m^2$ corresponds to $\varphi = \pi + i  \phi$.}
    \label{fig:behaviors_j0}
\end{figure}

Let us also recall that in the limit of large imaginary angles, we get
\be
\Gamma_{\text{cusp}}(g, - i \varphi) = {\Gamma_{\text{cusp}}^{\infty}(g) \over 2} \varphi+ ..., ~~~ \varphi \to \infty \ . 
\label{eq:light_like_cusp_phi}
\ee
This captures the behavior of the Regge trajectory for large $s$, namely we get
\be
\lim_{|s| \to \infty} j_0(s) = - {\Gamma_{\text{cusp}}^{\infty}(g) \over 2} \log \left( {-s \over m^2} \right) \ . 
\label{eq:light_like_cusp}
\ee
In particular, we see that at large negative $s$ the Regge intercept gets arbitrarily negative, however, it does so logarithmically, which reflects the less fine-tuned nature of the amplitude compared to the four-point amplitude that describes scattering of strings in flat space. A typical finite coupling shape of the leading Regge trajectory is shown in Figure \ref{fig:behaviors_j0}.

Let us next describe properties of $j_0(s)$ in the complex $s$-plane \cite{Mandelstam:1974fq}. From the explicit perturbative results, we see that we can write a once-subtracted dispersion relation with a unitarity cut starting from $4m^2$
\be
j_0(s) = -1 + \int_{4m^2}^\infty {ds' \over \pi} {s \over s'} {{\rm Im}j_0(s') \over s'-s} \ ,
\ee
where $\lim_{s \to \infty} {\rm Im}j_0(s) = \pi {\Gamma_{\text{cusp}}(g) \over 2}$ and ${\rm Im}j_0(s) >0$ because resonances become unstable and go to the second sheet above the two-particle cut.
It is interesting to see how the linear Regge trajectory emerges at strong coupling. We can expand the Regge trajectory at low energies to get
\be
j_0(s) = -1 + s  \int_{4m^2}^\infty {ds' \over \pi}{{\rm Im}j_0(s') \over (s')^2}+ s^2 \int_{4m^2}^\infty {ds' \over \pi}{{\rm Im}j_0(s') \over (s')^3} + ... \ . 
\ee
We expect all the moments at strong coupling to be $\sim g$. Therefore if we take $g \to \infty$ while keeping $g s$ fixed only the linear term survives and we recover the linear Regge trajectory of the flat spacetime amplitude.

\begin{figure}[h!]
    \centering
    \includegraphics[width=0.9\linewidth]{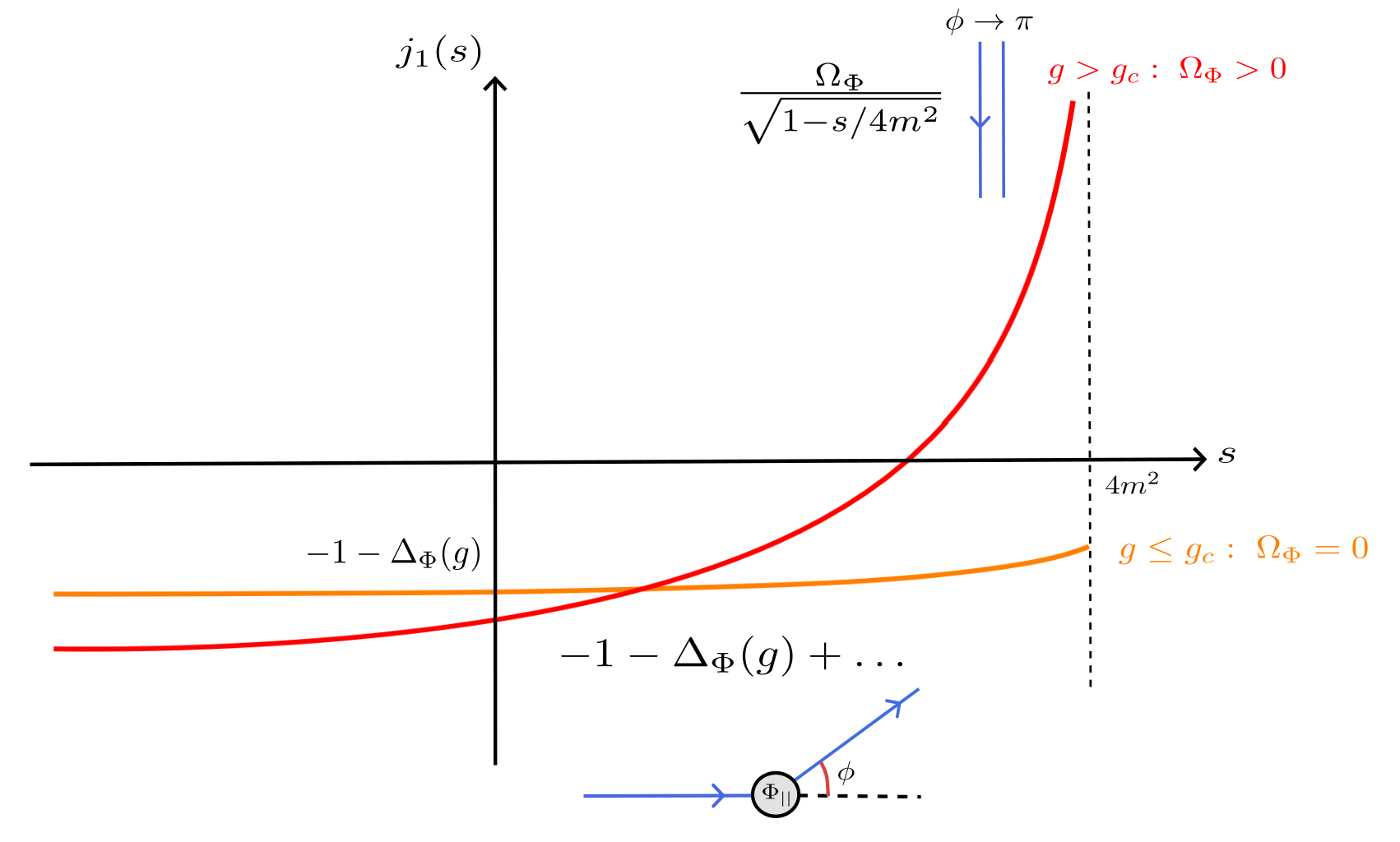}
    \caption{Summary of the different limits captured by $j_1(s) = -1-\Gamma_{\text{cusp},\Phi}(g,\phi)$, with $s = 4m^2 \sin^2(\phi/2)$.  Close to $s = 0$ (i.e. $\phi = 0$), $j_1(s)$ is related to the scaling dimension of the \textit{parallel} scalar inserted in the cusp $\Delta_{\Phi}(g)$, which can be computed using QSC \cite{Grabner:2020nis} and it is known at weak and strong coupling \eqref{eq:Delta_Phi_g}. When $s \to 4m^2$ (i.e. $\phi \to \pi$), the Wilson lines become anti-parallel and the subleading Regge trajectory diverges with residue $\Omega_{\Phi}$ \eqref{eq:Gamma_cusp_pi_Phi}. In \cite{Klebanov:2006jj} it was argued that $\Omega_{\Phi}(g) = 0$ for $g \leq g_c$, with $g_c = 1/4$ from the ladder approximation. Our integrability computation, reported in  \secref{sec:qsc_analysis_j1}, confirms this behavior and the critical coupling that we obtain is $g_c \approx 0.28$.}
    \label{fig:behavior_j1}
\end{figure}

Let us next discuss the first subleading Regge trajectory $j_1(s)$. In this case, the same argument used above for $j_0(s)$, suggests a relationship to the cusp anomalous dimension with a nontrivial insertion of a local operator at the cusp. It was indeed observed in \cite{Bruser:2018jnc} that the following relationship holds at weak coupling up to two loops
\begin{equation}
       j_1(s) =-1 -\Gamma_{\text{cusp},\Phi}(\phi), \qquad s = 4m^2 \sin^2 \frac{\phi}{2} ,
\label{eq:j1_and_gamma_cusp}
\end{equation}
where $\Gamma_{\text{cusp},\Phi}(g,\phi)$ is the cusp anomalous dimension with an insertion of the so-called \emph{parallel} scalar. $\Gamma_{\text{cusp},\Phi}(g,\phi)$ can be computed using integrability methods as well \cite{Grabner:2020nis}.
To understand the behavior of the subleading Regge trajectory we consider the limit $\phi \to \pi$
\begin{equation}
    \lim_{\phi \to \pi} \Gamma_{\text{cusp},\Phi}(\phi) = - {\Omega_{\Phi}(g) \over \pi - \phi} \, .
\label{eq:Gamma_cusp_pi_Phi}
\end{equation}
It was argued in \cite{Klebanov:2006jj}, based on the study of excitations of a conformal gauge theory flux tube, that
\be
\Omega_{\Phi}(g) = 0, ~~~ g \leq g_c ,
\label{eq:Omega1}
\ee
and $\Omega_{\Phi}(g)>0$ for $g > g_c$. Our integrability calculation is consistent with this expected behavior and we get that $g_c \approx 0.28$.
The relevant shape of the Regge trajectory is shown in \figref{fig:behavior_j1}.

\subsection{The Regge/cusp correspondence}
\label{sec:sublRegge}

Let us discuss next an extension of the Regge/cusp correspondence to subleading trajectories. It is convenient to consider two cases separately, $\phi = 0$ and $\phi \neq 0$, since the symmetries of the problem are different.\footnote{Recall that $s=4m^2 \sin^2(\phi/2)$, and $\phi=0$ corresponds to forward scattering.}  For the case $\phi =0$, we have a straight line, and the associated one-dimensional defect CFT has been studied extensively with various methods, see e.g. \cite{Giombi:2017cqn, Ferrero:2023znz, Ferrero:2023gnu, Cavaglia:2021bnz} and references therein. It exhibits $OSp(4^*|4)$ superconformal symmetry, and local operators are organized in the corresponding supermultiplets. Its bosonic subgroup includes $SO(3)$ rotations orthogonal to the line, as well as $SO(5)$ residual R-symmetry. In addition, it exhibits parity \cite{Cavaglia:2023mmu}, which will be naturally connected to the crossing symmetry of the amplitude. 

Focusing on the forward amplitude at $\phi=0$, we propose the following generalization of the Regge/cusp correspondence
\be \label{eq:jn_forward}
j_n(0)=- \Delta_{O_n} - 1 ,
\ee
where $O_n$ are the parity-even lowest weight scalar primaries ${\cal L}^{\Delta}_{0,[0,0]}$  \cite{Ferrero:2023znz}, and $j_n(0)$ are the Regge trajectories that contribute to the forward amplitude. ${\cal L}^{\Delta}_{0,[0,0]}$ refers to operators that belong to long superconformal multiplets of $OSp(4^*|4)$ which are neutral under $SO(3) \times SO(5)$. As we will demonstrate in the upcoming sections, we find nontrivial support for this picture both at weak and strong coupling.

As we turn on a non-zero angle $\phi \neq 0$, the symmetry of the problem changes. The Maldacena-Wilson line with a cusp is still invariant under dilatations, so that by the operator-state correspondence, the eigenstates of the cusp Hamiltonian $H_{\text{cusp}}$ can be created by local operators at the cusp. In addition, we have $SO(2)$ rotations in the plane orthogonal to the cusp, and the same $SO(5)$  R-symmetry. As we review in \appref{app:parity}, it also has several discrete symmetries. Since the conformal symmetry is broken, there is no reason for primaries and descendants to have trivially related scaling dimensions. A natural generalization of the correspondence is then 
\be
\label{eq:reggecusp}
j_n(s)=- E_{\text{cusp},n} (\phi) - 1 , ~~~ s = 4 m^2 \sin^2(\phi/2) \ , 
\ee
where $E_{\text{cusp},n}$ are excited states of the cusp Hamiltonian, which are parity-even, see \appref{app:parity}, and neutral $SO(2) \times SO(5)$. Of course, $E_{\text{cusp},n}(\phi)$ is nothing but $\Gamma_{\text{cusp},O_n}(\phi)$ by the operator-state correspondence. 

Once $\phi \neq 0$, there is no reason to restrict to $SO(3)$ singlet operators of the $\phi=0$ theory, because spinning $SO(3)$ representations admit singlet restrictions to the residual $SO(2)$ symmetry. Indeed, we will show that \emph{additional Regge trajectories} appear at $\phi \neq 0$, beyond those associated with cusp insertions of parity-even operators $O_n \in {\cal L}^{\Delta}_{0,[0,0]}$, both at weak and strong coupling. We expect that these trajectories are sourced  by \emph{spinning operators}: the ones that transform nontrivially under the rotations in the cusp plane, or $SO(3)$ in the straight line limit. This explains why they decouple as we set $\phi=0$, when only $SO(3)$ singlets contribute to the forward amplitude. However, with the current knowledge of the amplitude, we cannot yet probe their precise microscopic nature, and we leave this problem to future work.

\begin{figure}[h!]
    \centering
    \includegraphics[width=0.8\linewidth]{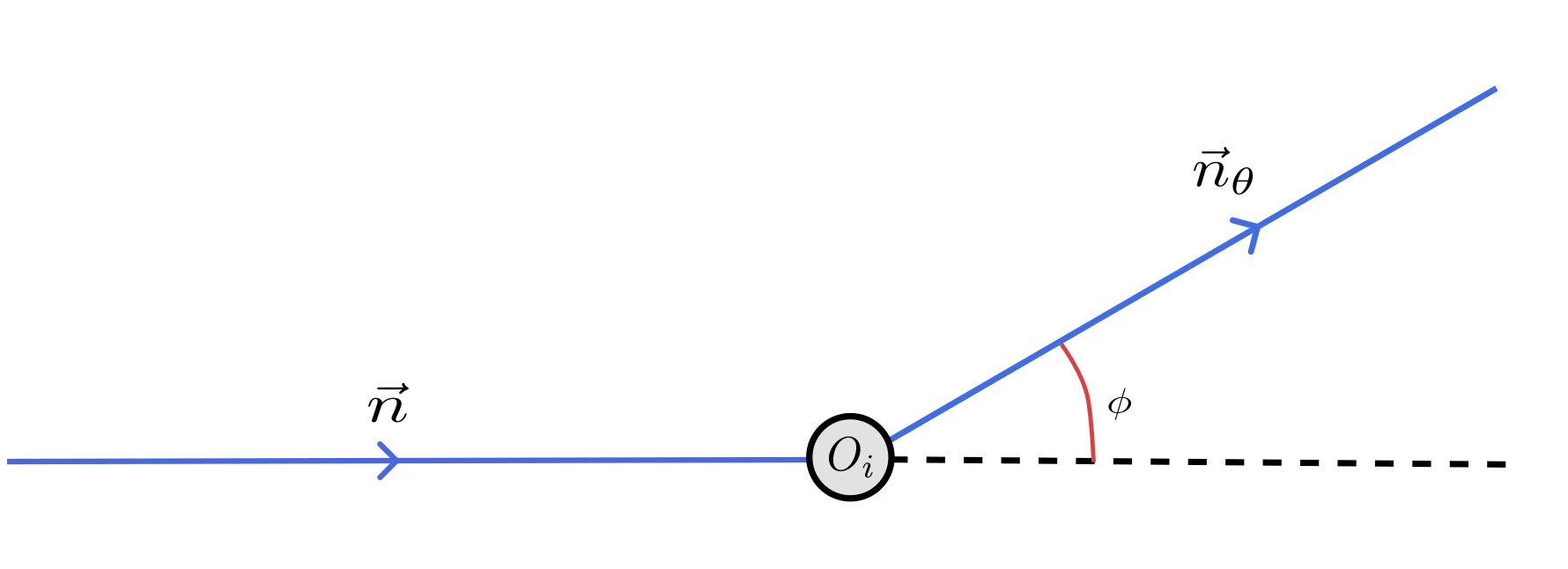}
    \caption{A Maldacena-Wilson line with the cusp angle $\phi$ and an insertion of operator $O_i$. In the more general case the coupling of the lines with the scalars is parametrized by two unit vectors $\vec{n}, \vec{n}_{\theta}$ in the R-symmetry space, such that $\vec{n} \cdot \vec{n}_{\theta} = \cos \theta$. In the case analyzed in this paper $\theta = 0$ \eqref{eq:n_all_aligned}.}
    \label{fig:cusped_WL}
\end{figure}

\section{Regge trajectories from integrability}
\label{sec:integrability}

In this section, we use integrability to compute the leading and the subleading Regge trajectories $j_0(s)$ and $j_1(s)$ for $0 \leq s \leq 4 m^2$ at finite coupling $g$. We then explore the generalization of the Regge trajectories/defect CFT correspondence beyond the first two Regge trajectories.

In the last decade, the integrability program has developed a powerful method called the Quantum Spectral Curve (QSC) to numerically compute the planar spectrum of local operators in $\mathcal{N} = 4$ SYM for any value of the coupling $g$ \cite{Gromov:2013pga}.
The QSC is based on the Q-system difference equations, which are a generalization of the Baxter difference equations that take into account the $psu(2,2|4)$ global symmetry of $\mathcal{N} = 4$ SYM \cite{Gromov:2017blm}. In these equations, the unknowns are the Q-functions: the $\mathbf{Q}$- and $\mathbf{P}$-functions that encode the $AdS_5$ and the $S^5$ dynamics respectively. The Q-functions depend on the spectral parameter $u$ and have a known, nontrivial analytic structure that depends on the 't Hooft coupling $g$. The set of Q-system equations combined with gluing conditions that specify the monodromies around the branch points of the Q-functions gives the QSC. The QSC is believed to have a discrete set of solutions and from the asymptotic behavior of the Q-functions it is possible to extract quantum numbers of the operators such as, for example, the scaling dimensions $\Delta$ of local operators in ${\cal N}=4$ SYM. %

Remarkably, the same system of equations admits solutions which describe other observables in the theory. What is relevant for the present paper is that by changing the boundary conditions (and the analytic structure) of the $Q$-functions appropriately, QSC can be used to compute the cusp anomalous dimension of the Maldacena-Wilson line $\Gamma_{\text{cusp}}(g,\phi,\theta)$ \cite{Gromov:2015dfa,Gromov:2016rrp}, see \figref{fig:cusped_WL} (the picture has an operator insertion, here we refer to the usual cusp). It can also be used to compute CFT data of the 1d defect CFT on the straight line \cite{Grabner:2020nis,Cavaglia:2021bnz,Cavaglia:2022qpg}. Finally, in \cite{Grabner:2020nis,Cavaglia:2021bnz} the authors used QSC to derive numerical data for the anomalous dimension of a Maldacena-Wilson line with operator insertions at the cusp. These are precisely the observables that we need to compute $j_0(s)$ and $j_1(s)$ at finite coupling via the relations \eqref{eq:j0_and_gamma_cusp} and \eqref{eq:j1_and_gamma_cusp}.

\subsection{Regge trajectories from Quantum Spectral Curve (QSC)}
\label{sec:QSC}

The observables we are interested in are $\Gamma_{\text{cusp}}(g,\phi)$ and $\Gamma_{\text{cusp},\Phi}(g,\phi)$ with $\theta = 0$. 
The $\mathbf{P}$ and $\mathbf{Q}$ functions in this case obey the same set of equations as in the case of local single-trace operators. The difference between the case of $\Gamma_{\text{cusp}}(g,\phi)$ and the one of local operators is two-fold, as explained in \cite{Gromov:2015dfa}.
First, the boundary conditions at large spectral parameter $u$ are modified to include the angle-dependence. Specifically, the leading asymptotics take the form $\mathbf{P} \sim e^{\pm \theta u}, \mathbf{Q} \sim e^{\pm \phi u}$. Second, the gluing conditions must also be adjusted, since the analytic structure of the $Q$-functions differs from the case of local operators. This generalization applies to insertions of orthogonal scalars at the cusp and, with further modifications, to the case of a parallel scalar \cite{Grabner:2020nis}. In this work, we are interested in the latter case.

It is important to emphasize that, because the QSC admits multiple solutions, one must carefully identify the physically relevant one. To establish the correct starting point for solving the QSC at general $(g,\phi)$ numerically, one exploits special limits, such as weak coupling or the BPS limit, where analytic solutions can be obtained. These analytic inputs then serve as seeds for an iterative algorithm that allows one to numerically solve the QSC for generic values of the coupling and the angles.

The procedure to numerically solve QSC is explained in detail, for example, in \cite{Gromov:2015dfa} and we will summarize it in the following. 
First, after having extracted the exponential and leading powerlike asymptotic, the $\mathbf{P}$-functions can be expressed as a convergent power series in $1/x(u)$, with $x(u)$ the so-called Zhukovsky variable. The $\mathbf{P}$-functions can then be approximated by the truncated series  $\sum_{n=0}^{\verb|ChPW|}{c_n \over x(u)^{2n}}$, with cutoff \verb|ChPW| and unknown coefficients $c_n$. These coefficients are the main objective of the algorithm: solving QSC equations reduces to minimizing a specific quantity which is a function of $c_n$'s.
More precisely, we use the gluing conditions on the cut $[-2g,2g]$ for the $\mathbf{Q}$-functions to find $c_n$'s. In fact, the function $\mathbb{S}$ to minimize is built from the gluing conditions of the $\mathbf{Q}$-functions. Defining $\Tilde{\mathbf{Q}}$ the analytic continuation of the $\mathbf{Q}$-functions on the second sheet and, for convenience, $\mathbf{q}_i(u) = \mathbf{Q}_i(u)/\sqrt{u}$ with $i = 1, \dots, 4$, the function $\mathbb{S}$ is \cite{Gromov:2015dfa}
\begin{equation}
    \mathbb{S} = \sum_{k = 1}^{N_{points}} |\tilde{\mathbf{q}}_1(u_k)-\mathbf{q}_1(-u_k)|^2 + |f_k-\hat{f}|^2\,, \qquad f_k = \frac{\tilde{\mathbf{q}}_4(u_k)-\mathbf{q}_4(-u_k)}{\textbf{q}_1(-u_k) \sinh(2 \pi u_k)}\,,
\end{equation}
where $\{u_k, \hspace{1mm} k = 1, \dots N_{points}\}$ is a Chebyshev grid of points on the cut and $\hat{f}$ is the average of all $f_k$.
To minimize $\mathbb{S} = |\vec{\mathbb{F}}|^2$, at each step we approximate $\vec{\mathbb{F}}(\dots + \eps) = \vec{\mathbb{F}}(\dots) + \eps \hspace{1mm}\partial  \vec{\mathbb{F}}$ (iterative Levenberg–Marquardt algorithm). The minimization procedure gives a result for the $c_n$'s and $\Gamma_{\text{cusp}}$ can be computed from them. The precision of $\Gamma_{\text{cusp}}$ is of order $\sqrt{\mathbb{S}}$: for later purposes we don't need more than $10$ digits of precision on $\Gamma_{\text{cusp}}$, so we decide to consider data with $\mathbb{S} < 10^{-20}$. To reach this precision one has to carefully choose the values of the parameters, for example \verb|ChPW|, which depend on the specific values of $g$ and $\phi$.

\begin{figure}[h!]
    \centering
    \includegraphics[width=0.9\linewidth]{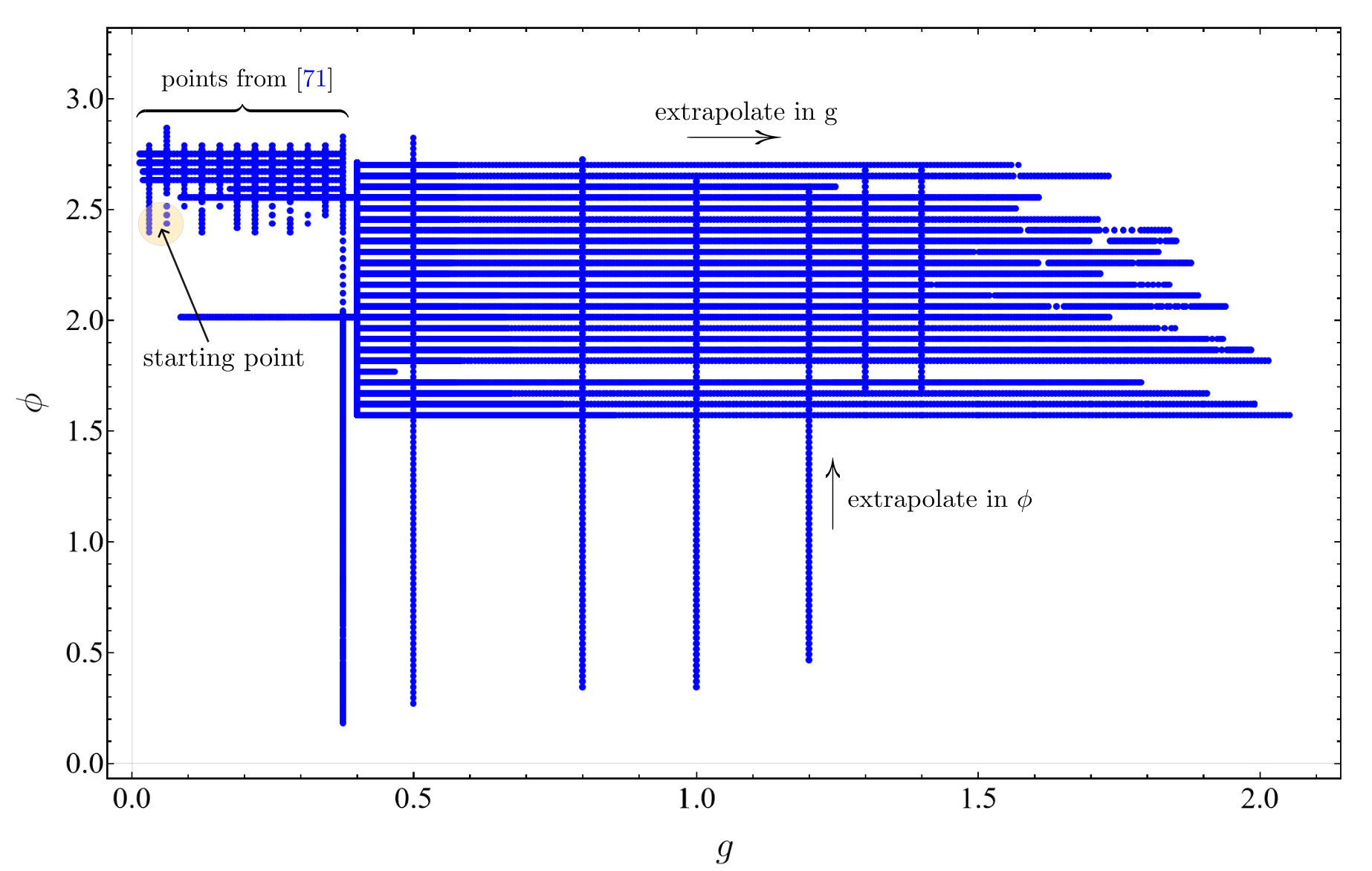}
    \caption{QSC data point for $\Gamma_{\text{cusp}}(g,\phi)$ as a function of $(g,\phi)$. Here we summarize our workflow: in the left corner, at small $g$, the QSC is initialized using the perturbative solution. Then we move in $\phi$ at fixed $g$ or in $g$ at fixed $\phi$.}
    \label{fig:workflow_j0}
\end{figure}

To practically implement the method described above, first we need to analytically solve QSC at small coupling and generic values of $\phi$. From these points, one can then extrapolate the solution moving in $\phi$ at fixed $g$ (or in $g$ at fixed $\phi$). 
Our workflow is depicted in \figref{fig:workflow_j0}, which represents all the QSC data points that we have for $\Gamma_{\text{cusp}}(g,\phi)$ as a function of $(g,\phi)$.

As stated in the figure, all the points for $g \leq 3/8$ (on the left of the long vertical line) were provided to us by Nikolay Gromov \cite{Gromov_QSC}, as well as the code to numerically compute these points. Effectively, we computed all the points on the right of the long vertical line ($g > 3/8$) by extrapolating in $g$ or in $\phi$. On the left corner, in the yellow dot, we indicate the starting point of the QSC algorithm, given by the initialization of the analytical perturbative solution. 
To extrapolate the result of $\Gamma_{\text{cusp}}$ for different values of $g$ or $\phi$, we consider the solution at the nearest previous points and interpolate them to get a first approximation for the coefficients $c_n$ of the new point. These coefficients are then tested and submitted to the minimization procedure.

As showed in figure \figref{fig:workflow_j0}, we computed values of $\Gamma_{\text{cusp}}(g,\phi)$ at different values of $\phi$ in $[0,\pi]$ and of $g$ in $[0,2]$. The precision on the value of $\Gamma_{\text{cusp}}$ obtained is at least of $\sim 10^{-10}$. The generation of these data was computationally expensive, especially at stronger couplings and close to $\phi \approx \pi$. For this reason we stopped around $g = 2$ but, as explained in the next section, for $g > 2$ we could generate data using a global fit and the strong-coupling result of $\Gamma_{\text{cusp}}(g,\phi)$.

This discussion can be equally applied for the computation of  $\Gamma_{\text{cusp},\Phi}$, as explained in \cite{Grabner:2020nis}. Also in the case of one parallel insertion, we compute the value of $\Gamma_{\text{cusp},\Phi}$ for different $\phi$ in $[0,\pi]$ and $g$ in $[0,2]$, with a precision of $\sim 10^{-10}$ at least. \\

In the following, we review known properties of the observables $\Gamma_{\text{cusp}}$ and $\Gamma_{\text{cusp},\Phi}$, which are related to $j_0(s)$ and $j_1(s)$ using \eqref{eq:j0_and_gamma_cusp} and \eqref{eq:j1_and_gamma_cusp} respectively. We then present the results obtained using the QSC and extrapolation of our results to arbitrary $g$ and $\phi$.
We divide the discussion into two sections: one for the leading trajectory $j_0(s)$, and one for the subleading trajectory $j_1(s)$.

\subsection{Leading Regge trajectory $j_0(s)$}

\paragraph{Review of known results}\leavevmode\newline
Let us start by reviewing the known properties of $j_0(s)$ or, equivalently, of $\Gamma_{\text{cusp}}(g,\phi)$ \eqref{eq:j0_and_gamma_cusp}. 
To remark the dependence of $j_0$ on the cusp angle $\phi$ we will use the following notation 
\begin{equation}
    \mathrm{j}_0(\phi) \equiv j_0(s(\phi))\,, \quad s = 4m^2 \sin^2(\phi/2)
\end{equation}
where we don't stress the $g$ dependence for simplicity.
In the following, we will discuss the $g$ and $\phi$ dependence of the leading trajectory. 
In \figref{fig:behaviors_j0} we summarized the different behaviors of $j_0(s)$ as a function of $s$ and we emphasized how these can be captured by different limits of the Wilson line with a cusp. 

First of all, we discuss the dependence of $\mathrm{j}_0(\phi)$ on $\phi$ at fixed $g$, expanding the description of \figref{fig:behaviors_j0}.
Notice that, since $\Gamma_{\text{cusp}}(g,\phi)$ is invariant under $\phi \to -\phi$ \cite{Correa:2012nk}, $\mathrm{j}_0(\phi)$ is an even function of $\phi$.
One can then discuss different limits of $\Gamma_{\text{cusp}}$ depending on the values of the angles $\phi$ and $\theta$. A limit of particular interest is the so-called BPS limit $\phi \to \pm \theta$ in which the Wilson line becomes $\frac{1}{2}$-BPS and the cusp anomalous dimension takes the form \cite{Cavaglia:2022qpg}
\begin{equation}
    \Gamma_{\text{cusp}}(g,\phi,\theta) = \frac{\cos\phi-\cos\theta}{\sin\phi} \Gamma^{(1)}(g,\phi) + \left(\frac{\cos\phi-\cos\theta}{\sin\phi}\right)^2 \Gamma^{(2)}(g,\phi) + \mathcal{O}((\cos\phi-\cos\theta)^3)\,,
\end{equation}
with 
\begin{equation}
    \Gamma^{(1)}(g,\phi) = \frac{2 \phi g}{\sqrt{\pi^2-\phi^2}} \frac{I_2(4 g \sqrt{\pi^2-\phi^2})}{I_1(4 g \sqrt{\pi^2-\phi^2})}\,, \quad \Gamma^{(2)}(g,\phi) = \phi^2 \mathbb{C}(\phi)\,.
\end{equation}
The $I_n(x)$ are modified Bessel functions of the first kind, while $\mathbb{C}(g)$ is called \textit{curvature} function. It can be written as a double contour integral using the QSC formalism \cite{Cavaglia:2022qpg}.
In our case, since $\theta = 0$, the BPS-limit corresponds to the limit $\phi \to 0$. In this regime, expanding the expressions above and using \eqref{eq:j0_and_gamma_cusp}, we obtain
\begin{equation}
\mathrm{j}_0(\phi) = -1 + \phi^2 B(g) + \phi^4 B_4(g) + \mathcal{O}(\phi^6)\,, 
\label{eq:small_phi_j0}
\end{equation}
where $B(g)$ is the Bremsstrahlung function, while $B_4(g)$ is a combination of Bessel functions and the curvature function
\begin{equation}
 B(g) = \frac{g}{\pi} \frac{I_2(4 \pi g)}{I_1(4 \pi g)}\,, \qquad B_4(g) = \left(\frac{B(g)}{12}+2B(g)^2-\frac{\mathbb{C}(g)}{4}-\frac{2g^2}{\pi^2} \frac{I_3(4\pi g)}{I_1(4 \pi g)}  \right) \\,
 \label{eq:bremss}
\end{equation}
$B(g)$ is exactly known at any coupling from supersymmetric localization \cite{Correa:2012at} and integrability \cite{Correa:2012hh}
while $\mathbb{C}(g)$ is known at weak and strong coupling. It takes the form \cite{Cavaglia:2022qpg}
\begin{equation}
    \mathbb{C}(g) = \begin{cases}
        4g^4 - \left( 24 \zeta_3 + \frac{16 \pi^2}{3} \right) g^6 + \mathcal{O}(g^8)\,, & \text{if } \hspace{1mm} g \ll 1 \ , \\
        \frac{(2\pi^2-3)g}{6\pi^3} + \frac{5-24\zeta_3-4 \pi^2}{32 \pi^4} + \mathcal{O}(1/g)\,, & \text{if } \hspace{1mm} g \gg 1 \ .
    \end{cases}
\label{eq:Curvature}
\end{equation}
Another interesting limit is $\phi \to \pi$ in which the two lines are anti-parallel. In this limit, the cusp anomalous dimension diverges as in equation \eqref{eq:Gamma_cusp_pi}, where $\Omega(g)$ is related to the quark-antiquark potential $V(r) = \Omega(g)/r$. The residue $\Omega(g)$ is known up to three loops at weak coupling \cite{Erickson:1999qv, Correa:2012nk} and at one loop at strong coupling \cite{Maldacena:1998im,Chu:2009qt, Forini:2010ek}
\begin{equation}
    \Omega(g) =
\begin{cases}
    4 \pi g^2 + \mathcal{O}(g^4), & \text{if } \hspace{1mm} g \ll 1 \\
     \frac{\pi (4 \pi g + a_1)}{4 K(1/2)^2} + \mathcal{O}(g^{-1}), & \text{if } \hspace{1mm} g \gg 1
\end{cases}
\label{eq:Omega_g}
\end{equation}
with $a_1 = -1.33459530528$. 
This quantity has also been studied in \cite{Gromov:2016rrp} using QSC. %

In the discussion above, we have considered $\phi$ as an Euclidean angle: it is possible to analytically continue to the Lorentzian cusp angle $\varphi = i \phi$. The limit $\varphi \to \infty$ is particularly relevant and was studied in \cite{Korchemsky:1987wg, Correa:2012nk}. In this regime, $\Gamma_{\text{cusp}}(g,\phi) \propto \varphi \Gamma_{\text{cusp}}^{\infty}(g)$  as in \eqref{eq:light_like_cusp_phi}, where $\Gamma_{\text{cusp}}^{\infty}(g)$ is the cusp anomalous dimension of a light-like Wilson line \cite{Korchemskaya:1992je}. Its expression is known up to four loops analytically \cite{Bern:2006ew,Cachazo:2006az}, but it can also be computed numerically for any value of $g$ using integrability \cite{Beisert:2006ez, Benna:2006nd}.
\\

Next, we can focus on the $g$ dependence at fixed cusp angle $\phi$. The perturbative expansion of $\mathrm{j}_0(\phi)$ has been computed, up to three loops, in \cite{Bruser:2018jnc} from the Regge limit of the amplitude, and it can be nicely expressed as a sum of harmonic polylogarithms \cite{Remiddi:1999ew}. Here we report its one-loop expression
\begin{equation}
    \mathrm{j}_0^{\text{weak}}(\phi) = -1 + 2 g^2 \phi \tan\left( \frac{\phi}{2}\right)+ \mathcal{O}(g^4) \,,
\end{equation}
The three-loop expansion of $\mathrm{j}_0(\phi)$ agrees with the one of $\Gamma_{\text{cusp}}(g,\phi)$ through equation \eqref{eq:j0_and_gamma_cusp}:
\be
\mathrm{j}_0(\phi) = - 1 - \Gamma_{\text{cusp}}(g,\phi), ~~~ s = 4m^2 \sin^2 \phi/2 \ . 
\ee
Previous results for the weak coupling expansion of $\Gamma_{\text{cusp}}(g,\phi)$ can be found for example in \cite{Makeenko:2006ds, Correa:2012nk, Henn:2019swt}, and we report them in \appref{app:perturbative_result_database}.

At strong coupling, the cusp anomalous dimension was originally computed in \cite{Drukker:1999zq} using the $AdS$/CFT correspondence and by relating the Wilson loop VEV $\langle W[C] \rangle$ to the area of the minimal surface bounded by $C$ spanned by the dual string. The leading trajectory $j_0(s)$ can be also computed \cite{Espindola:2016afe} as the angular momentum of a rotating string ending on a D3-brane in $AdS$. In the limit of large angular momentum $j \gg 1$ one can consider a classical rotating string, without having to quantize the system. In \cite{Espindola:2016afe} the authors express $j_0(s)$ at strong coupling only as a function of the parameter $\tilde{z}_0$, which is related to the maximum position of the rotating string profile in $AdS$
\begin{equation}
\mathrm{j}_0^{\text{strong}}(\phi) +1 = \frac{\pi g}{\Tilde{z}_0 \sqrt{1+\Tilde{z}_0^2} } {}_2 F_1 \left( \frac{1}{2}, \frac{3}{2},2,-\frac{1}{1+\Tilde{z}_0^2} \right) + \mathcal{O}(g^0)  \,,
\label{eq:j0_strong}
\end{equation}
where $\mathcal{O}(g^0)$ is a function of $s$ that goes to $0$ when $\phi \to 0$. The angle $\phi$ and the parameter $\Tilde{z}_0$ are related by 
\begin{equation}
 \phi(\Tilde{z}_0) = \pi + \frac{2 \sqrt{\Tilde{z}_0^2+1}
   \left(K\left(\frac{1}{\Tilde{z}_0^2+2}\right)-\left(\Tilde{z}_0^2+1\right) \Pi
   \left(-\Tilde{z}_0^2|\frac{1}{\Tilde{z}_0^2+2}\right)\right)}{\Tilde{z}_0
   \sqrt{\Tilde{z}_0^2+2}} ,
\end{equation}
where $\tilde z_0 \to 0$ corresponds to $\phi \to \pi$; $\tilde z_0 \to \infty$ corresponds to $\phi \to 0$; finally, $\tilde z_0 \to \pm i \sqrt{2}$ captures the Lorentzian regime $\phi \to \mp i \infty$. The function $K(x)$ is the complete elliptic integral of the first kind, while $\Pi(x|k)$ is of the third kind.

\paragraph{Global fit of the QSC data for $j_0(s)$}\leavevmode\newline
In this section, we describe our procedure to perform a global fit of $j_0(s)$ using the QSC data and the known results described in the previous section. Our method takes inspiration from the so-called conformal-Pad\'{e} approximation \cite{Dunne:2025wbq}, extensively used and described in section \ref{sec:bounds_using_pade_model}.

The idea of the conformal-Pad\'{e} approximation---that is a combination of  a Pad\'{e} approximation and a conformal map---comes from the fact that the weak coupling expansion of $\Gamma_{\text{cusp}}(g,\phi)$ has radius of convergence of $\frac{1}{4}$, due to a singularity at $g^2 = -\frac{1}{16}$. We can map the cut complex $g^2$ plane to the interior of the unit disk in the $z$ plane
\be
16 g^2 = {4 z \over (1-z)^2} ~~~ \longleftrightarrow ~~~ z = {\sqrt{1+16 g^2}-1 \over \sqrt{1+16 g^2}+1} \ . 
\ee
Under this map, the branch point $16 g^2=-1$ is mapped to $z=-1$, $g = \infty$ is mapped to $z=1$, and the cut along $16 g^2 < - 1$ is mapped to the boundary of the unit circle $|z|=1$. This method allows for improving the naive Pad\'{e} approximation in $g$ by using the $z$ variable instead and going back to $g$, see \cite{Dunne:2025wbq}.

In order to fit the leading Regge trajectory, we choose the following ansatz
\begin{equation}
    \mathrm{j}^{\text{ansatz}}_0(g,\phi) = -1 + \frac{\Omega(g)}{\Omega^{\text{strong}}} J(\phi) + \mathcal{J}(g,\phi)\, .
\label{eq:frankenstein_j0}
\end{equation}
Here $-1$ is a protected value of the trajectory at $\phi=0$. The second term $\frac{\Omega(g)}{\Omega^{\text{strong}}} J(\phi)$ is a known function, and it is constructed such that it correctly captures the strong coupling limit of the trajectory, as well as the finite coupling ${1 \over \pi - \phi}$ singularity. The last term $\mathcal{J}(g,\phi)$ parametrizes the non-singular $\phi$ dependence at finite coupling, and it is unknown.
We parametrize the unknown function $\mathcal{J}(g,\phi)$ as follows
\begin{equation}\label{eq:bigJ_gphi}
    \mathcal{J}(g,\phi) = \frac{a_1(g)(1-x) + a_2(g) (1-x)^2 + a_3(g) (1-x)^3}{1+a_4(g)(1-x) + a_5(g) (1-x)^2 + a_6(g) (1-x)^3}\,, \qquad x \equiv \left( \phi/\pi\right)^2\,,
\end{equation}
with unknown functions of the coupling parametrized as
\begin{equation}
    a_{i}(g) \equiv \frac{a_{i,0} + a_{i,1} z + a_{i,2} z^2 + a_{i,3} z^2}{1 + a_{i,4} z + a_{i,5} z^2 + a_{i,6} z^3}\,, \quad z = \frac{\sqrt{16 g^2 + 1}-1}{\sqrt{16 g^2 + 1}+1} \ .
\end{equation}
The Pad\'{e} ansatz was chosen to balance stability and flexibility: it uses a number of free parameters small enough to avoid overfitting relative to the available QSC data points, yet large enough to capture the nontrivial structure of $\mathrm{j}_0(g,\phi)$.

Notice that $\mathcal{J}(g,\phi)$ is ${\cal O}(g^0)$ in the strong coupling limit $g \to \infty$, while when $\phi \to \pi$
\begin{equation}
    \mathcal{J}(g,\phi) = \mathcal{O}(\phi-\pi) ,
\end{equation}
namely we assume that the constant term is absent (see footnote 40 in \cite{Kravchuk:2024qoh} and also \cite{Cuomo:2024psk}).\footnote{Our QSC data support the claim that the term $(\phi-\pi)^0$ is absent both at weak and strong coupling.}
Let us analyze more closely the second term in the ansatz \eqref{eq:frankenstein_j0}. It contains the quark-antiquark potential $\Omega(g)$, which can be derived as a function of $g$ by combining weak and strong coupling results \eqref{eq:Omega_g} in a conformal-Pad\'{e} approximation with ansatz
\begin{equation}
    \Omega(g) = \frac{1}{1-z} \frac{\omega_0 + \omega_1 z + \omega_2 z^2 + \omega_3 z^3}{1 + \omega_4 z + \omega_5 z^2 + \omega_6 z^3}\,, \quad z = \frac{\sqrt{16 g^2 + 1}-1}{\sqrt{16 g^2 + 1}+1} \ .
\end{equation}
The unfixed coefficients are fitted using the QSC data of \cite{Gromov:2016rrp}. 

\noindent The second term in ansatz \eqref{eq:frankenstein_j0} also contains $\Omega^{\text{strong}} = \frac{\pi^2}{K(1/2)^2}$ such that in the limit $g \to \infty$ we have $\frac{\Omega(g)}{\Omega^{\text{strong}}} = g + \mathcal{O}(g^0)$. The function $J(\phi)$ is simply given by the leading strong-coupling result for $\mathrm{j}_0(\phi)$ \eqref{eq:j0_strong}
\begin{equation}
    J(\phi) %
    = \frac{\pi}{\Tilde{z}_0(\phi) \sqrt{1+\Tilde{z}_0(\phi)^2} } {}_2 F_1 \left( \frac{1}{2}, \frac{3}{2},2,-\frac{1}{1+\Tilde{z}_0(\phi)^2} \right)\,,
\label{eq:J_strong}
\end{equation}
such that, when $g \to \infty$, the strong-coupling result for the leading Regge trajectory is correctly reproduced.

We describe now the constraints coming from the small $\phi$ expansion \eqref{eq:small_phi_j0} of $\mathrm{j}_0(\phi)$.
Expanding the ansatz \eqref{eq:frankenstein_j0} up to order $\mathcal{O}(\phi^6)$ we get 
\begin{equation}
    \mathrm{j}^{\text{ansatz}}_0(g,\phi) = -1 + \frac{\Omega(g)}{\Omega^{\text{strong}}} \left(\frac{\phi^2}{\pi} + \frac{5 \phi^4}{8 \pi^3} \right)+ \frac{a_1(g) + a_2(g) + a_3(g)}{1 + a_4(g) + a_5(g) + a_6(g)} + q_1(g) \phi^2 + q_2(g) \phi^4 +  \mathcal{O}(\phi^6)
\end{equation}
with $q_1(g)$ and $q_2(g)$ explicit but lengthy functions of the parameters $a_i(g)$. Matching order by order in $\phi$ the expansion above with the one given in \eqref{eq:small_phi_j0}, we can fix $3$ of the $g$-dependent coefficients $a_i(g)$. In order to fix the $\phi^4$ term we need to know $B_4(g)$ as an exact function of the coupling. As done for $\Omega(g)$, we do a conformal-Pad\'{e} approximation of $\mathbb{C}(g)$ using the weak and strong-coupling expansions \eqref{eq:Curvature}: 
\begin{equation}
    \mathbb{C}(g) = \frac{1}{1-z} \frac{\beta_0 + \beta_1 z + \beta_2 z^2 + \beta_3 z^3}{1 + \beta_4 z + \beta_5 z^2 + \beta_6 z^3}\,, \quad z = \frac{\sqrt{16 g^2 + 1}-1}{\sqrt{16 g^2 + 1}+1}\,,
\end{equation}
and then we use the definition \eqref{eq:bremss} to fix $B_4(g)$.
After imposing these constraints, by expanding $\mathcal{J}(g,\phi)$ at small $g$ and fixed $\phi$, we notice that 
\begin{equation}
    \mathcal{J}(g,\phi) = (\dots ) a_{1,0}  + \mathcal{O}(g^2)\,.
\end{equation}
Since at weak coupling we know that $\mathcal{J}(g,\phi)$ should start with $g^2$, we get that $a_{1,0} = 0$.
Imposing all the constraints, 
we are left with $ 3 \times 7 -1 = 20$ independent numerical parameters that we have to determine.
For this purpose, we use the Python function \verb|scipy.optimize.least_squares|, which finds the best solution for the independent parameters by minimizing the cost function $F$
\begin{equation}
    F = \frac{1}{2} \sum_{i,k = 1}^{N_{\text{points}}} (\mathrm{j}_0^{\text{ansatz}}(g_i,\phi_k) - \mathrm{j}_0^{\text{QSC}}(g_i,\phi_k))^2\,.
\end{equation}
The optimization process stops when the variation of the cost function at each step reaches a given tolerance, parametrized by $\verb|ftol|$: $dF < \verb|ftol| \cdot F$. In our runs, this parameter was fixed to be $\verb|ftol| = 1.5 \cdot 10^{-6}$. 

This minimization method proves to be very efficient, as it correctly reproduces the QSC data and allows to generate $\mathrm{j}_0(\phi)$ for any value of $(g,\phi)$. 
However, we warn that our result for $\mathrm{j}_0^{\text{ansatz}}(g,\phi)$ should not be used to predict any quantity other than the mass of the bound states. The chosen fitting function is not optimal because some parameters are correlated, leading to large error bars on the parameters. Nevertheless, we verified that variations of the parameters within the error bars have a negligible effect on the mass of the bound states, which are the main observable of interest. 

\begin{figure}[h!]
    \centering
    \includegraphics[width=0.9\linewidth]{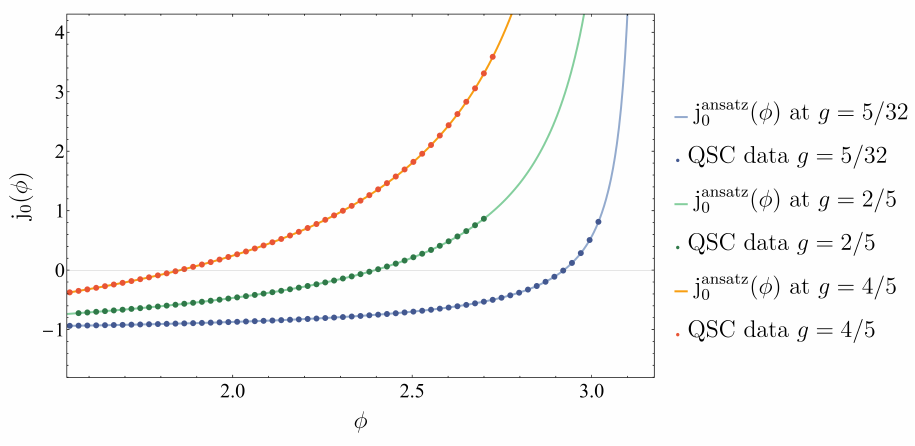}
    \caption{Comparison between $\mathrm{j}_0^{\text{ansatz}}(\phi)$ and QSC data for $g = \{5/32, 2/5, 4/5\}$ as a function of the cusp angle $\phi$.}
    \label{fig:j0_frankenstein_plot}
\end{figure}

The best values for the parameters in the ansatz are given in a Mathematica notebook attached to the submission, while in \figref{fig:j0_frankenstein_plot} we compare the QSC data with the result of the optimization procedure. 

\begin{figure}[h!]
    \centering
    \includegraphics[width=0.8\linewidth]{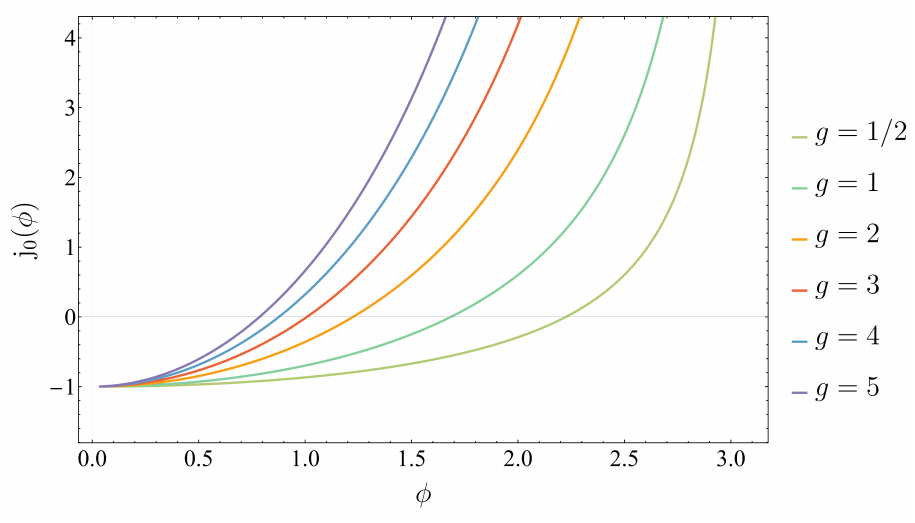}
    \caption{Plot of $\mathrm{j}_0(\phi)$ for higher values of $g$. The results are obtained from the $\mathrm{j}_0^{\text{ansatz}}(g,\phi)$.}
    \label{fig:j0_plot_strong}
\end{figure}

Finally, in \figref{fig:j0_plot_strong} we compare the shape of the leading Regge trajectory for different values of the coupling $g$. Notice that as the coupling increases the position of the first bound state $m_{\text{gap}}^2(g)$, defined by $j_0(m_{\text{gap}}^2(g)) = 0$, moves towards $\phi = 0$, as expected from eq. \eqref{eq:expansion_mgap}. We postpone a more detailed discussion about $m_{\text{gap}}^2(g)$ in section \secref{sec:mgap_section}.

\subsection{Subleading Regge trajectory $j_1(s)$}
\label{sec:subleading_trajectory}
\paragraph{Review of known results}\leavevmode\newline
The subleading Regge trajectory $j_1(s)$ is related to the cusp anomalous dimension $\Gamma_{\text{cusp},\Phi}(g,\phi)$ of the Maldacena-Wilson line with a \textit{parallel} scalar insertion at the cusp \eqref{eq:j1_and_gamma_cusp}. As done for the case of $j_0(s)$ we summarize the known properties of this observable as a function of $g$ and $\phi$ and we denote
\begin{equation}
    \mathrm{j}_1(\phi) \equiv j_1(s(\phi))\,, \quad s = 4m^2 \sin^2(\phi/2)\,.
\end{equation}
Compared to the case of $j_0(s)$, less results are available for $\Gamma_{\text{cusp},\Phi}(g,\phi)$. First of all, we discuss the dependence of $\mathrm{j}_1(\phi)$ on $\phi$ at fixed $g$. Similar to $\mathrm{j}_0(\phi)$, the function $\mathrm{j}_1(\phi)$ is even in $\phi$, as $\Gamma_{\text{cusp},\Phi}(g,\phi)$ is invariant under $\phi \to -\phi$.
In the BPS-limit $\phi \to 0$, the observable $\Gamma_{\text{cusp},\Phi}(g,\phi = 0) \equiv \Delta_{\Phi}(g)$ is equivalent to the anomalous dimension of a \textit{parallel} scalar in the one-dimensional defect CFT \cite{Drukker:2006xg, Giombi:2017cqn}
\begin{equation}
    \mathrm{j}_1(\phi) = -1-\Delta_{\Phi}(g) + \mathcal{O}(\phi^2) \ .
\label{eq:j1_intercept}
\end{equation}
This scaling dimension $\Delta_{\Phi}(g)$ can be computed perturbatively at weak coupling \cite{Alday:2007he,Bruser:2018jnc} or at strong coupling using $AdS_2$/CFT$_1$ \cite{Giombi:2017cqn} or analytic bootstrap \cite{Ferrero:2021bsb}
\begin{equation}
    \Delta_{\Phi}(g) =
\begin{cases}
    1 + 4 g^2 + \mathcal{O}(g^4), & \text{if } \hspace{1mm} g \ll 1 \\
    2 - \frac{5}{4 \pi g} + \mathcal{O}(g^{-2}), & \text{if } \hspace{1mm} g \gg 1
\end{cases}
\label{eq:Delta_Phi_g}
\end{equation}

This observable has also been widely studied using QSC \cite{Grabner:2020nis}. Notice that the value of $\Delta_{\Phi}(g)$ goes from $1$ at weak coupling to $2$ at strong coupling. This will be relevant for the discussion in 
\secref{sec:descendant trajectories}.

The anti-parallel line limit $\phi \to \pi$ is also interesting and it was studied in \cite{Klebanov:2006jj}. There, in the infinite quark mass limit, it was argued that for $g  \leq g_c$, where $g_c$ is a critical coupling, the quark-antiquark flux tube is not stiff enough to support any excited state. When the coupling increases and $g > g_c$, excited states start appearing in the spectrum. This translates into the fact that for $g > g_c$ the cusp anomalous dimension with the parallel insertion, which gives the first excited states, diverges as in \eqref{eq:Gamma_cusp_pi_Phi}, while for $g \leq g_c$, the residue $\Omega_{\Phi}(g) = 0$. We will present evidence supporting this picture below.

Now, we can discuss the $g$ dependence at fixed $\phi$. The perturbative expansion of $\mathrm{j}_1(\phi)$ has been computed up to three loops in \cite{Bruser:2018jnc} from the Regge limit of the amplitude\footnote{See \appref{app:perturbative_result_database}, for the result up to three loops.}
\begin{equation}
    \mathrm{j}^{\text{weak}}_1(\phi) = -2 - 4 g^2 + \mathcal{O}(g^4)\,.
\end{equation}
Notice that the one-loop anomalous dimension does not depend on the angle, as expected in a planar theory. 
On the other hand, the expression for $\Gamma_{\text{cusp},\Phi}(g,\phi)$ at weak coupling has been derived in \cite{Bruser:2018jnc} up to two loops, so the conjectured relation \eqref{eq:j1_and_gamma_cusp} has been checked up to this order. 
At strong coupling, the only known result is at $\phi = 0$, while an expression with general $\phi \neq 0$ has not been studied. 
Nevertheless, from the dual string theory picture, since the insertion of $\Phi$ is light and the leading order result comes from the classical string worldsheet, it follows that the leading strong-coupling result should be the same for $j_0(s)$ and $j_1(s)$, namely $\lim_{g \to \infty} j_1(s)/j_0(s) = 1$.

\begin{figure}[h!]
    \centering
    \includegraphics[width=\linewidth]{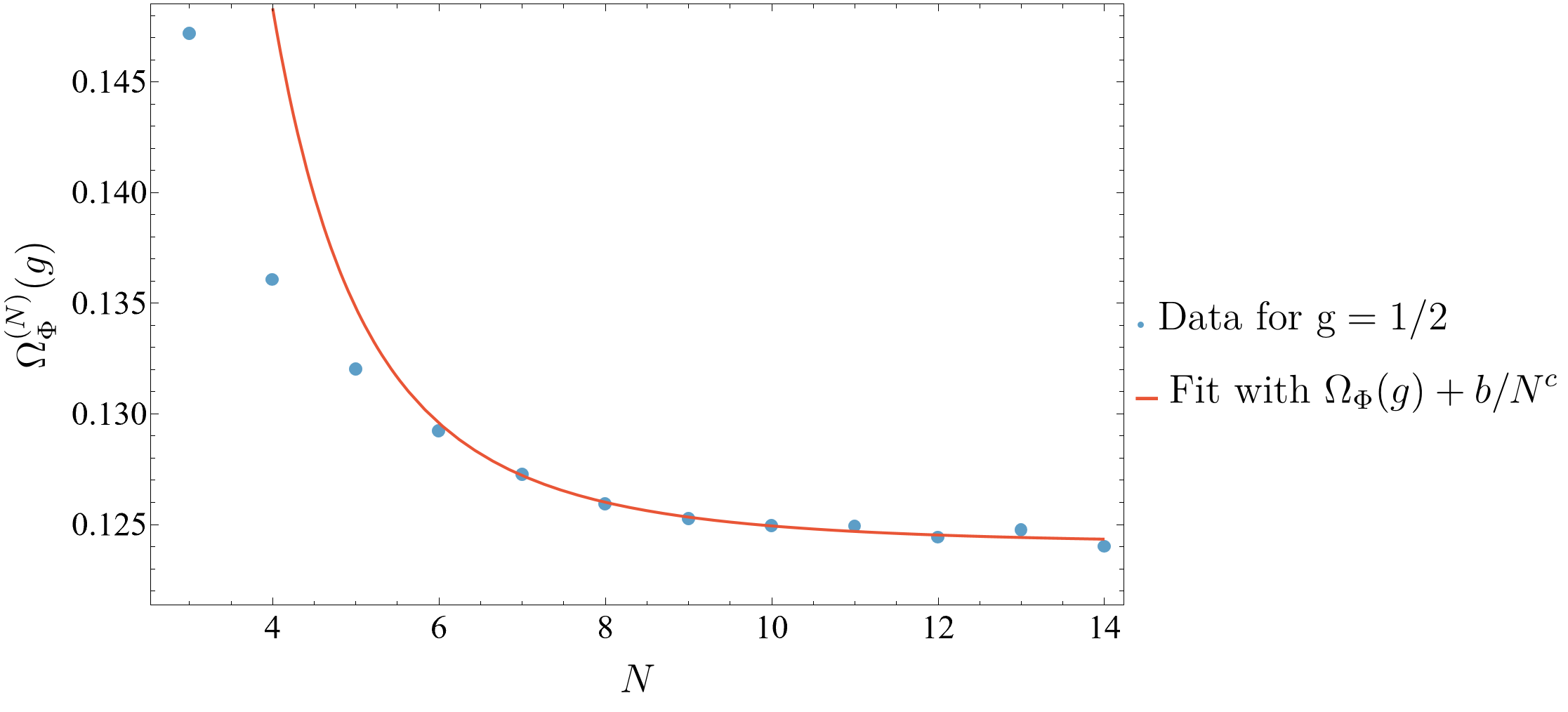}
    \caption{Values of $\Omega_{\Phi}^{(N)}(g)$ for $g = 1/2$ as a function of the order $N$ of the fit \eqref{eq:fit_j1}. In order to extrapolate the value of $\Omega_{\Phi}(g)$ at $N \to \infty$ we perform a fit with the function $\Omega_{\Phi}(g) + \frac{b}{N^c}$, with $\Omega_{\Phi}(g), b, c$ fitting parameters. The results of $\Omega_{\Phi}(g)$ for different values of $g$ are reported in \tabref{tab:tab_residue_j1}. For the specific fit in the plot we have used all the points with $N \geq 5$.}
    \label{fig:extrapolate_residue}
\end{figure}

\paragraph{Derivation of $\Omega_{\Phi}(g)$ from QSC data}\leavevmode\newline
\label{sec:residue_j1_QSC}In this section we analyze the QSC data obtained for $\mathrm{j}_1(\phi)$ in order to compute the value of $\Omega_{\Phi}(g)$ for different couplings.
We construct the following ansatz:
\begin{equation}
    \mathrm{j}^{\text{fit}}_1(\phi) = -1-\Delta_{\Phi}(g) + 2 \pi \Omega_{\Phi}(g) \left( \frac{1}{(\pi-\phi)(\pi+\phi)}-\frac{1}{\pi^2} \sum_{k = 0}^{N} \left( \frac{\phi}{\pi} \right)^{2k} \right) +  \sum_{k = 1}^N d_k \phi^{2k}\,.
\label{eq:fit_j1}
\end{equation}
The intercept $\Delta_{\Phi}(g)$ is the only quantity that we fix in the fitting function \eqref{eq:fit_j1}. To derive it as a function of $g$ we combine weak and strong coupling expansions in a conformal-Pad\'{e} approximation with ansatz
\begin{equation}
    \Delta_{\Phi}(g) = \frac{\delta_0 + \delta_1 z + \delta_2 z^2 + \delta_3 z^3}{1 + \delta_4 z + \delta_5 z^2 + \delta_6 z^3}\,, \quad z = \frac{\sqrt{16 g^2 + 1}-1}{\sqrt{16 g^2 + 1}+1}
\end{equation}
The unfixed coefficients are fitted using QSC data of \cite{Grabner:2020nis}.\\
All the other parameters in \eqref{eq:fit_j1} are left free. The order $N$ of the fitting function \eqref{eq:fit_j1} is also a parameter: by using increasing values of $N$ at fixed $g$ we obtain different results for $\Omega_{\Phi}^{(N)}(g)$. %
By plotting these values as a function of the order $N$, we observe that they converge to a value that we extrapolate, as shown in \figref
{fig:extrapolate_residue} for $g = 1/2$. The extrapolated value at $N \to \infty$ is our best determination of $\Omega_{\Phi}(g)$.
\\
To perform the extrapolation at $N \to \infty$,\footnote{For increasing values of $g$ we observe that for $N > 10$ the result of $\Omega_{\Phi}^{(N)}(g)$ from \eqref{eq:fit_j1} becomes unreliable. The reason is that at strong coupling it becomes more and more difficult to compute QSC data close to $\phi = \pi$, which is the relevant region for the determination of the residue  $\Omega_{\Phi}(g)$. However, by considering values of $N = 1, \dots, 10$, we still see a plateau in the value of $\Omega_{\Phi}^{(N)}(g)$ and we use the same procedure to extrapolate the result at $N \to \infty$.} we fit the values of $\Omega_{\Phi}^{(N)}(g)$ as functions of $N$ with 
\begin{equation}
    \Omega^{(N)}_{\Phi}(g) = \Omega_{\Phi}(g) + \frac{b}{N^c}\,,
\label{eq:fit_residue_N}    
\end{equation}
where $\Omega_{\Phi}(g), b$ and $c$ are the fitting parameters. The results for $\Omega_{\Phi}(g)$ obtained using this procedure are shown in \tabref{tab:tab_residue_j1}. We observe that, by using different ansatz for \eqref{eq:fit_residue_N}, the results of $\Omega_{\Phi}(g)$ differ from the ones reported in \tabref{tab:tab_residue_j1} by $1 \% - 4 \%$ depending on the value of the coupling, which serves us as a rough error estimate for our procedure.

\begin{table}[h!]
\centering
\resizebox{1\textwidth}{!}{%
\begin{tabular}{|c|c|c|c|c|c|c|c|c|c|c|c|}
\hline
$g$ & $5/32$ & $1/4$ & $5/16$ & $2/5$ & $1/2$ & $3/5$ & $7/10$ & $4/5$ & $1$ & $6/5$ & $7/5$ \\
\hline
$\Omega_{\Phi}(g)$ & 
$\simeq 0$ & 
$\simeq 0$ & 
$\simeq 5 \cdot 10^{-3}$ & 
$4.2 \cdot 10^{-2}$ & 
$0.12$ &
$0.25$ & 
$0.41$ & 
$0.59$ & 
$1.01$ & 
$1.46$ & 
$1.97$ \\
\hline
\end{tabular}}
\captionof{table}{Values of the residue $\Omega_{\Phi}(g)$ of $\mathrm{j}_1(\phi)$ in $\phi = \pi$ from the extrapolation to $N\to \infty$ of the fitting function \eqref{eq:fit_j1}. The error estimate is $1\%-4\%$.}\label{tab:tab_residue_j1}
\label{tab:omega_phi}
\end{table}

\begin{figure}[h!]
    \centering
    \includegraphics[width=0.9\linewidth]{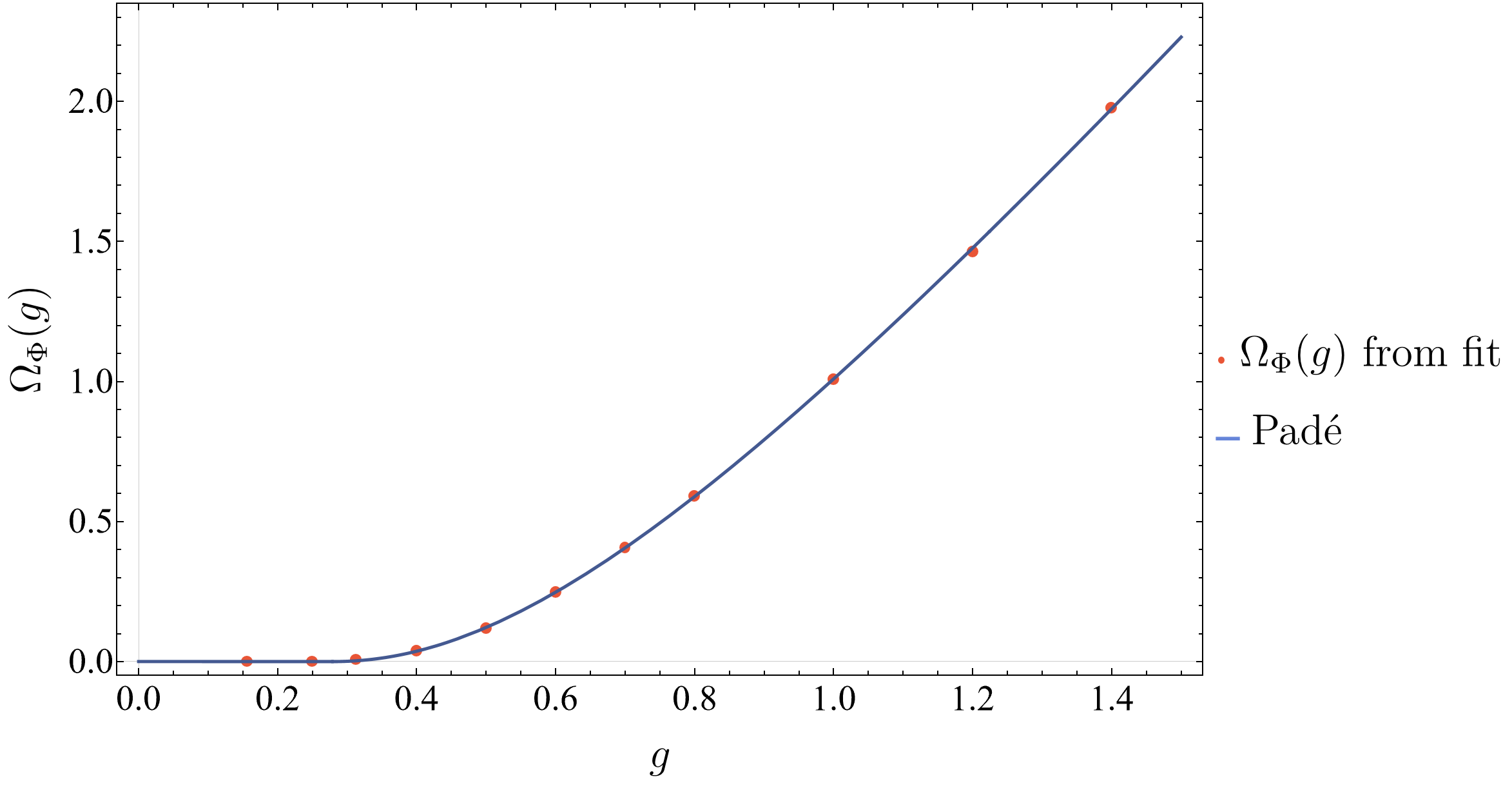}
    \caption{Plot of the residue of $\mathrm{j}_1(\phi) \simeq {\Omega_{\Phi}(g) \over \pi - \phi}$ for different values of $g$. For $g < g_{c} \approx 0.28$ the residue is effectively zero, while it increases for $g \geq g_{c}$. The orange dots were obtained using extrapolation of the QSC results at $\phi = \pi$. }
    \label{fig:residue_j1}
\end{figure}

After having determined the values of the residue from the extrapolation in $N$, we make a conformal-Pad\'{e} approximation ansatz for $\Omega_{\Phi}(g)$, by assuming that $\Omega^{\text{strong}} = \Omega_{\Phi}^{\text{strong}}$:
\begin{equation}
    \Omega_{\Phi}(g) = \theta(g-g_c) \frac{1}{1-z} \frac{\rho_0 + \rho_1 z + \rho_2 z^2 + \rho_3 z^3}{1 + \rho_4 z + \rho_5 z^2 + \rho_6 z^3}\,, \quad z = \frac{\sqrt{16 g^2 + 1}-1}{\sqrt{16 g^2 + 1}+1}\,,
\end{equation}
and we minimize the parameters, including the critical coupling $g_c$, on the values found above. We plot the result with the data of \tabref{tab:tab_residue_j1} in \figref{fig:residue_j1}. The value of the critical coupling that we find from the minimization is $g_c \approx 0.28$, which is close to the ladder-approximation value of $1/4$, see \cite{Klebanov:2006jj}.

\paragraph{Global fit of the QSC data for $j_1(s)$}\leavevmode\newline
\label{sec:qsc_analysis_j1}In this section, as done for $j_0(s)$, we describe our procedure to perform a global fit of $j_1(s)$ using the QSC data and the known results described in the previous section. 
We consider the following ansatz for the subleading trajectory
\begin{equation}
\begin{aligned}
\mathrm{j}^{\text{ansatz}}_1(g,\phi) = &-1-\Delta_{\Phi}(g) + \frac{\Omega_{\Phi}(g)}{\Omega_{\Phi}^{\text{strong}}} J_1(\phi) + \frac{a_0(g) + a_1(g)(1-x) + a_2(g) (1-x)^2 + a_3(g) (1-x)^3}{1+a_4(g)(1-x) + a_5(g) (1-x)^2 + a_6(g) (1-x)^3}\,,
\label{eq:frankenstein_j1}
\end{aligned}
\end{equation}
with $x \equiv \left( \phi/\pi\right)^2$ and all the coefficients are functions of $g$ as follows 
\begin{equation}
    a_i(g) \equiv \frac{a_{i,0} + a_{i,1} z + a_{i,2} z^2 + a_{i,3} z^2}{1 + a_{i,4} z + a_{i,5} z^2 + a_{i,6} z^3}\,, \quad z = \frac{\sqrt{16 g^2 + 1}-1}{\sqrt{16 g^2 + 1}+1} .
\end{equation}

Let us now enumerate all the known ingredients of \eqref{eq:frankenstein_j1}. The intercept $\Delta_{\Phi}(g)$ and the quark-antiquark potential $\Omega_{\Phi}(g)$ have been derived for any $g$ using Pad\'{e} in the previous section. The function $J_1(\phi)$ is precisely the same as $J(\phi)$ in \eqref{eq:J_strong}. We denote the remaining term in the ansatz \eqref{eq:frankenstein_j1} as $\mathcal{J}_1(g,\phi)$ and we discuss how to fix some of its parameters. In the $\phi \to 0$ limit, up to $\mathcal{O}(\phi^6)$
\begin{equation}
    \begin{aligned}
        \mathrm{j}_1^{\text{ansatz}}(g,\phi) = &-1-\Delta_{\Phi}(g) + \frac{\Omega_{\Phi}(g)}{\Omega^{\text{strong}}} \left(\frac{\phi^2}{\pi} + \frac{5 \phi^4}{8 \pi^3} \right)+ \frac{a_0(g) + a_1(g) + a_2(g) + a_3(g)}{1 + a_4(g) + a_5(g) + a_6(g)} + \\ &+ q_1(g) \phi^2 + q_2(g) \phi^4 +  \mathcal{O}(\phi^6)\,,
    \end{aligned}
\end{equation}
with $q_1(g)$ and $q_2(g)$ some lengthy functions of the parameters $\{ a_0(g), \dots ,a_6(g) \}$. Matching the order $\phi^0$ with \eqref{eq:j1_intercept}, we can fix $a_0(g)$, while at orders $\phi^2$ and $\phi^4$, since the coefficients are not known as an exact function of $g$, we can just use the weak and strong-coupling expansion of $\mathrm{j}_1(\phi)$. This fixes $8$ parameters. In total we are left with $34$ independent parameters, that can be fixed using the function \verb|least_squares| in Python, with $\verb|ftol| = 1.5 \cdot 10^{-3}$.

\begin{figure}[h!]
    \centering
    \includegraphics[width=0.95\linewidth]{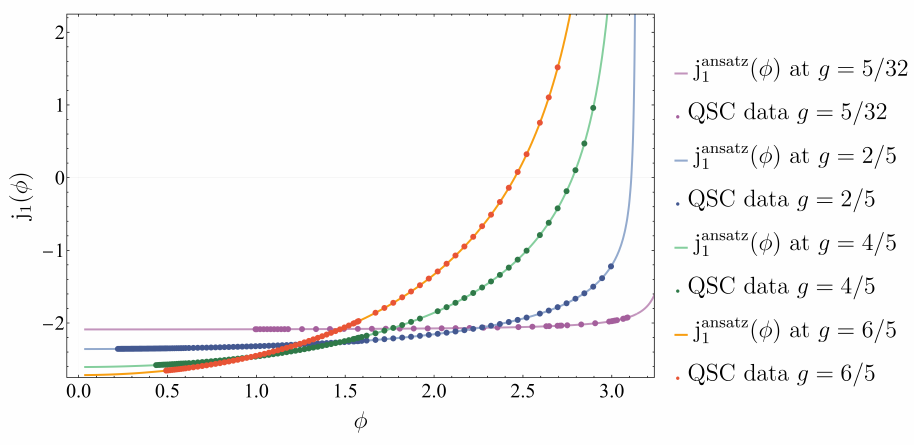}
    \caption{Comparison between $\mathrm{j}_1^{\text{ansatz}}(\phi)$ and QSC data for $g = \{5/32, 2/5, 4/5, 6/5\}$ as a function of the cusp angle $\phi$. As computed above, the $g = 5/32$ trajectory does not diverge at $\phi = \pi$.}
    \label{fig:j1_frankenstein_plot}
\end{figure}

The best values for the parameters in the ansatz are given in the attached Mathematica notebook, while in \figref{fig:j1_frankenstein_plot} we compare the QSC data with the result of the optimization procedure. We find that the ansatz $\mathrm{j}_1^{\text{ansatz}}(\phi)$ works well compared to the QSC data.
Also in this case, we warn that the result of $\mathrm{j}_1^{\text{ansatz}}(g,\phi)$ is reliable just for computing the masses of the bound states and should not be trusted to make other predictions.

\noindent In the next sections we analyze the Regge/cusp correspondence at subleading orders.

\subsection{Beyond the first two trajectories}
\label{sec:descendant trajectories}

So far, we have only discussed the behavior of the leading Regge trajectory $j_0(s)$ and the first subleading Regge trajectory $j_1(s)$ at finite coupling. In this section, we would like to explore the correspondence between Regge trajectories and Wilson lines with insertions of local operators at the cusp beyond the first two trajectories. 

\begin{figure}[h!]
    \centering
    \includegraphics[width=\textwidth]{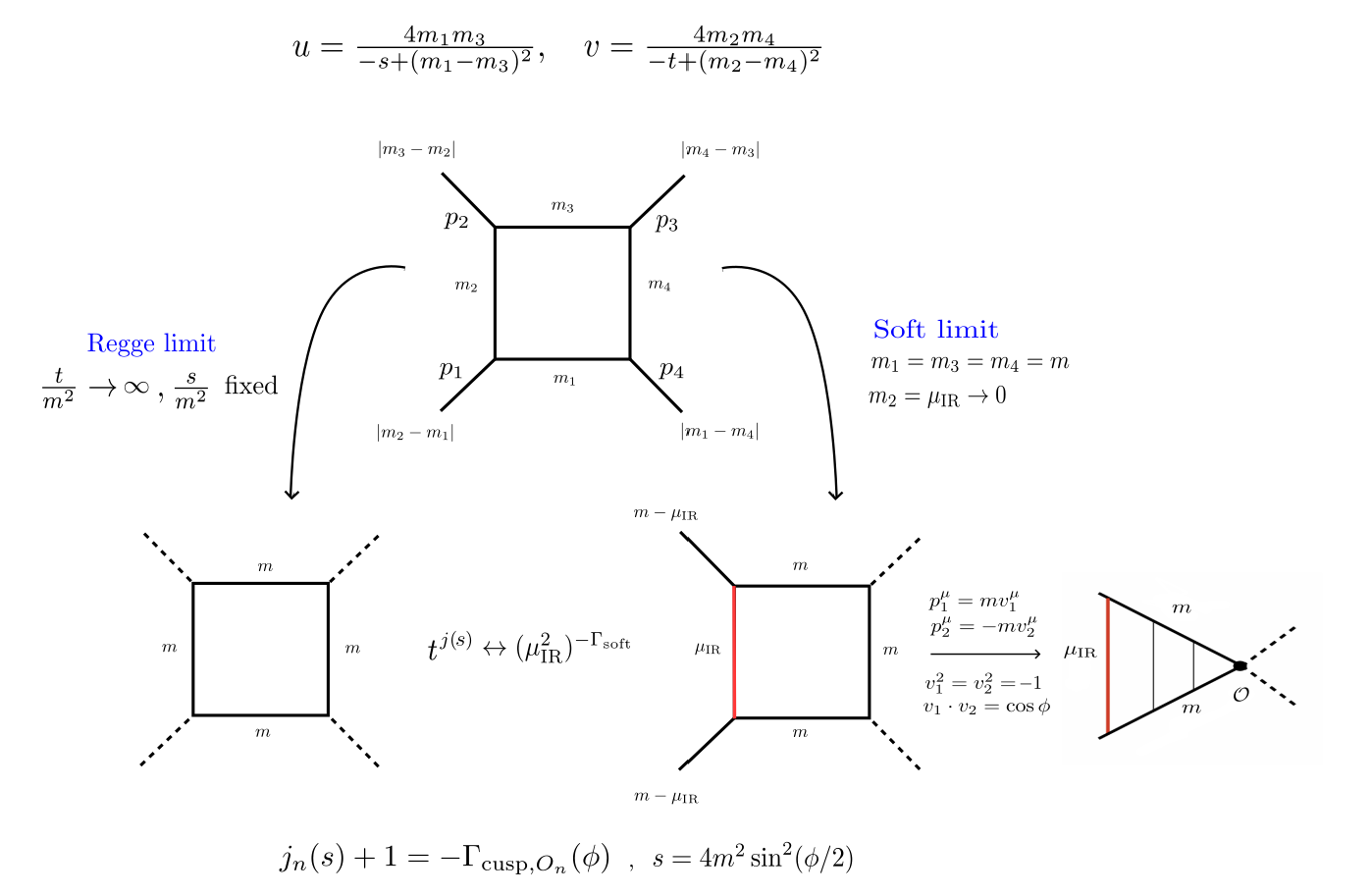}
    \caption{Summary of the argument for the Regge/cusp correspondence. Thanks to dual conformal invariance, the Regge limit and the soft limit are equivalent. By matching the behavior of the amplitude in the two limits it is possible to get the equivalence \eqref{eq:reggecusp}.}
    \label{fig:Cusp_and_regge_correspndence}
\end{figure}

First, we review the basic argument that relates the Regge limit to the soft limit. The argument can be summarized in \figref{fig:Cusp_and_regge_correspndence}. Thanks to dual conformal invariance, the limit $v \to 0$ and $u$ fixed corresponds both to the Regge limit ($t/m^2 \to \infty$, $s/m^2$ fixed) and to the soft limit (one of the masses $m_2 = \mu_{\text{IR}} \to 0$). These two limits are then equivalent, and it is possible to equate the behaviors of the amplitude. On one hand, the Regge limit is controlled by the Regge trajectories $j_n(s)$ and the amplitude goes as $\sim t^{j_0(s) +1} + \dots$ where the $\dots$ denote the contribution of subleading trajectories. In the soft limit, on the other hand, the amplitude is IR divergent $\sim (\mu_{\text{IR}}^2)^{-\Gamma_{\text{soft}}}$ and the divergence is controlled by a soft anomalous dimension. The IR divergences of amplitudes are connected to the UV divergences of the Wilson line correlators \cite{Becher:2009cu,Gardi:2009qi,Catani:1998bh,Korchemskaya:1992je}. For a two-particle heavy quark form factor that arises in our case, this connection was established in \cite{Korchemsky:1992xv}, where it was shown that the corresponding soft anomalous dimension is given by the cusp anomalous dimension of a Wilson line \cite{Polyakov:1980ca,Korchemsky:1987wg,Maldacena:1998im}.

The leading and the first subleading Regge trajectories are captured by an insertion of parity-even $SO(2)\times SO(5)$ singlet operators at the cusp.\footnote{We discuss discrete symmetries of the Maldacena-Wilson cusp in \appref{app:parity}.} As we explore the subleading effects in $m^2/t$, we expect to observe broadly two types of physical effects. First, we expect that the Wilson line picture acquires soft or `recoil' corrections due to the finite mass of  W-bosons, as described, for example, in the heavy-quark effective theory (HQET) \cite{Eichten:1989zv,Georgi:1990um,Falk:1990pz}. Second, we can have other operators inserted at the cusp. In the language of Regge theory, we expect that the first type of corrections to be important for the corrections to the three-point functions.
As for the new operators insertion at the cusp, they are capturing excited energy eigenstates of the $H_{\text{cusp}}$ Hamiltonian. As we discussed in Section \ref{sec:sublRegge}, it is natural to expect that the correspondence between the eigenstates of the cusp Hamiltonian and the Regge trajectories continues beyond the first two trajectories. Below, we check it explicitly at weak and strong coupling.

\subsubsection{Checks of the Regge/cusp correspondence}
To analyze the Regge/cusp correspondence \eqref{eq:reggecusp}, we follow \cite{Bruser:2018jnc} and introduce, as in \eqref{eq:Regge_limit}, the variables $Y = e^{i \tilde{\theta}_s}$ and $\xi = -i \tan(\phi/2)$ and express the Mandelstam variables $s$ and $t$ as functions of $(Y,\xi)$ as follows
\begin{equation}
    t = -\frac{\xi^2 m^2 (Y-1)^2}{Y}\,, \qquad s = -\frac{4 \xi^2 m^2}{1-\xi^2}\,,
\end{equation}
Also, for later convenience, we introduce the following notation for the Maldacena-Wilson line \cite{Bruser:2018jnc}
\begin{equation}
    W[C] = \text{Tr} \, \mathcal{P} \exp \left( i \, g_{\text{YM}} \int_C d\tau \hspace{1mm} A^{\mu}(x + v \,\tau ) \dot{x}_{\mu}(\tau)  + g_{\text{YM}} \int_C d\tau \hspace{1mm} \Phi_{||}(x +v \, \tau) |\dot{x}(\tau)| \right)\,.
\end{equation}
For a Maldacena-Wilson line with a cusp the contour $C$ is made of two straight lines $C_1$ and $C_2$ meeting at the cusp $x$ and forming a cusp angle $\phi$
\begin{equation}
    C_1 = \{ \tau v_1^{\mu}| \tau \in [-\infty,0] \}, \quad C_2 = \{ \tau v_2^{\mu}| \tau \in [0,\infty] \}, \quad v_1 \cdot v_2 = \cos(\phi)\,,
\end{equation}
with $v_1^2 = v_2^2 = 1$. Notice that there is a symmetry $v_1 \leftrightarrow -v_2$.

\paragraph{Weak coupling}\mbox{}\\
The Regge expansion of the amplitude $M(s,t)$ \eqref{eq:Regge_limit}, related to $f(s,t)$ through \eqref{eq:f_and_M_rel}, can be organized as
\begin{equation}
  \lim_{Y \to 0} \frac{1+Y}{1-Y} M(Y,\xi) = \sum_n r_n(s) Y^{-j_n(s)-1}\,,
\label{eq:regge_expansion}
\end{equation}
where $j_n(s)$ are the Regge trajectories that we would like to analyze and $r_n(s)$ are the corresponding residues (see below equation (B.11) in \cite{Bruser:2018jnc} for the precise definition). The RHS can be computed using the Regge expansion of the amplitude, which is known up to three loops \cite{Bruser:2018jnc}. The LHS is expanded at small $g$ after the assumption that a certain number of Regge trajectories contributes at each order in $Y$. By matching the two sides it's possible to derive loop by loop the expression of the trajectories.

The matching with the leading Regge trajectory has been analyzed in \cite{Bruser:2018jnc}, where they could check the relationship with $\Ga_{\text{cusp}}(g,\phi)$ up to three loops. Similarly, the first subleading trajectory was computed up to three loops (see the precise expression in \appref{app:perturbative_result_database}), finding perfect agreement with $\Ga_{\text{cusp},\Phi}(g,\phi)$.

Let us now study the sub-subleading order. The Regge limit of the amplitude at this order reads \cite{Bruser:2018jnc} (here we just show the leading $\log(Y)$ contributions at each loop order)
\begin{equation}
\begin{aligned}
     \lim_{Y \rightarrow 0}  \frac{1+Y}{1-Y} M(Y,\xi)\Big{|}_{Y^2} &=  Y^2 \Big( 2 + 8 g^2 \frac{\xi^2-1}{\xi^2} \log(Y) -8 g^4 \frac{(2\xi^4-6\xi^2+5)}{\xi^2(1-\xi^2)} \log^2(Y) + \\ &-64 g^6 \frac{\xi^4-4\xi^2+5}{3 \xi^2(1-\xi^2)}  \log^3(Y) \Big) + \mathcal{O}(g^8) \ .
\label{eq:Regge_limit_Y2}
\end{aligned}
\end{equation}
We consider different scenarios for the number of trajectories that contribute at this order. 
\begin{itemize}
    \item one trajectory is inconsistent with the equations above, 
    \item two trajectories 
    \begin{equation}
    \lim_{Y \to 0} \frac{1+Y}{1-Y} M(Y,\xi)\Big{|}_{Y^2} = r_{2,1}(s) Y^{-j_{2,1}(s)-1} + r_{2,2}(s) Y^{-j_{2,2}(s)-1} \, ,
    \end{equation}
    give the following solution 
    \begin{equation}
    \begin{aligned}
        j_{2,i}(s) = -3 + 2 g^2 \left( -1 + \frac{3}{\xi^2-2} \pm \sqrt{1+ \frac{2}{\xi^2-1} + \frac{9}{(\xi^2-2)^2}}  \right)\,, \quad i = 1,2\,.
    \end{aligned}
    \label{eq:one_loop_j2_s}
    \end{equation}
    This expression for the $j_{2,i}(s)$ has singularities at locations $s \neq 4m^2$, which would lead to extra singularities at unphysical locations in the amplitude. For this reason we discard this option.
    \item to explore the solutions with higher number of Regge trajectories we apply the following strategy: if we naively continue the correspondence between the Regge trajectories and the Wilson line with local operator insertions at the cusp, we find two new local operators at this level $O_{\pm} \equiv \sum_{i=1}^5 \Phi_{\perp}^i \Phi_{\perp}^i \pm \sqrt{5} (\Phi_{||})^2$. 
    In addition, there are four potential spinning trajectories related to the insertion of the following operators at the cusp \cite{Cooke:2017qgm,Agmon:2020pde}: $F_{\mu \nu} v_1^{\mu} v_2^{\nu}$, $(v_1 \pm v_2)^{\mu} D_{\mu} \Phi_{||}$, $F^{\mu \nu} v_1^{\rho} v_2^{\sigma} \epsilon_{\mu \nu \rho \sigma}$. However, as we show in \appref{app:parity}, only two out of four are parity-even. Therefore, we expect the solution to contain four trajectories and only two out of four to contribute in the forward limit.
    Making this assumption, we focus on the leading part of \eqref{eq:Regge_limit_Y2} in the forward $\phi \to 0$ limit and we solve it for two scalar trajectories to find
    \begin{equation}
    j_{2,\pm}(0) = -\Delta_{O_{\pm}} -1, \qquad \Delta_{O_{\pm}} = 2 + g^2 (5 \pm \sqrt{5}) + \mathcal{O}(g^4)\, ,
\label{eq:regge_and_cusp_2}
\end{equation}
    in agreement with the Regge/cusp correspondence applied to the sub-subleading Regge trajectories. The RHS of \eqref{eq:regge_and_cusp_2} in fact does not depend on $\phi$ at one loop, as expected in a planar theory. Assuming that the correspondence continues to hold for $j_{2,i}(s)$ away from $s = 0$, we find that it is easy to arrange for a pair of spinning trajectories, such that the four-trajectory (or higher number of trajectories) solution exists and the amplitude exhibits right analyticity properties for any $s$.\footnote{For three Regge trajectories in this case we have the same analyticity problem as we observed for the two-trajectory solution.} 
    In fact, let us quickly show that as long as the Regge trajectories have the correct analyticity properties the equations that we are solving do not generate extra singularities. Consider an example with four Regge trajectories. The set of equations that we are solving is given by
    \be
    \label{eq:eqorig}
    \sum_{i=1}^4 r_i (j_i)^k = f_k , ~~~ k=0,1,2,3 ,
    \ee
    where we suppressed the dependence on $s$ and we assume that $j_i$ and $f_k$ only have physical non-analyticities. These can be trivially solved for $r_i$. The residues $r_k$ exhibit singularities at unphysical locations $j_a = j_b$. We would like to check that these cancel in the amplitude, which amounts to showing that
    \be
    \sum_{i=1}^4 r_i (j_i)^k , ~~~  k \geq 0 ,
    \ee
    is regular at these locations. For $k=0,1,2,3$ the cancellation of singularities is guaranteed due to \eqref{eq:eqorig}. The way it happens is that given $a$ and $b$ there is a pair of residues that exhibit a pole $r_{a,b}\sim \pm {1 \over j_a-j_b}$ with residues equal up to a sign. It is clear then that as $j_a \to j_b$, the combination $r_a j_a^k  + r_b j_b^k$ is regular for any $k$.\\
    It would be very interesting to test the four-trajectory (or higher  number of trajectories) picture further by calculating the Coulomb branch amplitudes at higher loops and comparing it to the explicit calculations of  anomalous dimensions of the spinning operators.
    
\end{itemize}

Let us next push the analysis above to the next order in the small $g$ expansion. On the integrability side, we have the following result
\cite{Cavaglia:2022qpg}
\begin{equation}
    \Gamma_{\text{cusp},O_{\pm}}(0) = 2 + g^2 (5 \pm \sqrt{5}) + g^4 (-18 \pm 8 \sqrt{5}) + \mathcal{O}(g^6)\,.
\label{eq:Gamma_cusp2_two_loop}
\end{equation}
The two-loop values of $j_{2,i}(s)$ appear in the subleading $Y^2 \log(Y)$ terms in the small $g$ expansion of the amplitude \eqref{eq:regge_expansion}. In particular, they contribute at orders $g^2 Y^2$, $g^4 Y^2 \log(Y)$ and $g^6 Y^2 \log^2(Y)$ terms which in the small $\xi$ expansion read
\begin{equation}
    Y^2 \left( g^2 \frac{4(-1+4 \log(\xi))}{\xi^2} + g^4 \log(Y) \frac{80 (1+2 \log(\xi))}{\xi^2} + g^6 \log^2(Y) \frac{8 (73 + 80 \log\xi))}{\xi^2}  \right) + \mathcal{O}(1/\xi)\,.
\label{eq:subleading_Y2}
\end{equation}
As discussed above, in the $\xi \to 0$ limit, only two Regge trajectories contribute. Their expressions and their residues in the $\xi \to 0$ limit are
\begin{equation}
    \begin{aligned}
        &j_{2,1}(s) = -3 + g^2 (-5 -\sqrt{5}) + \mathcal{O}(g^4)\,,\qquad r_{2,1}(0) = -\frac{4}{\sqrt{5} \xi^2} + \mathcal{O}(\xi^0) + \mathcal{O}(g^2)\,,  \\
        &j_{2,2}(s) = -3 + g^2 (-5 +\sqrt{5}) + \mathcal{O}(g^4)\,,\qquad r_{2,2}(0) = +\frac{4}{\sqrt{5} \xi^2}+ \mathcal{O}(\xi^0) + \mathcal{O}(g^2)\,.
    \end{aligned}
\end{equation}
The unknowns appearing in the LHS of \eqref{eq:regge_expansion}, which has to be matched with \eqref{eq:subleading_Y2}, are $r_{2,i}^{\text{1-loop}}$ and $j_{2,i}^{\text{2-loop}}$ and we can solve for three of them as functions of $r_{2,2}^{\text{2-loop}}$. In particular we get 
\begin{equation}
    r_{2,1}^{\text{1-loop}}(0) = -r_{2,2}^{\text{1-loop}}(0) + \frac{4(-1+4 \log(\xi))}{\xi^2}\,.
\end{equation}
By assuming the following small $\xi$ structure 
\begin{equation}\label{eq:r221loopStructure}
    r_{2,2}^{\text{1-loop}}(0) = \frac{b}{\xi^2} + \frac{c \log(\xi)}{\xi^2} + \mathcal{O}(\xi^0)\,,
\end{equation}
suggested by the equation above, we are left with two unknown parameters $b,c \in \mathbb{R}$ to compute. One more equation can be obtained by considering that the $Y \to 0$ limit in \eqref{eq:regge_expansion} has to be considered as a limit from above $Y \to 0^+$. Comparing it with the limit from below $Y \to 0^-$ gives the constraint
\begin{equation}
    r^*_n(s) Y^{-j_n(s)-1} = r_n(s) e^{-2\pi i (-j_n(s)-1)} Y^{-j_n(s)-1}\,,
\end{equation}
that translates to
\begin{equation}
    r^*_n(s) = r_n(s) e^{-2\pi i j_n(s)}\,, \qquad \forall \hspace{2mm} n \hspace{2mm} \text{separately} \ .
\label{eq:residues_and_Regge_traj}
\end{equation}
Expanding the above equation perturbatively, it is easy to find some relations between the Regge trajectories and the residues. At leading order in $g$ for instance $(r_n^{(\text{tree-level})}(s))^* = r_n^{(\text{tree-level})}(s)$ so $r_n^{(\text{tree-level})}(s) \in \mathbb{R}$. At the next order
\begin{equation}
    \text{Im}(r_n^{(\text{1-loop})}(s)) = - \pi \hspace{1mm} r_n^{(\text{tree-level})}(s) \hspace{1mm} j_n^{(\text{1-loop})}(s)\,,
\label{eq:constraint_r1}
\end{equation}
which can be applied to compute $c$ in \eqref{eq:r221loopStructure} (since $\xi = -i \tan(\phi/2))$:
\begin{equation}
    \text{Im}(r_{2,2}^{(\text{1-loop})}) = -\frac{\pi c}{2 \xi^2} = -\pi \left( \frac{4}{\sqrt{5} \xi^2} \right) \left( -5 + \sqrt{5} \right)\, \hspace{3mm} \Rightarrow \hspace{3mm} c = 8(1-\sqrt{5})\,.
\end{equation}
Using the ansatz for $r_{2,2}^{\text{2-loop}}$ with the value of $c$ above we get the following result for the two-loop Regge trajectories
\begin{equation}
    \begin{aligned}
        &j_{2,1}^{(\text{2-loop})}(0) = \frac{1}{4} \left(82 + 50 \sqrt{5} + 5 b \right)+\mathcal{O}\left(\xi^1\right)\,, \\
        &j_{2,2}^{(\text{2-loop})}(0) = \frac{1}{4} \left(62 - 50 \sqrt{5}-5b\right)+\mathcal{O}\left(\xi ^1\right)\,.
    \end{aligned}
\end{equation}
We don't have enough data to compute $b$ independently, but we can check that, by imposing 
\begin{equation}
    j_{2,1}^{(\text{2-loop})}(0) \equiv 18 - 8 \sqrt{5} \quad \Rightarrow \quad b = -2-\frac{82}{\sqrt{5}}\,,
\end{equation}
then automatically $j_{2,2}^{(\text{2-loop})}(0) = 18 + 8 \sqrt{5}$ as expected from \eqref{eq:Gamma_cusp2_two_loop}. We can conclude that the available perturbative data for the amplitude suggest that the relationship in \eqref{eq:regge_and_cusp_2} holds also at two loops.

As a final comment, to further check the Regge trajectories/cusp anomalous dimension correspondence at subleading orders (starting from $j_3$ and so on) and weak coupling, higher loop corrections of the amplitude are needed. Eventually, one could try to match results for the trajectories with the available perturbative data at $\phi = 0$ (see Appendix D of \cite{Cavaglia:2022qpg}). \\
A trivial consistency check, though, is to note that at weak coupling, the Regge trajectories are equidistant and we get
\be
j_n(0)= -(n+1) + \cO(g^2)\,,
\ee
which matches the expected free spectrum of local operators, schematically $\Phi^n$, on the Maldacena-Wilson line. These checks of the conjecture at weak coupling are reinforced by the strong-coupling arguments, presented in the next section.

Let us briefly comment on the structure of the forward limit at finite coupling, which also clarifies the appearance of the ${1 \over \xi^2}$ singularity in the calculation above. A given Regge trajectory contributes to the amplitude in the forward limit as follows
\be
M(s/m^2,t/m^2) \sim r\left(s\right) (s/t)^{-j-1} .
\ee
For $f(s,t) \sim M(s,t)/(st)$ to stay finite as $s \to 0$ requires that the residue function $r(s) \sim ({m^2 \over s})^{-j-2}$. This is precisely the ${1 \over \xi^2}$ singularity observed in the $Y^2$ analysis above. This behavior also implies that if we go further in the Regge expansion, the residue functions $r_i(s)$ will become more singular. We indeed tested that at the order $Y^3$, the leading singularity in the residue function is $1/\xi^4$.

\paragraph{Strong coupling}\mbox{}\\
At strong coupling, two different checks can be done. 
The more straightforward one involves the comparison between the spectrum of the super-Veneziano amplitude \eqref{eq:super_Veneziano} and the strong-coupling results for the scaling dimensions of the operators mentioned above. Expanding the strong-coupling Regge trajectories at small $s$ we get
\be
j_n^{\text{strong}}(s)=-(2n+1)+ {g \over \pi} {s \over m^2} + \mathcal{O}(s^2) \ ,~~~ g \gg 1 . 
\ee
This, if the conjecture is true, translates to the fact that the scaling dimensions of the $O_n$ operators \eqref{eq:jn_forward} are even $\Delta_{O_n, g = \infty} = 2n$ at strong coupling.
This is indeed correctly reproduced by the spectrum of the \emph{parity-even} operators $\mathcal{L}^{\Delta}_{0,[0,0]}$. Recall that to the states on the 1d defect CFT a parity charge can be assigned $\mathbb{Z}_2^-$ \cite{Cavaglia:2023mmu}, see \appref{app:parity}. In the case of states belonging to $\mathcal{L}^{\Delta}_{0,[0,0]}$ it is given by
\begin{equation}
    \mathbb{Z}_2^- = (-1)^{\Delta_{g = \infty}} \ .
\label{eq:parity_strong_coupling}
\end{equation}
In our conjecture, we therefore see that only states that have parity $+$ under $\mathbb{Z}_2^-$ appear in the flat space limit of the amplitude.
It means that, at strong coupling, the parity-even states in $\mathcal{L}^{\Delta}_{0,[0,0]}$ are also the ones with even $\Delta_{g = \infty}$ and this is consistent with the values of the Regge trajectories of the super-Veneziano amplitude at $g \to \infty$. 

A more detailed check for subleading Regge trajectories would involve matching the actual number of states at each level $n$. In \cite{Ferrero:2023znz}, the authors provide a strong-coupling counting of the degeneracy of the long $\{0,[0,0]\}$ states independently from their length (see Table 1), while in \cite{Cavaglia:2021bnz} there is a weak-coupling counting. A similar matching has recently been done for local operators on the leading and the subleading Regge trajectories in \cite{Julius:2024ewf}. We leave this analysis for the future.

Another important comment is related to the $1/g$ correction to the Regge trajectories. At $\phi = 0$ the spinning trajectories decouple and
we have $j_n(0) = -\Delta_{O_n}-1$ \eqref{eq:jn_forward}, with $O_n$ parity-even operators in $\mathcal{L}^{\Delta}_{0,[0,0]}$. 
In \cite{Ferrero:2021bsb} the $1/g$ correction to the scaling dimension of these operators has been computed using analytic bootstrap methods. It has a universal form
\begin{equation}
\label{eq:LOcorr}
\gamma^{(1/g)}_{{O}} = -\frac{\Delta_{g = \infty}(\Delta_{g = \infty}+3)}{8 \pi}\,,
\end{equation}
that coincides with the integrability-based result of \cite{Grabner:2020nis}.
Operators that are degenerate at $g = \infty$ remain degenerate at first order. The strong-coupling expansion of the $n$-th Regge trajectory $j_n(0)$ (since the degeneracy is not lifted we consider $j_n$ as a single trajectory) reads
\begin{equation}
\label{eq:leadingcorrectionstrong}
    \begin{aligned}
     &j_n(0) = -(2n+1) + \frac{n (2n+3)}{4 \pi g}+ \mathcal{O}(g^{-2}) \ .
    \end{aligned}
\end{equation}
This universal correction to the intercept of the Regge trajectories is important for the discussion in \secref{sec:Worldsheet bootstrap}, where it will be reproduced using the worldsheet bootstrap for the strong-coupling amplitude.

\subsection{Computing $m_{\text{gap}}^2(g)$ from QSC}
\label{sec:mgap_section}

Another relevant quantity that can be computed using QSC data is the position of the first spin $0$ bound state on the leading trajectory, denoted as $m_{\text{gap}}(g)$.
Knowing the form of $\mathrm{j}_0(\phi)$ for different values of $g$ we can evaluate $m_{\text{gap}}(g)$ by solving the equation $j_0(m_{\text{gap}}^2) = 0$. The weak-coupling expansion for $m_{\text{gap}}(g)$ is known \cite{Caron-Huot:2014lda}, while the strong-coupling one can be derived from the small $\phi$ expansion in \cite{Cavaglia:2022qpg}, taking into account the $B_4(g) \phi^4$ term. The result takes the form
\begin{equation}
\begin{aligned}
    & m_{\text{gap}}(g) = 2 m \Big( 1 - 2 \pi ^2 g^4-16 \pi ^2 g^6 \left(2 \log \left(8 \pi  g^2\right)-1\right) + ... \Big) \ , \qquad g \ll 1 \ , \\
    &m_{\text{gap}}(g) = \frac{m \sqrt{\pi}}{\sqrt{g}}\left(1-\frac{(\pi^2+3)}{24 \pi g} + \dots \right) \ , \qquad g \gg 1 \ .
\end{aligned}
\label{eq:expansion_mgap}
\end{equation}

From the analysis of the QSC data we get the following behavior for $m_{\rm gap}(g)$: as expected, the QSC data smoothly interpolate between the weak and strong-coupling expansions. 
\begin{figure}[h!]
    \centering
    \includegraphics[width=0.9\linewidth]{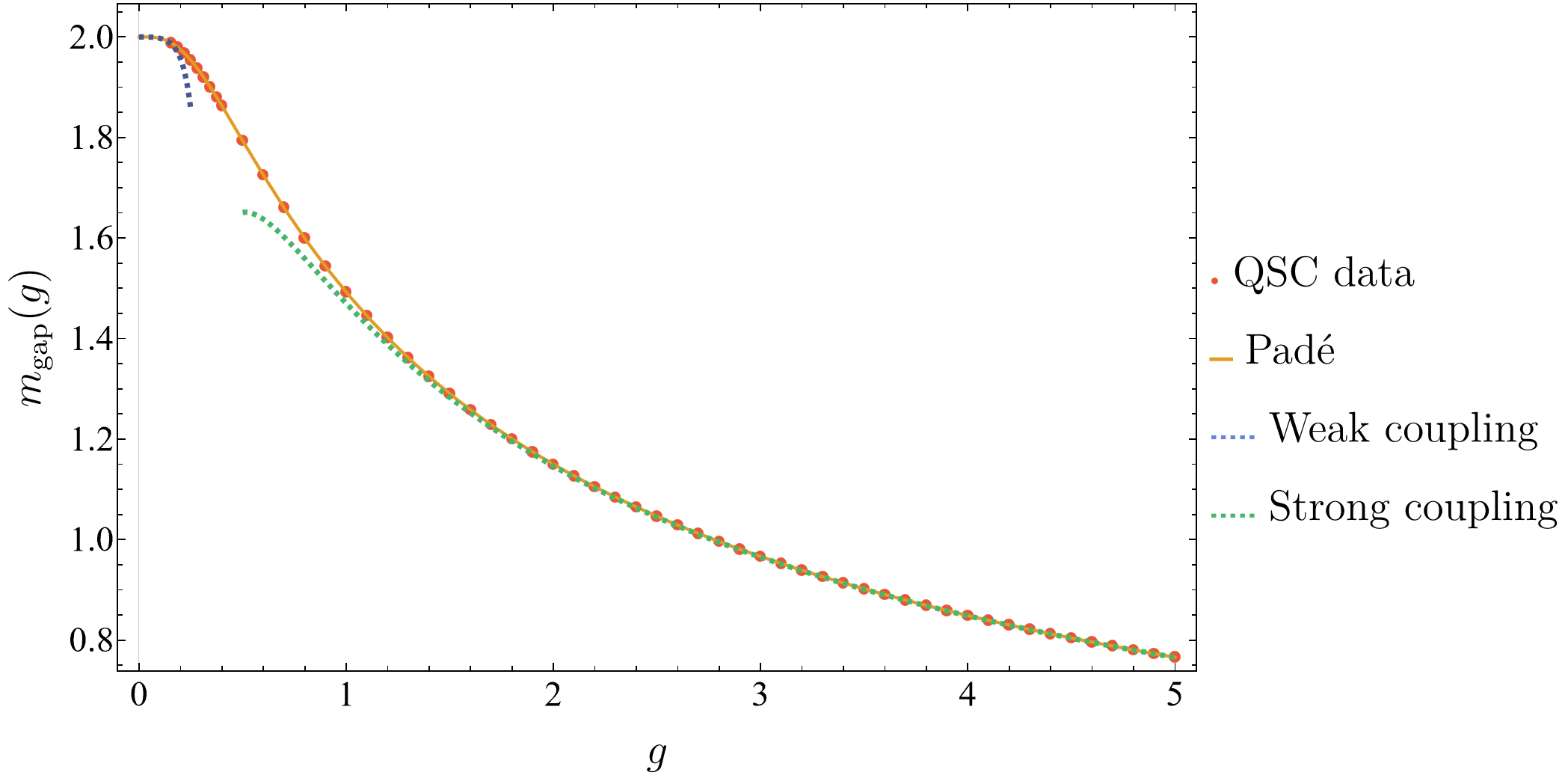}
    \caption{Plot of $m_{\text{gap}}$ as a function of $g$. Data from QSC interpolate the weak-coupling and strong-coupling expansions \eqref{eq:expansion_mgap}.}
    \label{fig:mgap_plot}
\end{figure}

As done for other quantities, we can use the QSC data, with the weak and strong-coupling expansions, to perform a conformal-Pad\'{e} approximation of $m_{\text{gap}}(g)$
\begin{equation}
\label{eq:mgappade}
    m_{\text{gap}}(g) = \sqrt{1-z} \hspace{1mm} \frac{\mu_0 + \mu_1 z + \mu_2 z^2 + \mu_3 z^3}{1 + \mu_4 z + \mu_5 z^2 + \mu_6 z^3}\,, \quad z = \frac{\sqrt{16 g^2 + 1}-1}{\sqrt{16 g^2 + 1}+1}.
\end{equation}
Some coefficients are fixed by using the leading and subleading orders at weak and strong coupling \eqref{eq:expansion_mgap}, the others are fitted using the QSC data and the three-loop result for $m_{\text{gap}}(g)$.  The result of this fit is presented in \figref{fig:mgap_plot} and available in the attached Mathematica notebook.

\section{Worldsheet bootstrap}
\label{sec:Worldsheet bootstrap}

In this section we study the scattering amplitude at strong coupling and compute the first $1/g$ correction to the leading-order result. In order to find the result, we make a number of assumptions: first, we assume that this correction admits a world-sheet representation, with an integrand given by a linear combination of harmonic polylogarithms of weight up to three, as in similar studies \cite{Alday:2023jdk,Alday:2023mvu,Alday:2024yax,Alday:2024ksp}. In addition, we assume that, to this order, only the states that already appear at the leading flat space result contribute to the answer. We then find that the underlying worldsheet correlator is completely determined. We check that the final result predicts mass shifts that are consistent with previous results from analytic bootstrap and integrability.

\subsection{Worldsheet ansatz}

At strong coupling, the leading low-energy behavior of the amplitude is controlled by the flat-space tree-level string scattering
\be
f(s,t) =-{g^2 \over \pi^2 m^4} \frac{\Gamma \left(-\frac{g s}{m^2 \pi }\right) \Gamma \left(-\frac{g t}{m^2 \pi}\right)}{\Gamma \left(1-\frac{g s}{m^2 \pi }-\frac{g t}{m^2 \pi }\right)} + \dots \ , ~~~ g \gg 1 \ , 
\label{eq:super_VenezianoWS}
\ee
where we consider the regime $g \gg 1$ with $(\frac{g s}{m^2 \pi },\frac{g t}{m^2 \pi })$  kept fixed. Our goal is to compute the first $1/g$ correction to this formula. It is convenient to introduce the dimensionless Mandelstam variables 

\begin{equation}
S=\frac{g s}{m^2 \pi },~~~T=\frac{g t}{m^2 \pi } \ ,
\end{equation}
which remain finite at strong coupling. We furthermore rescale the amplitude by an overall factor, so that
\begin{equation}\label{eq:hatf_def}
\hat f(S,T) \equiv {\pi^2 m^4 \over g^2}f \left({m^2 \pi \over g} S, {m^2 \pi \over g} T \right),
\end{equation}
where $\hat f(S,T)$ is a dimensionless amplitude. At strong coupling, it admits the expansion
\begin{equation}
\hat f(S,T)=\hat f^{(0)}(S,T) + \frac{1}{g}\hat f^{(1)}(S,T)+\cdots \ ,
\label{eq:strongcouplingcorrection}
\end{equation}
where from \eqref{eq:super_VenezianoWS} it follows that
\be
\hat f^{(0)}(S,T) = - {\Gamma(-S)\Gamma(-T) \over \Gamma(1-S-T)} \ . 
\ee
Let us comment that while \eqref{eq:strongcouplingcorrection} is a very natural form for the leading correction, we do not know the systematic structure of the strong-coupling expansion. For example, we do not know if it involves $\log g$ terms, which are known to appear at weak coupling, see \eqref{eq:expansion_mgap}.

The amplitude $\hat f(S,T)$ must satisfy crossing symmetry and admit an $SO(4)$ partial wave decomposition. More precisely,
we have the following dispersive representation for $\hat f(S,T)$
\begin{equation}\label{eq:hatf_dispRep}
\hat f(S,T) = -\sum_{J,n} \frac{C_J(n)}{S-M_{n,J}^2} \tilde P_J \left(1+2\frac{T}{M_{n,J}^2}- \frac{\pi T}{2g} \right) + \hat f_{cut}(S,T)\,.
\end{equation}
Unitarity implies
\be
C_J(n) \geq 0 \ . 
\ee
At leading order, the result can be written using the worldsheet representation as
\begin{equation}
\hat f^{(0)}(S,T) =\frac{1}{S+T} \int_0^1 \frac{x^{-S}(1-x)^{-T}}{x(1-x)} dx\,.
\end{equation}
The spectrum of intermediate states at leading order can be obtained from the partial-wave decomposition. We find
\begin{equation}
C^{(0)}_J(n) \neq 0~~ \text{for}~~~n=J+1,J+3,J+5,\cdots,
\end{equation}
with $J=0,1,2,\cdots$ and
\begin{equation}
M_{n,J}^2= n + \frac{1}{g} \gamma_{n,J}+\cdots,
\label{eq:mass_bs_correction}
\end{equation}
so that $n$ labels the excitation level. In terms of \figref{fig:behaviors_j0}, these are bound states close to $s=0$, where the Regge trajectory is approximately linear. We assume that the leading correction $\hat f^{(1)}(S,T)$ is fully captured by these states. For later reference, we also record the three-point couplings of the leading Regge trajectory $j_0$, where $n=J+1$,
\begin{equation}
C^{(0)}_J(J+1) =(J+1)^{J+1}\frac{2^{-2J}}{\Gamma (J+2)},
\end{equation}
as well as the three-point couplings in the next two trajectories
\begin{eqnarray}
C^{(0)}_J(J+3) &=&\frac{2^{-2 J-4} (J+1)^2 (J+3)^J (J+5)}{3 \Gamma (J+3)},\\
C^{(0)}_J(J+5) &=&\frac{2^{-2 J-9} (J+1)^2 (J+5)^J \left(5 J^3+103 J^2+735 J+1773\right)}{45 \Gamma (J+4)}. \label{degenerateC}
\end{eqnarray}
$AdS$-curvature corrections to the flat-space limit of CFT correlators have been constructed in various backgrounds, including in \cite{Alday:2023jdk,Alday:2023mvu} for closed-string amplitudes and in \cite{Alday:2024yax,Alday:2024ksp} for open-string amplitudes. In all these cases, the curvature corrections admit a worldsheet representation analogous to the flat-space one, but with additional insertions of polylogarithms. 
Specifically, the closed-string case involves single-valued polylogarithms, while the open-string case involves harmonic polylogarithms.
Motivated by these examples, we look for corrections of the form
\begin{equation}
\hat f^{(1)}(S,T) = \int_0^1 \frac{x^{-S}(1-x)^{-T}}{x(1-x)} h(S,T;x) dx,
\end{equation}
where $h(S,T;x)$ is a linear combination of harmonic polylogarithms $H_{w}(x)$, of weight up to three and with letters $\{ 0,1\}$, producing branch points at $x=0$ and $x=1$.\footnote{See \appref{sec:HPL} for their definition.} More precisely, we assume
\begin{equation}
h(S,T;x)= \sum_{w} \frac{P_w(S,T)}{(S+T)^2} H_{w}(x),
\end{equation}
where $P_w(S,T)$ are polynomials of degree three.\footnote{This assumption is very strong, and is based on experience with similar cases. One could consider, for instance, analytic functions $P(x)$ in the worldsheet, such as polynomials.} This correction must satisfy several conditions. We start with the following two
\begin{itemize}
\item It must be symmetric under $S \leftrightarrow T$ and $x \leftrightarrow 1-x$, ensuring $f^{(1)}(S,T)=f^{(1)}(T,S)$
$$h(S,T;x)=h(T,S;1-x) \ . $$
\item The non-analytic behavior near $x=0$ must take the form
$$h(S,T;x)= \left(c_0+c_1 x+ c_2 x^2 + \cdots \right)\log x +\text{regular}.$$
In other words, terms proportional to $\log^2 x$ and $\log^3 x$ must be absent. The reason for this is that $f^{(1)}(S,T)$ develops only double poles at $S=1,2,\cdots$, but no higher-order poles.\footnote{This follows directly from substituting \eqref{eq:mass_bs_correction} into \eqref{eq:hatf_dispRep} and performing the large-$g$ expansion.} Crossing symmetry then implies the same structure around $x=1$. 
\end{itemize}
These conditions lead to the following basis. First we introduce the symmetric and anti-symmetric combinations
\begin{equation}
H^{\pm}_w(x) = \frac{1}{2} \Big( H_w(x) \pm H_w(1-x) \Big).
\end{equation}
In terms of those 
\begin{eqnarray*}
h(S,T;x) &=&  \frac{P^+_{001}(S,T)}{(S+T)^2}  H^+_{001}(x)+ \frac{P^+_{010}(S,T)}{(S+T)^2}  H^+_{010}(x)+ \frac{P^-_{001}(S,T)}{(S+T)^2}  H^-_{001}(x)+ \frac{P^-_{010}(S,T)}{(S+T)^2}  H^-_{010}(x) \\
&&+ \frac{P^+_{01}(S,T)}{(S+T)^2}  H^+_{01}(x)+ \frac{P^-_{01}(S,T)}{(S+T)^2}  H^-_{01}(x)\\
& & + \frac{P^+_{0}(S,T)}{(S+T)^2}  H^+_{0}(x)+ \frac{P^-_{0}(S,T)}{(S+T)^2}  H^-_{0}(x)+\frac{P^+(S,T)}{(S+T)^2}.
\end{eqnarray*}
Here $P^+_w(S,T)$ are symmetric and $P^-_w(S,T)$ are anti-symmetric polynomials. The next set of constraints arises from the partial wave decomposition at order $1/g$, and the expected structure of corrections to the bound-state masses

\begin{itemize}
\item Double poles at $S=n$ correspond to leading-order bound states acquiring mass corrections
\begin{equation}
\label{doublepoles}
\hat f^{(1)}(S,T)  = -\sum_{J,n} \frac{\gamma_{n,J} C^{(0)}_J(n)}{(S-n)^2}\tilde P_J \left(1+\frac{2T}{n} \right) + \text{single order poles} \ .
\end{equation}
\item Bound states on the leading Regge trajectory, which correspond to $n=J+1$, $\gamma_{0}(J) \equiv \gamma_{J+1,J}$ have the mass shift
\begin{equation}
\label{dimleading}
\gamma_{0}(J)=(J+1) \left(-\frac{\left(15+2 \pi ^2\right) J}{24 \pi }-\frac{3+\pi ^2}{12 \pi }\right) .
\end{equation}
This expression can be derived from the small-angle expansion of the cusp anomalous dimension at strong coupling \eqref{eq:small_phi_j0}.
\end{itemize}
Note that starting from the subleading Regge trajectory intermediate operators are degenerate at strong coupling, so that the sum over $J,n$ should also include a sum over species, omitted here in order to ease the notation. Another condition arising from the partial wave decomposition is the following 
\begin{itemize}
\item Intermediate operators with $n=J+2$, which are absent at strong coupling, should also be absent at order $1/g$. This follows both from the spectrum of excitations of the Maldacena-Wilson line at strong coupling, and from the flat-space string spectrum, see \appref{eq:fspspectrum}.
\end{itemize}
This last condition, together with (\ref{doublepoles}) and (\ref{dimleading}), fixes the polynomials $P^{\pm}_{w}(S,T)$ up to a handful of coefficients. In addition, we have the following requirements. 
\begin{itemize}
\item The pole at $S=0$ and the constant term do not receive $1/g$ corrections
\begin{equation}
\hat f(S,T)=-\frac{1}{S \, T} + {\pi^2 \over 6} + {\cal O} (S,T) \ .
\end{equation}
\item The correction $\hat f^{(1)}(S,T)$ must remain regular in the limit $S+T \to 0$, since a pole in this limit would correspond to an unphysical $u$-channel singularity. This condition is nontrivial because our ansatz contains a double pole at $S+T=0$. 

\item The Regge behavior of $\hat f^{(1)}(S,T)$ 
should not be more power-law singular than that of the leading-order solution
\begin{equation}
\hat f^{(0)}(S,T) \sim \Gamma(T) (-S)^{T-1},~~~|S| \gg 1, \text{finite $T$} \ .
\end{equation}
This follows from the fact that the $1/g$ corrections to the Regge trajectories lead to the $(\log S)^{\#}$ terms in the amplitude.
\end{itemize}

\subsection{Bootstrapping the solution}

These conditions lead to an over-constrained system of equations for the coefficients of the polynomials $P^{\pm}_{w}(S,T)$, which admits a unique solution. In terms of classical polylogarithms, the solution takes the form

\begin{eqnarray}
h(S,T;x)&=& \frac{3 S (S+2 T)}{4 \pi  (S+T)}\text{Li}_3(x)+\frac{3 T (2 S+T)}{4 \pi  (S+T)}\text{Li}_3(1-x)-\frac{3 (S+T)}{4 \pi } \zeta(3) \nonumber\\
&&-\frac{S (S+2 T)}{4 \pi  (S+T)}\text{Li}_2(x) \log (x)-\frac{T (2 S+T)}{4 \pi  (S+T)}\text{Li}_2(1-x) \log (1-x) \nonumber\\
&& -\zeta(2)\left(\frac{S^2}{2 \pi  (S+T)} \log x +\frac{T^2}{2 \pi  (S+T)} \log(1-x)\right)  \\
&& +\frac{S-T}{4 \pi  (S+T)}\text{Li}_2(x)+\frac{T-S}{4 \pi  (S+T)}\text{Li}_2(1-x)-\frac{S^2-4 S T+T^2}{4 \pi  (S+T)^2} \zeta(2)\nonumber\\
&& \frac{S}{4 \pi  (S+T)^2}\log (x)+\frac{T}{4 \pi  (S+T)^2}\log (1-x)-\frac{1}{4 \pi  (S+T)^2} \ . \nonumber
\end{eqnarray}
Importantly, the rational function multiplying a given harmonic polylogarithm is a homogenous polynomial divided by $(S+T)^2$, and the degree of the polynomial is always equal to the weight of the harmonic polylogarithm in question. In this counting, $\zeta(n)$ is assigned weight $n$. 
This structure ensures that the low-energy expansion has uniform transcendental weight, as shown below.

Having obtained a candidate for the $1/g$ correction, we can now make concrete predictions that can be tested by independent methods. 
We begin by evaluating the mass corrections for bound states on the subleading and sub-subleading Regge trajectories. We obtain

\begin{eqnarray}
\gamma_{1}(J)&=&-\frac{2 \pi ^2 J^3+15 J^3+22 \pi ^2 J^2+136 J^2+78 \pi ^2 J+353 J+90 \pi ^2+300}{24 \pi  (J+5)},\label{dimsubleading}\\
\gamma_{2}(J)&=&-\frac{\left(15+2 \pi ^2\right) J^2}{24 \pi }-\frac{4 \left(127 J^2+2720 J+11007\right)}{25 \pi  \left(5 J^3+103 J^2+735 J+1773\right)}-\frac{\left(101+20 \pi ^2\right) J}{24 \pi }\nonumber \\
& & -\frac{2093+625 \pi ^2}{300 \pi }. \nonumber
\end{eqnarray}
As noted above, the bound states on the subleading trajectories are degenerate, so the mass corrections obtained from our solution are weighted averages
\begin{equation}
\langle \gamma \rangle = \frac{\sum_I \gamma_I C^{(0)}_I}{\sum_I C^{(0)}_I}  \ ,
\end{equation}
where the sums run over all species 
degenerate in mass (and spin) at leading order in the strong-coupling expansion, and $\sum_I C^{(0)}_I$ denotes the corresponding leading-order three-point coupling. The poles at $J+5=0$ and $\left(5 J^3+103 J^2+735 J+1773\right)=0$ in \eqref{dimsubleading} 
clearly signal the degeneracy.
They arise from dividing by $\sum_I C^{(0)}_I$, given in (\ref{degenerateC}) for the first degenerate Regge trajectories. Finally, 
all Regge trajectories exhibit the same asymptotic behavior at large spin 
\begin{equation}
\gamma_{n}(J) = \left(-\frac{5}{8 \pi }-\frac{\pi }{12}\right) J^2 + \cO(J) \ .
\end{equation}
It is expected since this correction can be traced back to the classical string solution \eqref{eq:j0_strong}, which controls the limit $g \to \infty$, $J/g$ - fixed and is universal for all trajectories.

Next, we look at the Wilson coefficients in the low-energy expansion of $f^{(1)}(S,T)$. To this end, we introduce two building blocks
\begin{equation}
\Phi_2(S,T) = \int_0^1 \frac{x^{-S}(1-x)^{-T}}{x(1-x)} \text{Li}_2(x) dx,~~~\Phi_3(S,T) = \int_0^1 \frac{x^{-S}(1-x)^{-T}}{x(1-x)} \text{Li}_3(x) dx \ .
\end{equation}
Their low-energy expansions can be obtained following \cite{Alday:2024ksp,Alday:2025bjp,Baune:2025hfu}
\begin{eqnarray}
\Phi_2(S,T) &=&-\frac{\pi ^2}{6 T }-\zeta (3)+\frac{7}{360} \pi ^4 (S-T) \\
&&+\frac{1}{6} \left(\pi ^2 \zeta (3) \left(3 S^2+5 S T+T^2\right)-3 \zeta (5) \left(9 S^2+16 S T+9 T^2\right)\right)+\cdots \nonumber\\
\Phi_3(S,T) &=&-\frac{\zeta (3)}{T }-\frac{\pi ^4}{360}+\frac{1}{2} \zeta (5) (11 S+7 T)-\frac{1}{3} \pi ^2 \zeta (3) (S+T)\\
&&+\frac{1}{2} \zeta (3)^2 \left(S^2-3 S T-3 T^2\right)+\frac{\pi ^6 \left(24 S^2+166 S T+85 T^2\right)}{45360} + \cdots \nonumber
\end{eqnarray}
The proposed $1/g$ correction can be expressed in terms of $\Phi_2(S,T),\Phi_3(S,T)$, the leading-order amplitude, and their derivatives. Its low-energy expansion takes the form
\begin{eqnarray}
24 \pi \hat f^{(1)}(S,T)&=& 18 \zeta(4)(S+T)+12 \zeta(2)\zeta (3) (S-T)^2\\
& &+15 \zeta (5) \left(S^2+4 S T+T^2\right) + \cdots \nonumber
\end{eqnarray}
Finally, for the correction to the three-point coupling on the leading Regge trajectory we find
\begin{eqnarray}
C^{(1)}_J(J+1) =C^{(0)}_J(J+1) \frac{57 J^2-36 (J+1)^2 \zeta (3)-8 \pi ^2 J+99 J-8 \pi ^2+24}{48 \pi } \ . \nonumber
\end{eqnarray}
\subsection{Consistency checks against known results}

We now assume that the Regge/cusp correspondence holds for all Regge trajectories. %
We can use this assumption to generate an infinite set of predictions for the leading $1/g$ correction to the amplitude derived above.

To obtain the desired prediction, recall that on the Wilson-line side there are two types of $SO(2) \times SO(5)$ singlet operators: (i) parity-even scalar insertions $O_n \in \mathcal{L}_{0,[0,0]}^{\Delta}$, which are neutral under $SO(3)$ in the forward limit, and (ii) spinning operators transforming non-trivially under $SO(3)$ in the forward limit. 
As before, we assume that the spinning trajectories decouple in the forward limit.
Moreover, as discussed in \eqref{eq:leadingcorrectionstrong}, the leading strong-coupling correction to the scaling dimension of parity-even primary operators $O_n \in \mathcal{L}_{0,[0,0]}^{\Delta}$ does not lift the degeneracy and is known explicitly. 
Taken together, these two observations lead to the prediction
\be
\label{eq:predictionIA}
\langle \gamma_{n}(-2n-1) \rangle = - {n(2n+3) \over 4 \pi} \ ,
\ee
where the argument $J=-2n-1$ corresponds to the straight Wilson line.

Remarkably, the amplitude correction obtained above satisfies \eqref{eq:predictionIA}. More precisely, we have explicitly checked that
\begin{equation}
\gamma_{1}(-3)=-\frac{5}{4\pi},~~~\gamma_{2}(-5)=-\frac{7}{2 \pi },~~~\gamma_{3}(-7)=-\frac{27}{4 \pi },~~~\gamma_{4}(-9)=-\frac{11}{\pi }.
\end{equation}
We emphasize that in deriving the correction to the amplitude in this section, only the leading Regge trajectory data was used as input.

A further remark concerns the microscopic origin of the subleading-trajectory degeneracy.
Away from the forward limit, and in contrast to weak coupling, our solution exhibits the appearance of an additional Regge trajectory at the first subleading order, besides the $SO(3)$-singlet one $j_1(s)$. 
A natural candidate for this additional spinning trajectory is the insertion of the displacement operator at the cusp. Its presence is also consistent with the counting of the flat-space spectrum, see \appref{eq:fspspectrum}, and it would be interesting to check this prediction.

\subsection{Towards the monodromy relations}
Tree-level color-ordered open string amplitudes in flat space satisfy monodromy relations \cite{osti_4155317}. Indeed, one can explicitly check
\begin{equation}
e^{-i \pi S} \hat f^{(0)}(S,U) + \hat f^{(0)}(S,T) + e^{i \pi T} \hat f^{(0)}(T,U)=0.
\end{equation}
In \cite{Alday:2025cxr} a generalization of the monodromy relations for open string theory amplitudes in $AdS$, as defined in the setup of \cite{Alday:2024yax,Alday:2024ksp}, was proposed. The present case is somewhat different. While in  \cite{Alday:2024yax,Alday:2024ksp} one considers the scattering of open strings on a D7-brane wrapping $AdS_5 \times S^3$, here we are considering the scattering of open strings on a D3-brane at fixed radial distance on $AdS$. In particular, in the present case we have an extra scale $m$, which breaks conformal invariance. On the other hand, in both cases the amplitude at strong coupling admits a world-sheet representation with identical structure. Hence, we will look for a deformation to the monodromy relations along the lines of that proposed in \cite{Alday:2025cxr}. It is convenient to introduce the following building blocks, see \cite{Alday:2025bjp}, labelled by words $w$, 
\begin{equation}
J_w(S,T)= \int_0^1 \frac{x^{-S}(1-x)^{-T}}{x(1-x)} H_w(x)dx.
\end{equation}
In terms of these, the amplitude at zeroth order is simply $\hat f^{(0)}(S,T)=\frac{1}{S+T}J_e(S,T)$ while at order $1/g$ it takes the form
\begin{eqnarray}
\hat f^{(1)}(S,T) &=& \frac{S (S+2 T)}{S+T}(J_{010}(S,T) - J_{001}(S,T))+ \frac{T (T+2 S)}{S+T}(J_{101}(S,T) - J_{110}(S,T)) \nonumber \\
&& - 3 \frac{S^2}{S+T} \zeta(3) J_e(S,T)
-2\frac{\zeta(2)}{S+T} S^2 J_0(S,T) +4 \frac{\zeta(2)}{S+T} S T J_1(S,T) \\
&&+ \frac{S-T}{S+T} (J_{10}(S,T)-J_{01}(S,T)) + \frac{S}{(S+T)^2} J_0(S,T)+ \frac{T}{(S+T)^2} J_1(S,T) \nonumber\\
&&+\frac{-2 S (S-2 T)\zeta(2)  -1}{(S+T)^2}J_e(S,T) \,.\nonumber
\label{f1vsJ}
\end{eqnarray}
To order $1/g$ we find the amplitude $\hat f^{(0)}(S,T)+ \frac{1}{g} \hat f^{(1)}(S,T)+\cdots$ satisfies the modified monodromy relation

\begin{equation}
e^{ i \pi (-S+K_0)} e^{-\frac{\pi}{4 g} S^2 \partial_U} \hat f(S,U) + \hat f(S,T) + e^{i \pi (T-K_0)}  e^{-\frac{\pi}{4 g} T^2 \partial_U} \hat f(T,U)=0.
\end{equation}
Here $K_0$ is the operator controling the monodromy around zero, and its action on the building blocks is given by
\begin{equation}
K_0 J_{wa}(S,T) = J_w(S,T) \delta_{a0},~~~K_0 J_{e}(S,T)=0.
\end{equation}
The operators $e^{-\frac{\pi}{4 g} S^2 \partial_U},e^{-\frac{\pi}{4 g} T^2 \partial_U}$ are absent in the $D7-$branes case. Note that in terms of the un-rescaled Mandelstam variables $s,t$:
$$\frac{\pi}{4 g} S^2 \partial_U = \frac{1}{4m^2} s^2 \partial_u,~~~\frac{\pi}{4 g} T^2 \partial_U = \frac{1}{4m^2} t^2 \partial_u .$$
So that these new terms presumably appear due to the existence of a new scale/cut in the amplitude. It would be interesting to understand why the monodromy relations take precisely this form.

\subsection{Contribution of the $AdS$ states}

In the calculation above, we have assumed that only the flat-space states contribute to the leading correction. In this subsection, we examine the consistency and implications of this assumption. A convenient starting point is the fixed-$t$ dispersion relation
\be
\label{eq:dispagain}
f(s,t) +{1 \over s t}= \oint {ds' \over 2 \pi i} {1 \over s'-s} \left(f(s',t) +{1 \over s' t}  \right) = \int_{m_{gap}^2}^{\infty} {d s' \over \pi} {f_s(s',t) \over s'-s} ,
\ee
where, as usual, the right-hand side receives contributions from both the bound states and the two-particle cut. 
The mass gap decreases at strong coupling as $m_{gap}^2 \sim {1 \over g}$. We refer to a state $X$ as a \emph{flat-space state} if its mass vanishes in the strong-coupling limit, $\lim_{g \to \infty} m_X(g) = 0$ at fixed W-boson mass $m$. All states considered in this section are of this type. On the other hand, we refer to a state $X$ as \emph{AdS state} if $\lim_{g \to \infty} m_X(g) > 0$. 

We introduce a fixed small parameter $\epsilon$ and consider the contribution of the $AdS$ states to the dispersive representation of $\hat f(S,T)$, defined in \eqref{eq:hatf_def},
\be
\hat f_{AdS}(S,T) = \left( {\pi m^2 \over g} \right)^2 \int_{\epsilon m^2}^{\infty} {d s' \over \pi} {f_s(s',{\pi m^2 \over g} T) \over s'-{\pi m^2 \over g} S}.
\ee
The flat-space limit in $AdS$ corresponds to the large-$g$ expansion of this expression. From the point of view of a low-energy observer, in this limit the $AdS$ states become infinitely heavy, with a characteristic mass scale $\Lambda_{{AdS}}^2 = {\epsilon m^2 \over g}$. Consequently, we expect their contribution to be captured by a polynomial in $\left({S \over \Lambda_{{AdS}}^2},{T \over \Lambda_{{AdS}}^2}\right)$. The flat-space low-energy expansion can be written as
\be\label{eq:fADS_moment_exp}
\hat f_{{AdS}}(S,T) &= \left( {\pi m^2 \over g} \right)^2 \int_{\epsilon m^2}^{\infty} {d s' \over \pi} {f_s(s',0) \over s'} + \left( {\pi m^2 \over g} \right)^3 S \int_{\epsilon m^2}^{\infty} {d s' \over \pi} {f_s(s',0) \over (s')^2} \nn \\
&+ \left( {\pi m^2 \over g} \right)^3 T \int_{\epsilon m^2}^{\infty} {d s' \over \pi} {\partial_t f_s(s',0) \over s'}+\left( {\pi m^2 \over g} \right)^4 S T \int_{\epsilon m^2}^{\infty} {d s' \over \pi} {\partial_t f_s(s',0) \over (s')^2} + \dots \ ,  
\ee
where each of these moment integrals is itself a function of $g$. For the flat-space low-energy expansion to be well defined, the moments containing an additional $\partial_t$ derivative must not be enhanced at large $g$. Otherwise, the heavy-state contributions would fail to produce a polynomial expansion order by order in $1/g$. However, this is not something we can prove from the first principles.\footnote{The $AdS$ part of the amplitude discussed here is not necessarily crossing-symmetric on its own. The only applicable first-principle constraint is the positivity of the Hankel matrix of moments (see, e.g., \cite{Bellazzini:2020cot}), which is not sufficient to establish this statement.}

Assuming that the heavy states indeed organize into a well-defined polynomial expansion in $\left({S \over \Lambda_{{AdS}}^2},{T \over \Lambda_{{AdS}}^2}\right)$, the leading term in the expansion above is a constant
\be
\label{eq:adsa00}
\hat a_{0,0}^{{AdS}}(\epsilon,g) = \left( {\pi m^2 \over g} \right)^2 \int_{\epsilon m^2}^{\infty} {d s' \over \pi} {f_s(s',0) \over s'} \ . 
\ee
The non-renormalization theorem, together with the results of this section, implies that
\be
\label{eq:WIstrongpr}
\lim_{g \to \infty}{\hat a_{0,0}^{{AdS}}(\epsilon,g)  \over g} = 0 .
\ee
It would be very interesting to test this prediction.  In this context, let us note that it was conjectured in \cite{Bruser:2018jnc} that, at fixed $g$,
\be
\label{eq:crosssection}
\lim_{s \to \infty}  s f_s(s,0) = {\pi \over m^2} B(g) \ , 
\ee
where $B(g)$ is the Bremsstrahlung function given in \eqref{eq:bremss}. Its strong-coupling expansion reads
\be
B(g) = {g \over \pi} \left( 1 - {3 \over 8 \pi g} + {3 \over 128 \pi^2 g^2} + \dots  \right) \ .
\ee
Taken together, equations \eqref{eq:adsa00}, \eqref{eq:WIstrongpr}, and \eqref{eq:crosssection} imply that the asymptotic regime, where $f_s(s,0) \sim {B(g) \over s}$, can only set in at energies that diverge as $g \to \infty$.

\begin{figure}[h!]
    \centering
    \includegraphics[width=0.9\linewidth]{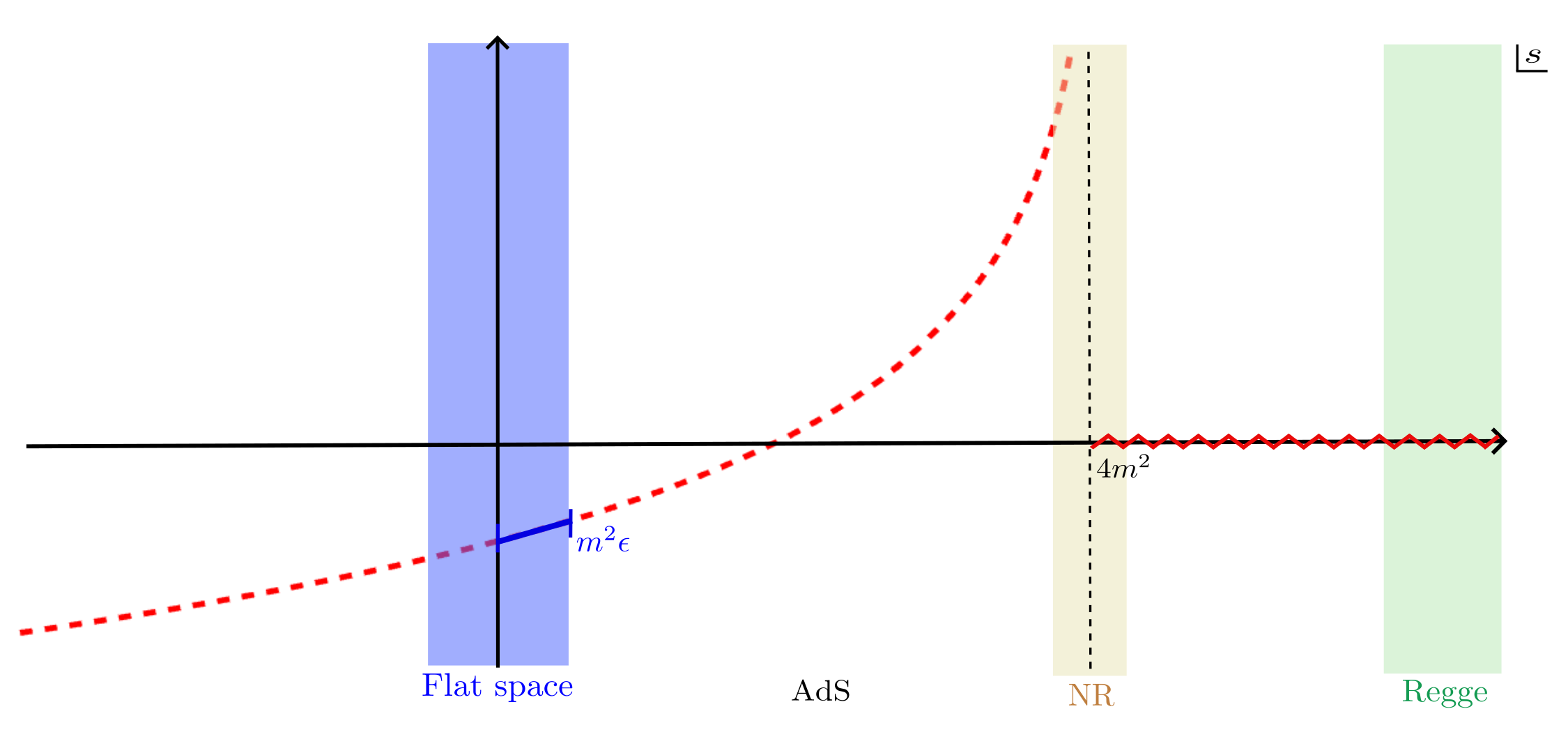}
    \caption{Contribution of different states to the dispersion relation \eqref{eq:dispagain}. Blue: close to $s = 0$ flat-space states contribute. There are two other regions that we have control over: close to the two-particle threshold (yellow), where the non-relativistic EFT expansion applies, and at $s \to \infty$ (green), where the behavior of $f_s(s,0)$ in $g$ is fixed by the Bremsstrahlung function $B(g)$, see \eqref{eq:crosssection}.}
    \label{fig:regimes_AdS_flat}
\end{figure}

Equation \eqref{eq:WIstrongpr} can also be tested by focusing on the contribution of states near the two-particle threshold, where the non-relativistic EFT expansion applies, see \appref{app:residue} and \figref{fig:regimes_AdS_flat}.
To apply these results, we must ensure that the Coulomb potential used in their derivation remains reliable. The leading correction to this potential arises from the term $V_{1/m}$ in equation (4.57) of \cite{Beneke:2013jia}. Its contribution is suppressed relative to the Coulomb potential by a factor of $\Omega(g) v$ in the non-relativistic limit $v \ll 1$, where $\Omega(g)$ is the strength of the Coulomb potential. At strong coupling we have $\Omega(g) \sim g$, and to ensure that the corrections is suppressed requires $g \sqrt{1-4m^2/s} \ll 1$. Applying the results of \appref{app:residue} within the dispersion relation then yields results consistent with \eqref{eq:WIstrongpr}.

\section{Bootstrapping the amplitude at finite coupling}\label{sec:bootstrapping_setup}

In \secref{sec:setup}, we reviewed the known properties of the amplitude $f(s,t)$, which we now use to derive constraints on various observables.
The basic idea is the same as in \cite{Adams:2006sv,Tolley:2020gtv,Caron-Huot:2020cmc,Arkani-Hamed:2020blm,Caron-Huot:2021rmr}, which were recently been applied to weakly-coupled stringy amplitudes in \cite{Chiang:2023quf,Berman:2023jys,Haring:2023zwu,Albert:2024yap,Berman:2024wyt,Berman:2025owb,Huang:2025icl}. 
The amplitude considered in this work exhibits distinctive properties --- notably enhanced unitarity: $SO(4)$ for bound states, $SO(1,3)$ for scattering states --- which make some of our results qualitatively different from previous analyzes.

Assuming that the Regge/cusp correspondence extends to all Regge trajectories, we can relate the spectral gap between the leading and  subleading trajectories to the energy of the first excited state of the cusp Hamiltonian. Both at weak and strong coupling, the lightest excited state is given by the insertion of $\Phi_{||}$ at the cusp. At intermediate coupling, $\Phi_{||}$ remains the first excited state in the forward limit, as seen in Figure 2 of \cite{Cavaglia:2021bnz}. Away from the forward limit, $\phi \neq 0$, determining the gap requires analyzing the spectrum of spinning Regge trajectories. This information is not available at present. In the absence of this data, we will assume assume that the spinning trajectories do not cross the $\Phi_{||}$ trajectory at intermediate coupling, and we can therefore use the integrability results of the previous section to define the gap, see \figref{fig:spectrumSchematic}.

In the planar $\mathcal{N}=4$ SYM CFT, an analogous bootstrap problem was studied recently in \cite{Caron-Huot:2024tzr}. In that work, instead of the amplitude on the Coulomb branch, the authors bootstrapped the four-point conformal correlation function of half-BPS operators. A crucial ingredient in their analysis was supersymmetric localization, which is not currently available in our framework.

\subsection{Dispersion relations and sum rules}\label{sec:DR_and_SR}

Our starting point is the unsubtracted dispersion relation \eqref{eq:Disprel}, valid for $t< m_{\text{gap}}^2$. By unitarity, the discontinuity of the amplitude can be decomposed into partial waves with non-negative coefficients. Dual conformal invariance further distinguishes two regimes. For bound states, $s < 4m^2$, the amplitude admits an $SO(4)$ partial-wave expansion \eqref{eq:fsInSO4PW}. For scattering states, $s \geq 4m^2$, it instead admits an $SO(1,3)$ harmonic decomposition \eqref{eq:fsInSO31PW}. Substituting these decompositions into the dispersion relation gives
\begin{align}
\label{eq:fSRIntegral} 
    f(s,t)+\frac{1}{st} 
    &= \frac{1}{\pi}\int_{m_{\rm gap}^2}^{4m^2} dz \frac{f_s(z,t)}{z-s} + \frac{1}{\pi}\int_{4m^2}^{\infty} dz \frac{f_s(z,t)}{z-s}\\
    &= \frac{1}{\pi}\int_{m_{\rm gap}^2}^{4m^2} \frac{dz}{z} \sum_{J=0}^\infty c_J(z)\frac{\PJfive(\cos\tilde\theta_z)}{z(z-s)}  + \frac{1}{\pi}\int_{4m^2}^{\infty} \frac{dz}{z} \int_{0}^{\infty} d\nu c_\nu(z) \frac{\Omega_\nu(\cosh\tilde\theta_z)}{z(z-s)}\, , 
    \nonumber
\end{align}
where recall that $\cos \tilde\theta_z = 1+{2t \over z} - {t \over 2 m^2}$, and analogously $\cosh\tilde\theta_z$ when $z \geq 4m^2$.

\begin{figure}[h!]
    \centering
    \includegraphics[width=0.6\linewidth]{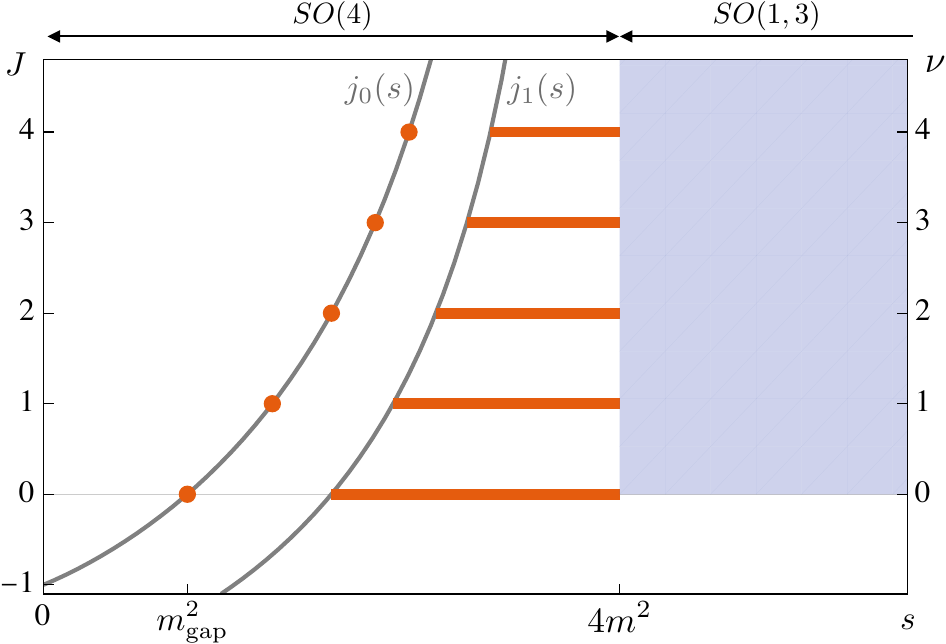}
    \caption{Spectrum constraints in the bootstrap. For $z < 4m^2$, the spectral density $c_J(z)$ from the $SO(4)$ decomposition \eqref{eq:cJDecomposition} is used, while for $z > 4m^2$, the $SO(1,3)$ decomposition with the spectral density $c_\nu(z)$ applies.}
    \label{fig:spectrumSchematic}
\end{figure}

On the right-hand side of \eqref{eq:fSRIntegral}, we have separated the contributions of the bound and scattering states. To derive bounds on observables, it is important to know the masses (or positions) of the lightest states for each spin. In 
\secref{sec:integrability}, we used integrability to compute the leading $j_0(s)$ and first subleading $j_1(s)$ Regge trajectories. These trajectories predict the location of the bound states via the relation $j_i(m_{i,J}^2)=J$. This relation allows us to restrict the spectrum, as illustrated in 
\figref{fig:spectrumSchematic}.
Consequently, the $SO(4)$ partial waves $c_J(z)$ in \eqref{eq:fSRIntegral} take the form
\begin{equation}\label{eq:cJDecomposition}
    c_J(z) = \pi \lambda^2_{0,J} m_{0,J}^2\delta(z-m_{0,J}^2) + c_J'(z) \theta(z-\msubJ{J}^2)\,,
\end{equation}
where $\theta(x)=1$ for $x\geq 0$ and is zero otherwise. The first term corresponds to the leading Regge trajectory, while the second term encodes the contribution of all other states starting from the subleading Regge trajectory.

In \eqref{eq:fSRIntegral}, we have introduced $\mgap \equiv m_{0,J=0}^2$ corresponding to the mass of the lightest bound state. The coefficient $\lambda_{0,J}$ denotes the three-point coupling between the external states and the spin-$J$ bound states of mass $m_{0,J}$. Indeed, a $SO(4)$ spin-$J$ particle exchange contributes to the amplitude as\footnote{At strong coupling, the three-point coupling $\lambda_{0,J}^2$ is trivially related to the residue $C_J(J+1)$ defined in \eqref{eq:hatf_dispRep}.}
\begin{equation}
    f(s,t)\underset{s\to m_{0,J}^2}{\sim} -\frac{\lambda_{0,J}^2  \PJfive(\cos\tilde\theta_s)}{m_{0,J}^2(s-m_{0,J}^2)}\,,
\label{eq:f_close_to_bs}
\end{equation}
and leads to the first term in \eqref{eq:cJDecomposition}. For $z\geq m_{1,J}^2$, we do not make further spectrum assumptions.\footnote{If additional Regge trajectories $j_i(z)$ were accessible through integrability, the spectral ansatz in \eqref{eq:cJDecomposition} could be refined accordingly.} 

It is convenient to introduce a compact bracket notation for the averages appearing in \eqref{eq:fSRIntegral}
\begin{align}
\label{eq:spdpwBS}
    \avgBS{\dots} &= \frac{1}{\pi}\int_{m_{\rm gap}^2}^{4m^2} \frac{dz}{z} \sum_{J=0}^\infty c_J(z)(\dots)\,,\\
    \avgCut{\dots} &= \frac{1}{\pi}\int_{4m^2}^{\infty} \frac{dz}{z} \int_{0}^{\infty} d\nu c_\nu(z)(\dots)\,,
    \label{eq:spdpwSS}
\end{align}
where $c_J(z)$ is further given by \eqref{eq:cJDecomposition}. These brackets denote non-negative averages over the bound-state and scattering-state spectral densities, respectively. The dispersion relation \eqref{eq:fSRIntegral} then takes the form
\begin{equation}\label{eq:fSumRule}
    f(s,t) =-\frac{1}{st}+\avgBS{\frac{ \PJfive(\cos\tilde\theta_z)}{z(z-s)}} + \avgCut{\frac{\Omega_\nu(\cosh\tilde \theta_z)}{z(z-s)}} \,.
\end{equation}
Finally, note that in the limit of large $W$-boson mass, $m \to \infty$, the cosine of the $SO(4)$ angle $\cos \tilde \theta_s$ reduces to the familiar expression
\begin{equation}\label{eq:usual_cos}
\cos\tilde\theta_s\underset{m\to \infty}{\longrightarrow} 1+\frac{2t}{s}\,.
\end{equation}
In this limit, the sum rules simplify to 
\begin{equation}\label{eq:SRatInfiniteMass}
    f(s,t) =\frac{-1}{st}+\frac{1}{\pi}\int_{m_{\rm gap}^2}^{\infty} \frac{dz}{z} \sum_{J=0}^\infty c_J(z)\frac{\PJfive\!\left(1+\frac{2t}{z}\right)}{z(z-s)}\,.
\end{equation}
In other words, we recover the standard dispersion relations for scattering in \emph{five-dimensional} flat space. Equivalently, one can send $s,t,\mgap \to 0$ while keeping their ratios fixed and $m^2$ finite, leading to the same result. The latter limit is relevant at strong coupling. Thus, the limiting dispersion relation \eqref{eq:SRatInfiniteMass}  provides a clear holographic interpretation.

In the formulas above, we used the enhanced unitarity that follows from dual conformal invariance. Nevertheless, the amplitude also obeys the standard unitarity relation \eqref{eq:usualPW}. 
This in turn leads to the standard sum rules 
\begin{align}
\label{eq:usualsumrule}
    f(s,t)+\frac{1}{st} 
    &= \frac{1}{\pi}\int_{m_{\rm gap}^2}^{\infty} dz \frac{f_s(z,t)}{z-s} = \frac{1}{\pi}\int_{m_{\rm gap}^2}^{\infty} \frac{dz}{z} \sum_{J=0}^\infty \rho_J(z)\frac{P_J\left(1+2t/z\right)}{z(z-s)}  \\
    &\equiv \avgSTD{\frac{P_J\left(1+2t/z\right)}{z(z-s)}}\,. \nonumber
\end{align}
We emphasize that \eqref{eq:fSRIntegral} automatically implies the standard dispersion relation \eqref{eq:usualsumrule}, whereas the converse is not true.

\paragraph{Wilson coefficients}\mbox{}\\
Wilson coefficients are low-energy observables obtained by expanding the amplitude $f(s,t)$ in a Taylor series around small $s$ and $t$
\begin{align}
    f(s,t) &= -\frac{1}{st} + \sum_{0 \leq \ell \leq n/2} a_{n,\ell} (s^{n-\ell} t^\ell + s^\ell t^{n-\ell})\frac{1}{1+\delta_{n-\ell,\ell}} \\ &= -\frac{1}{st} + a_{0,0} + a_{1,0} (s+t) + a_{2,0} (s^2 + t^2) + a_{2,1} s t +  \dots\,.
\label{eq:low_enery_expansion}
\end{align}
By expanding the dipersive representation of the amplitude \eqref{eq:fSRIntegral}, we obtain the following integral representation for these coefficients
\begin{equation}\label{eq:anlSR}
    \begin{split}
        a_{n,\ell}&=\avgBS{\frac{1}{z^{2+n}}\left(\frac{z}{z_{5d}}\right)^\ell\frac{4^\ell  \Gamma (J+\ell+2) }{(J+1) \Gamma (2 \ell+2) \Gamma
   (J-\ell+1)}}\\
   &+\avgCut{\frac{1}{z^{2+n}} \left(\frac{z}{\tilde z_{5d}}\right)^\ell \frac{4^\ell\prod_{k=1}^\ell(k^2+ \nu^2)}{ \Gamma
   (2 \ell+2)} }\,,
    \end{split}
\end{equation}
where we introduced the effective five-dimensional kinematic variables $z_{5d}=\frac{z}{1 - z/(4m^2)}$ and $\tilde z_{5d}=\frac{z}{z/(4m^2) - 1}$ to streamline the notation.
In this representation, the positivity of all coefficients becomes manifest
\begin{equation}
    a_{n,\ell}\geq 0\,.
\end{equation}
The same conclusion follows from standard unitarity, which yields
\begin{equation}
    a_{n,\ell}=\avgSTD{\frac{1}{z^{2+n}}\frac{ \Gamma (J+\ell+1)}{\Gamma (\ell+1)^2 \Gamma (J-\ell+1)}} \geq 0\,.
\end{equation}
As discussed above, the enhanced dual conformal unitarity imposes a stronger constraint. A distinctive feature of the sum rules \eqref{eq:anlSR} is the presence of the $(z/z_{5d})^\ell$ factor, which as we will see in \secref{sec:analyticBounda10}, leads to nontrivial lower bounds.

\paragraph{Tree-level pole}\mbox{}\\
So far, by writing the sum rules for Wilson coefficients we have assumed that the amplitude admits an unsubtracted dispersion relation \eqref{eq:fSRIntegral} for $t \leq \mgap$. However, for $t < 0$, the amplitude instead satisfies an anti-subtracted sum rule, since $\lim_{|s|\to\infty} s f(s,t) = 0$. This follows from the fact that the leading Regge trajectory satisfies $j_0(t) < -1$ in this region. It implies that we can write a sum rule for the tree-level pole \eqref{eq:unsubtractedDR}
\begin{equation}
\label{eq:sumrulepole}
    \frac{1}{-t} = \avgBS{\frac{ \PJfive(\cos\tilde\theta_z)}{z}} + \avgCut{\frac{\Omega_\nu(\cosh\tilde \theta_z)}{z}}\,, ~~~ t<0 \ .
\end{equation}

This sum rule diverges in the forward limit. To make use of it, we smear the sum rule with a wave packet $f(q)$ supported in $0<q<q_0$, where $t=-q^2$ \cite{Caron-Huot:2021rmr}. 
More precisely, we define the functional 
\begin{equation}
    \Phi_f[g(q)]\equiv \int_0^{q_0}\frac{dq}{q_0}f(q) g(q) \,,
\end{equation}
where $q_0$ can be chosen below the two-particle cut $q_0<2m$. In practice, we take $q_0=m_{\text{gap}}$ and defer the study of the dependence on $q_0$ to future work. If we apply the functional to the sum rule \eqref{eq:sumrulepole}, we  obtain 
\begin{equation}\label{eq:tree-levelPoleSumRuleSmeared}
\begin{split}
      \Phi_f[q^{-2}] \equiv\int_0^{q_0}{dq\over q_0} \frac{f(q)}{q^2} &= \avgBS{ \frac{1}{z}\int_0^{q_0}{dq\over q_0} f(q)\, \PJfive(1-\frac{2q^2}{z_{5d}})}\\
     &+\avgCut{ \frac{1}{z}\int_0^{q_0}{dq\over q_0}  f(q)\, \Omega_\nu(1+\frac{2q^2}{\tilde z_{5d}})}\,.
\end{split}
\end{equation}
There is freedom in the choice of the smearing function $f(q)$. We will restrict ourselves to $f(q)$ which are finite degree polynomials
\begin{equation}\label{eq:functionalPoleWithbasis}
    f(q) = \sum_{k=1}^{k_{\rm max}}c_k f_k(q)\,,~~\text{with}~~f_k(q) = \frac{k }{k+1}\left(\frac{q}{q_0}\right)^2-\left(\frac{q}{q_0}\right)^3 +\frac{1}{k+1}\left(\frac{q}{q_0}\right)^{k+3}\, .
\end{equation}
This basis is chosen so that, for large $z$, the integrand $\int_0^{q_0}\,dq\, f_k(q)\,\Omega_\nu(1 + 2q^2/\tilde z_{5d})$ remains sign-definite at large $\nu$.\footnote{In contrast, in the strict $g=\infty$ limit, the choice ensuring positivity at large impact parameters is  $f_k^{\infty}(q)= (q/q_0)-(q/q_0)^{2+k}$, see \cite{Caron-Huot:2021rmr}.}
We discuss this choice in more detail in \appref{sec:numerical_truncations_details}.

Let us comment on the crucial role of dual conformal invariance for usefulness of the ${1 \over t}$ pole sum rule in four dimensions. Suppose we only knew that $j_0(t) < -1$ for $t < 0$, without invoking the $SO(4)/SO(1,3)$ unitarity structure. In that case, the functional for the pole would exist but could not be made positive for all four-dimensional impact parameters \cite{Caron-Huot:2021rmr}. %
This difficulty is avoided once the enhanced $SO(4)/SO(1,3)$ unitarity implied by dual conformal invariance is taken into account.

\paragraph{Crossing symmetry and null constraints}\mbox{}\\
Crossing symmetry, $f(s,t) = f(t,s)$, is not manifest in the dispersion relations and therefore imposes additional constraints on the spectral densities. Enforcing this symmetry, we obtain the first \emph{master null constraint}, denoted $\mathbb{X}$,
\begin{equation}\label{eq:nullmasterX}
\begin{split}
    0=\mathbb{X}(s,t) &\equiv \avgBS{\frac{\PJfive\(1+ \frac{2t}{z_{5d}}\)}{z(z-s)} - \frac{\PJfive\(1+ \frac{2s}{z_{5d}}\)}{z(z-t)}}\\
    &+\avgCut{\frac{\Omega_\nu\(1- \frac{2t}{\tilde z_{5d}}\)}{z(z-s)}- \frac{\Omega_\nu\(1- \frac{2s}{\tilde z_{5d}} \)}{ z(z-t)}}\,.
\end{split}
\end{equation}
Expanding this relation around $s,t = 0$ generates an infinite sequence of polynomial constraints, which we denote $\chi_{n,\ell} = 0$
\begin{equation}\label{eq:nullX}
    0= \sum_{n,\ell} \chi_{n,\ell} s^{n-\ell} t^{\ell}\, \ .
\end{equation}
Analogously to the Wilson coefficients \eqref{eq:anlSR}, a closed-form expression for each $\chi_{n,\ell}$ can be obtained. For example, the first null constraint is
\begin{equation}
    0=\chi_{1,0} = \avgBS{\frac{1}{z^3}\left(1- \frac{2J(J+2)}{3}\frac{z}{z_{5d}}\right)} + \avgCut{\frac{1}{z^3}\left(1- \frac{2(1+\nu^2)}{3}\frac{z}{\tilde z_{5d}}\right)}\,.
\end{equation}
Higher-order null constraints $\chi_{n,\ell}=0$ follow in the same way and encode crossing symmetry in the dispersive language.

There is a second set of null constraints. It arises when the dispersion relation is written at fixed $u$ rather than fixed $t$
\begin{align}
    f(s,t=-s-u) + \frac{1}{st}&= \frac{1}{\pi} \int_{\mgap(g)}^\infty\frac{dz}{z}f_s(z, -z-u)\left(\frac{1}{z-s} + \frac{1}{z-t}\right) \nonumber \\
    &=\avgBS{\frac{\tilde P_J\(1-\frac{2(u+z)}{z_{5d}}\)}{z}\left(\frac{1}{z-s} + \frac{1}{z-t}\right)} \\
    &+ \avgCut{\frac{\Omega_\nu\(1+\frac{2(u+z)}{\tilde z_{5d}}\)}{z}\left(\frac{1}{z-s} + \frac{1}{z-t}\right)} \ .  \nonumber
\end{align}
Equating this fixed-$u$ representation with the fixed-$t$ dispersive representation \eqref{eq:fSumRule} yields the second \emph{master null constraint}, $\mathbb{Y}$,
\begin{align}\label{eq:nullY_master}
    0=\mathbb{Y}(s,t) &= \avgBS{\frac{\PJfive\(1+ \frac{2t}{z_{5d}}\)}{z(z-s)} -\frac{\tilde P_J\(1-\frac{2(u+z)}{z_{5d}}\)}{z}\left(\frac{1}{z-s} + \frac{1}{z-t}\right) }\\
    &+\avgCut{\frac{\Omega_\nu\(1- \frac{2t}{\tilde z_{5d}}\)}{z(z-s)} -\frac{\Omega_\nu\(1+\frac{2(u+z)}{\tilde z_{5d}}\)}{z}\left(\frac{1}{z-s} + \frac{1}{z-t}\right)}\,. \nonumber
\end{align}
These constraints can likewise be expanded at small $s,t$ to produce a series of relations $\mathcal{Y}_{n,\ell}=0$. We will not make use of them in the present analysis, and leave their detailed exploration to future work.

\paragraph{Observables}\mbox{}\\
The Wilson coefficients $a_{n,\ell}$ defined above are dimensionful quantities. To compare amplitudes across different couplings, it is convenient to introduce dimensionless combinations. For each coupling $g$, it is natural to work in the units where we set $\mgap=1$. We further normalize by $a_{0,0} = \frac{g^2}{6m^4}$, which is one-loop exact by the non-renormalization theorem and provides a natural measure of the coupling. In this way, we define dimensionless ratios
\begin{equation}
\label{eq:observabledimless}
    \bar a_{n,\ell} \equiv \frac{a_{n,\ell}}{a_{0,0}}m_{\text{gap}}^{2n}\,.
\end{equation}
In this normalization, the rescaled coefficients $\bar a_{n,\ell}$ remain $\mathcal{O}(1)$ at all couplings. At weak coupling, where $a_{n,l}\sim\frac{g^2}{m^{2(2+n)}}$ and $\mgap \sim 4m^2$, we indeed find $\bar a_{n,\ell}\sim 1$. At strong coupling, using $a_{n,\ell}\sim \left(\frac{1}{\mgap}\right)^{2+n}$ and $\mgap \sim \frac{\pi m^2}{g}$, we reach the same conclusion. 

As a second observable, we consider the amplitude itself evaluated at $s,t\leq 0$. In this case, we will bound the following dimensionless ratio
\begin{equation}
\label{eq:funcitself00}
    \bar{f}(s,t) \equiv \frac{f(s,t) + \frac{1}{s t}}{a_{0,0}}\,,
\end{equation}
where we subtracted the tree-level pole so that $\bar f(0,0)=1$. More generally, we consider derivatives of the amplitude
\begin{equation}
\label{eq:funcitself}
    \bar{f}_{n,\ell}(s,t) \equiv \frac{\partial_s^n\partial_t^{n-\ell}(f(s,t) + \frac{1}{s t})}{a_{0,0}}m_{\rm gap}^{2 n}\,,
\end{equation}
which satisfy $\bar f_{n,\ell}(0,0)= \bar a_{n,\ell}$.

Finally, we also study the residues of the leading Regge trajectory, which we express as dimensionless ratios
\begin{equation}
\label{eq:threepointcouplingdless}
    \bar \lambda_{0,J}^2 = \frac{\lambda_{0,J}^2}{a_{0,0} m_{\text{gap}}^4}\,.
\end{equation}
In this normalization, the residue of the tree-level massless pole is fixed to $\frac{1}{a_{0,0} m_{\text{gap}}^4}$ at any coupling. This relation provides an additional sum rule on the spectral density. We will impose this relation as an additional constraint in our bootstrap analysis.

\subsection{Bounds on observables}

In the previous subsection we reviewed the dispersion relations and described how unitarity and crossing symmetry appear in the dispersive sum rules. We also defined dimensionless observables of interest. Here we formulate \emph{the bootstrap problem} and explain how to use the sum rules to constrain these observables.

\subsubsection{Bootstrap problems}

Let us consider a dispersive observable, denoted by $\ODisp$. We call an observable \emph{dispersive} if it can be expressed as an integral over the discontinuity of the amplitude,
\begin{equation}
  \ODisp = \avgBS{\ODisp^{SO(4)}(z,J)}+\avgCut{\ODisp^{SO(1,3)}(z,\nu)} \,,
\end{equation} 
where, on the right-hand side, we used that the discontinuity of the amplitude decomposes into a sum over non-negative partial-wave spectral densities \eqref{eq:spdpwBS} and \eqref{eq:spdpwSS}. In our case, these correspond to the $SO(4)$ sector for the bound states, with $c_J(s)\geq 0$, and the $SO(1,3)$ sector for the scattering states above the two-particle threshold, with $c_{\nu}(s) \geq 0$.
The dispersive observables discussed so far include: the Wilson coefficients $a_{n,\ell}$ \eqref{eq:anlSR}, the ${1 \over t}$ pole \eqref{eq:sumrulepole} and its smeared version $\Phi[q^{-2}]$ \eqref{eq:tree-levelPoleSumRuleSmeared}; and, of course, the amplitude itself and its derivatives \eqref{eq:funcitself}. Crossing symmetry is imposed using the null constraints $\chi_{n,\ell}=0$ \eqref{eq:nullmasterX}. Finally, the support of the bound-state spectral density $c_J(s)$, see \eqref{eq:cJDecomposition} and  \figref{fig:spectrumSchematic}, encodes our knowledge of the leading and the subleading Regge trajectories. 

We now turn to how this information can be used to derive bounds on dimensionless observables. The simplest question we can ask is: what are the minimal and maximal allowed values of $\bar a_{n,\ell}$? We can also consider pairs of Wilson coefficients and determine the allowed region in the $(\bar a_{n,\ell}, \bar a_{n',\ell'})$ plane. This can be achieved by fixing, say, $\bar a_{n',\ell'}$ and bounding $\bar a_{n,\ell}$. The same strategy naturally extends to higher-dimensional spaces of observables. In some situations, it is useful to impose inequalities of the form $ \ODisp_i\in [\ODisp_{i}^{\min}, \ODisp_{i}^{\max}]$, for instance when $\ODisp_i$ is known to lie within an uncertainty range. %
We will use such partial information in \secref{sec:bounds_using_pade_model}, where the Pad\'{e} model is considered. 

All these questions can be concisely summarized in what is called the \emph{primal problem}.
\begin{mdframed}[frametitle={Primal Problem},frametitlealignment=\centering,backgroundcolor=white, leftmargin=1cm, rightmargin=1cm, topline=true,bottomline=true, leftline=false, rightline=false] 
\vspace{-0.4cm}
\begin{align} 
&\underset{\text{in } \{c_J(s), c_\nu(s)\}}{\text{Maximize}} && \pm \ODisp\nonumber\\
& \text{subject to}  && a_{0,0}=1\nn\\ 
& && \ODisp_e = \ODisp_e^{\text{fixed}}\,,   &&e \in E \nn\\
& && \ODisp_{i}^{\min}\leq \ODisp_i\leq \ODisp_{i}^{\max}\,, &&i \in I \label{eq:Primal2}\\
& && \chi_{n,\ell}=0\,,&&(n, \ell)\in \cN \nn\\
& && c_J(s)\geq 0 \,\, &&J=\mathbb{Z}_{\geq 0}\,, \, s=m_{0,J}^2 ~ \text{and} ~ \msubJ{J}^2 \leq s< 4m^2  \nn \\
& && c_\nu(s)\geq 0 \,\, && \nu\geq 0 \,,~ s \geq  4m^2\nn
\end{align}
\end{mdframed}
Here and below we set $\mgap=1$ for convenience. 
The first line defines the primal objective $\pm \ODisp$, where the two signs correspond to maximizing or minimizing the observable of interest.
Fixing $a_{0,0}=1$ conveniently normalizes the amplitude, consistent with the definition of our dimensionless observables \eqref{eq:observabledimless}.
Next, observables in the set $E$ are fixed, while those in the set $I$ are constrained to lie within prescribed intervals.
The $\chi_{n,\ell}$ with $(n,\ell)\in \cN$ denote a finite subset of imposed null constraints, which we truncate in practice to $n\leq n_{\text{max}}$ and $\ell=0,...,\lfloor\frac{n+1}{2}\rfloor$, i.e. to all non-trivial $\ell$. 

In the primal formulation, the optimization variables are the spectral densities $c_J(s)$ and $c_\nu(s)$, which parametrize the imaginary part of the amplitude. 
To perform the optimization, one must choose a parameterization of $c_J(s)$ and $c_\nu(s)$.
The larger the function space considered, the greater the range of the observables that can be explored, thus, \emph{ruling in} possible values.\footnote{This approach has been pursued in a similar context in \cite{Li:2023qzs,Haring:2023zwu, Eckner:2024ggx}.} 
In this work we do not pursue this (primal) approach; instead we turn to its \emph{dual} formulation. In the dual formulation, our goal is to \emph{rule out} parts of the parameter space. 
A key advantage of the dual formulation is that such exclusions are rigorous at any finite level of truncation. Given a primal problem, there is a standard way to arrive at the corresponding dual problem. 
This is done by constructing the Lagrangian and integrating out the primal variables. For the primal problem defined above, the corresponding dual reads, see \appref{app:detailPrimalDual}.
\begin{mdframed}[frametitle={Dual problem},frametitlealignment=\centering,backgroundcolor=white, leftmargin=0cm, rightmargin=0cm, topline=true,bottomline=true, leftline=false, rightline=false] 
\vspace{-0.4cm}
\begin{align} 
&\underset{\text{in } \cD=\{\Lambda, f_e, u_i, d_i, \kappa_{n,\ell} \}}{\text{Minimize}} && \bar \ODisp_{\pm} = \Lambda +\sum_{e\in E} f_e \ODisp_e^\text{fixed} + \sum_{i\in I} (u_i \ODisp_i^{\max}+ d_i \ODisp_i^{\min})  \nonumber\\
& \text{subject to}  &&\cP_\pm^{SO(4)}(z,J)\geq 0 \,,~~~~~ \text{for} \,\, J=\mathbb{Z}_{\geq 0}\,,  z= m_{0,J}^2  \text{ and }  \msubJ{J}^2\leq z< 4m^2 \nn \\[10pt]
& &&\cP_\pm^{SO(1,3)}(z,\nu) \geq 0 \,,~~~\,\text{for} \,\, \nu\geq 0 \,, z\geq 4m^2\nn \\[10pt]
& \text{and} &&  u_i\geq 0\,,~~ d_i\leq 0 \,, ~~~~~~\,\, i\in I\,\label{eq:dualWilsonCoefMax}\\[10pt]
& \text{where}  &&\cP_\pm(z,\cdot)=\Lambda a_{0,0}(z,\cdot)\mp \ODisp(z,\cdot)+\sum_{(n,\ell)\in \cN}\kappa_{n,\ell}\chi_{n,\ell}(z,\cdot) \nn \\
& &&~~~~~~~~~~\,
+ \sum_{e\in E} f_e \ODisp_e(z,\cdot) +\sum_{i\in I} (u_i+d_i) \ODisp_i(z,\cdot) \,.
\nn
\end{align}
\end{mdframed}
In the first line, we defined the dual objective $\bar \ODisp_{\pm}$, which is related to the primal objective through the weak duality relation, see e.g. \cite{Guerrieri:2021tak},
\begin{equation}
    \max_{(c_J(z), c_\nu(z))} (\pm \cA) \leq \min_{\cD} \bar \ODisp_{\pm}\,, 
\end{equation}
where the collection $\cD$ denotes the set of \emph{dual variables}. This inequality expresses the standard weak duality: the maximal value of the primal objective never exceeds the minimal value of the dual. It illustrates the complementary ‘ruling-in’ versus ‘ruling-out’ nature of the primal and dual formulations. 
The sign constraints on $u_i$ and $d_i$ follow from the positivity of the Lagrange multipliers associated with the inequality bounds in the primal problem.
In the final line, we defined the dual functional $\cP_\pm(z,\cdot)$, which acts on both the $SO(4)$ and $SO(1,3)$ partial waves, and $(\cdot)$ stands for $J$ and $\nu$ respectively.

Above, we described the primal/dual problems where the objective is a dispersive observable $\cA$. Similarly, one can also maximize the three-point coupling $\bar \lambda_{0,J}^2$. The corresponding primal and dual formulations are presented in \appref{sec:3pt_primal_dual_problems}.\footnote{This is analogous to OPE-coefficient maximization in the CFT bootstrap literature  \cite{Caracciolo:2009bx}.}

\subsubsection{Truncation of constraints on the dual functional}
\label{sec:truncation_constraints_dual}
The dual formulation can be written conveniently in vector form, since all constraints are linear in the dual variables~$\cD$. Explicitly, the dual objective takes the form 
\begin{equation}
  \bar \ODisp_{\pm} = \cD \cdot \vec{v}_{\bar \ODisp_{\pm}},
\end{equation}
where $\vec v_{\bar \ODisp_{\pm}}$ collects the coefficients multiplying the dual variables in the objective function
\begin{equation}
  \vec{v}_{\bar \ODisp_{\pm}} = (1, \vec{\cA}^{\rm fixed}, \vec{\cA}^{\rm max}, \vec{\cA}^{\rm min}, \vec{0}) \, .
\end{equation}
Similarly, the dual functional $\cP_\pm(z,\cdot)$ can be expressed as a scalar product between the dual variable vector $(\pm1,\cD)$ and the kinematic vector $\vec v(z,\cdot)$ 
\begin{equation}
  \cP_{\pm}(z, \cdot) = (\pm 1, \cD)\cdot \vec{v}(z, \cdot) \, .
\end{equation}
This vectorized formulation allows for efficient numerical implementation using standard linear-programming methods. 
Up to this point, we have described the exact, infinite-dimensional problem that we wish to solve. 
In practice, the dual functional $\cP(z,\cdot)$ is subject to infinitely many positivity constraints, which cannot be imposed directly in numerical computations, hence a truncation must be introduced.
When the problem involves a single continuous variable on a half-line and the constraints are polynomial, the positivity conditions can be handled using semidefinite programming, as implemented in the solver SDPB \cite{Simmons-Duffin:2015qma, Landry:2019qug} that we employ.\footnote{In such cases, positivity can be verified using rational arithmetic, see e.g.\ \cite{Kravchuk:2021akc}.} In our problem, the $SO(4)$ bound-state region involves the variables $(J,z)$, with $J$ discrete and $z$ continuous. In the $SO(1,3)$ scattering-state region, both variables $(\nu,z)$ are continuous. Thus, in both cases, a truncation must be applied to at least one of the variables, and particular care must be taken to control the asymptotic regimes.
We will now explain, separately for the two regions, the truncation procedure and the treatment of asymptotic regimes, followed by an explanation of the adaptive-refinement strategy. Detailed implementation are presented in \appref{sec:numerical_truncations_details}.

 \paragraph{$SO(4)$ region (bound states):}\mbox{}\\
We now discuss the truncation strategy and asymptotic control in the $SO(4)$ (bound-state) sector. For the leading trajectory $j_0$, the constraints are pointwise in spin and must be truncated in number. We impose them for $J=0,1,\ldots,J_{\rm Large}$ with $J_{\rm Large}=\cO(10^3)$.
For the subleading trajectories, the dual functional $\cP^{SO(4)}(z,J)$ becomes a polynomial (or rational function) of $z$ at fixed spin $J$.\footnote{At fixed spin we have $\cP^{SO(4)}(z,J) = \tfrac{P(z,J)}{Q(z,J)}$ with $Q(z,J)>0$ over the support of $z$. Hence the positivity condition $\cP^{SO(4)}(z,J)\geq 0$ is equivalent to the polynomial constraint $P(z,J)\geq 0$.}
To simplify the analysis, we use the mapping
\begin{equation}
    z_{5d}= \frac{z}{1-\frac{z}{4m^2}}\,,
\end{equation}
which transform the interval $z\in[m_{1,J}^2,4m^2]$ onto the half-line $z_{5d}\geq m_{1,J}^2\big/\big(1-\tfrac{m_{1,J}^2}{4m^2}\big)$. This leaves us with a truncation in $J$. We impose the constraints for $J=1,\ldots,J_{\rm max}$ with $J_{\rm max}=\cO(50)$, and add additional sparse constraints up to $J_{\rm Large}$.

\begin{figure}
    \centering
\includegraphics[width=0.8\linewidth]{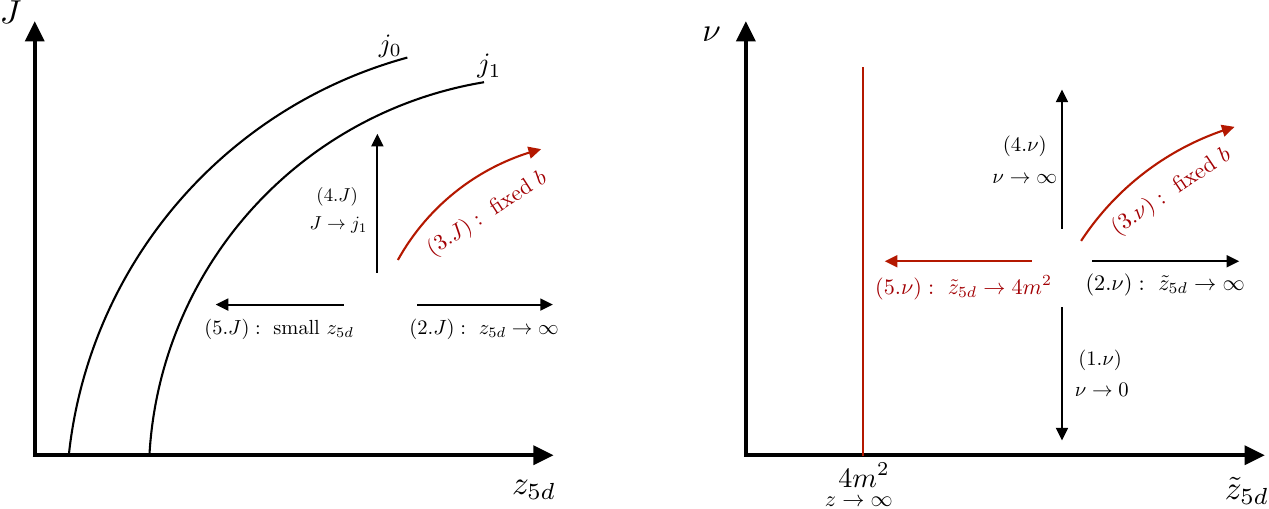}
    \caption{Asymptotic regions in the parameter space for both the $SO(4)$ and $SO(1,3)$ sectors. Regions where additional constraints are applied are highlighted in red.}
\label{fig:j_x_nu_x_regions}
\end{figure}

The following asymptotic regimes must be analyzed and controlled:
\begin{enumerate}
    \renewcommand{\theenumi}{(\roman{enumi})}
    \renewcommand{\labelenumi}{\theenumi}
    \item Large $z_{5d}$ at fixed $J$ (region $2.J$ in \figref{fig:j_x_nu_x_regions});
    \item Small $z_{5d}$ at fixed $J$, or equivalently large $J$ at fixed $z_{5d}$, both approaching the subleading trajectory $j_1(s)$ (regions $4.J$ and $5.J$);
    \item The regime where both $z_{5d}$ and $J$ are large (region $3.J$).
\end{enumerate}
In the first two cases, nothing special occurs in the sum rules, so we now focus on the third. To study the region of simultaneously large $z_{5d}$ and $J$, we introduce the impact parameter $b$ by taking $z_{5d},J\gg 1$ while keeping fixed the ratio\footnote{This choice of impact parameter is natural since  $\cos\tilde\theta_z = 1+\frac{2t}{z_{5d}} $, see \eqref{eq:cosWith5dEnergy}.}
\begin{equation}\label{eq:impact_b_def}
    b = \frac{2J}{\sqrt{z_{5d}}}\,.
\end{equation}
 The subleading trajectory behaves as $j_1\sim \sqrt{z_{5d}}$, which imposes a bound $b\leq b_*$. In impact-parameter space, both the pole sum rule $\Phi_f[q^{-2}]$ and the amplitude $f(s,t)$ exhibit oscillatory behavior, since they are evaluated away from the forward limit. Schematically they take the form 
\begin{equation}
    A(b)+ B(b)\cos\phi(b)+ C(b)\sin\phi(b).
\end{equation}
Here $A(b)$, $B(b)$, and $C(b)$ are smooth functions capturing the slowly varying envelope of the oscillations. Consequently, ensuring positivity in this regime requires adding explicit constraints at fixed values of the impact parameter, see \appref{sec:numerical_truncations_details} for details.

\paragraph{$SO(1,3)$ region (scattering states):}\mbox{}\\
We now turn to the truncation in the $SO(1,3)$ (scattering-state) sector, where the analysis is complicated by the presence of two continuous variables. In this region, there are two continuous variables, $(z,\nu)$ (or equivalently $(\tilde z_{5d},\nu)$). While the sum rules for the Wilson coefficients $a_{n,\ell}$ \eqref{eq:anlSR} and the null constraints $\chi_{n,\ell}$ \eqref{eq:nullX} are polynomial (or rational) in these variables, the sum rules for the pole $\Phi[q^{-2}]$ and the amplitude $f(s,t)$ are non-polynomial.

We therefore distinguish two cases. First, if these non-polynomial sum rules are not included in the dual functional, we can follow the same strategy as in the $SO(4)$ region, except that we must also discretize the continuous spin variable $\nu\geq 0$. Second, if the non-polynomial sum rules are included, we may either approximate them by polynomials or discretize the energy variable $\tilde z_{5d}$.
In both cases, the spin variable $\nu$ is discretized up to $\nu\leq \nu_{\rm Large}$, using a grid that becomes denser as $\nu\to 0$ (corresponding to the region $(1.\nu)$ in \figref{fig:j_x_nu_x_regions}). When discretizing $\tilde z_{5d}$, we employ a Chebyshev grid. Further details on the discretization grids are provided in \appref{sec:numerical_truncations_details}.

We now discuss the asymptotic regimes in the $SO(1,3)$ sector. For large $\tilde z_{5d}$ at fixed $\nu$ (region $(2.\nu)$ in \figref{fig:j_x_nu_x_regions}) and large $\nu$ at fixed $\tilde z_{5d}$ (region $4.\nu$), nothing special occurs. In the double-scaling limit $\tilde z_{5d},\nu\gg 1$, the dual functional in impact-parameter space (region $3.\nu$) reduces to that of the $SO(4)$ sector, except that the positivity condition must be imposed for all $b\geq 0$. Finally, in the limit $\tilde z_{5d}\to 4m^2$ (corresponding to $z\to\infty$) at fixed $\nu$ (region $5.\nu$), the pole sum rule dominates, as it is defined through an anti-subtracted dispersion relation \eqref{eq:sumrulepole}. Since this sum rule oscillates in $\nu$, we add explicit constraints on the dual functional in the limit $\lim_{z\to\infty} z \cP^{SO(1,3)}(z,\nu)$ to control these oscillations.  Ensuring positivity of this sum rule at large $\nu$, fixes the choice of basis functions $f_k$ in \eqref{eq:functionalPoleWithbasis} (see \appref{sec:numerical_truncations_details} for details). If the pole sum rule is omitted, we instead impose constraints on $\lim_{z\to\infty} z^2 \cP^{SO(1,3)}(z,\nu)$. 

These truncation and asymptotic-control conditions ensure that the numerical implementation of the dual constraints remains stable across all kinematic regimes.

\paragraph{Adaptive refinement:}\mbox{}\\
When solving the dual bootstrap problem \eqref{eq:dualWilsonCoefMax} with a fixed truncation (i.e. a given choice of grids in $(z_{5d},J)$, $(\tilde z_{5d},\nu)$, and $b$), we obtain a candidate dual vector $\cD_0$, which depends on this choice. We then identify regions where the dual functional $\cP_0(z,\cdot)$ is negative. If such regions are found, we add additional sampling points to the grids and re-solve the dual bootstrap problem with these extra constraints, yielding a new candidate dual vector $\cD_1$. This procedure is iterated until the negative regions are numerically negligible. Further implementation details are provided in \appref{sec:adptative_refinement}.

\subsection{Analytic bound on $\bar a_{1,0}$ at weak coupling $g\to0$}\label{sec:analyticBounda10}

We now illustrate the power of the dual formulation by deriving an explicit bound on the first Wilson coefficient $\bar a_{1,0}$ in the weak-coupling limit.

We start by considering the simplest problem, which consists of minimizing $\bar a_{1,0}$ subject to a single null constraint. We further take the weak-coupling limit $g\to 0$, where the gap is $\mgap= 4m^2$ and the spectrum is reduced to the $SO(1,3)$ sector, see \figref{fig:spectrum_comparison}. In this case, the dual problem \eqref{eq:dualWilsonCoefMax} simplifies to 
\begin{align}
    \min_{\text{in}~\kappa}(\Lambda)~~\text{s.t.}~~\cP^{SO(1,3)}(x,\nu)=\Lambda  (x+1)+1+ \kappa  \left(1-\frac{2}{3} \left(\nu ^2+1\right)
   x\right)\geq 0\,,  ~~ \forall x\geq 0\,,
\end{align}
where we defined the dimensionless variable $z = 4m^2(1+x)$ and multiplied the entire equation by an overall positive factor of $z^3$ to remove denominators. 

The optimal solution is 
\begin{equation}\label{eq:analytic_dual_sol}
    \kappa=  -\frac{3}{5}\,,~~ \Lambda = -\frac{2}{5} \,, ~~ \cP(x,\nu) = \frac{2}{5}\nu^2 x \geq 0\,,
\end{equation}
which in turn implies that 
\begin{equation}
    \bar a_{1,0}\Big|_{g=0} = \frac{a_{1,0}}{a_{0,0}}4m^2 \geq \frac{4}{10}\,.
\end{equation}
The minimal value coincides exactly with the one-loop result in $\mathcal{N}=4$ SYM. This shows that in the weak-coupling limit the $\mathcal{N}=4$ amplitude is \emph{extremal}, sitting at a special point in the space of low-energy observables. 
Furthermore, from the zeros of the dual functional, we can read off the possible spectrum of the amplitude.
In this case, the allowed spectrum is $x=0$ or $\nu=0$. Comparing with the one-loop amplitude, we find that it has support only at $\nu=0$, showing that the bootstrap correctly identified the one-loop amplitude.

Note that the bound derived here is fully rigorous. Since we are working in the dual formulation, any truncation in the size of the dual vector $\cD$ (i.e. in the number of null constraints imposed) leads to rigorous exclusion regions. Moreover, it is straightforward to verify that the dual functional \eqref{eq:analytic_dual_sol} is non-negative, so the subtleties discussed in \secref{sec:truncation_constraints_dual} do not apply.

This result is to be contrasted with the result obtained using $SO(3)$ partial waves. This problem has been largely studied in the literature (see e.g. \cite{Berman:2023jys} in the context of $\cN=4$ SYM) and leads to $\bar a_{1,0}\geq 0$. In this case, the vanishingly small observable can be interpreted as arising from an amplitude whose spectrum is pushed far into the UV. When using $SO(1,3)$ partial waves with the $SO(1,3)$ scattering angle \eqref{eq:SO31angle}, this is no longer possible, and the resulting bound becomes non-trivial.

\section{Bootstrap bounds on finite-coupling observables}\label{sec:Universal_bounds}
In the previous sections, we analyzed the scattering amplitude in the weak- and strong-coupling regimes and used integrability to compute part of the spectrum at finite coupling. 
Our next goal is to understand the observables defined above at finite coupling $g$.
In particular, in this section, we derive the bootstrap bounds on various observables, such as the Wilson coefficients and the three-point coupling. %
We obtain these bounds by solving the dual problem \eqref{eq:dualWilsonCoefMax}.
This extends the analytic bound derived in \secref{sec:analyticBounda10} to a broader class of observables.

\subsection{Limiting regimes}

Before turning to finite coupling, it is instructive to examine the bounds on Wilson coefficients in the weak- ($g \to 0$) and strong-coupling ($g \to \infty$) limits, where the position of the $\cN = 4$ amplitude is known.
\begin{figure}[h!]
    \centering
\includegraphics[width=\linewidth]{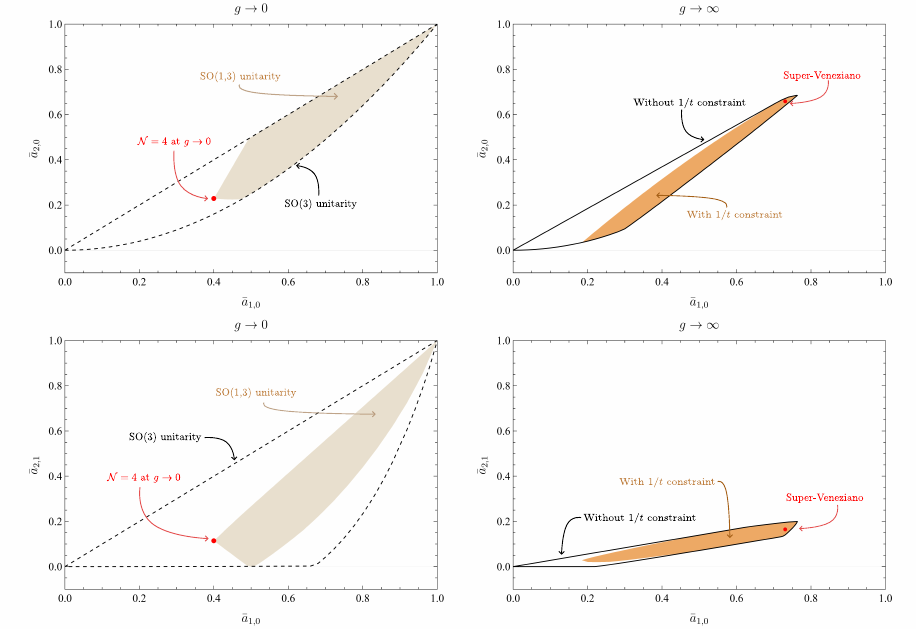}
    \caption{Exclusion plots for $(\bar{a}_{1,0}, \bar{a}_{2,0})$ and $(\bar{a}_{1,0}, \bar{a}_{2,1})$ at the extremal couplings $g \to 0$ and $g \to \infty$. 
    In the $g\to0$ plots (left panels) we highlight the difference between results obtained with (light brown) and without (black dashed) dual conformal invariance.
    The 1-loop box amplitude \eqref{eq:weak1loop} is located at the bottom-left kink.
    In the $g \to \infty$ plots (right panels) we show the bounds with (orange) and without (black) the pole constraints. Both sets of bounds are obtained using $SO(4)$ partial waves. The super-Veneziano amplitude lies close to the upper-right kink.}
    \label{fig:a10_a20_g=0}
\end{figure}
The bounds are shown in \figref{fig:a10_a20_g=0}. In particular, we show exclusion plots for $(\bar{a}_{1,0}, \bar{a}_{2,0})$ and $(\bar{a}_{1,0}, \bar{a}_{2,1})$ at the extremal couplings $g \to 0$ and $g \to \infty$. The plots were generated with $n_{\rm max}=12$ and, when the pole sum rule was imposed, $k_{\rm max}=6$ . Increasing the size of the dual functional had a negligible impact on the results. Unless stated otherwise, all bounds shown below use these parameters. The $g \to 0$ bounds are first obtained using $SO(3)$ partial waves (black dashed), and then refined by imposing dual conformal invariance, corresponding to $SO(1,3)$ partial waves (light brown). Notice that, also for $(\bar a_{1,0}, \bar a_{2,1})$, the one-loop amplitude lies at the lower-left kink, as expected from the analysis in the previous section. For the $g \to \infty$ bounds, we use $SO(4)$ partial waves and compare results with (in orange) and without (in black) the pole constraints. The pole constraints shift the lower-kink from $(0,0)$, the 1-loop result, to a finite value. The super-Veneziano amplitude lies close to the upper-right kink.

\begin{figure}[h!]
    \centering
    \includegraphics[width=\linewidth]{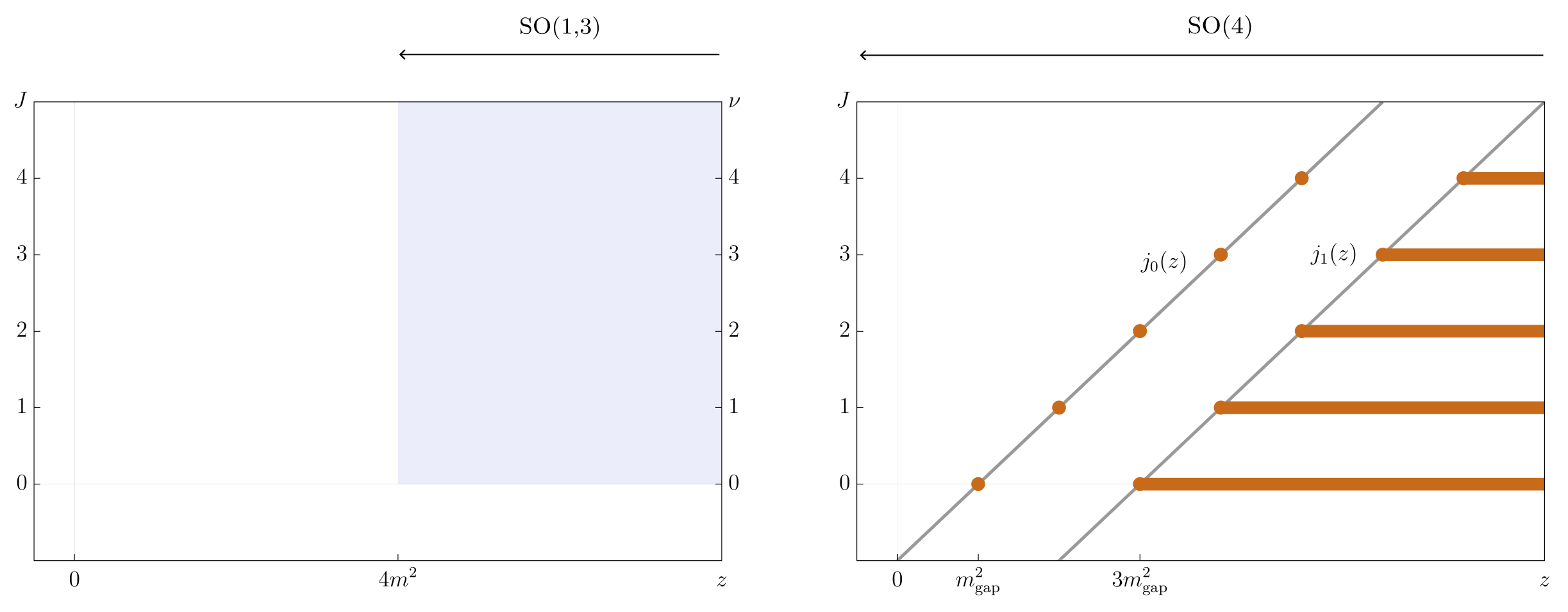}
    \caption{Spectra at weak and strong coupling. Left: $g \to 0$, with no bound states and only a two-particle cut. Right: $g \to \infty$, exhibiting $SO(4)$ bound states on linear Regge trajectories. The right panel also illustrates the spectral assumptions used to generate the bounds in this limit, shown in \figref{fig:a10_a20_g=0}.}
    \label{fig:spectrum_comparison}
\end{figure}

The $g \to 0$ and $g \to \infty$ are two limiting regimes governed by very different physical dynamics. 
In particular, the spectra differ drastically between the two limits, as shown in \figref{fig:spectrum_comparison}. In the $g\to0$ limit, the amplitude contains only a two-particle cut with no bound-state contributions, whereas at $g\to\infty$ it consists solely of $SO(4)$ bound states lying on linear Regge trajectories.

It is clear from \figref{fig:a10_a20_g=0} that the lower-left kink, where the one-loop amplitude lies, is cut-dominated, while the upper-right kink, close to the super-Veneziano point, is bound-state dominated. 
We therefore expect the finite-$g$ amplitude, which smoothly interpolates between these two regimes, to move from the lower-left to the upper-right kink as $g$ increases.

\begin{figure}[h!]
    \centering
\includegraphics[width=\linewidth]{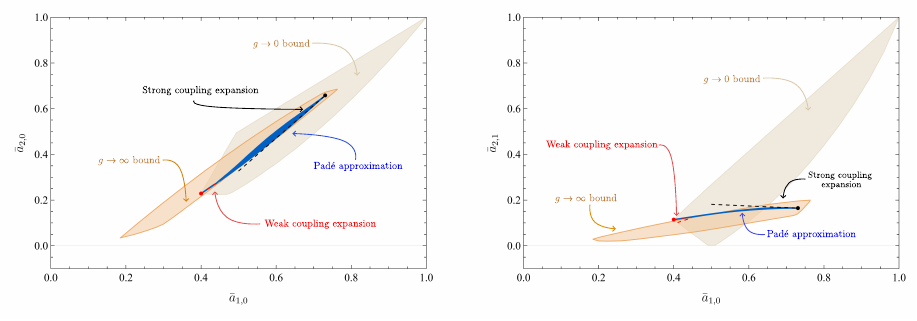}
    \caption{Comparison of the bounds at $g \to 0$ (light brown) and $g \to \infty$ (dark brown), together with the weak-coupling (red) and strong-coupling (black) expansions for $(\bar{a}_{1,0},\bar{a}_{2,0})$ and $(\bar{a}_{1,0},\bar{a}_{2,1})$. 
    A Pad\'{e} approximation (blue, with $\mu = 2$, see \eqref{eq:pade_band_def}) provides a possible prediction for the Wilson coefficients, smoothly interpolating between the two extremal regimes.
    The weak-coupling expansion (dashed red) is shown for $g\leq 0.2$, while the strong-coupling expansion (dashed black) corresponds to $g \geq 0.4$.}
    \label{fig:exp1}
\end{figure}

To gain intuition, we construct Pad\'{e} approximants that interpolate between the weak- and strong-coupling results. The construction is explained in detail in the next section. %
These approximants serve as non-rigorous priors, or heuristic expectations, for the finite-coupling amplitude, as illustrated in \figref{fig:exp1}. We now turn to a quantitative analysis of the bootstrap bounds at finite coupling.

\subsection{Wilson coefficients}

In this section we present exclusion plots for pairs of Wilson coefficients at several values of $g$, ranging from weak to strong coupling.
The first Wilson coefficients appearing in the low-energy expansion of the amplitude \eqref{eq:low_enery_expansion}, after the non-renormalized $a_{0,0}$, are $a_{1,0}$, $a_{2,0}$, and $a_{2,1}$. We first bound $\bar{a}_{1,0}$ as a function of $g$ for $0 \leq g \leq 7$, and then study the two-dimensional exclusion plots for $(\bar{a}_{1,0} ,\bar{a}_{2,0} )$ and  $(\bar{a}_{1,0} ,\bar{a}_{2,1} )$ with $1/4 \leq g \leq 3$. As explained in the previous section, the dimensionless ratios $\bar{a}_{n,\ell}$ are defined so as to remain ${\cal O}(1)$ at any coupling. The bounds at $g = 1/4 <g_c$ (where $g_c\approx 0.28$, see \figref{fig:residue_j1}) are obtained by imposing that the bound-state spectrum follows only the leading Regge trajectory, as detailed in \secref{sec:subleading_trajectory}. For $g \geq g_c$, the corresponding spectrum constraint is shown in \figref{fig:spectrumSchematic}.

\begin{figure}[t!]
    \centering
    \includegraphics[width=0.8\linewidth]{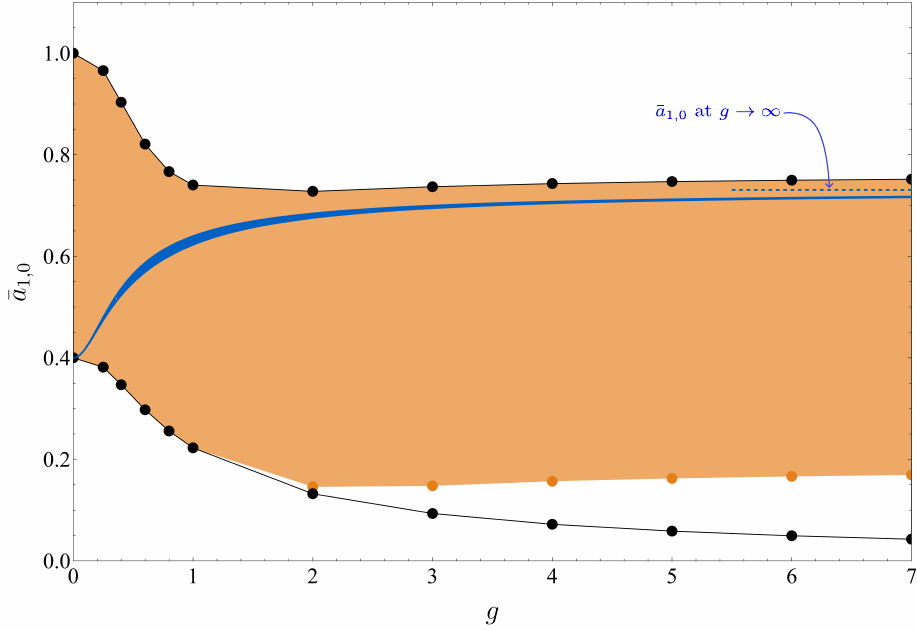}
    \caption{%
    Exclusion plot for $\bar a_{1,0}$ as a function of the coupling $g$.
    Black: bounds derived without imposing the pole constraints.
    Orange: bounds with pole constraints included.
    Blue: Pad\'{e} band of $\bar a_{1,0}(g)$ from \secref{sec:Pade model for Wilson coefficients} with $\mu = 1$, see \eqref{eq:pade_band_def}.
    The asymptotic $g \to \infty$ value is indicated by the dashed blue line.}
    \label{fig:a10vsg}
\end{figure}

We begin by discussing the bounds on $\bar a_{1,0}$ shown in \figref{fig:a10vsg}.
Figure \ref{fig:a10vsg} compares the bounds obtained without the pole constraints (black) and with them included (orange, using $k_{\text{max}}=6$). 
As the coupling increases, the pole constraints have an increasingly pronounced effect on the lower bound of $\bar a_{1,0}$.
It is straightforward to verify that, without the pole constraints, the minimum of $\bar a_{1,0}$ coincides with the one-loop result for all values of $g$.\footnote{Recall that $\bar a_{1,0}|_{1-\text{loop}} = \frac{2}{5}\frac{\mgap}{4 m^2}$, which is not constant in $g$, see \secref{sec:mgap_section}.} 
The change in the lower bound of $\bar a_{1,0}$ upon adding the pole constraint $\Phi[q^{-2}]$ \eqref{eq:tree-levelPoleSumRuleSmeared} indicates that, as the ratio between the residue of the tree-level pole and the non-renormalized $a_{0,0}$ is varied, the one-loop amplitude ceases to be an allowed solution.
For the upper bound, no difference is observed between the cases with and without pole constraints.
This behavior is expected: using standard unitarity in $d=10$, \cite{Albert:2024yap} showed that the upper bound on $\bar a_{1,0}$ is independent of the allowed pole residue.
In blue we show the Pad\'{e} band obtained from the conformal-[2,2] Pad\'{e} approximation described in \secref{sec:Pade model for Wilson coefficients}.
As explained in \secref{sec:analyticBounda10}, at $g \to 0$, %
the lower bound on $\bar a_{1,0}$ coincides with its value in $\cN=4$ SYM. 
At finite coupling this equality no longer holds, but the constructed Pad\'{e} band remains consistent with the bootstrap bounds.
Moreover, for $g \gtrsim 5$ the upper bound approaches the strong-coupling result given by the super-Veneziano amplitude together with the $1/g$ correction of \secref{sec:Worldsheet bootstrap}.

\begin{figure}[h!]
    \centering
\includegraphics[width=\linewidth]{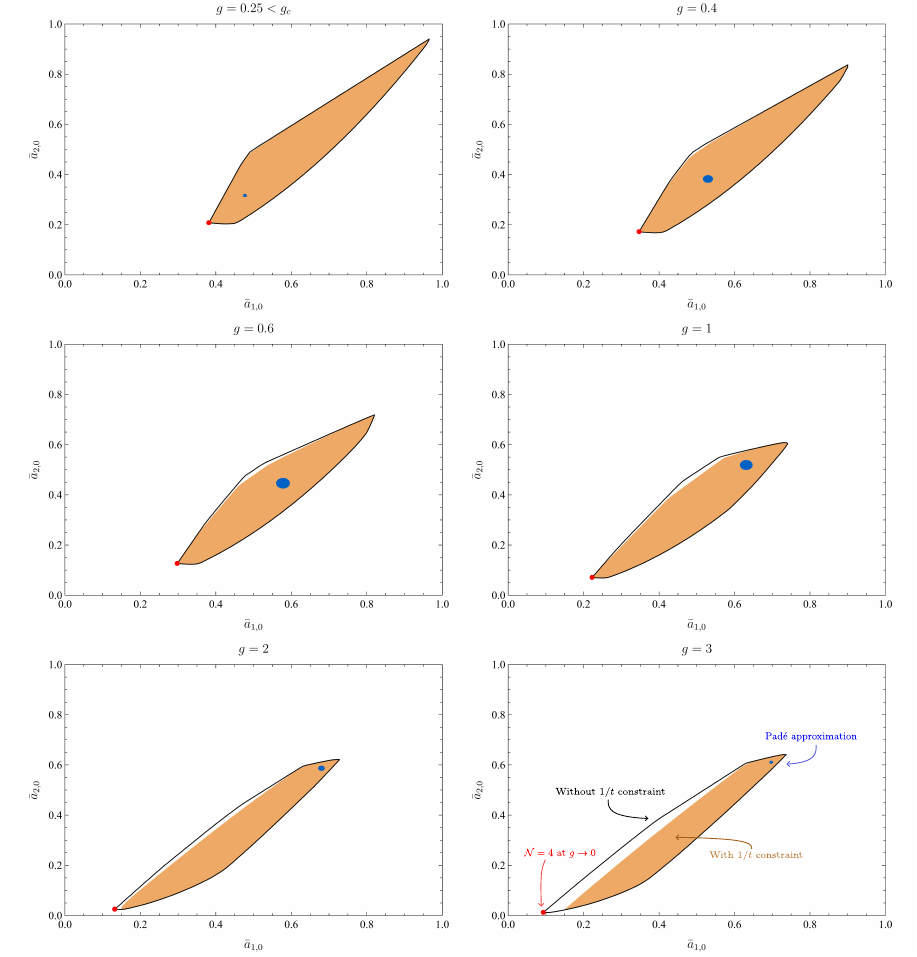}
    \caption{
    Exclusion plot for $(\bar a_{1,0},\bar a_{2,0})$ at various values of the coupling $g$.
    The black contour shows the allowed region obtained without the pole constraints, while the orange contour includes them. The red dot marks the one-loop box amplitude, corresponding to the $g \to 0$ limit. The blue ellipse indicates the Pad\'{e}-model prediction with $\mu = 2$, see \eqref{eq:pade_band_def}.}
    \label{fig:a20vsa10}
\end{figure}

\begin{figure}[h!]
    \centering
\includegraphics[width=\linewidth]{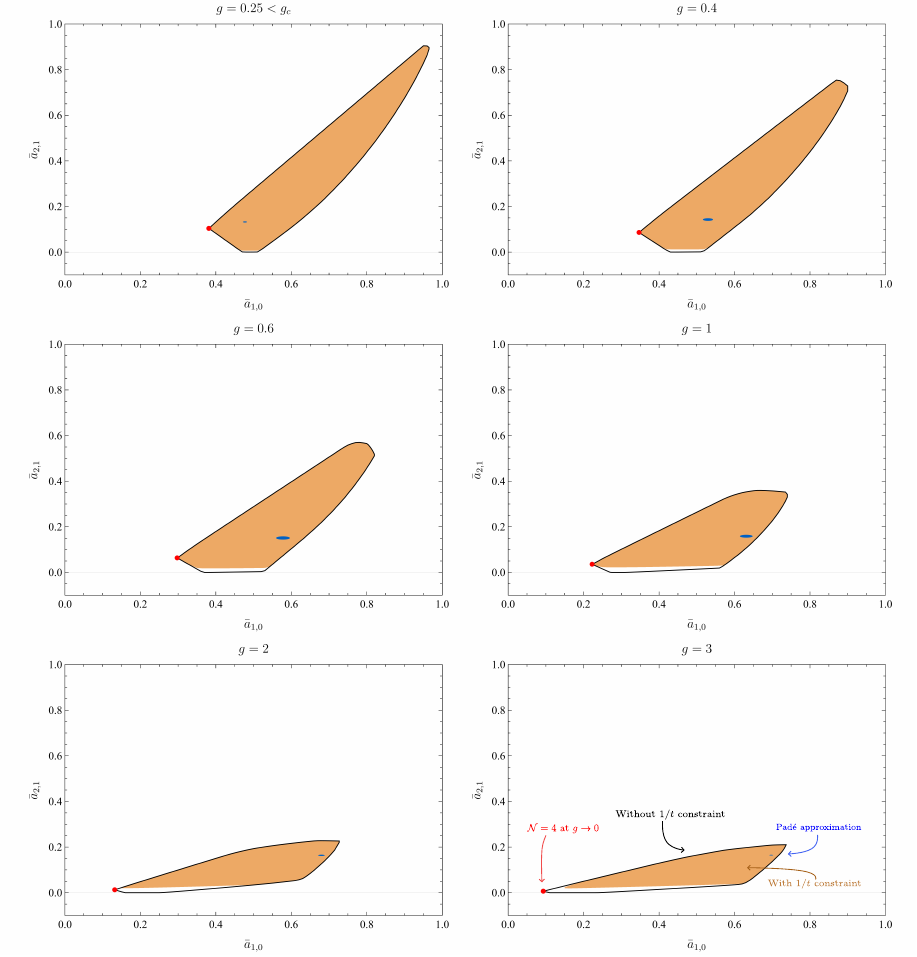}
    \caption{
    Exclusion plot for $(\bar{a}_{1,0},\bar{a}_{2,1})$ at various values of the coupling $g$.
    The black contour shows the allowed region obtained without the pole constraints, while the orange contour includes them. The red dot marks the one-loop box amplitude, corresponding to the $g \to 0$ limit. The blue ellipse indicates the Pad\'{e}-model prediction with $\mu=2$, see \eqref{eq:pade_band_def}.}
    \label{fig:a21vsa10}
\end{figure}

We now discuss the bounds on the regions $(\bar a_{1,0},\bar a_{2,0})$ and $(\bar a_{1,0},\bar a_{2,1})$, shown respectively in \figref{fig:a20vsa10} and \figref{fig:a21vsa10}, in comparison with \figref{fig:a10vsg}.
Each plot displays the bounds derived without pole constraints (black contour) and with them included (orange, $k_{\text{max}}=6$).
It is useful to emphasize the main differences between the two cases. In the plots without the pole constraints, the kink at the minimum corresponds to the one-loop box, as in the case of $g \to 0$ discussed in \secref{sec:analyticBounda10}. 
This indicates that the bound states do not contribute significantly, since the one-loop box is cut-dominated (see \figref{fig:a10_a20_g=0}).
When the pole constraints are imposed, the lower kink shifts away from the one-loop value, as already observed in the \figref{fig:a10vsg}.
This signal that the extremal amplitude contains more structure.
The upper-right part of the allowed region remains essentially unaffected by the pole constraints.
The shape of the Regge trajectories, and hence the imposed spin constraints, also have a strong impact on the upper-right region of the bounds.
In particular, the point $(1,1)$ is excluded in both plots. This point would correspond to an amplitude of the form $\frac{1}{(s-1)(t-1)}$, featuring an infinite tower of spins at $m_{\text{gap}}^2=1$. Imposing the spin constraints that each spin-$J$ bound state satisfies $z\geq m_{0,J}^2>m_{\text{gap}}^2$ rules out this amplitude. 
As a final and important remark, all plots also display (in blue) the prediction from the conformal-Pad\'{e} approximation of \secref{sec:Pade model for Wilson coefficients}, with $\mu = 2$, see \eqref{eq:pade_band_def}.
As the coupling increases, the Pad\'{e} estimate  moves from the lower-left corner (the one-loop box) to the upper-right corner, close to the super-Veneziano point at strong coupling. 
Moreover, we observe that the Pad\'e approximation is always included in the projective allowed region.

\subsection{Three-point coupling $\bar{\lambda}_{0,0}^2$}

We now turn to a different observable, namely the three-point coupling of the first bound state $\bar{\lambda}_{0,0}^2$, studied as a function of the coupling in the range $0 \leq g \leq 7$. This bound is presented later in \secref{sec:bounds_using_pade_model}  in order to be compared with the same bound derived using the Pad\'{e} model (see black/orange curves in \figref{fig:res00_vs_g_allTogether}). 
Here we focus on the bound on $\bar\lambda_{0,0}^2$ as a function of $\bar a_{1,0}$ at fixed intermediate coupling $g=1$.

\begin{figure}[h!]
    \centering
    \includegraphics[width=0.7\linewidth]{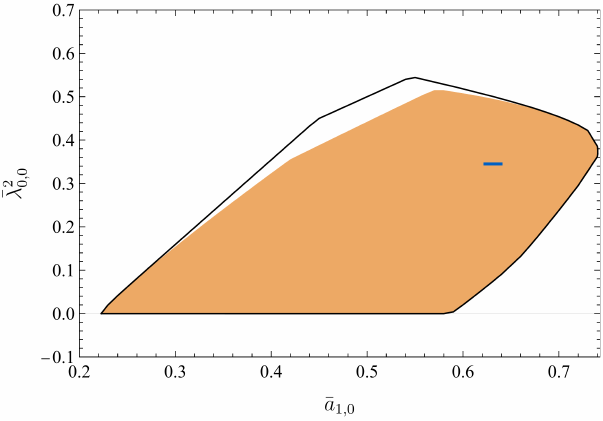}
    \caption{%
    Exclusion plot for $\bar\lambda_{0,0}^2$ as a function of the Wilson coefficient $\bar a_{1,0}$ at fixed intermediate coupling $g=1$.
    Black: bounds obtained without pole constraints.
    Orange: bounds with pole constraints included.
    Blue: Pad\'{e}-model prediction with $\mu = 1$, see \eqref{eq:pade_band_def}. The size in the $\bar \lambda_{0,0}^2$ direction has no meaning as we did not create an error estimate for this observable.}
    \label{fig:resvsa10}
\end{figure}

The bound, shown in \figref{fig:resvsa10}, exhibits an interesting structure.
A nontrivial lower bound appears at large values of $\bar a_{1,0}$, both without pole constraints (black) and with them included (orange).
This behavior is expected, since $\bar a_{1,0}$ represents the ratio between two moments of the integrated cross section. Increasing $\bar a_{1,0}$ forces the cross section to become larger near $z = \mgap$. Because there is a gap to the next bound state, the resulting lower bound on the three-point coupling $\bar \lambda_{0,0}^2$ becomes nontrivial.
As expected at this value of $g$, close to the minimum of the $\bar a_{1,0}$ corresponding to the one-loop box, the allowed value of the residue $\bar \lambda_{0,0}^2$ shrinks to $0$. In blue the Pad\'{e} prediction of \secref{sec:Pade model for Wilson coefficients}.

\section{Pad\'{e} model}
\label{sec:bounds_using_pade_model}

In Section~\ref{sec:integrability}, we used the conformal-Pad\'{e} approximation to model several observables at intermediate coupling~$g$.
Comparison with explicit integrability data showed excellent agreement. 
The effectiveness of Pad\'{e} interpolation has also been observed in planar~${\cal N}=4$~SYM, for example, in \cite{Banks:2013nga,Costin:2020hwg,Costin:2020pcj,Dunne:2025wbq}, and it has a long history in both physics and mathematics~\cite{Baker_Graves-Morris_1996}. Beyond its empirical success, the convergence of diagonal Pad\'{e} approximants is supported by rigorous mathematical results.\footnote{For analytic functions with finitely many branch points, the convergence domain and rate are established by Stahl’s theorem~\cite{stahl1997convergence}. In this context one proves convergence in capacity, which allows for a countable set of exceptional or ``bad'' points. The situation relevant here, where both weak- and strong-coupling expansions are available, is covered by the two-point generalization of Buslaev~\cite{buslaev2013convergence}, see also \cite{yattselev2021convergence}.} 
These considerations motivate us to apply the same strategy to the Wilson coefficients, modeling them with Pad\'{e} approximants. 
At this stage it is important that at least two nontrivial orders are known from both the weak- and strong-coupling sides. Intuitively, the leading term fixes the asymptotic value, while the first subleading correction determines the slope (or derivative) of the interpolant.
In the following section, we test the internal consistency of this Pad\'{e} construction using the S-matrix bootstrap and use it to make predictions for other observables.
The weak and strong-coupling expressions for the first few Wilson coefficients are listed in \appref{app:weak_and_strongCoupling}. The complete set of $a_{n,\ell}$, $n = 1,\dots,5$ can be found in the attached Mathematica notebook.

\subsection{Pad\'{e} model for Wilson coefficients}
\label{sec:Pade model for Wilson coefficients}
We use the weak- and strong-coupling data to construct a finite-coupling model for the observables, interpolating smoothly between the two regimes. We first outline the general procedure and then present results for several Wilson coefficients.

Consider an observable $ \cA(g)$ whose behavior is known both at weak and at strong coupling
\be
\label{eq:pertexp}
 \cA(g) &= \cA_{0,0}+g^2 \cA_{0,1} + g^4 \cA_{0,2} +g^6 \cA_{0,3}  + ... \ , ~~~ g \ll 1 , \cr 
 \cA(g) &= \cA_{\infty,0} + {\cA_{\infty,1} \over g} + ... \ .
\ee
In practice, we apply this construction to the dimensionless ratios defined in \eqref{eq:observabledimless}.

To construct the finite-coupling model, we follow the conformal-mapping procedure described, for example, in \cite{Dunne:2025wbq}. We map the cut complex $g$-plane to the interior of the unit disk in the $z$-plane
\be
16 g^2 = {4 z \over (1-z)^2} ~~~ \longleftrightarrow ~~~ z(g) = {\sqrt{1+16 g^2}-1 \over \sqrt{1+16 g^2}+1} \ . 
\ee
The location of singularities $g = \pm {i \over 4}$ is inferred from the analytic properties of the cusp anomalous dimension. For real coupling $0 \leq g < \infty$, the map covers $0 \leq z(g) < 1$.

We then model the finite-coupling dependence of the observable with a Pad\'{e}-type ansatz
\be
\label{eq:ansatzPadeWC}
\cA_{[2,2]}(g) = {b_0 + b_1 z + b_2 z^2   \over 1 + c_1 z + c_2 z^2} \Big|_{z=z(g)} \ . 
\ee
This ansatz corresponds to a diagonal $[n,n]$ Pad\'{e} approximant, where the numerator and denominator have equal degree. Here we set $n=2$ which precisely has five parameters that we can match to the perturbative expansion \eqref{eq:pertexp}.

A few subtleties must be addressed. First, after fixing the parameters to their perturbative values, the Pad\'{e} approximant $\cA_{[2,2]}(g)$ may develop a spurious zero–pole pair within the physical domain $0 \leq z(g) \leq 1$. Such pairs are known as Froissart doublets \cite{froissart1973approximation}.
If the pole is located at $z=z_*$, its residue is typically tiny, owing to a nearby zero, and it can be safely removed without affecting the overall fit or perturbative behavior of the ansatz.

To achieve this we consider
\be
\tilde \cA_{[2,2]}(g) = \cA_{[2,2]}(g) - {{\rm Res}_{z=z_*}a_{[2,2]}(z) \over z-z_*} {z^3 (1-z)^2 \over z_*^3 (1-z_*)^2} \Big|_{z=z(g)} .
\ee
This procedure yields a regular finite-coupling model for $\cA(g)$. 

A second subtlety arises because the weak-coupling expansion of $m_{\text{gap}}$ contains logarithmic terms $\log g$, see \eqref{eq:expansion_mgap}, which are not captured by the Pad\'{e} ansatz above.
We have dealt with this problem in two ways. First, we used the Pad\'{e} approximation for the $m_{\text{gap}}$ \eqref{eq:mgappade} which does not contain logarithms. Second, we rewrite $  \bar a_{n,\ell} = \left( \frac{a_{n,\ell}}{a_{0,0}} \tilde m_{\text{gap}}^{2n} \right) {m_{\text{gap}}^{2n} \over \tilde m_{\text{gap}}^{2n}}$, where $\tilde m_{\text{gap}}$ is a simple function that we choose to mimick the qualitative behavior of $m_{\text{gap}}$ but which does not contain logarithms. We apply the Pad\'{e} approximation to the expression in parentheses, and finally multiply by the known ratio ${m_{\text{gap}}^{2n} \over \tilde m_{\text{gap}}^{2n}}$ obtained from integrability. 
We used the difference between the two methods to estimate the uncertainty in the finite-coupling ansatz,\footnote{We also tried other methods to estimate the uncertainty. For instance, we extended the perturbative expansion by one additional order to build a $[3,3]$ Pad\'{e} ansatz and then varied the unknown coefficients based on the observed behavior of the perturbative series to generate the band. This procedure is more involved but it leads to similar results.} yielding a Pad\'{e} band
\begin{equation}\label{eq:pade_band_def}
    \cA_\mu^{\rm Band}(g) = \cA^{\rm mid}(g) \pm \mu\cdot   \Delta \cA(g)/2 \ , 
\end{equation}
where $\cA^{\rm mid}(g)$ is the average of the two Pad\'{e} constructed and $\Delta \cA(g)$ is their difference. We also included a scale factor $\mu$ that can be used to rescale the size of the band.

\subsubsection{Consistency checks of the Pad\'{e} model}\label{sec:consistency_pade}

\begin{figure}[h!]
    \centering
    \includegraphics[width=\linewidth]{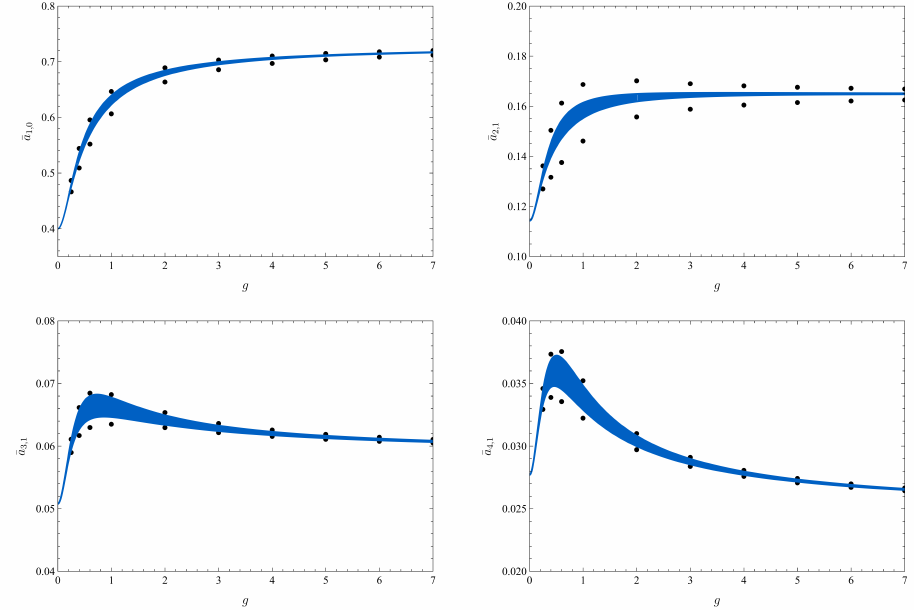}
    \caption{Consistency of the Pad\'{e} bands of \secref{sec:Pade model for Wilson coefficients} with the bootstrap bounds derived by imposing that all the Wilson coefficients $\bar{a}_{n,l}$ with $n = 1,\dots,5$ except the one that we are bounding have to be inside the Pad\'{e} bands. Notice that the bootstrap bounds are perfectly consistent with the bands and reproduce the expected shape. Here, we do not rescale the Pad\'{e} bands and take $\mu=1$, see \eqref{eq:pade_band_def}.}
    \label{fig:consistency_pade}
\end{figure}

When building the Pad\'{e} model for the Wilson coefficients, we used the expansion at weak and strong coupling. It is thus natural to ask whether the bands obtained at finite coupling are compatible with the bootstrap constraints. To answer this question, for a given set of Wilson coefficient, we impose that at fixed coupling $g$, all but one are constrained to be inside the bands provided by Pad\'{e} and bound the unfixed one. In this way we can check the consistency of the Pad\'{e} bands with first-principle constraints. In this work, we consider the coefficients $a_{n,\ell}$ for $n=1,...,5$ and $\ell \leq [n/2]$. This amounts to considering 11 dispersive observables. In \figref{fig:consistency_pade}, we compare the bootstrap bounds (shown in black, $n_{\text{max}} = 10$, no pole constraints) with the Pad\'{e} bands derived in \secref{sec:Pade model for Wilson coefficients} for several representative Wilson coefficients, namely $\bar{a}_{1,0}$, $\bar{a}_{2,1}$, $\bar{a}_{3,1}$ and $\bar{a}_{4,1}$. We find that the bootstrap bounds are perfectly consistent with the Pad\'{e} bands and accurately reproduce the expected shape.

\subsection{Bootstrap bounds using the Pad\'{e} model}
In this section, we make use of the Pad\'{e} bands on the 11 Wilson coefficients mentioned in the previous section to derive bounds on other observables. 
\subsubsection{Three-point coupling $\bar \lambda_{0,0}^2$}
\begin{figure}
    \centering
    \includegraphics[width=0.9\linewidth]{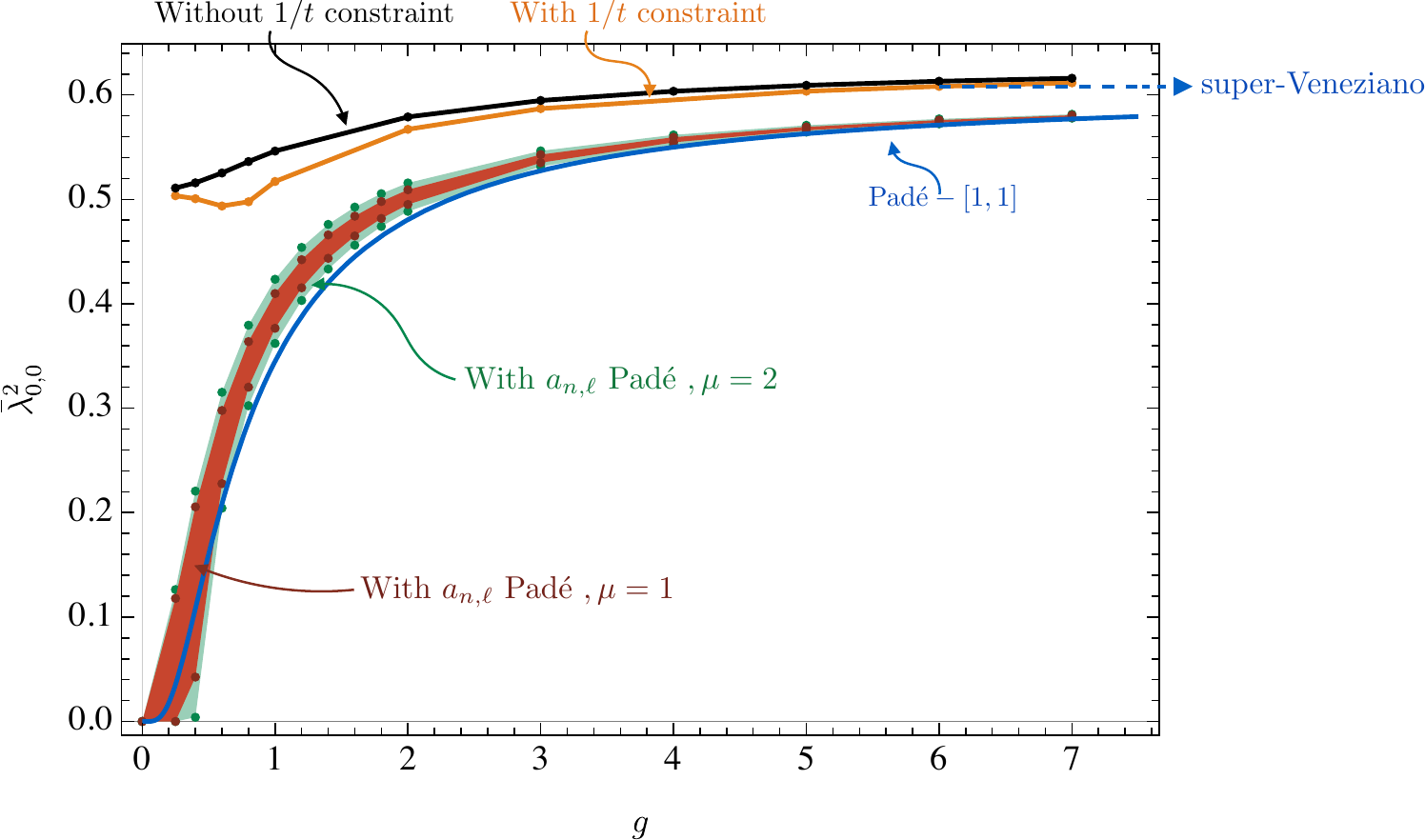}
    \caption{Bound on the residue  $\bar \lambda_{0,0}^2$ at finite coupling. In the black and orange curves, we show the upper bound without $\bar a_{n,\ell}$ Pad\'{e} constraints. In red/green, we show the bands obtained the $\bar a_{n,\ell}$ Pad\'{e} bands. In blue, we compare with the conformal-Pad\'{e}-[1,1] \eqref{eq:lambda_pade} which asymptotes to super-Veneziano.}
    \label{fig:res00_vs_g_allTogether}
\end{figure}

First we compute the two-sided bounds on the leading three-point coupling $\bar \lambda_{0,0}^2$, defined in \eqref{eq:f_close_to_bs} and \eqref{eq:threepointcouplingdless}. The leading weak- and strong-coupling expansions of this observable take the form 
\begin{equation}
\begin{aligned}
\bar{\lambda}_{0,0}^2 &= 48\pi^4 g^6 + \ldots \ , \qquad && g \ll 1 \ ,\\
\bar{\lambda}_{0,0}^2 &= \frac{6}{\pi^2}
 - \frac{9\zeta(3) + \pi^2 - 9}{2\pi^3}\,\frac{1}{g}
 + \ldots \ , \qquad && g \gg 1  \ , 
\end{aligned}
\label{eq:lambda_asymptotics}
\end{equation}
where the weak-coupling calculation can be found in \appref{app:residue}, and the strong-coupling calculation is straightforward given the explicit formulas for the amplitude presented in \secref{sec:Worldsheet bootstrap}. We can easily construct a simple interpolating function to model the residue at finite coupling
\be
\label{eq:lambda_pade}
(\bar{\lambda}_{0,0}^2)_{[1,1]} = z^3 {b_0+b_1 z \over 1+c_1 z} \ , ~~~ z = {\sqrt{1+16 g^2}-1 \over \sqrt{1+16 g^2}+1} \ , 
\ee
where the three unknown coefficients are fixed by matching to \eqref{eq:lambda_asymptotics}. We \emph{do not} expect this approximation to be very accurate due to the lack of knowledge of higher-order corrections at weak coupling. It is therefore particularly interesting to compare \eqref{eq:lambda_pade} to the bootstrap bounds. 
This comparison is shown in \figref{fig:res00_vs_g_allTogether} for $g \leq 7$.
In this figure, we first show the upper bound on the residue obtained without imposing the Padé bands. These bounds are shown both with (orange) and without (black) the pole constraints $\Phi[q^{-2}]$. As expected from previous works, 
the upper bound does not follow the Pad\'{e}–[1,1] behavior of \eqref{eq:lambda_pade} (blue curve). This case was studied using usual partial waves in $d=10$ in  \cite{Berman:2024wyt, Albert:2024yap}, where the authors showed that the residue has a finite upper bound for any gap between the spin-0 bound state and the rest of the spectrum. This is in contrast to the vanishing residue as $g \to 0$ in the $\mathcal{N}=4$ SYM amplitude.

In \figref{fig:res00_vs_g_allTogether}, we also present the bootstrap bands obtained for $\bar\lambda_{0,0}^2$, assuming that the eleven Wilson coefficients $\bar a_{n,\ell}$ with $n \leq 5$ lie within their respective Pad\'{e} bands.
The bounds are shown for two choices of scale for the error, $\mu = 1$ (red) and $\mu = 2$ (green). The bands are obtained without imposing the pole constraints, in order to simplify the numerical problem. Adding the pole constraint would not change the overall picture, since the constraints on the Wilson coefficients have a stronger effect on the result. 

We see that our bounds predict that the higher-order corrections will drive the expected form to slightly higher values at intermediate coupling. 
We can use the bootstrap bands to predict the next correction to the three-point coupling. To do so, we assume that at order $1/g^2$ there is no correction proportional to $\log g / g^2$.\footnote{Such a term could, in principle, be present. Recall that at weak coupling, $m_{\text{gap}}$ contains a $\log g$ term at order $\mathcal{O}(g^6)$, see \eqref{eq:expansion_mgap}.} Then, we construct a Pad\'{e}-[2,2] approximant containing two unknown parameters. These free parameters are then bounded using the bootstrap bands, requiring that the resulting Pad\'{e} approximant remains entirely within them. Using this procedure, we obtain
\begin{equation}
    (\bar{\lambda}_{0,0}^2)|_{1/g^2} \in [-0.14, 0.01]~ \text{for } \mu=1\,, \qquad  (\bar{\lambda}_{0,0}^2)|_{1/g^2} \in [-0.18, 0.35]\, \text{for } \mu=2\,,
\end{equation}
which in particular excludes the $1/g^2$ expansion of the Pad\'{e}-[1,1] given by $ (\bar{\lambda}_{0,0}^2)_{[1,1]}|_{1/g^2}\approx -0.23$. %
It would be interesting to test this prediction with higher-order explicit calculations. 

\subsubsection{The amplitude}
\label{sec:Bounds on the amplitude itself}

We next apply the same method to investigate the behavior of the amplitude itself at finite $s$ and $t$. To guide our expectations, we recall that in \cite{Henn:2024qwe} it was observed that the perturbative amplitude on the Coulomb branch exhibits \emph{complete monotonicity}. Here we briefly review this property and use our results to test it at finite coupling.

To introduce this concept, it is convenient to introduce the dimensionless amplitude \eqref{eq:funcitself00}
\be
\bar f(s, t) = \frac{f(s,t) + \frac{1}{s t}}{a_{0,0}} \ . 
\ee
We can set $\mgap=1$ and consider its partial derivatives $\partial_s^n \partial^{n-\ell}_t \bar f(s,t)$ denoted in \eqref{eq:funcitself} as $\bar f_{n, \ell}(s,t)$.

Following \cite{Henn:2024qwe}, we say that $\bar f(s,t)$ is \emph{completely-monotonic} (CM) if it satisfies
\be
\label{eq:CMcondition}
\bar f_{n, \ell}(s,t) \geq 0, ~~~ s,t \leq 0 , ~~~ \forall n, \ell \geq 0 \ .
\ee
While bounds of the type \eqref{eq:CMcondition} naturally follow from the study of dispersion relations for $s,t \geq 0$, here positivity appears in the region $s,t \leq 0 $ for which no first-principle argument for positivity exists.

In this context, the Bernstein-Hausdorff-Widder-Choquet theorem states that a function is CM if and only if it admits the following representation
\be
\label{eq:CMtheorem}
\bar f(s, t) = \int_0^\infty d\mu(x,y) e^{x s+y t} ,
\ee
where $d\mu(x,y) \geq 0$ is a non-negative measure. CM functions have a natural connection to positive geometry \cite{Mazzucchelli:2025gyg}, and as a consequence to the program of expressing scattering amplitudes as certain generalized volumes \cite{Hodges:2009hk,Arkani-Hamed:2013jha,Arkani-Hamed:2014dca,Arkani-Hamed:2017mur}.

It is easy to check that the four-point amplitude is CM at weak coupling $g \to 0$. The easiest way to see it is to notice that the one-loop amplitude satisfies the unsubtracted Mandelstam representation with positive double spectral density, see \eqref{eq:oneloopf},
\be
f(s,t)+{1 \over s t} \sim \int_0^\infty {ds' dt' \over \pi^2} {\rho_{\text{1-loop}}(s',t') \over (s'-s)(t'-t)} \ , ~~~ \rho_{\text{1-loop}}(s',t') \geq 0 \ . 
\ee
In \cite{Henn:2024qwe} it was checked that the two- and three-loop corrections $\bar f^{(L)}(s,t)$ are completely-monotonic when multiplied by an additional factor $(-1)^{L+1}$, where $L$ is the loop order. In particular, it means that the two-loop correction violates the CM property as we defined it. The alternation of positivity signs with loop order is a feature shared by many observables \cite{Dixon:2016apl,Henn:2024qwe}.

At strong coupling instead we get the super-Veneziano amplitude and we can write the following integral representation
\be
-{\Gamma(-s)\Gamma(-t) \over \Gamma(1-s-t)} + {1 \over s t} = \int_0^1 d x d y \theta(x+y-1) x^{-s-1} y^{-t-1} ,
\ee
which makes complete monotonicity manifest as well because derivatives in $s$ or $t$ bring manifestly positive factors $-\log x$, $- \log y$. In the formula above, we combined ${1 \over s t} = \int_0^1 dx dy x^{-s-1} y^{-t-1}$, and 
\be
-{\Gamma(-s)\Gamma(-t) \over \Gamma(1-s-t)} &= {1 \over s+t} \int dx x^{-s-1}(1-x)^{-t-1} 
=- \int_0^1 d\alpha dx \alpha^{-1-s-t} x^{-s-1}(1-x)^{-t-1} \nonumber \\
&=- \int_0^1 d\tilde x d\tilde y \tilde x^{-s-1} \tilde y^{-t-1} \theta(1-\tilde x-\tilde y) ,
\ee
where in the last line we performed a change of variables $\tilde x = \alpha x, \tilde y = \alpha (1-x)$. We have also checked that the first subleading correction at strong coupling found in \secref{sec:Worldsheet bootstrap} \emph{does not} obey complete monotonicity.

To summarize, at weak coupling, $g \ll 1$, complete monotonicity appears to be satisfied order-by-order in perturbation theory, however, it is sign-alternating. At strong coupling, $g \gg 1$, the leading super-Veneziano amplitude satisfies it, but the stringy correction found in the paper does not. This leaves open an interesting possibility that complete monotonicity is a true property of the scattering amplitude at finite coupling. 
We notice that at high energies --- both in the Regge limit $-s \gg \mgap$ with fixed $t$, and in the fixed-angle regime $-s,-t \gg \mgap$ with fixed $s/t$ --- the amplitude satisfies $f(s,t) \ll {1 \over s t}$ for $s,t < 0$. In the Regge limit, it follows from the fact that $j_0(t)<-1$ for $t<0$, and from \eqref{eq:GMADS} and \eqref{eq:GMregime} at fixed angle. Consequently, at high energies $\bar f(s,t) \sim {1 \over s t}$, and complete monotonicity trivially follows.
A complete proof of \eqref{eq:CMcondition} at finite coupling for general $s,t \leq 0$ is beyond our present reach, but it can be tested numerically for the first few derivatives in selected regions of the $(s,t)$ plane at fixed coupling. This is carried out next. %

\begin{figure}[h!]
    \centering
    \includegraphics[width=0.9\linewidth]{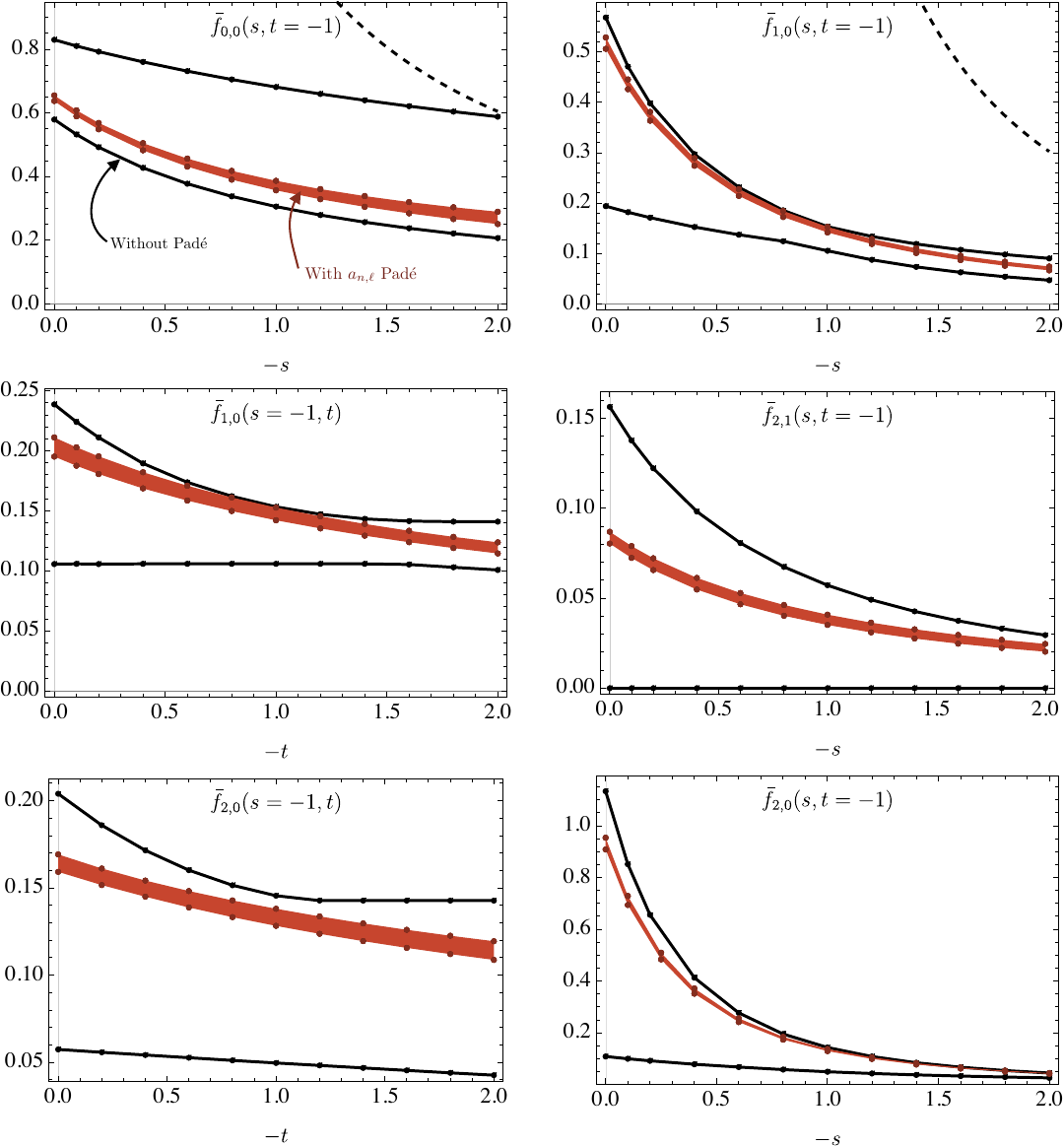}
    \caption{Two-sided bounds on the amplitude and its derivatives at $g=1$, we set $\mgap=1$. In all cases we fixed $s\text{ or } t=-1$ and and varied the other Mandelstam. In black, we present the two-sided bootstrap bounds obtained without using the Pad\'{e} bands and in red, the two-sided bootstrap bounds, where the Wilson coefficients were forced to lie within their Pad\'{e} bands (with scale $\mu=1$).  In the top row, we used the black dashed curve to show the contribution of $1/st$ to the observables. }
    \label{fig:amplitude_vs_s}
\end{figure}

\begin{figure}
    \centering
    \includegraphics[width=0.8\linewidth]{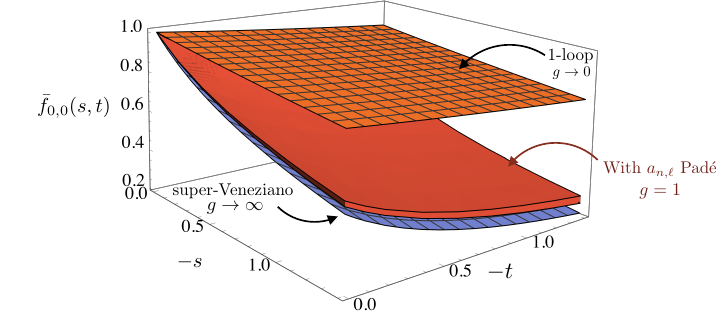}
    \caption{Two-sided bounds on the dimensionless amplitude $\bar f_{0,0}(s,t)$, see \eqref{eq:funcitself00}, at $g=1$, setting $\mgap = 1$. The resulting region is shown as the red solid volume, obtained using Padé bands for the Wilson coefficients (with scale $\mu = 1$). The orange surface corresponds to the one-loop box \eqref{eq:weak1loop}, and the purple surface to the super-Veneziano amplitude \eqref{eq:super_Veneziano}. Recall that $\bar f_{0,0}(0,0) = 1$ by definition \eqref{eq:funcitself}.}
    \label{fig:f_vs_st}
\end{figure}

The results for the first derivatives ($\bar f_{n,\ell}$ with $n\leq 2,\, \ell=0,1$) at finite coupling $g = 1$ are shown in \figref{fig:amplitude_vs_s}, where one of the Mandelstam variables is fixed to either $s = -1$ or $t = -1$, with $\mgap = 1$. In \figref{fig:f_vs_st}, we show the result for the amplitude $\bar f_{0,0}$ in the region $s, t < 0$. These bands were obtained with $n_{\text{max}} = 8$, omitting the pole constraint. We verified that increasing $n_{\rm max}$ and including the pole constraint does not significantly affect the results. In \figref{fig:f_vs_st}, we also compare the results for the amplitude with the one-loop box amplitude \eqref{eq:weak1loop} and the super-Veneziano amplitude \eqref{eq:super_Veneziano}. The  $g = 1$ amplitude (shown as the red volume) lies between the two surfaces. 

In both \figref{fig:amplitude_vs_s} and \figref{fig:f_vs_st}, using as an input the Pad\'{e} model for the Wilson coefficients, we get sharp two-sided bounds on the amplitude. Remarkably, we find that the behavior of the finite coupling scattering amplitude is consistent with \eqref{eq:CMcondition}.  It suggests that complete monotonicity could be a true property of the Coulomb branch amplitude at finite coupling. We leave further explorations of this surprising aspect of the finite-coupling scattering amplitude to future work. %

\section{Conclusions and open directions}
\label{sec:conclusions}

Scattering on the Coulomb branch of the planar ${\cal N}=4$ SYM defines a one-parameter family of consistent, stringy S-matrices in four dimensions labelled by the 't Hooft coupling $g$.\footnote{Recall that $\lambda = g_{\text{YM}}^2 N_c = (4 \pi g)^2$.} 
In this paper we have studied the simplest matrix element of this S-matrix, the four-point amplitude, viewed as a function of $g$.

The amplitude exhibits a remarkably rich structure: nonlinear Regge trajectories, a non-degenerate spectrum of bound states, an ionization threshold where the bound states accumulate, and a two-particle cut.\footnote{In a confining gauge theory, such as QCD, the dual geometry caps off at a finite radial distance. This removes the ionization threshold and the two-particle cut associated with the $AdS$ horizon from the excitation spectrum of the string.}  At weak coupling, the relevant physical degrees of freedom are the quarks and gluons of the gauge theory. At strong coupling, a dual string description emerges which captures both flat-space string scattering at low energies and string scattering in $AdS$ at high energies. As such, this amplitude beautifully interpolates between the one-loop box and the super-Veneziano amplitude.

Several remarkable properties of the amplitude allowed us to make quantitative progress at finite coupling. First, %
the spectrum of Regge trajectories $j_n(s)$ maps to the energy spectrum of a Maldacena-Wilson line with a local operator insertion at the cusp, $E_n(\phi)$, see \eqref{eq:reggecusp} for the precise relation. The latter can be computed using integrability at any $g$. We used this Regge/cusp correspondence to determine in detail the shapes of the leading and the first subleading Regge trajectory, see \figref{fig:j0_frankenstein_plot} and \figref{fig:j1_frankenstein_plot} , for which the connection was already established. We provide some additional nontrivial checks for the correspondence both at weak and strong coupling. Second, dual conformal invariance leads to \emph{enhanced unitarity}, see \secref{sec:enhunit}, which in turn leads to stronger S-matrix bootstrap bounds. At strong coupling, the enhanced unitarity becomes manifestly holographic, describing partial wave expansion in five spacetime dimensions. 

Using the worldsheet bootstrap, we computed the leading correction to the Veneziano amplitude at strong coupling in \secref{sec:Worldsheet bootstrap}. This builds on recent progress in computing \emph{AdS amplitudes}  and constitutes the first application of these methods to a flat-space scattering amplitude.

We continued by performing the S-matrix bootstrap analysis of various observables at finite coupling $g$ using enhanced unitarity into the standard framework of dispersion relations in \secref{sec:bootstrapping_setup}. There, the value of the coupling $g$ is specified in two ways: first by incorporating our knowledge of the first two Regge trajectories; second by fixing the ratio between the tree-level pole and the constant $a_{0,0}$, see \eqref{eq:protectedopen}, fixed by non-renormalization theorem.{\footnote{Thanks to enhanced unitarity, we get a nontrivial effect from the $1/t$-pole sum rule in $d=4$.} We performed a detailed quantitative analysis of the leading Wilson coefficients at various values of $g$. 

At weak coupling, the amplitude approaches the cusp of the exclusion plot, with the one-loop result being extremal, see \figref{fig:a10_a20_g=0}. In this regime, the dual string is too floppy to support nontrivial excitations, and string scattering does not probe the fine structure of the dual $AdS$ geometry, see also \cite{Klebanov:2006jj}. The amplitude displays a two-particle cut, encoding the physics of the $AdS$ horizon, and a single Regge trajectory of bound states corresponding to the hydrogen-like spectrum. As the coupling increases beyond the critical value $g_c \approx 0.28$, the subleading Regge trajectory begins to contribute to the bound-state spectrum, see \figref{fig:behavior_j1}. At infinite coupling $g \to \infty$, the amplitude is approximated by the super-Veneziano amplitude at low energies, characterized by infinitely many equidistant, linear trajectories. At strong coupling, in agreement with the previous studies, the amplitude lies close to the boundary in the allowed region but does not saturate any bound, see \figref{fig:a10_a20_g=0}. We then performed the S-matrix bootstrap analysis of the Wilson coefficients at finite coupling, see \figref{fig:a10vsg}, \figref{fig:a20vsa10}, and \figref{fig:a21vsa10}.

Although the spectrum of bound states undergoes a structural change, we expect coarse-grained observables such as the Wilson coefficients, which represent moments of the amplitude discontinuity, to vary smoothly with $g$. This smoothness suggests that combining weak- and strong-coupling perturbative data can yield an accurate finite-coupling model for the Wilson coefficients.\footnote{Recall that at weak-coupling results are available up to three loops, while at strong coupling, including the correction derived in the present paper, two terms in the expansion are known.} To estimate an expected value of various observables, we combined the weak- and strong-coupling results to construct a finite-coupling Pad\'{e} model for the Wilson coefficients, which we subsequently confronted with S-matrix bootstrap constraints in \secref{sec:consistency_pade}. We have found that the finite-coupling Pad\'{e} model is fully consistent with the bootstrap constraints. 
Using this model for the Wilson coefficients, we obtained the two-sided bounds on the spin-0 three-point coupling at finite coupling $g$, see \figref{fig:res00_vs_g_allTogether}.
We also explored the finite-$g$ scattering amplitude, see \figref{fig:amplitude_vs_s} and \figref{fig:f_vs_st} and observed that it is compatible with the property of complete monotonicity discussed in \cite{Henn:2024qwe}.

Even though in this paper we have made some progress toward understanding the scattering amplitude on the Coulomb branch of ${\cal N}=4$ SYM at finite coupling, many fascinating open questions remain:
\begin{itemize}
\item \textbf{Regge/cusp correspondence.} We presented evidence that the Regge/cusp correspondence observed in \cite{Henn:2010bk,Caron-Huot:2014gia, Bruser:2018jnc} for the leading and the subleading Regge trajectories continues to all trajectories.  It will be very important to test and possibly refine this correspondence further, as well as understand deeper its origins and possible generalizations to other theories. For example, performing a four-loop calculation at weak coupling would allow us to further resolve the degeneracy of the sub-subleading trajectories. 
\item \textbf{Spinning cusp insertions.} On the Wilson loop side, we are naturally led to consider insertions of spinning operators at the cusp. This goes beyond existing results in the literature. Computing the scaling dimensions of these operators and investigating whether integrability techniques can be applied to them remains an open problem.
\item \textbf{Anti-parallel limit and QSC.} In this paper, we explore the $\phi \to \pi$ limit of the first subleading trajectory using the Quantum Spectral Curve and predict a transition in its behavior at a critical coupling $g_c$. In this limit, the numerical QSC methods become challenging, and the value of $g_c$ was determined by fitting data at intermediate values of $\phi$. It would be interesting to explore alternative methods in the anti-parallel limit that would allow one to compute the critical coupling for the subleading trajectories directly. The anti-parallel limit is physically interesting because one can consider the fusion of the two lines and formulate an effective field theory (EFT) \cite{Cuomo:2024psk, Kravchuk:2024qoh}. 
\item \textbf{Beyond the subleading strong‑coupling term.} A promising avenue is to compute $1/g^2$ corrections to the strong coupling result. Although the result to order $1/g$ has the same structure as curvature corrections to $AdS$ amplitudes in other contexts, in which the second order correction was computed, the details of the problem are very different. In the present paper we are considering a bona-fide amplitude, as opposed to the transform of a correlator. This should make the problem simpler. On the other hand, there is an extra scale, which should make the problem harder. It would be also interesting to understand the systematics of the $1/g$ expansion, and whether non-analytic terms, containing $\log g$ can arise. 
\item \textbf{Monodromy relations.} To order $1/g$ we have found that the result satisfies a deformed version of  monodromy relations. This deformation is different to the one recently found in the context of $AdS$ amplitudes \cite{Alday:2025cxr}, and presumably arises due to the extra scale in the present problem. If one can understand this deformation on physical grounds, one could hope to derive an all-order version of the monodromy relations.
\item \textbf{Smeared tree‑level sum rule.} In the smeared sum rule for the tree-level pole, the integration over momenta is performed with an upper limit $q_0 < 2m$. In this work, when performing the S-matrix bootstrap in \secref{sec:Universal_bounds}, we set $q_0 = m_{\rm gap}$ and it would be interesting to study how the bounds depend on $q_0$. This dependence is expected to be particularly relevant at strong coupling, where the constraint has a stronger impact and the allowed range of $q_0$ becomes larger.
\item \textbf{Fixed‑$u$ dispersion relations.} By writing a fixed-$u$ dispersion relation, crossing symmetry leads to a second set of null-constraints \eqref{eq:nullY_master}, see also \cite{Albert:2022oes}. The sum rules obtained contains Legendre and $SO(1,3)$ harmonics not evaluated in the forward limit and we did not imposed them in this work. They could be treated similarly as the sum rule for the amplitude and it would be interesting to study their effect.  
\item \textbf{Complete monotonicity beyond perturbation theory.} We have observed that complete monotonicity observed in the perturbative studies of scattering amplitudes is consistent with our finite $g$ results. In perturbative studies, complete monotonicity naturally arises from positive geometry and the notion of the amplitude as a generalized volume in an appropriate space.  In the context of Coulomb-branch scattering, this was recently discussed in \cite{Arkani-Hamed:2023epq}. It would be fascinating if this notion could be made precise at finite $g$ in the current context. Interpreting complete monotonicity as a genuine property of the $\mathcal{N}=4$ SYM amplitude leads to an infinite set of sum rules which can, in principle, be imposed within the bootstrap framework.
\item \textbf{Integrability with D3‑brane boundary conditions.} It has been recently shown, that the D3-brane boundary condition preserves integrability of the string sigma-model \cite{Demjaha:2025axy}. One-point functions of non-BPS single-trace operators on the Coulomb branch were calculated using integrability at finite 't Hooft coupling in \cite{Coronado:2025xwk}. It would be very interesting to see if one could apply integrability methods directly to the scattering amplitudes on the Coulomb branch of ${\cal N}=4$ SYM. Still, as was the case for the four-point conformal correlation function in the planar limit of ${\cal N}=4$ SYM \cite{Caron-Huot:2024tzr}, it could be that the S-matrix bootstrap provides a simpler and more direct way to access the amplitude.
\item \textbf{Nonrelativistic EFT.} Close to the two-particle threshold $s = 4m^2$, the scattering amplitude is effectively described by non-relativistic scattering of massive particles, see \appref{app:residue}. A natural further step is to develop a systematic EFT expansion of the amplitude in this limit. 
\item \textbf{Other external states and higher-point generalizations.} We restricted ourselves to the analysis of the $2 \to 2$ scattering of the massless mesons $\chi$. We also limited our consideration to the simplest possible choice of the R-symmetry polarizations of external scalars. It would be interesting to extend our analysis to scattering amplitudes with general external states, see \cite{Flieger:2025ekn}, as well as to consider the first bound state as an external particle.\footnote{In this context, the work of \cite{Caron-Huot:2021usw} establishes the relationship between a four-point amplitude on the Coulomb branch of massive external particles (and massless internal) and the octagon. 
It has been recently extended to five-point amplitudes in \cite{Belitsky:2025bgb}.
} 
On the bootstrap side, the latter amounts to considering a mixed system of amplitudes, which is expected to lead to stronger bounds.
It would be also very interesting to generalize the analysis of the present paper to higher-point Coulomb-branch amplitudes, see e.g. \cite{Craig:2011ws,Kiermaier:2011cr,Herderschee:2019dmc,Bork:2022vat,MdAbhishek:2023nvg}.\footnote{The six-point one-loop amplitude on the
moduli space of ABJM was calculated in \cite{Bianchi:2015tba}.} It might be a good testing ground for the S-matrix bootstrap techniques beyond four external points. 
\item \textbf{Amplitude/Wilson‑loop duality on the Coulomb branch.} There is a well-understood correspondence between the gluon amplitudes and null Wilson loops in the planar ${\cal N}=4$ SYM \cite{Alday:2007he,Drummond:2007aua,Brandhuber:2007yx}. However, the generalization of this correspondence to scattering on the Coulomb branch is not well understood. In particular, it is not clear what is the Wilson loop dual of the four-point amplitude on the Coulomb branch, see \cite{Belitsky:2021huz}. Recently the correspondence between the Wilson loops and light-like limits of the correlators have been generalized to non-conformal theories in \cite{Chen:2025ffl}. It would be interesting to understand if similar techniques can be applied to the Coulomb branch scattering amplitudes. %
\item \textbf{Beyond the planar limit.} It would be very interesting to analyze how the methods used in the paper change if we consider $1/N$ corrections. A fascinating aspect of this problem is that $1/N$ corrections will probe the closed string sector. Perhaps the simplest first step in this direction will be the calculation of the first nontrivial non-planar correction, which first appear at four loops. A more ambitious goal would be to combine the S-matrix bootstrap methods with supersymmetric localization techniques on the Coulomb branch. For the finite $N_c$ conformal correlators in ${\cal N}=4$ SYM, the conformal bootstrap and the supersymmetric localization were used, for example, in \cite{Chester:2021aun,Chester:2023ehi}.\footnote{Supersymmetric localization was also used in the planar limit to constrain the four-point correlation function of half-BPS operators \cite{Caron-Huot:2024tzr}.}  
\end{itemize}

We hope that future work will shed light on many of these interesting questions.

\acknowledgments

We are very grateful to Nikolay Gromov for many discussions, collaboration, and help with the integrability aspect of the project. We thank Johannes Henn for many discussions concerning the properties of Coulomb-branch scattering amplitudes. We also thank Benjamin Basso, Victor Gorbenko, Luca Griguolo, Andrea Guerrieri, Patrick Hager, Julius Julius, Petr Kravchuk, Shota Komatsu, Marco Meineri, Joao Penedones, Prashanth Raman, Luigi Tizzano, Gherardo Vita, Zahra Zahraee for illuminating discussions. This project has received funding from the European Research Council (ERC) under the European Union’s Horizon 2020 research
and innovation program (grant agreement number 949077). The work of L.F.A. is partially supported by the STFC grant ST/T000864/1. KH is supported by the European Research Council (ERC) under the European Union’s Horizon 2020 research and innovation programme (grant agreement No 101115511) and by the Simons Collaboration on Celestial Holography.

\appendix

\section{String mass in $AdS_5 \times S^5$}
\label{app:stringmass}

Consider a static open string stretched between a pair of D3-branes in $AdS_5 \times S^5$. We work in the Poincare patch. We would like to compute the string mass and match it to the gauge theory prediction. The equations of motions and the Virasoro constraints can be found, for example, in \cite{Tseytlin:2010jv}. 

We work in the static gauge $\tau=t$. We parameterize the string solution by its position across $S^5$ given by $n^i(\sigma)$, such that $n(\sigma) \cdot n(\sigma) = 1$. It also has a nontrivial dependence on the radial direction $z(\sigma)$. The Virasoro constraints and the equation of motion for $z(\sigma)$ imply that
\be
\label{eq:Vconstraint}
{1-z'(\sigma)^2 \over z(\sigma)^2} = n'(\sigma) \cdot n'(\sigma) = \Omega^2 ,
\ee
where $x' = \partial_\sigma x$ and $\Omega$ is a constant. The free equation of motion for $n(\sigma)$ leads to the following solution
\be
n^i(\sigma) = n_1^i \cos \Omega \sigma + n_0^i \sin \Omega \sigma,
\ee
where $n_0^2 = n_1^2 = 1$ and $n_0 \cdot n_1 = 0$, which together guarantee that \eqref{eq:Vconstraint} holds. We choose $n_0^i$ such that it rotates $n_1^i$ in the desired direction so that the final position of the string is $n_2^i = n^i(\sigma_f)$. In this way we get
\be
\label{eq:rotangle}
n_1 \cdot n_2 = \cos \Omega \sigma_f .
\ee
The solution for $z(\sigma)$ takes the form
\be
z(\sigma) = {\sin (\Omega \sigma+\sigma_0) \over \Omega} \ .
\ee
We assume that $z_1=z(0)$ and $z_2=z(\sigma_f)$ which together with \eqref{eq:rotangle} fixes $\sigma_0$ and $\Omega$.

To evaluate the mass, we compute the on-shell action, and we get that
\be
m &=T R_{AdS}^2 \int_0^{\sigma_f} {d \sigma \over z(\sigma)^2} \nn \\
&=2g {\sqrt{1 - (n_1 \cdot n_2)^2} \over z_1 z_2} {1 \over \Omega} = 2g \sqrt{\Big({n_1 \over z_1} - {n_2 \over z_2} \Big)^2} = \sqrt{(m_1 n_1 - m_2 n_2 )^2} \ ,
\ee
which matches \eqref{eq:mesonmass}. In the expression above, we used the standard $AdS$/CFT dictionary $T R_{AdS}^2 = 2 g$, and identified the radial position to the gauge theory variables as follows
\be
z_i = {2 g \over m_i} \ . 
\ee

\section{Extended unitarity}
\label{app:extunit}

In this section, we present the derivation of the generalized unitarity conditions used in the main text of the paper. We consider $AdS_5$ parametrized as follows
\be
- (Y^{-1})^2 - (Y^0)^2 + \sum_{i=1}^4 (Y^i)^2 = - 1 ,
\ee
where we set $R_{AdS}=1$. We use it as a T-dual space for our scattering process. In other words, a scattering amplitude is characterized by four points $Y_i$. Let us now explain it more precisely. We can introduce the following coordinates in this $AdS$
\be
Y^{\mu} &= {y^\mu \over r} , ~~~ \mu = 0,1,2,3 \ , \nn \\
Y^{-1}+Y^{4} &= {1 \over r}, ~~~ Y^{-1}-Y^{4} = {r^2 + y^2 \over r} , 
\ee
where for $y^2 = - (y^0)^2 + \sum_{i=1}^3 (y^i)^2$ we use mostly-plus signature. Dual conformal invariance is the statement that the amplitude is invariant under $SO(4,2)$ that acts linearly on $Y^A$. 

Let us next connect to the kinematics of the four-point amplitude. We have
\be
\label{eq:momenta}
p_i^\mu = y_{i+1}^{\mu}-y_{i}^{\mu} ,
\ee
whereas the radial coordinate is related to VEVs of the scalar fields in the dual theory
\be
r_i = m_i \ . 
\ee
The invariant scalar product takes the form
\be
Y_i \cdot Y_j +1 = - {(r_i-r_j)^2 + (y_i - y_j)^2 \over 2 r_i r_j} \ .  
\ee
The on-shell condition for the scattering process takes therefore the form
\be
Y_i \cdot Y_{i+1} +1 =- {(m_i-m_{i+1})^2 + p_i^2 \over 2 m_i m_{i+1}}  = 0,
\ee
where we used that $p_i^2 = - (m_i-m_{i+1})^2$. We have however two nontrivial products
\be
\label{eq:invariants}
Y_1 \cdot Y_3 +1 = - {(m_1-m_3)^2 + (p_1 + p_2)^2 \over 2 m_1 m_3}  = -{(m_1-m_3)^2 - s \over 2 m_1 m_3}  \ , \\
Y_2 \cdot Y_4 +1 = - {(m_2-m_4)^2 + (p_2 + p_3)^2 \over 2 m_2 m_4}  = -{(m_2-m_4)^2 - t \over 2 m_2 m_4} .
\ee
Dual conformal invariance of the amplitude implies that it is only the function of the invariant products
\be
M \Big( (Y_1 \cdot Y_3) ,  (Y_2 \cdot Y_4) \Big) .
\ee
Let us recall various symmetry generators
\be
J_{MN} = Y_{M} \partial_{Y^N} - Y_{N} \partial_{Y^M} \ .
\ee
The usual $SO(1,3)$ Lorentz invariance of the amplitude is encoded in $J_{\mu \nu}$. Translation invariance is generated by 
\be
J_{4,\mu} - J_{-1,\mu} = \partial_\mu = P_\mu \ ,
\ee
and is obvious because of \eqref{eq:momenta}. There are more non-trivial generators which act as
\be
J_{-1,4} &= r \partial_r + y^{\mu} \partial_\mu = D~~~~(\text{dilatation}), \\
J_{4,\mu} + J_{-1,\mu} &= 2 y_{\mu} (y_\nu \partial^\nu +r \partial_r) - (y^2 + r^2) \partial_\mu = K_\mu~~~~(\text{special conformal}) \ . 
\ee
Dilatations are somewhat obvious because we do not have any scale in the theory so it makes sense that things only depend on dimensionless ratios. Special conformal transformations are however nontrivial.

\subsection{$s$-channel scattering}
Let us consider scattering in the $s$-channel center-of-mass frame \cite{Bruser:2018jnc}. Using spatial translation invariance and $SO(3)$ rotations we set $\vec y_1 = \vec y_3 =0$. In this frame we have
\be
p_2^0+p_3^0 = y_4^0 - y_2^0  = 0 \ . 
\ee
Using time translation invariance we can set $y_2^0 = y_4^0 = 0$. We would like next to impose that all external particles are massless, which is simply the statement that
\be
r_i = r \ . 
\ee
We thus have the following vectors $Y_1^A(\sqrt{s}/2,\vec 0,r)$,$Y_2^A(0,-\vec p_2,r)$, $Y_3^A(-\sqrt{s}/2,\vec 0,r)$, $Y_4^A(0,-\vec p_4,r)$,
where we used the notation $Y^A(y^0, \vec y, r)$ and recall that $| \vec p_2 | = | \vec p_3 | ={1 \over 2} \sqrt{s}$ . This corresponds to the physical momenta
\be
p_1^\mu &= (-{\sqrt{s} \over 2}, \vec p_1), ~~~p_2^\mu = (-{\sqrt{s} \over 2}, -\vec p_1), \nn \\
p_3^\mu &= ({\sqrt{s} \over 2}, -\vec p_4), ~~~p_4^\mu = ({\sqrt{s} \over 2}, -\vec p_4),
\ee
where $|\vec p_1|=|\vec p_4|={\sqrt{s} \over 2}$ so that $p_i^2=0$. With this choice we have $t=-(p_2+p_3)^2 =- {s \over 2} - 2 \vec p_2 \cdot \vec p_3$.

Physically, $r=m$, where $4m^2$ is the location of the two-particle threshold. The amplitude, of course, only depends on the ratios ${s \over m^2}$ and ${t \over m^2}$, see \eqref{eq:invariants}. In particular, it is convenient to switch to the dimensionless variables $(s,t)=r^2 (\tilde s, \tilde t)$ such that the amplitude is $M(\tilde s, \tilde t)$.

With this choice of frame, we find that there are two possible little groups for $s$-channel scattering, depending on the scattering energy: $SO(4)$ for the bound states $s<4m^2$, and $SO(1,3)$ for the scattering states $s\geq 4m^2$. We describe them below. 

\paragraph{Little group ${SO(4)}$}\mbox{}\\
In the frame define above, we have
\be
Y_1^{4} = Y_3^{4} ={1 \over 2} \left( {1 \over r} - {r^2 - {s \over 4} \over r} \right) . 
\ee
We can set it to zero $0$ by choosing 
\be
r = {1 \over \sqrt{1-{\tilde s \over 4} }} \ , 
\ee
where the solution only exists for $\tilde s<4$ which is the location of the bound states. 

In this frame $SO(4)$ that rotates $Y^i$ with $i=1,2,3,4$ leaves $Y_1^{A}$ and $Y_3^{A}$ invariant. In particular, under this $SO(4)$ we have $r_1 = r_3 =  {2 \over \sqrt{4-\tilde s}}$. On the other hand, $r_2$ and $r_4$ will change. Recall that the mass of external states is
\be
m_i = | r_{i+1} - r_{i} | \ ,
\ee
which means that under $SO(4)$ we stay in the following subsector
\be
\label{eq:so4orbit}
m_1 = m_2, ~~~ m_3 = m_4 \ . 
\ee

We now consider a space of scattering amplitude with masses that satisfy \eqref{eq:so4orbit}, which thanks to dual conformal invariance are all described by the same function $M$. We can consider intermediate states labelled by $0<\tilde s<4 m^2$ (total energy) and $J_{SO(4)}$ in this space. The scattering amplitude between such states is diagonal because $SO(4)$ is a symmetry of the $S$-matrix. Let us compute the scattering angle
\be
\cos \tilde \theta_s = {\sum_{i=1}^{4} Y_2^i Y_4^i \over \sqrt{(\sum_{i=1}^{4} Y_2^i Y_2^i) (\sum_{i=1}^{4} Y_4^i Y_4^i)}} = 1+{2 \tilde t \over  \tilde s} - {\tilde t \over 2} \ . 
\ee

We can therefore decompose the amplitude as
\be
\label{eq:pwgen}
M \Big( (Y_1 \cdot Y_3) ,  (Y_2 \cdot Y_4) \Big)  = \sum_{J=0}^\infty \tilde f_J(s) \PJfive(\cos \tilde \theta_s), ~~~ \cos \tilde \theta_s = 1+{2 t \over  s} - { t \over 2 m^2} ,
\ee
where we restored the dependence on the mass for the case $m_i = m$. In this equation, $\tilde f_J(s)$ are the $SO(4)$  partial waves, and $\PJfive(x)$ are the $SO(4)$ Gegenbauer polynomials, see \eqref{eq:PJ_s04}.

We next can interpret the imaginary part of the amplitude as the probability of creating a given $SO(4)$ multiplet, which therefore should be non-negative
\be
\label{eq:genunit}
{\rm Im} f_J(s) > 0 , ~~~ 0< s < 4 m^2\,,
\ee
and to obtain \eqref{eq:fsInSO4PW} we define $c_J(s) = s^2\Im \tilde f_J(s)$.

\paragraph{Little group ${SO(1,3)}$}\mbox{}\\
As we go in the regime $s >4 m^2$, we can instead impose $Y_1^{-1}=Y_3^{-1}=0$ through
\be
{1 \over 2} \left( {1 \over r} + {r^2 - {s \over 4} \over r} \right) = 0.
\ee
This leads to
\be
r = {2 \over \sqrt{\tilde s -4}} . 
\ee
For $\tilde s >4$ it is therefore possible to set instead $Y_1^{-1}=Y_3^{-1}=0$ and the little group becomes $SO(1,3)$ instead of $SO(4)$.

The scattering angle in this case becomes
\be
\label{eq:hyperangle}
\cosh \tilde \theta_s = { Y_2^{-1} Y_4^{-1} - \sum_{i=1}^3 Y_2^i Y_4^i \over \sqrt{\left( (Y_2^{-1})^2 - \sum_{i=1}^3 (Y_2^i)^2\right)\left( (Y_4^{-1})^2 - \sum_{i=1}^3 (Y_4^i)^2\right)}} \ . 
\ee
This is consistent with the fact, in this case, the little group becomes $SO(1,3)$ which is a group of isometries of $H_{3}$ (whereas $SO(4)$ naturally acted on $S^3$). Using the scattering angle \eqref{eq:hyperangle} and restoring the $m$ dependence, we get 
\be
\cosh \tilde \theta_s = 1+{2 t \over  s} - { t \over 2 m^2} \geq 1 ,~~ s  \geq 4m^2 \ , 
\ee
which we can think of as a geodesic distance on $H_3$. The natural expansion is in terms of harmonic functions, see e.g. Appendix 4.C in \cite{Penedones:2007ns},
\be
\label{eq:pwgenHF}
M(s,t)  = \int_{0}^{\infty} d \nu \tilde f_\nu (s) \Omega_{\nu}(\cosh \tilde \theta_s) \ ,
\ee
where $\Omega_{\nu}(x)$ is the $SO(1,3)$ harmonic function and $\tilde f_\nu (s)$ the corresponding $SO(1,3)$ partial waves.
We can now formulate unitarity as
\be
{\rm Im} \tilde f_{\nu}(s) \geq 0 \ ,
\ee
and we obtain \eqref{eq:fsInSO4PW} after defining $c_\nu(s) = s^2 \Im \tilde f_\nu(s)$.

\section{Nonexistence of the Mandelstam representation at finite coupling}
\label{sec:nomandelstam}

In \eqref{eq:mandelstamRepFiniteg}, we asked about the possibility that the Mandelstam representation holds at finite coupling. If it were to be true, it would imply that the $SO(4)$ region (i.e. the bound state region) satisfies crossing symmetry independently as the cut part is explicitly crossing-symmetric in the Mandelstam representation. This is precisely what happens in perturbation theory at weak coupling. In other words, the Mandelstam representation yields  
\begin{equation}\label{eq:X_crossing_SO4_only}
    0\overset{?}{=}\avgBS{\frac{\PJfive\(1+ \frac{2t}{z_{5d}}\)}{z(z-s)} - \frac{\PJfive\(1+ \frac{2s}{z_{5d}}\)}{z(z-t)}}\,,
\end{equation}
where we used the notation $\avgBS{\cdot}$ defined in \eqref{eq:spdpwBS}.
In particular, it implies a subset of constraints $\chi_{n,0}^{SO(4)}\overset{?}{=}0$, where their explicit form is
\begin{equation}
    \chi_{n,0}^{SO(4)} = \avgBS{\frac{1}{z^{n+2}}\(1-\left(\frac{z}{z_{5d}}\right)^n\frac{(-1)^n(-J)_n (J+2)_n}{  n! \left(\frac{3}{2}\right)_n}\)}\, \ .
\end{equation}
For any finite $g$, these constraints cannot be satisfied with the allowed spectrum, as we can always find an $n$ such that the entries are sign definite for all $J$ with $z\geq m_{0,J}^2$. Thus, the $SO(4)$ crossing constraint \eqref{eq:X_crossing_SO4_only} cannot be satisfied, see \figref{fig:example_no_doubleDisc}.  

\begin{figure}[h!]
    \centering
    \includegraphics[width=0.94\linewidth]{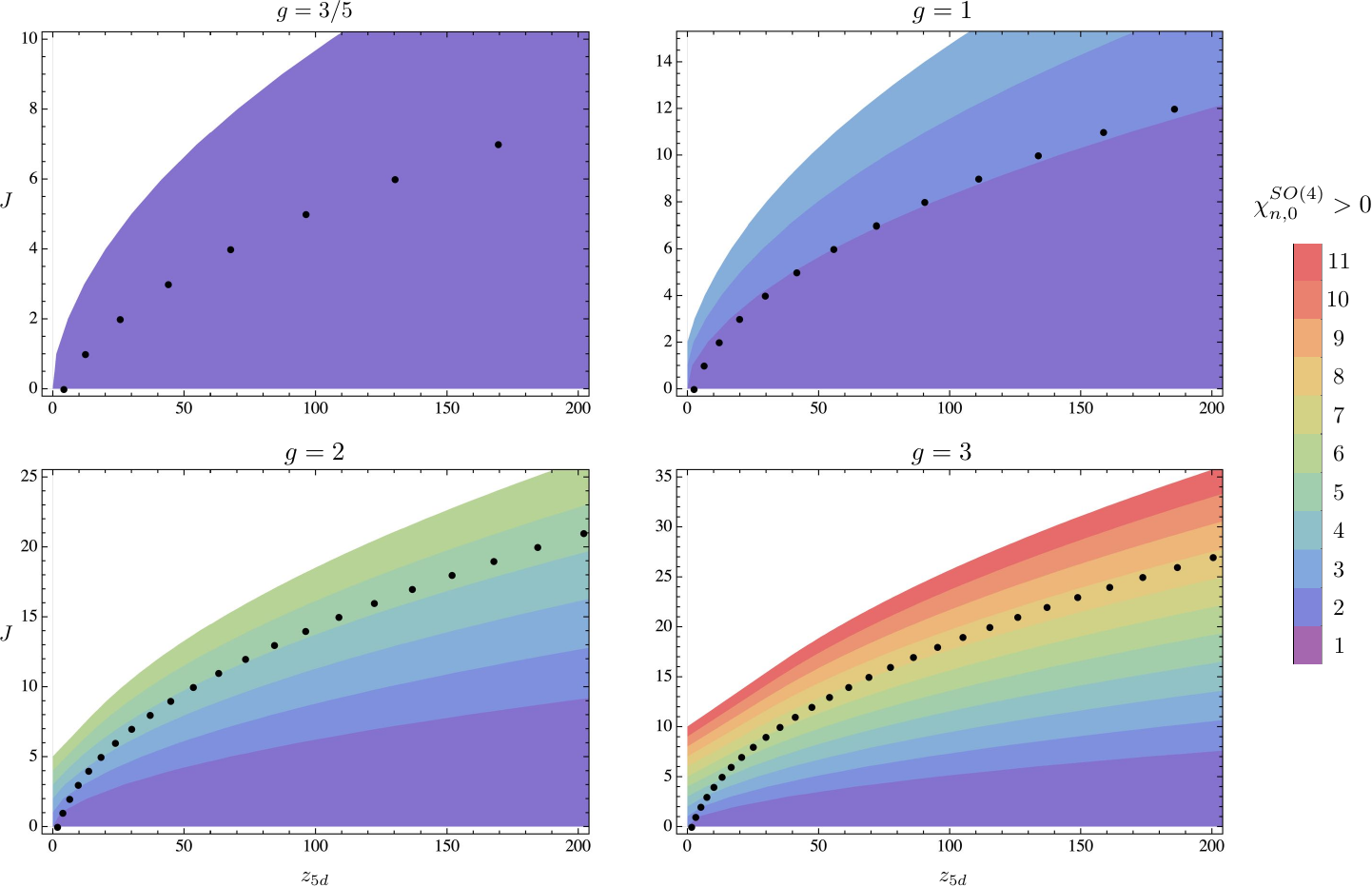}
    \caption{Examples of null constraints that cannot be satisfied on the allowed spectrum. The black dots represent the leading trajectory $j_0$, and the allowed spectrum is on the right of $j_0$. The filled colours represent when the sum rule for the null constraint $\chi_{n,0}^{SO(4)}(z_{5d}, J)$ is positive.}
\label{fig:example_no_doubleDisc}
\end{figure}

Intuitively, we expect the problem to arise from the following simple fact. To write the Mandelstam representation we require that the amplitude is polynomially bounded in the $(s,t)$ cut-plane. We know that at finite coupling in the Regge limit $t \to \infty$ and $s \to 4m^2$ we have $f(s,t) \sim t^{\Omega(g)/\sqrt{1-s/4m^2}}$ which is not polynomially bounded. On the other hand, if we expand the amplitude in perturbation theory, it is polynomially bounded at every order, though we expect higher and higher order singularities close to $s=4m^2$, which can be treated by introducing higher and higher order subtractions in the Mandelstam representation, see \eqref{eq:Imf_from_G_NR}.

\section{Discrete symmetries of the Maldacena-Wilson cusp}
\label{app:parity}

The Maldacena-Wilson cusp exhibits two  $\mathbb{Z}_2$ symmetries. To describe them, it is convenient to introduce a pair of orthogonal unit vectors, see Figure \ref{fig:cuspplane},
\be
n_{\pm}^{\mu} = {v_2^\mu \pm v_1^\mu \over |v_1 \pm v_2|} \ .
\ee
We have two extra coordinates in the plane transverse to the cusp for which we can use orthogonal coordinates $(w,z)$.
\begin{figure}[h!]
    \centering
\includegraphics[width=0.5\linewidth]{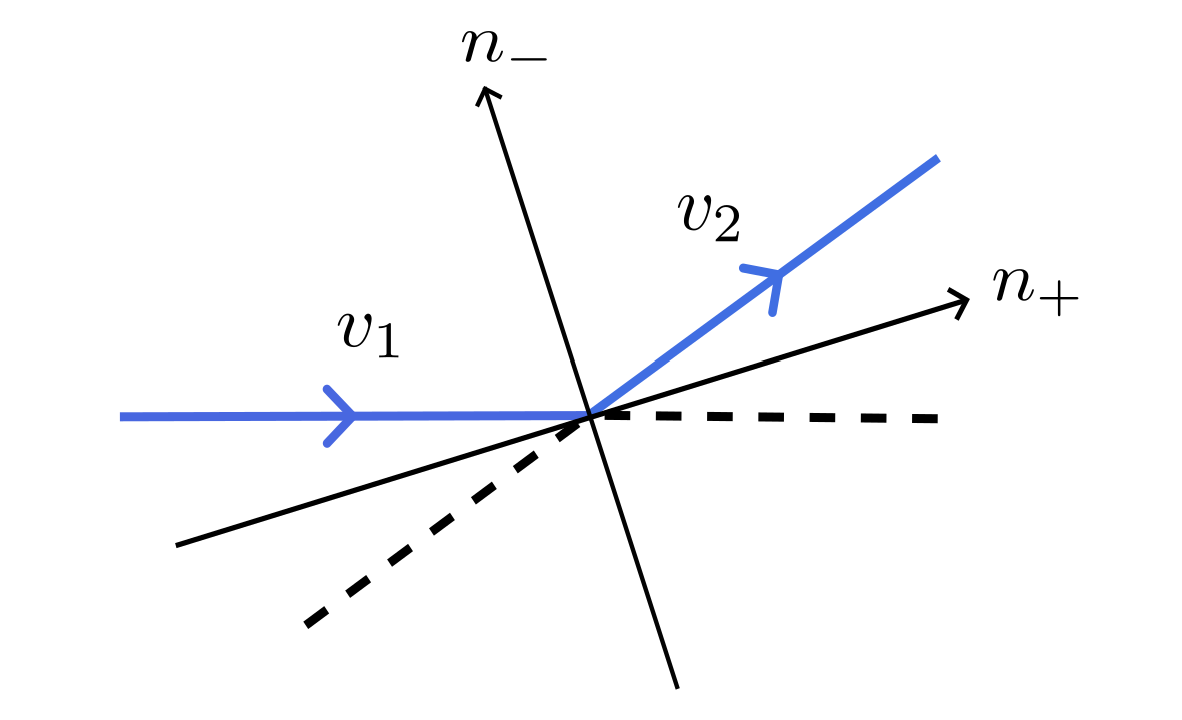}
    \caption{Definition of the orthogonal unit vectors $n_{\pm}$ w.r.t. the Maldacena-Wilson line with a cusp. Reflections around these two vectors are used to classify the discrete symmetry properties of local operators.}
    \label{fig:cuspplane}
\end{figure}
The discrete symmetries then are related to the reflections around $n_{\pm}^{\mu}$.

Let us first consider reflection around $n_+^\mu$, which acts as $n_-^\mu \to -n_-^\mu$ in the cusp plane and trivially in the transverse plane. It maps $H_{\text{cusp}}(\phi)$ to $H_{\text{cusp}}(-\phi)$ and therefore all energy eigenvalues obey $E_{n}(\phi)=E_{n}(-\phi)$. We can then map the cusp back to itself by combining it with the rotation by $\pi$ in the $(n_-,w)$-plane
\be
\mathbb{Z}_2^\perp :  R_{n_-,w}^\pi \circ P_{n_-}=P_{w} \ . 
\ee
It acts as $w \to -w$ in the transverse plane, therefore we can call this symmetry transverse parity. In this paper, we are only interested in operators that are invariant under the transverse $SO(2)$ rotations, and therefore, they are characterized by charge $\pm 1$ under $\mathbb{Z}_2^\perp$. 

The second symmetry is a finite-$\phi$ version of the parity transformation defined in \cite{Cavaglia:2023mmu}. In this case we combine charge conjugation $C: \cO \to -\cO^T$, where $\cO$ are fundamental fields represented as $N \times N$ matrices \eqref{eq:funda_fields},  
the R-symmetry flip $P_{\Phi_{||}}:\Phi_{||} \to -\Phi_{||}$, and reflection around $n_-^{\mu}$, which acts as $n_+^\mu \to -n_+^\mu$ in the cusp plane and trivially in the transverse plane. The first two transformation reverse the orientation of the line, while the third one maps it back to itself
\be
\mathbb{Z}_2^+ :  P_{n_+} \circ P_{P_{\Phi_{||}}} \circ C  \ . 
\ee
At the level of local operators this transformation exchanges $v_1^{\mu} \to - v_2^{\mu}$ and $v_2^{\mu} \to - v_1^{\mu}$ which acts as crossing $p_1 \leftrightarrow p_2$ at the level of the scattering amplitude. Let us list the discrete properties of the operators with bare dimension $\Delta \leq 2$ considered in the paper.
\begin{table}[ht]
  \centering
  \label{tab:parity}
\begin{tabular}{ |c|c|c| } 
 \hline
 ${O}$ & $\mathbb{Z}_2^+$ & $\mathbb{Z}_2^\perp$ \\ 
 \hline
 $1$ & $+$ & $+$ \\ 
 \hline
 $\Phi_{||}$ & $+$ & $+$ \\  
 \hline
 $\Phi_{\perp}^i \Phi_{\perp}^i + \sqrt{5} (\Phi_{||})^2$ & $+$ & $+$ \\ 
 \hline
  $\Phi_{\perp}^i \Phi_{\perp}^i - \sqrt{5} (\Phi_{||})^2$ & $+$ & $+$ \\ 
 \hline
 $F_{\mu \nu} n_+^{\mu} n_-^{\nu}$ & $+$ & $+$ \\
 \hline
 $n_-^\mu D_{\mu} \Phi_{||}$ & $+$ & $+$ \\
 \hline
 $F^{\mu \nu} n_+^{\rho} n_-^{\sigma} \eps_{\mu \nu \rho \sigma}$ & $-$ & $-$ \\
 \hline
 $n_+^\mu D_{\mu} \Phi_{||}$ & $-$ & $+$ \\
 \hline
\end{tabular}
\caption{Quantum numbers of the low-lying transverse $SO(2)$-invariant operators under the discrete symmetries of the cusp.}
\label{tab:opers}
\end{table}

Focusing on $\mathbb{Z}_2$-even operators we get four operators at $\Delta=2$, see Table \ref{tab:opers},  two of them are spinning in the sense that in the forward limit $v_2^{\mu} \to v_1^\mu$ they transform nontrivially under the emergent $SO(3)$ symmetry. This is the counting used in the main part of the paper. For the straight line case, the discrete symmetries were discussed recently in \cite{Gabai:2025hwf}. Our symmetry $\mathbb{Z}_2^+$ is called $\mathbf{CT}$ there, whereas the second symmetry considered there, i.e. $\mathbf{R}$, acts as $(n_-,w,z)\to -(n_-,w,z)$, and it is given by $\mathbf{R} = R_{n_-,z}^\pi \circ P_{w}$. This transformation is not a symmetry of the cusp, or, equivalently, when $\phi \neq 0$. The displacement operator is $\mathbf{R}$-odd, however, for us it decomposes into a $SO(2)$-vector which we discard, and $SO(2)$-scalar which is $\mathbb{Z}_2^{\perp}$-even.

\section{Perturbative results database}

\subsection{Regge trajectories}
\label{app:perturbative_result_database}

In this section we collect the perturbative data available up to three loops for $j_0(s)$ and $j_1(s)$. We follow \cite{Correa:2012nk,Bruser:2018jnc}, and parametrize the result in terms of 
\begin{equation}
    x  = e^{i \phi}\,, \qquad \xi = \frac{1-x}{1+x} = -i \tan(\phi/2)\,.
\label{eq:conventions_angle}
\end{equation}
Recall that $s$ and $\phi$ are related by $s = 4 m^2 \sin^2(\phi/2)$.
The result for $j_0(s)$ is 
\begin{equation}
\begin{aligned}
    j_0(s) = &-1 + 2 g^2 \xi \log(x) + \frac{4}{3}g^4 \Big[ -\xi \log(x) (\pi^2 + \log(x)^2) + \xi^2 (\log(x)^3 + 6 \log(x)^2 \log(1-x^2) + \\ &+ 6 \log(x) \text{Li}_2(x^2)  + 3 \log(x) \text{Li}_2(1-x^2) - 3 \text{Li}_3(x^2) + 3 \zeta_3 )  \Big] + g^6 \Big[ -\frac{4}{3} \xi \log(x) (\pi^2 + \log(x)^2)^2 + \\ & + \xi^2 ( 8 H_{1,1,1,2} + 6 H_{1,2,1,1} - \frac{4}{15} (10 \pi^2 \log(x)^3 + 11 \log(x)^5 - 20 \log(x)^4 \log(1-x^2) + \\
    &- 40 \log(x)^3 \text{Li}_2(x^2) + 10 \pi^2 \log(x) \text{Li}_2(1-x^2) + 60 \log(x)^2 \text{Li}_3(x^2) - 60 \log(x) \text{Li}_4(x^2) + \\
    &+ 30 \text{Li}_5(x^2) - 90 \log(x)^2 \zeta_3 - 30 \zeta_5 )  ) + \xi^3 (8 H_{1,1,1,2} + 8 H_{1,2,2} + 8 H_{1,1,1,2} + 4 H_{1,1,2,1} + \\
    &+ 4 H_{1,2,1,1} - \frac{8}{5} \log^5(x) )\Big]\,,
\end{aligned}
\end{equation}
where the harmonic polylogarithms $H_n$ \cite{Remiddi:1999ew} are functions of $1-x^2$, see \appref{sec:HPL} for a brief review.
For $j_1(s)$ 
\begin{equation}
    \begin{aligned}
        \rm j_1(\phi) &= -2-4g^2 + g^4 \Big[-\frac{1}{\xi} \left(4 \zeta_2 H_1 + \frac{H_1^3}{6}\right)-8\xi H_1 + 2 H_1^2 + 16 (\zeta_2 + 1)   \Big] + \\
        &+ g^6 \Big[ 32 H_{1,2}-16(4 \zeta_2-6 \zeta_3 + 15 \zeta_4 + 8) - 12 (2 \zeta_2 + 1) H_1^2 - \frac{5 H_1^4}{6} + \frac{8 H_1^3}{3} + \\
        &+ \xi \left( 48 (\zeta_2 + 2 ) H_1 + \frac{14 H_1^3}{3} - 16 H_1^2 - 64 H_2 \right) - 8 \xi^2 H_1^2 + \\
        & + \frac{1}{\xi} \left( -8 H_{1,1,2} + \frac{2}{3}(4 \zeta_2 +1) H_1^3 + 8 (2 \zeta_2 - 3 \zeta_3 + 11 \zeta_4) H_1 + \frac{H_1^5}{20} - \frac{H_1^4}{6} \right) \Big].
    \end{aligned}
\end{equation}

The residue functions $r_0(s)$ and $r_1(s)$ computed in \cite{Bruser:2018jnc} read (here we just show the result up to one loop, but the three-loop expressions can be found in the ancillary Mathematica notebook of \cite{Bruser:2018jnc})
\begin{equation}
    \begin{aligned}
        &r_{0}(s) = 1 + g^2 \xi (\pi^2 - 4 H_{-2}(x) - 4 H_{2}(x)) +  \mathcal{O}(g^4)\,,\\
        &r_{1}(s) = 2 + 8 g^2 (2 H_{-1}(x) + 2 H_1(x) -1) + \mathcal{O}(g^4)\,.
    \end{aligned}
\end{equation}

\subsection{Wilson coefficients}\label{app:weak_and_strongCoupling}

In this appendix we report the weak- and strong-coupling expansions of the Wilson coefficients appearing in the low-energy expansion \eqref{eq:low_enery_expansion} of $f(s,t)$. At weak coupling we used the method explained in detail in Appendix C of \cite{Bruser:2018jnc} to expand the differential equations describing the amplitude in the $s/m^2, t/m^2 \ll 1$ limit. Up to three loops we obtained the following expansions (here we show just the first few, the others are reported in the attached Mathematica notebook):
\begin{equation}
    \begin{aligned}
        &a_{0,0} = \frac{g^2}{6 m^4}\,,\qquad a_{1,0} = \frac{g^2}{60 m^6}+\frac{g^4}{12 m^6}-\frac{g^6}{3 m^6}\,,\\
        &a_{2,0} = \frac{g^2}{420 m^8}+\frac{g^4}{45 m^8}-\frac{g^6}{24 m^8}\,,\qquad a_{2,1} = \frac{g^2}{840 m^8}+\frac{g^4}{180 m^8}\,,\\
        &a_{3,0} = \frac{g^2}{2520 m^{10}}+\frac{53 g^4}{10080 m^{10}}+\frac{g^6}{4320 m^{10}}\,, \qquad a_{3,1} = \frac{g^2}{7560 m^{10}} + \frac{g^4}{1260 m^{10}}+ \frac{g^6}{2160 m^{10}}\,, \\
    \end{aligned}
\end{equation}
At strong coupling, expanding \eqref{eq:super_Veneziano} and adding the correction of \secref{sec:Worldsheet bootstrap} we get 
\begin{equation}
    \begin{aligned}
        &a_{1,0} = \frac{g^3 \zeta (3)}{\pi ^3 m^6}+\frac{g^2}{120 m^6}\,, \qquad a_{2,0} = \frac{g^4}{90 m^8}+g^3 \left(\frac{\zeta (3)}{12 \pi ^3 m^8}+\frac{5 \zeta (5)}{8 \pi ^5 m^8}\right)\,, \\
        &a_{2,1} = \frac{g^4}{360 m^8} + g^3 \left(\frac{5 \zeta (5)}{2 \pi ^5 m^8}-\frac{\zeta (3)}{6 \pi ^3 m^8}\right)\,, \quad a_{3,0} = \frac{g^5 \zeta (5)}{\pi ^5 m^{10}} + g^4 \left(\frac{3 \zeta (3)^2}{8 \pi ^6 m^{10}}+\frac{47}{22680 m^{10}}\right)\,,\\
        &a_{3,1} = \frac{g^5 \left(2 \zeta (5)-\frac{\pi ^2 \zeta (3)}{6}\right)}{\pi ^5 m^{10}}+ g^4 \left(\frac{3 \zeta (3)^2}{4 \pi ^6 m^{10}}-\frac{13}{18144 m^{10}}\right)\,.
    \end{aligned}
\end{equation}

\subsection{Harmonic polylogarithms}\label{sec:HPL}
The harmonic polylogarithms are defined in \cite{Remiddi:1999ew}. They are identified by a set of $w$ indices of values $\{0,\pm 1\}$, grouped into $\vec{m}_w$ and are recursively defined as
\begin{equation}
    H(\vec0_w,x) = \frac{1}{w!} \ln^w(x)\,, \qquad H(\vec{m}_w,x) = \int_0^x \hspace{1mm} dx' \hspace{1mm} f(a,x') H(\vec{m}_{w-1},x') \quad \text{if } \hspace{1mm}\vec{m}_w \neq \vec0_w \,.
\end{equation}
with $a = m_w$ the leftmost component of $\vec{m}_w$. The function $f$ can take the following values
\begin{equation}
    f(0,x) = 1/x, \qquad f(1,x) = 1/(1-x), \qquad f(-1,x) = 1/(1+x)\,.
\end{equation}
Explicitly, for $w = 1$ we get 
\begin{equation}
    H(0,x) = \ln(x)\,, \qquad H(1,x) = -\ln(1-x)\,, \quad H(-1,x) = \ln(1+x)\,.
\end{equation}
Notice that above we have used the shortening convention of the indices in $\vec{m}_w$: proceeding from right to left, all the zeros in $\vec{m}_w$ are removed by increasing the absolute value of the closest non-zero index on the right by 1, so for example
\begin{equation}
    H_{1,2}(x) = H(1,0,1,x), \qquad H_{1,1,2}(x) = H(1,1,0,1,x)\,.
\end{equation}

\section{Flat-space string spectrum}
\label{eq:fspspectrum}

For the analysis of \secref{sec:Worldsheet bootstrap}, it is important to know the spectrum of bound states in the flat space limit. To do so, we start from the $10$d state counting of \cite{Hanany:2010da}, see formula (3.19) in that paper. The string states are labelled using irreducible representations of $SO(9)$ and we consider the symmetric traceless ones, i.e. with the Dynkin labels $[l,0,0,0]_9$, since only these will appear in the four-point amplitude of interest. To classify the $SO(4)$ multiplets, we decompose the $SO(9)$ irreducible representations to $SO(4) \times SO(5)$ and we take into account just the $SO(5)$ singlets, since these are the states that can appear in the scattering process that we consider. 
In the $SO(4) \times SO(5)$ notation, the states that we consider are labelled as $[J,J;0,0]$ and we denote them with $(n,J)$ where $n$ is the mass level and $J$ is the spin. $n$ is related to the power of $q^\#$ in the expansion of \cite{Hanany:2010da} as $n = \# + 1$. For the first few mass levels the decomposition gives
\begin{itemize}
    \item at $q^0: (1,0)$ state;
    \item at $q: [1,0,0,0]_9 \to [1,1;0,0]$, so a $(2,1)$ state; 
    \item at $q^2: [2,0,0,0]_9 \to [0,0;0,0] + [2,2;0,0]$ so a $(3,0)$ and a $(3,2)$ state;
    \item \dots
\end{itemize}

In the \figref{fig:superstring_spectrum} we show the result of this computation and the corresponding leading and subleading linear trajectories. The small numbers on the states denote their degeneracy which naively comes from counting. 

\begin{figure}[h!]
    \centering
    \includegraphics[width=0.75\linewidth]{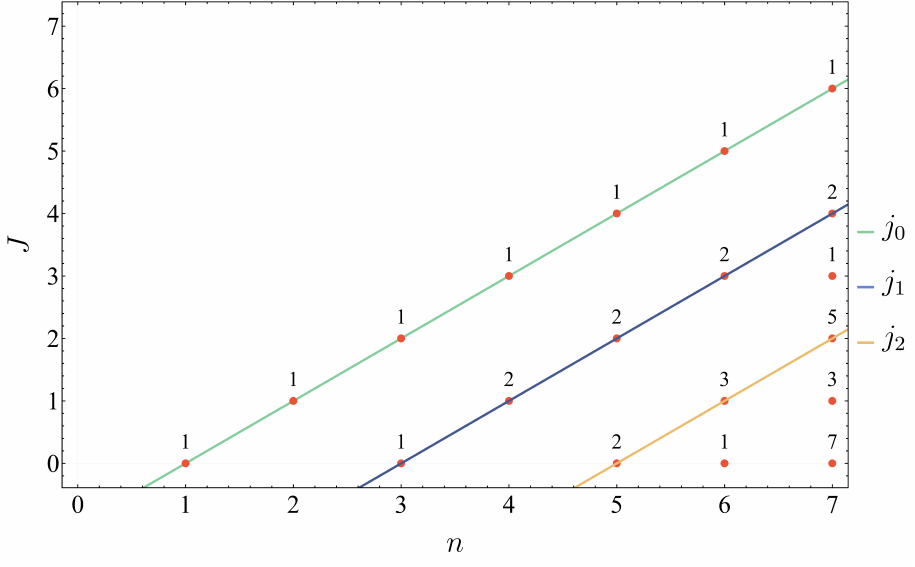}
    \caption{Regge trajectories and the spectrum of bound states in flat space from the decomposition of irreducible representations in $10$d. The numbers on top of the states denote degeneracy.}
    \label{fig:superstring_spectrum}
\end{figure}

\section{Leading-order weak-coupling result for $\lambda_{0,J}^2$}
\label{app:residue}

Recall that the scattering amplitude $f(s,t)$ describes a process in which external particles produce a pair of massive $W$-bosons which subsequently interact with each other. Close to the two-particle threshold, $s \to 4m^2$, the W-bosons become non-relativistic, and we expect that the scattering amplitude admits a one-body effective description \cite{Todorov:1970gr,Correia:2024jgr}. 
At leading order, the W-bosons interact via a Coulomb potential, and the effective description allows one to re-sum the series of ladder (Born) diagrams that dominate in the nonrelativistic limit. This was used in \cite{Bruser:2018jnc}, together with the results of \cite{Beneke:2013jia}, to predict the near-threshold behavior of the Coulomb branch amplitude for $s>4m^2$. Here we extend the analysis of \cite{Bruser:2018jnc} to $s<4m^2$, and use the results of \cite{Beneke:2013jia} to calculate the leading-order weak-coupling result for the three-point coupling $\lambda_{0,J}^2$.

In \cite{Schwinger:1964zzb}, Schwinger computed the Green's function of the nonrelativistic Coulomb potential in momentum space\footnote{We follow the conventions of \cite{Beneke:2013jia}, where the usual factor of $1/(2 \mu)$ in the non-relativistic propagator ${\vec p^2 \over 2 \mu}$ is replaced by $1/m$ since in our case the reduced mass $\mu$ is $m/2$. Recall that in the two-body problem $\mu = {m_1 m_2 \over m_1+m_2}$ and in our case $m_1 = m_2 = m$.}
\begin{equation}
\begin{aligned}
    \tilde G(\vec{p},\vec{p}',E) = &-\frac{(2 \pi)^3 \delta^{(3)}(\vec{p}-\vec{p}')}{E-\vec{p}^2/m} + \frac{1}{E-\vec{p}^2/m} \frac{16 \pi^2 g^2}{(\vec{p}-\vec{p}')^2} \frac{1}{E-\vec{p}'^2/m} + \\
    &+ \frac{1}{E-\frac{\vec{p}^2}{m}} \left(\int_0^1 dt \frac{16 \pi^2 g^2 \lambda \hspace{1mm} t^{-\lambda}}{(\vec{p}-\vec{p}')^2 t - \frac{m}{4E} (E-\frac{\vec{p}^2}{m})(E-\frac{\vec{p}'^2}{m})(1-t)^2}\right) \frac{1}{E-\frac{\vec{p}'^2}{m}}\,,
\label{eq:coulomb_eq}
\end{aligned}
\end{equation}
with 
\begin{equation}
    E = \sqrt{s}-2m\,, \qquad \lambda = \frac{4 \pi g^2}{2\sqrt{-E/m}}\,.
\end{equation}
$\tilde G(\vec{p},\vec{p}',E)$ has a cut for $E > 0$ and a set of bound states for $E < 0$. We would like to compute the bound-state residues of the Green's function and then relate them to the residues of the scattering amplitude $f(s,t)$. At weak coupling, the bound states of $f(s,t)$ are localized close to $s = 4m^2$ and they can be described as two particles interacting through a nonrelativistic Coulomb potential.\footnote{At finite coupling, we can focus on the nonrelativistic regime, where the description described here applies, by considering the large spin limit $J \gg 1$.} 
In \cite{Beneke:2013jia}, the residues are computed from the Fourier transform $G(\vec{r},\vec{r}',E)$ of \eqref{eq:coulomb_eq}, evaluated at $\vec{r} = \vec{r}' = 0$,
\begin{equation}
    G(0,0,E) = \frac{m^2}{4 \pi} \left( -\sqrt{\frac{E}{m}} - 4 \pi g^2 \left( \frac{1}{2} \ln\left(\frac{-4 m E}{\mu^2}\right)-\frac{1}{2} + \gamma_E + \Psi(1-\lambda) \right) \right)\,,
\end{equation}
where $\Psi$ is the digamma function.  Expanding around its poles at $E_n = -\frac{(4 \pi)^2 m g^4}{4 n^2}$ with $n = 1,2,\dots$ we get
\begin{equation}
    G(0,0,E) = -\frac{8 \pi^2 m^3 g^6}{n^3 (E-E_n)} + \mathcal{O}((E-E_n)^0)\,.
\end{equation}
By expressing $s(E) = (2m + E)^2$ it is easy to derive the expansion of the Green function around the poles in $s$ at locations
\begin{equation}
    s_n = (2m + E_n)^2 \equiv m_{0,J}^2, \quad J \equiv n-1\,.
\end{equation}
It reads
\begin{equation}
    G(0,0,s) = -\frac{8 \pi^2 m^3 g^6}{n^3} (2 m_{0,J}) \frac{1}{s-m_{0,J}^2} + \mathcal{O}((s-m_{0,J}^2)^0)\,,
\end{equation}
with $m_{0,J}=2m$ at the order in $g$ we are working at.
In \cite{Beneke:2013jia} the authors also present the expression of the imaginary part of the nonrelativistic Coulomb Green function above threshold $E > 0$: 
\begin{equation}
    \text{Im} G(0,0,E) = \sum_{n = 1}^{\infty} 8 \left(\frac{m \pi g^2}{n} \right)^3 \delta(E-E_n) + \theta(E) \frac{m^2 \pi g^2}{1-e^{-4 \pi^2 g^2/\beta_u}}\,,
\end{equation}
with $\beta_u = \sqrt{1-4m^2/s} \approx \sqrt{E/m}$ at threshold $s = 4m^2$. In order to match the non-relativistic Coulomb Green function with the amplitude $f(s,t)$ at weak coupling we can compare the imaginary parts at one loop and in the limit $s \to 4m^2$. We get
\begin{equation}\label{eq:Imf_from_G_NR}
    \text{Im} f(s,t) = \frac{\pi^2 g^2}{m^6} \text{Im} G(0,0,s) =  \sum_{n = 1}^{\infty} \frac{32 \pi^5 g^8}{m^2 n^3} \delta(s-m_{0,J}^2) + \theta(s-4m^2) \frac{\pi^3 g^4}{m^4(1-e^{-\frac{4 \pi^2 g^2}{\beta_u}})}\,.
\end{equation}
Finally we can match the expansion of $f(s,t)$ close to $s \to m_{0,J}^2$ from \eqref{eq:f_close_to_bs} with the residue of $f(s,t)$ computed above
\begin{equation}
    f(s,t)\underset{s\to m_{0,J}^2}{\sim} -\frac{\lambda_{0,J}^2  \PJfive(\cos\tilde\theta_s)}{m_{0,J}^2(s-m_{0,J}^2)} = -\frac{32 \pi^4 g^8}{m^2 (J+1)^3 (s-m_{0,J}^2)}\,.
\end{equation}
Since $\PJfive(\cos\tilde\theta_s) = 1$ for $s = 4m^2$ we can obtain the leading-order weak-coupling result for the residues $\lambda_{0,J}^2$ on the leading Regge trajectory
\begin{equation}
    \lambda_{0,J}^2 = \frac{128 \pi^4 g^8}{(J+1)^3}\,,
\end{equation}
where we have used that $n \equiv J+1$. Moreover, the formula \eqref{eq:Imf_from_G_NR} predicts that the four-loop correction to the scattering amplitude contains a pole $f(s,t) \sim -{32 \pi^4 \zeta(3) g^8 \over m^2(s-4m^2)}$.

In the discussion above, we observed that in the weak-coupling regime the non-relativistic limit of the scattering amplitude is governed by scattering in a Coulomb potential. This immediately implies the existence of hydrogen-like bound states and allows one to compute the corresponding three-point couplings. It is important, however, to point out a difference between interactions of the hydrogen-like states discussed here and the usual hydrogen atoms in QED.\footnote{This difference was already emphasized in \cite{Maldacena:2012sf}, see footnote 18 in that paper.} In the present case, interactions between different particles, either massless particles considered in the bulk of the paper, or massive bound states discussed here, are controlled by the same string coupling $g_s \sim g_{\text{YM}}^2$ which we take to be parametrically small in the planar limit. In particular, it means that the scattering amplitudes between the higher-spin bound states are parametrically small in the planar limit. Such a weak-coupling limit does not exist for the hydrogen-hydrogen scattering in QED. In this case, if we send the electromagnetic strength $g \to 0$, we expect that the effective interaction strength between a pair of hydrogen atoms grows because it is controlled by the Bohr radius, which grows as $1/g^2$. The resulting inter-atomic van der Waals interaction is, however, a non-planar effect which is absent in our case, see e.g. \cite{Feinberg:1970zz}.

\section{Details of the bootstrap problem}

In this section, we will provide details on the numerical implementation of the bootstrap problem at finite coupling.\footnote{We refer the reader to \cite{Caron-Huot:2021rmr, Albert:2024yap} for the special case of $g\to \infty$.}

\subsection{Sum rule for the pole}\label{sec:SRPole}
For the tree-level pole, we explained in \secref{sec:DR_and_SR} that a smeared sum rule can be written. Here we provide details and explain the choice of basis \eqref{eq:functionalPoleWithbasis}.

We begin by computing the action of a monomial weight on the Legendre and hyperbolic harmonics
\begin{align}
   \phi^{SO(4)}_p &\equiv \Phi_{\left(\frac{q}{q_0}\right)^p}\[\PJfive(1-\frac{2q^2}{z_{5d}})\] =\frac{\, _3F_2\left(-J,J+2,\frac{p+1}{2};\frac{3}{2},\frac{p+3}{2};\frac{q_0^2}{z_{5 d}}\right)}{p+1} \label{eq:poleFunc_1SO4}\\
   \phi^{SO(1,3)}_p &\equiv \Phi_{\left(\frac{q}{q_0}\right)^p}\[\Omega_\nu(1+\frac{2q^2}{\tilde z_{5d}})\] =\frac{\, _3F_2\left(1-i \nu ,i \nu +1,\frac{p+1}{2};\frac{3}{2},\frac{p+3}{2};-\frac{q_0^2}{\tilde z_{5 d}}\right)}{p+1}\label{eq:poleFunc_1SO13}\,.
\end{align}
In the impact parameter space defined in \eqref{eq:impact_b_def}, the $SO(4)$ and $SO(1,3)$ structures degenerate and yield the common functional
\begin{equation}\label{eq:poleFunc_impact}
    \phi_p^{\infty} = \frac{{}_1F_2\left( \frac{1+p}{2} ;  \frac{3}{2}, \frac{3+p}{2}; -\frac{(bq_0)^2}{4}  \right)}{p+1}\,.
\end{equation}
Here the superscript $\infty$ indicates the impact-parameter limit. For integer $p$, all $\phi_p$ can be expressed in the form 
\begin{equation}
    \phi_p(\cdot) = A(\cdot) + B(\cdot)\cos\varphi(\cdot) +C(\cdot)\sin\varphi(\cdot)\,,
\end{equation}
where $(\cdot)= (z_{5d},J),(\tilde z_{5d},\nu) $ or $(b)$. This form will be important in the choice of basis that we will discuss next.

Because the sum rule for the pole  \eqref{eq:tree-levelPoleSumRuleSmeared} originates from an anti-subtracted dispersion relation, it behaves at large $z$ as $\cO(1/z)$, and thus dominates over the other sum rules, which decay at least as $\cO(1/z^2)$. At large $z$, the dual functional \eqref{eq:dualWilsonCoefMax} becomes
\begin{equation}
    \cP^{SO(1,3)}(z,\nu) = z^{-1}(\underbrace{\sum_k c_k \Phi_{f_k} \[\PJfive(1-\frac{2q^2}{4m^2})\] }_{\cP_{\text{LE}}^{SO(1,3)}(\nu)} ) + \cO(z^{-2})\,,
\end{equation}
and to solve the dual problem, we need to impose $\cP_{\text{LE}}^{SO(1,3)}(\nu)\geq 0$. This constraint will explain the choice of basis for the functional, which is constructed to ensure positivity at large $\nu$ and this is what we do next.

In order to impose positivity asymptotically, we start by considering the large $\nu$ limit of \eqref{eq:poleFunc_1SO13}
\begin{equation}
\begin{split}
 \phi_p^{SO(1,3)}(\tilde z_{5d}=4m^2,\nu) &= \frac{\, _3F_2\left(1-i \nu ,i \nu +1,\frac{p+1}{2};\frac{3}{2},\frac{p+3}{2};-X\right)}{p+1} \\
    &= -\frac{\cos \phi(\nu)}{4X}\frac{1}{\nu^2} +\frac{(p - 1)\sin \phi(\nu)\sqrt{1+X}}{8X^{3/2}}\frac{1}{\nu^3} \\
    &+ \frac{-2\delta_{p,2} + (p-1) ((p-1) X+p-2)\cos\varphi(\nu)}{16 X^2}\frac{1}{\nu^4}+ \cO(\nu^{-5})
    \end{split}
\end{equation}
where $\varphi(\nu)=2 \nu  \sinh ^{-1}\left(\sqrt{X}\right)$  and $X=\frac{q_0^2}{4m^2}<1$. This sum rule clearly oscillates in $\nu$. 
Therefore, we choose a basis such that the $1/\nu^2$ and $1/\nu^3$ terms vanish.
This can be done by choosing a basis element of the form $a_2(q/q_0)^2 + a_3(q/q_0)^3 + a_{n}(q/q_0)^n$. To  cancel the first two oscillating terms, we need
\begin{align}
    a_2 + a_3 + a_n &= 0, \\
    a_2 + 2a_3 + (n-1)a_n &= 0.
\end{align}
By choosing $a_3=1$, we obtain the basis $f(q)= \sum_{k\geq 1} c_k f_k(q)$ of \eqref{eq:functionalPoleWithbasis}, where  
\begin{equation}
    f_k(q) = \frac{k }{k+1}\left(\frac{q}{q_0}\right)^2-\left(\frac{q}{q_0}\right)^3 +\frac{1}{k+1}\left(\frac{q}{q_0}\right)^{k+3}\,.
\end{equation}
In the language of the dual formulation \eqref{eq:dualWilsonCoefMax}, the $c_k$ are dual variables.

In summary, the smeared pole sum rule for a basis element (which solves positivity at large $\nu$)  takes the form
\begin{equation}
\begin{split}
      \Phi_{f_k}[q^{-2}] =\frac{k}{2 (k+2) q_0^2}&= \avgBS{ \frac{1}{z}\left(\frac{k }{k+1}\phi^{SO(4)}_{2}-\phi^{SO(4)}_{3}+\frac{1}{k+1}\phi^{SO(4)}_{k+3}\right)} \\
      &+\avgCut{ \frac{1}{z}\left(\frac{k }{k+1}\phi^{SO(1,3)}_{2}-\phi^{SO(1,3)}_{3}+\frac{1}{k+1}\phi^{SO(1,3)}_{k+3}\right) }\,.
\end{split}
\end{equation}

\subsection{Numerical truncation of the constraints}\label{sec:numerical_truncations_details}
As explained in the definition of the dual problem \eqref{eq:dualWilsonCoefMax}, for a fixed dual functional, we must impose the following positivity constraints
\begin{equation}
\begin{split}\label{eq:positivity_constraints_app}
        \cP^{SO(4)}(z,J) &\geq 0 ~~\text{for}~ (z,J)\in (m_{0,J}^2, J)\cup (m_{1,J}^2<z<4m^2, J)\\
        \cP^{SO(1,3)}(z,\nu) &\geq 0 ~~\text{for}~ z\geq 4m^2 \,, \nu\geq 0\,.
\end{split}
\end{equation}
In practice, as discussed in \secref{sec:truncation_constraints_dual}, since this represents an infinite set of constraints, we discretize them. This appendix presents the details of this procedure. %
We then explain the adaptive refinement procedure that iteratively improves the grids to ensure \eqref{eq:positivity_constraints_app} are satisfied.

\paragraph{Bound-state region $z<4m^2$ -- fixed spin $J$}\mbox{}\\
In the bound state region, we distinguish the leading trajectory from the subleading trajectories (see \figref{fig:spectrumSchematic}). 
The leading-trajectory constraints are pointwise, we impose them for all $J$ up to $J_{\text{Large}}$
\begin{equation}
    \text{\textit{Leading trajectory:}}~~\cP^{SO(4)}(m_{0,J}^2,J)\geq 0 ~~\text{for}~~  J=0,1,\dots,J_{\text{Large}}
\end{equation}

For the subleading trajectories, whose positions are not fixed a priori, $z$ is treated as continuous. It is convenient to work with $z_{5d}=\frac{z}{1 - z/(4m^2)}$. At fixed $J$, the constraints are polynomial (or rational) and can be implemented efficiently in SDPB via the change of variables 
\begin{equation}
    z_{5d} = \frac{\msubJ{J}^2}{1-\frac{\msubJ{J}^2}{4m^2}} + x \equiv z_{th,J}+x
\end{equation}
imposed for $x\geq 0$. The pole sum rules \eqref{eq:poleFunc_1SO4}, as well as the amplitude sum rules, are polynomials of degree $\cO(J)$, which become numerically heavy at large spin. Thus, for $J\geq J_{\text{max}}$ we discretize the constraints in $x$. We use a Chebyshev grid\footnote{
At large spin, the hypergeometric argument is small and the sum rules can be approximated by lower-degree polynomials; nonetheless, a sparse discrete grid in $x$ is numerically advantageous.
}  
\begin{equation}\label{eq:chebGrid}
    x_k = C_{\text{Cheb}}\tan^2\left( \frac{\phi_k}{2} \right) \,,~~ \text{with} \,~\phi_k = \pi  \sin\left( \frac{k}{2n} \pi \right) , \quad k = 0, 2, \dots, N_{\text{Cheb}}-1\,.
\end{equation}

Furthermore, it is not necessary to include all spin constraints and above the spin threshold $J_{\text{max}}$ we select a subset using the grid
\begin{equation}
    \label{eq:listJLarge}
    \verb|list|\,J_{\text{Large}} = \verb|round|\left[\frac{(J_\text{max}+1) J_{\text{Large}} N_{\text{Large spin}}}{J_{\text{Large}}N_{\text{Large spin}} + (J_\text{max}+1 - J_{\text{Large}}) i}\right] , \quad i = 0, 1, \dots, N_{\text{Large spin}}-1
\end{equation}
Collecting everything together, we impose at fixed spin\footnote{Here and in the following, we slightly abuse notation by writing $\cP^{SO(4)}(z_{th,J} + x, J)$, which should be understood as shorthand for $\cP^{SO(4)}(z(z_{5d}), J)\big|{z_{5d} = z_{th,J} + x}$.} 
\begin{equation}
   \begin{array}{c}
        \textit{Subleading } \\
        \textit{trajectories }
    \end{array} :~~\begin{array}{l}\cP^{SO(4)}\(z_{th,J}+ x,J\)\geq 0 ~~\text{for}~~ x\geq 0\,, ~ J=0,1,\dots,J_{\text{max}}\\
    \cP^{SO(4)}\(z_{th,J} + x_k,J\)\geq 0 ~~\text{for}~~ x_k \in \eqref{eq:chebGrid}\,,\, J\in \verb|list|\,J_{\text{Large}}
    \end{array}\,.
\end{equation}
In doing this discretization, we introduced several parameters and typically, we used: $J_{\rm max}= \cO(50-100)\,,\,J_{\rm Large}= \cO(10^3)\,,\,N_{\text{Large spin}}=\cO(10^2)\,,\, C_{\text{Cheb}}=  \cO(1)$ and $N_{\text{Cheb}}=\cO(10^2)$.
To control the large-spin region, we also impose constraints at fixed impact parameter $b$ (see below).

\paragraph{Scattering region $z\geq4m^2$ -- fixed continuous spin $\nu$}\mbox{}\\
For $z\geq 4m^2$, both $\nu$ and $z$ are continuous. We discretize $\nu$ on a grid. We found it is necessary to use a denser grid close to $\nu=0$. We adopt a quadratic grid up to $\nu_{\rm max}$
\begin{equation}
    \verb|list|~\nu_{\rm max} = \left\{ \, \nu_{\max} \left(\frac{i}{N_\nu - 1}\right)^{2} \;\middle|\; i = 0,1,2,\dots, N_\nu - 1 \, \right\}\,.
\end{equation}
For larger $\nu$, we add a sparse list $\verb|list|~\nu_{\rm Large}$ constructed analogously to \eqref{eq:listJLarge}. Typically, we used $\nu_{\rm max}=\cO(50)\,,\,  N_\nu= C_\nu\cdot\,\nu_{\rm max}$ with $C_\nu = \cO(1)$ and $\nu_{\rm Large}=\cO(10^3)$.

For the energy variable we set
\begin{equation}
\tilde z_{5d} = 4m^2 + x,
\end{equation}
where $x \geq 0$.
Unlike the $(z,J)$ region, the sum rules evaluated away from $s,t\to 0$ are non-polynomial, for example the smeared sum rule for the pole \eqref{eq:poleFunc_1SO13}.
We are thus left with two options: either impose constraints on a discrete grid or approximate the sum rules by polynomials.\footnote{This is what one does in the CFT bootstrap, where conformal blocks are efficiently approximated by polynomials \cite{Kos:2013tga}, see also \cite{Chang:2025mwt} for a recent discussion.} In this work, we tested both approaches and we discuss the polynomial approximation in detail in \appref{sec:polynomial_constraints}. Since the required polynomial degree grows with $\nu$, we use a Chebyshev grid \eqref{eq:chebGrid} at large $\nu$. The same grid is also used at small $\nu$ when non-polynomial constraints are imposed. 

Summarizing, in this region we impose the following constraints
\begin{equation}
   \begin{array}{c}
        \textit{Scattering}\\
        \textit{region}
    \end{array} :~~\begin{array}{l}
        \verb|Poly|\left[\cP^{SO(1,3)}\(4m^2+ x,\nu\)\right]\geq 0 ~~\text{for}~~  x\geq 0 \,, ~ \nu\in \verb|list|~\nu_{\rm max}< \nu_{\rm poly}\\
    \cP^{SO(1,3)}\(4m^2+ x,\nu\)\geq 0 ~~\text{for}~~  x_k \in \eqref{eq:chebGrid}\,, ~ \nu\in (\verb|list|~\nu_{\rm max}\geq\nu_{\rm poly} \\
    \hspace{8cm}\cup~\verb|list|~\nu_{Large} )
    \end{array}\,.
\end{equation}
where $\nu_{\rm poly}$ denotes the threshold below which the polynomial approximation is applied. We emphasize that this represents the initial grid choice. These grids are subsequently refined through the adaptive-refinement procedure described in \appref{sec:adptative_refinement}.

\paragraph{Scattering region -- large energy $z\to\infty$ -- fixed continuous spin $\nu$}\mbox{}\\
As anticipated in \secref{sec:SRPole}, the first region requiring special attention is the large-$z$ limit (equivalently $\tilde z_{5d}\to 4m^2$) at fixed spin $\nu$ (region $5.\nu$ in \figref{fig:j_x_nu_x_regions}). The reason is that, at large energy, all sum rules decay as $\cO(1/z^2)$ or faster, except for the pole sum rules, which scales as $\cO(1/z)$. Accordingly, the dual functional has the asymptotic form
\begin{equation}
    \cP^{SO(1,3)}(z,\nu) = z^{-1}(\underbrace{\sum_k c_k \Phi_{f_k} \[\PJfive(1-\frac{2q^2}{4m^2})\] }_{\cP_{\text{LE}}^{SO(1,3)}(\nu)} ) + \cO(z^{-2})\,,
\end{equation}
and we must impose $\cP_{\text{LE}}^{SO(1,3)}(\nu)\geq 0$ for all $\nu\geq 0$.

Our basis guarantees that each element behaves as $1/\nu^4$ at large $\nu$. Furthermore, applying $f(q)= \sum_{k= 1}^{k_{\rm max}} c_k f_k(q)$ from \eqref{eq:functionalPoleWithbasis} yields, at order $1/\nu^4$, the constraint
\begin{equation}
    \sum_{k=1}^{k_{\rm max}} c_k + \cos\varphi(\nu) \sum_{k=1}^{k_{\rm max}}\frac{1+X}{2} k \,c_k \geq 0\,,
\end{equation}
with $X=q_0^2/4m^2$. This can be translated into linear constraints on the coefficients $c_k$
\begin{equation}\label{eq:asymptoticnu_threshold}
  \sum_{k=1}^{k_{\rm max}}c_k\geq 0\,,~~~ \sum_{k=1}^{k_{\rm max}} \(1-\frac{1+ \frac{q_0^2}{4m^2}}{2}k\)c_k\geq 0 \,,~~~ \sum_{k=1}^{k_{\rm max}} \(1+\frac{1+ \frac{q_0^2}{4m^2}}{2} k\)c_k\geq 0 \,.
\end{equation}

The choice of basis, together with the additional constraints \eqref{eq:asymptoticnu_threshold}, ensures positivity of $\cP_{\text{LE}}^{SO(1,3)}(\nu)$ asymptotically. In addition, we impose positivity on a grid and because the sum rule is oscillatory, we choose 
\begin{equation}
    \verb|list threshold |~\nu = \{\eps_\nu, \eps_\nu+ \delta_\nu, \dots, \eps_\nu + \left\lceil \frac{\nu_{\text{max}} - \eps_\nu}{\delta_\nu}-1\right\rceil \delta_\nu \} \cup \verb|list|~\nu_{\rm Large}\,.
\end{equation}

\paragraph{Impact-parameter space $b$}\mbox{}\\
To control the large-spin, large-$z_{5d},\tilde z_{5d}$ region, we define the natural impact-parameter variable $b$ via the double-scaling limit $J, z_{5d}\gg1$ 
\begin{equation}
    J= \frac{b \sqrt{z_{5d}}}{2}\,,
\end{equation}
and analogously for the $SO(1,3)$ variables. 
In this limit, the $SO(4)$ and $SO(1,3)$  sum rules degenerate and we must impose\footnote{Because of the spin constraints, the $SO(4)$ region only requires positivity for $b\leq b_*(g)$ whereas all $b\geq 0$ are required for the $SO(1,3)$ asymptotics.}
\begin{equation}
    \cP^{\infty}(b)\geq 0~~ \text{for}~~ b\geq 0\,.
\end{equation}

With the chosen functionals for the pole and the amplitude (see \secref{sec:SRPole}), this constraints takes the form \cite{Caron-Huot:2021rmr}
\begin{equation}
    \label{eq:PinImpactParameterSpace}
    \cP^{\infty}(b) = A(b) + B(b) \cos(b q_0) + C_1(b)\sin(b q_0)+  C_2(b)\sin(b\sqrt{-t_*})\,,
\end{equation}
where $A(b), B(b), C_1(b)$ and $C_2(b)$ are polynomials (or rational functions). The terms $B$ and $C_1$ originate from the pole sum rule \eqref{eq:poleFunc_impact}, while $C_2$ comes from the amplitude $f(s_*, t_*)$. Due to its oscillating nature, for small $b< b_{\text{max}}$, we impose this constraint pointwise on a grid $b\in \verb|list|~b$ 
\begin{equation}
    \verb|list|~b = \{\eps_b, \eps_b+ \delta_b, \dots, \eps_b + \left\lceil \frac{b_{\text{max}} - \eps_b}{\delta_b}-1\right\rceil \delta_b \}\,.
\end{equation}

To ensure positivity for large values of $b$, we enforce a set of polynomial matrix constraints that are strictly stronger than those in \eqref{eq:PinImpactParameterSpace}. To do so, we consider the general case with two frequencies and replace \eqref{eq:PinImpactParameterSpace} with
\begin{equation}\label{eq:Pb_general}
     P(b) = A(b) + B_1(b) \cos(\varphi_1(b)) + C_1(b)\sin(\varphi_1(b))+  
     B_2(b) \cos(\varphi_2(b)) + C_2(b)\sin(\varphi_2(b)) \,.
\end{equation}
To cast the expression into polynomial matrix form, we define the vectors $v_i = (\cos(\varphi_i/2), \sin(\varphi_i/2))$ and the matrices
\begin{equation}\label{eq:PiMatrix}
\mathbb{P}_i(b)=\begin{pmatrix}
t_i + B_i(b) & C_i(b) \\
C_i(b) & t_i - B_i(b)
\end{pmatrix}\,,~~~i=1,2\,.
\end{equation}
Rewriting \eqref{eq:Pb_general} as
\begin{equation}
    P(b) = v_1^T \mathbb{P}_1v_1 +v_2^T \mathbb{P}_2v_2 \Big|_{t_1 + t_2 = A(b)}\,,
\end{equation}
it follows that imposing the semidefinite constraints
\begin{equation}
    \mathbb{P}_i(b) \succeq 0\,,~~~  t_1+t_2= A(b)\,,~~~i=1,2\,,
\end{equation}
guarantees that $P(b) \ge 0$ and we impose it for $b\geq b_{\rm max}$. In this formulation, $t_1$ is commonly referred to as a \textit{slack} variable.

For large values of $b$, the functional is dominated by the contribution of the sum rules evaluated in the forward limit. Indeed, $a_{n,\ell}$ and $\chi_{n,\ell}$ scale as $\sim b^{2\ell}$, whereas $\Phi_{f_k}[q^{-2}] \sim 1/b^4$ and $f(s,t) \sim 1/b$. Consequently, in \eqref{eq:PinImpactParameterSpace}, $A(b) \sim b^{n_{\rm max}}$, while $B(b) \sim 1/b^4$, $C_1(b) \sim 1/b^5$, and $C_2(b) \sim 1/b$. Therefore, for sufficiently large $n_{\rm max}$ and $b_{\rm max}$, the constraint can be approximated by $A(b) \ge 0$. This is the approximation we adopt in practice when both sum rules are included.

\subsubsection{Adaptive refinement}\label{sec:adptative_refinement}
In the previous subsection, we described the initial grid used to impose the constraints \eqref{eq:positivity_constraints_app}.
Our goal, however, is to ensure that \eqref{eq:positivity_constraints_app} are satisfied over the entire domains. To achieve this, we employ an adaptive refinement procedure, which proceeds as follows:
\begin{enumerate}
    \item Choose an initial collection of grids, collectively denoted by  $\verb|grids_0|$. 
     \item Using this grid, run SDPB to obtain a candidate dual solution $\mathcal{D}_0$ to \eqref{eq:dualWilsonCoefMax}.
     \item Search for regions where the dual functional $\mathcal{P}$ in \eqref{eq:positivity_constraints_app} becomes negative in the $(x, J)$, $(x, \nu)$, and $b$ spaces. This step, detailed below, produces a list of negative points $\verb|list_negative_0|$, which are added to the grid to form $\verb|grids_1| = \verb|list_negative_0|+\verb|grids_0|$.
     \item Using $\verb|grids_1|$, run SDPB again to obtain an updated candidate solution $\mathcal{D}_1$ to the dual problem.
     \item Iterate this procedure until the remaining negative regions become negligible. A region is considered negligible if its size, in all directions ($x$, $J$, $\nu$, or $b$), falls below a chosen threshold (typically we choose $10^{-6}$).
\end{enumerate}
At the end of this procedure, we obtain a solution $\mathcal{D}_*$ that is positive except for negligible negative regions. We next investigate these regions in detail.\footnote{
A formal proof of positivity would require exact-arithmetic verification, which is nontrivial when the functionals are non-polynomial.
}

\paragraph{Negativity search}\mbox{}\\
We aim to identify all regions in which \eqref{eq:positivity_constraints_app} is violated by scanning the $(x, J)$, $(x, \nu)$, and $b$ spaces. Because the functionals are generally non-polynomial, this task is computationally challenging and must therefore be carried out numerically.

We begin by describing the procedure in the $(x, \nu)$ plane, which corresponds to the $SO(1,3)$ scattering region. For a given dual solution $\mathcal{D}$, $\mathcal{P}(x,\nu)$ defines a real-valued function over the first quadrant. We begin by testing positivity at the threshold $x = 0$. We partition the interval $[0, \nu_{\rm big}]$ into $N_{\rm th}$ subintervals $[\nu_i, \nu_{i+1}]$. Within each sub-interval, we minimize the function starting from its midpoint. This step is carried out at machine precision. If a negative minimum is found, we map the corresponding region and sample points within it, which we store in \verb|list_negative|. The intervals $[\nu_i, \nu_{i+1}]$ are chosen pseudo-randomly to ensure that at each SDP step, the minimization starts from a distinct initial point.

For $x>0$ we adopt a similar strategy. We divide the domain $[x_{\rm small}, x_{\rm big}] \times [0, \nu_{\rm big}]$ into $N_x \times N_\nu$ boxes. In each box, we perform a two-dimensional minimization and append points within negative regions to \verb|list_negative|. 

In the bound-state region $(x,J)$, we proceed analogously, using $N_x \times N_J$ boxes spanning $[x_{\rm small}, x_{\rm big}] \times [J_{\rm max}, J_{\rm big}]$. The functional is analytically continued in $J$, and we search for negative regions within each box. When a negative region is detected, we identify the region and retain only points corresponding to integer $J$ values in \verb|list_negative|.

Finally, the search in impact-parameter space $b$ is carried out analogously to the threshold case ($x=0$) in the scattering region. Here, we consider intervals in $[\eps_b, b_{\rm max}]$, noting that by the polynomial-matrix constraint the functional is known to be positive for $b \geq b_{\rm max}$.

To perform this search, we introduce various numerical parameters defining the number of boxes and the range of the space spanned. The choice of the number of boxes $N_i$ should be large enough to avoid missing negative regions within a given box (e.g. due to a local minima), and depends on the number of possible negative regions identified at each step. In practice, we used $N_i \sim \mathcal{O}(10^2)$. For the energy direction, we choose $x_{\rm small} \sim 10^{-4}$ and $x_{\rm big} \sim 10^{4}$. As explained above, we also include a point at $x = 0$. In the spin direction, we choose $J_{\rm big}$ to be of the same order as $J_{\rm Large} \sim \mathcal{O}(10^3)$. Similarly, we choose $\nu_{\rm big}$ to be larger but of the same order as $\nu_{\rm Large}$. Typically, we do not find negative regions near the upper thresholds of the search space. However, if that were to happen, we can increase the boundaries to ensure that such regions are not missed. 

\paragraph{Remark}\mbox{}\\
This refinement step is computationally expensive, as it requires running SDPB multiple times.
In practice, we typically solve the dual problem on a large fixed grid. For a selected subset of points we then performed the adaptive-refinement procedure to verify convergence of the objective function.

\subsubsection{Polynomial constraints}\label{sec:polynomial_constraints}
As discussed above, in the scattering region $SO(1,3)$, the sum rules evaluated away from $s,t\to 0$ are not polynomial. This follows directly from the form of the $SO(1,3)$ harmonics
\begin{equation}\label{eq:Omegainapp}
    \Omega_\nu(\cos\tilde\theta_z) = {}_2F_1\(1-i \nu, 1+i\nu;\frac{3}{2}; \frac{t}{\tilde z_{5d}} \)\,
\end{equation}
which are non-polynomial for $\nu\geq 0$ and $t\neq 0$. 

In this work, the non-forward sum rules  correspond to the smeared tree-level pole $\Phi[q^{-2}]$ and to the amplitude and its derivatives $f_{n,\ell}$. 
As an example, we now explain how \eqref{eq:Omegainapp} can be approximated by a polynomial in $1/{\tilde z_{5d}}$.
The basic idea is to approximate the function using Chebyshev polynomials.

We begin by mapping the semi-infinite interval $\tilde z_{5d}\geq 1$ to $[-1,1]$ via $x=-1+\frac{2}{\tilde z_{5d}}$, thereby defining a function $f(x)$ on $[-1,1]$. 
We then approximate $f(x)$ by a polynomial of degree $N$ 
\begin{equation}
    f_{\rm approx}^N(x)= \sum_{k=0}^{N} c_k T_k(x)\,,
\end{equation}
where $T_k(x)$ are the Chebyshev polynomials of the first kind. 
The coefficients $c_k$ are efficiently computed via a discrete cosine transform, by evaluating $f(x)$ at the Chebyshev nodes 
\begin{equation}
    c_k = \frac{2-\delta_{0k}}{N}\sum_{i=0}^N \cos\(k \theta_i\) f(x_i)\,,
\end{equation}
where $x_i=\cos\theta_i$ and $\theta_i=\frac{\pi  \left(i-\frac{1}{2}\right)}{N}$. Substituting back $x=-1+2/\tilde z_{5d}$, leads to a polynomial approximation of \eqref{eq:Omegainapp} in $1/\tilde z_{5d}$. The quality of the approximation is given by $N$. In practice, we observed that to have a reasonable approximation, $N$ scaled with $\nu$ and $|t|$ as shown in \figref{fig:chebychev_nu_example}. We used $N\sim \nu$ to have an error bounded by $|f-f_{\rm approx}|\lesssim 10^{-8}$. 

For the pole sum rule, which involves the hypergeometric function ${}_3F_2$ (see \eqref{eq:poleFunc_1SO13}), we proceed analogously. We first map the argument via $-(q_0^2/\tilde z_{5d}) = (x-1)/2$ and perform the Chebyshev approximation as described above.

It would be interesting if a better approximation, could be found. One possibility is to divide the energy range in intervals, and use separate polynomial fits in each. 

\begin{figure}
    \centering
\includegraphics[width=0.9\linewidth]{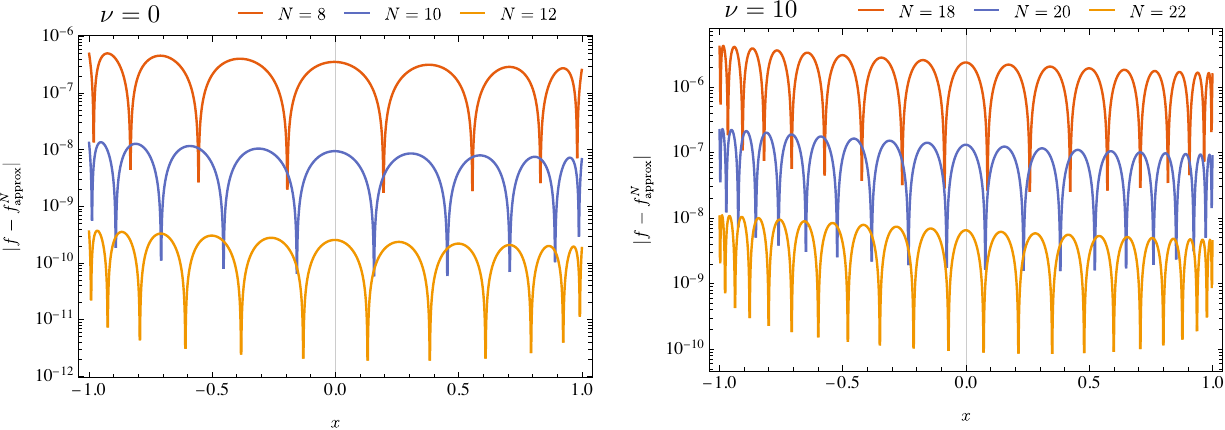}
    \caption{Error of the Chebyshev approximation to \eqref{eq:Omegainapp} at $t=-1$ and $\nu=0,10$. }
\label{fig:chebychev_nu_example}
\end{figure}

\subsection{Primal-dual formulation}\label{app:detailPrimalDual}
For a given linear primal problem, such as the one considered in this work, there exists a standard procedure to obtain its dual formulation (see e.g., \cite{Boyd_Vandenberghe_2004}).
This procedure is implemented by introducing a Lagrangian which for the primal problem \eqref{eq:Primal2} reads
\begin{equation}
\begin{split}
\cL &= \pm \cA  + \Lambda(1-a_{0,0}) + \frac{1}{\pi}\int_\mgap^{4m^2} \frac{dz}{z}\sum_{J=0}^\infty  \lambda_J(z) c_{J}(z) +\frac{1}{\pi}\int_{4m^2}^\infty \frac{dz}{z}\int_0^\infty d\nu  \lambda_\nu(z) c_{\nu}(z) \\
&- \sum_{{(n,\ell)}\in \cN} \kappa_{n,\ell} \chi_{n,\ell} + \sum_{e\in E} f_e( \cA_e^{\text{fixed}}-\cA_e) + \sum_{i\in I} u_{i}( \cA_i^{\max}-\cA_i) + d_i (\cA_i^{\min}- \cA_i)\,,
\end{split}
\end{equation}
where we introduced the Lagrange multipliers
$\Lambda, \kappa_{n,\ell}, f_j \in \mathbb{R}$,
$\lambda_{J,\nu}(z)\geq 0$, $u_i\geq 0$, and $d_i\leq 0$,
which we collectively denote by $D$. Clearly 
\begin{equation}
    \min_D \{\cL| \lambda_{J,\nu}(z)\geq 0, u_i\geq 0, d_i\leq 0  \} = \left\{\begin{array}{ll}
        \pm \cA & \text{if constraints } \eqref{eq:Primal2} \\
        -\infty &  \text{otherwise}
    \end{array}\right. \,.
\end{equation}
Using the standard \emph{min-max inequalities}, we can write 
\begin{equation}
\begin{split}
    \max_{c_J(z), c_{\nu}(z)} \pm \cA &= \max_{c_J(z), c_{\nu}(z)}\(\min_D \{\cL| \lambda_{J,\nu}(z)\geq 0, u_i\geq 0, d_i\leq 0  \}\)\\
    &~~\leq \min_D \{\bar \cA_\pm | \text{dual constraints}\}
\end{split}\,,
\end{equation}
where the dual objective $\bar{\cA}_\pm$ and its associated constraints are obtained by integrating out the primal variables. We now proceed to perform this integration explicitly.

We begin by expressing the observables through their dispersive representations,
\begin{equation}
\begin{split}
\cL &= \overbrace{\Lambda + \sum_{e\in E} f_e \cA_e^{\text{fixed}} + \sum_{i\in I} u_{i} \cA_i^{\max} + d_i \cA_i^{\min}}^{\bar \cA_\pm}\\
&- \frac{1}{\pi} \int_\mgap^{4m^2} \frac{dz}{z}\sum_J c_{J}(z) \Bigg\{ \cP^{SO(4)}_\pm(z,J) - \lambda_J(z)\Bigg\}\\
&- \frac{1}{\pi} \int_{4m^2}^\infty \frac{dz}{z}\int_{0}^\infty d\nu c_{\nu}(z) \Bigg\{\cP^{SO(1,3)}_\pm(z,\nu) - \lambda_\nu(z)\Bigg\}\,,
\end{split}
\end{equation}
where we defined
\begin{equation}
    \begin{split}
        \cP_\pm(z,\cdot)&=\Lambda a_{0,0}(z,\cdot)\mp \ODisp(z,\cdot)+\sum_{(n,\ell)\in \cN}\kappa_{n,\ell}\chi_{n,\ell}(z,\cdot) \nn \\
&
+ \sum_{e\in E} f_e \ODisp_e(z,\cdot) +\sum_{i\in I} (u_i+d_i) \ODisp_i(z,\cdot) \,.
    \end{split}
\end{equation}
Where we used the short-hand notation $(\cdot)$ for $J$ and $\nu$ respectively.
It is then straightforward to see that
\begin{equation}
    \max_{c_J(z), c_\nu(z)} \{\cL| c_J(z)\,, c_\nu(z)\geq 0\} = \left\{\begin{array}{ll}
        \bar \cA_\pm & \text{if } \cP_\pm(z,\cdot)\geq \lambda_{(\cdot)}(z)\geq 0\\
        \infty &  \text{otherwise}
    \end{array}\right. \,.
\end{equation}
Finally, by minimizing over the Lagrange multipliers $D$, we obtain the dual problem \eqref{eq:dualWilsonCoefMax}, whose dual variables are
$\cD = {\Lambda, \kappa_{n,\ell}, f_j,u_i, d_i}$.

\subsubsection{Three-point coupling optimization}\label{sec:3pt_primal_dual_problems}
The three-point coupling can also be bounded directly.
In contrast to the Wilson coefficient maximization, the residues are not a dispersive observable but are part of the primal variables.  In this case, the primal problem is 
\begin{mdframed}[frametitle={Primal Problem : three-point coupling},frametitlealignment=\centering,backgroundcolor=white, leftmargin=1cm, rightmargin=1cm, topline=true,bottomline=true, leftline=false, rightline=false] 
\vspace{-0.4cm}
\begin{align} 
&\underset{\text{in } \{c_J'(s), c_\nu(s)\}}{\text{Maximize}} && \pm \lambda_{0,\ell}\nonumber\\
& \text{subject to}  && a_{0,0}=1\nn\\ 
& && \ODisp_e = \ODisp_e^{\text{fixed}}\,,   &&e \in E \nn\\
& && \ODisp_{i}^{\min}\leq \ODisp_i\leq \ODisp_{i}^{\max}\,, &&i \in I \label{eq:Primal_3pt}\\
& && \chi_{n,\ell}=0\,,&&(n, \ell)\in \cN \nn\\
& && c_J(s)\geq 0 \,\, &&J=\mathbb{Z}_{\geq 0}\,, \, s=m_{0,J\neq{\ell}}^2  ~ \text{and} ~ \msubJ{J}^2 \leq s< 4m^2  \nn \\
& && c_\nu(s)\geq 0 \,\, && \nu\geq 0 \,,~ s \geq  4m^2\nn
\end{align}
\end{mdframed}
where we removed the spin $\ell$ from the constraints on the leading trajectory. Proceeding as in the previous section, the corresponding dual problem is 
\begin{mdframed}[frametitle={Dual Problem : three-point coupling},frametitlealignment=\centering,backgroundcolor=white, leftmargin=1cm, rightmargin=1cm, topline=true,bottomline=true, leftline=false, rightline=false] 
\vspace{-0.4cm}
\begin{align} 
&\underset{\text{in } \cD=\{\Lambda, f_e, u_i, d_i, \kappa_{n,\ell} \}}{\text{Minimize}} && \Lambda +\sum_{e\in E} f_e \ODisp_e^\text{fixed} + \sum_{i\in I} (u_i \ODisp_i^{\max}+ d_i \ODisp_i^{\min})  \nonumber\\
& \text{subject to}  &&\cP^{SO(4)}(m_0^2(\ell),\ell)=\pm 1\nn \\
& && \cP^{SO(4)}(z,J)\geq 0 \,,~~~~~ \text{for} \,\, J=\mathbb{Z}_{\geq 0}\,,  z= m_{0,J\neq \ell}^2  \text{ and }  \msubJ{J}^2\leq z< 4m^2 \nn \\[10pt]
& &&\cP^{SO(1,3)}(z,\nu) \geq 0 \,,~~~\,\text{for} \,\, \nu\geq 0 \,, z\geq 4m^2\nn \\[10pt] 
& \text{and} && u_i\geq 0\,,~~ d_i\leq 0 \,, \,\, i\in I\nn \\
& \text{where}  &&\cP(z,\cdot)=\Lambda a_{0,0}(z,\cdot)+\sum_{(n,\ell)\in \cN}\kappa_{n,\ell}\chi_{n,\ell}(z,\cdot) \nn \\
& &&~~~~~~~~~~\,
+ \sum_{e\in E} f_e \ODisp_e(z,\cdot) +\sum_{i\in I} (u_i+d_i) \ODisp_i(z,\cdot) \,.
\nn
\end{align}
\end{mdframed}
This formulation is used in this manuscript to derive bounds on the spin-0 residue of the leading trajectory, see \figref{fig:res00_vs_g_allTogether}.

\subsection{Convergence in $n_{\text{max}}$ for $\bar{\lambda}_{0,0}^2$}

In this subsection, we examine the convergence of the upper bound on the three-point coupling  $\bar{\lambda}_{0,0}^2$ with respect to the number of null constraints $n_{\text{max}}$, as a function of the coupling $g$. We observed that the three-point coupling had greater dependence on the truncation of the size of the dual functional than the Wilson coefficients. We remind the reader, however, that this truncation is rigorous and increasing $n_{\rm max}$ can only improve the bound. 
Here we focus on the case without pole constraints, although the same analysis applies when they are included.

The coupling is fixed at $g = 1$, and we maximize $\bar\lambda_{0,0}^2$. In \figref{fig:conv_nmax}, we present the results for several values of $n_{\text{max}} = 6, \ldots, 18$. The results indicate that the bound changes between $n_{\text{max}} = 6$ to $n_{\text{max}} = 16$, but stabilizes thereafter, with no observable difference between $n_{\text{max}} = 16$ and $n_{\text{max}} = 18$.

\begin{figure}[h!]
    \centering
    \includegraphics[width=0.7\linewidth]{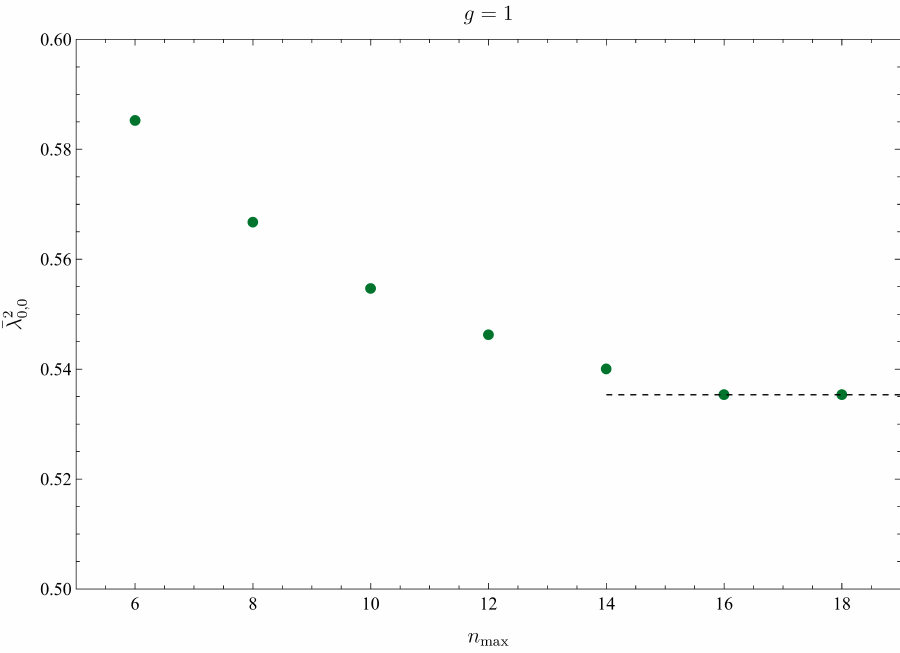}
    \caption{Maximum value of  $\bar{\lambda}_{0,0}^2$ at $g = 1$ for different numbers of null constraints $n_{\text{max}}$.}
    \label{fig:conv_nmax}
\end{figure}

In the bounds presented in the main text, we use $n_{\text{max}} = 12$, since the resulting difference in the bound is only a few percent, while the computational cost and run time increase substantially at higher $n_{\text{max}}$ due to the larger size of the functional, and the large number of $\nu$-points required for convergence.

\bibliographystyle{JHEP}
\bibliography{refs.bib}

\providecommand{\href}[2]{#2}\begingroup\raggedright\begin{thebibliography}{100}

\bibitem{Adams:2006sv}
A.~Adams, N.~Arkani-Hamed, S.~Dubovsky, A.~Nicolis and R.~Rattazzi, \emph{{Causality, analyticity and an IR obstruction to UV completion}}, \href{http://dx.doi.org/10.1088/1126-6708/2006/10/014}{\emph{JHEP} {\bf 10} (2006) 014}, [\href{https://arxiv.org/abs/hep-th/0602178}{{\tt hep-th/0602178}}].

\bibitem{Camanho:2014apa}
X.~O. Camanho, J.~D. Edelstein, J.~Maldacena and A.~Zhiboedov, \emph{{Causality Constraints on Corrections to the Graviton Three-Point Coupling}}, \href{http://dx.doi.org/10.1007/JHEP02(2016)020}{\emph{JHEP} {\bf 02} (2016) 020}, [\href{https://arxiv.org/abs/1407.5597}{{\tt 1407.5597}}].

\bibitem{Maldacena:2012sf}
J.~Maldacena and A.~Zhiboedov, \emph{{Constraining conformal field theories with a slightly broken higher spin symmetry}}, \href{http://dx.doi.org/10.1088/0264-9381/30/10/104003}{\emph{Class. Quant. Grav.} {\bf 30} (2013) 104003}, [\href{https://arxiv.org/abs/1204.3882}{{\tt 1204.3882}}].

\bibitem{Chowdhury:2019kaq}
S.~D. Chowdhury, A.~Gadde, T.~Gopalka, I.~Halder, L.~Janagal and S.~Minwalla, \emph{{Classifying and constraining local four photon and four graviton S-matrices}}, \href{http://dx.doi.org/10.1007/JHEP02(2020)114}{\emph{JHEP} {\bf 02} (2020) 114}, [\href{https://arxiv.org/abs/1910.14392}{{\tt 1910.14392}}].

\bibitem{Maldacena:1997re}
J.~M. Maldacena, \emph{{The Large N limit of superconformal field theories and supergravity}}, \href{http://dx.doi.org/10.4310/ATMP.1998.v2.n2.a1}{\emph{Adv. Theor. Math. Phys.} {\bf 2} (1998) 231--252}, [\href{https://arxiv.org/abs/hep-th/9711200}{{\tt hep-th/9711200}}].

\bibitem{Alday:2009zm}
L.~F. Alday, J.~M. Henn, J.~Plefka and T.~Schuster, \emph{{Scattering into the fifth dimension of N=4 super Yang-Mills}}, \href{http://dx.doi.org/10.1007/JHEP01(2010)077}{\emph{JHEP} {\bf 01} (2010) 077}, [\href{https://arxiv.org/abs/0908.0684}{{\tt 0908.0684}}].

\bibitem{Beneke:2013jia}
M.~Beneke, Y.~Kiyo and K.~Schuller, \emph{{Third-order correction to top-quark pair production near threshold I. Effective theory set-up and matching coefficients}},  \href{https://arxiv.org/abs/1312.4791}{{\tt 1312.4791}}.

\bibitem{Caron-Huot:2014gia}
S.~Caron-Huot and J.~M. Henn, \emph{{Solvable Relativistic Hydrogenlike System in Supersymmetric Yang-Mills Theory}}, \href{http://dx.doi.org/10.1103/PhysRevLett.113.161601}{\emph{Phys. Rev. Lett.} {\bf 113} (2014) 161601}, [\href{https://arxiv.org/abs/1408.0296}{{\tt 1408.0296}}].

\bibitem{Maldacena:2022ckr}
J.~Maldacena and G.~N. Remmen, \emph{{Accumulation-point amplitudes in string theory}}, \href{http://dx.doi.org/10.1007/JHEP08(2022)152}{\emph{JHEP} {\bf 08} (2022) 152}, [\href{https://arxiv.org/abs/2207.06426}{{\tt 2207.06426}}].

\bibitem{Gary:2009ae}
M.~Gary, S.~B. Giddings and J.~Penedones, \emph{{Local bulk S-matrix elements and CFT singularities}}, \href{http://dx.doi.org/10.1103/PhysRevD.80.085005}{\emph{Phys. Rev. D} {\bf 80} (2009) 085005}, [\href{https://arxiv.org/abs/0903.4437}{{\tt 0903.4437}}].

\bibitem{Penedones:2010ue}
J.~Penedones, \emph{{Writing CFT correlation functions as AdS scattering amplitudes}}, \href{http://dx.doi.org/10.1007/JHEP03(2011)025}{\emph{JHEP} {\bf 03} (2011) 025}, [\href{https://arxiv.org/abs/1011.1485}{{\tt 1011.1485}}].

\bibitem{Maldacena:2015iua}
J.~Maldacena, D.~Simmons-Duffin and A.~Zhiboedov, \emph{{Looking for a bulk point}}, \href{http://dx.doi.org/10.1007/JHEP01(2017)013}{\emph{JHEP} {\bf 01} (2017) 013}, [\href{https://arxiv.org/abs/1509.03612}{{\tt 1509.03612}}].

\bibitem{Caron-Huot:2025hmk}
S.~Caron-Huot, J.~Chakravarty and K.~Namjou, \emph{{Looking at bulk points in general geometries}}, \href{http://dx.doi.org/10.1007/JHEP06(2025)197}{\emph{JHEP} {\bf 06} (2025) 197}, [\href{https://arxiv.org/abs/2502.14963}{{\tt 2502.14963}}].

\bibitem{Klebanov:2006jj}
I.~R. Klebanov, J.~M. Maldacena and C.~B. Thorn, III, \emph{{Dynamics of flux tubes in large N gauge theories}}, \href{http://dx.doi.org/10.1088/1126-6708/2006/04/024}{\emph{JHEP} {\bf 04} (2006) 024}, [\href{https://arxiv.org/abs/hep-th/0602255}{{\tt hep-th/0602255}}].

\bibitem{Coon:1969yw}
D.~D. Coon, \emph{{Uniqueness of the veneziano representation}}, \href{http://dx.doi.org/10.1016/0370-2693(69)90106-3}{\emph{Phys. Lett. B} {\bf 29} (1969) 669--672}.

\bibitem{khuri1969derivation}
N.~Khuri, \emph{Derivation of a veneziano series from the regge representation}, {\emph{Physical Review} {\bf 185} (1969) 1876}.

\bibitem{Matsuda:1969zz}
S.~Matsuda, \emph{{Model for infinite Veneziano series}}, .

\bibitem{Mandelstam:1968czc}
S.~Mandelstam, \emph{{Veneziano formula with trajectories spaced by two units}}, \href{http://dx.doi.org/10.1103/PhysRevLett.21.1724}{\emph{Phys. Rev. Lett.} {\bf 21} (1968) 1724--1728}.

\bibitem{Cheung:2023adk}
C.~Cheung and G.~N. Remmen, \emph{{Stringy Dynamics from an Amplitudes Bootstrap}},  \href{https://arxiv.org/abs/2302.12263}{{\tt 2302.12263}}.

\bibitem{Cheung:2023uwn}
C.~Cheung and G.~N. Remmen, \emph{{Bespoke dual resonance}}, \href{http://dx.doi.org/10.1103/PhysRevD.108.086009}{\emph{Phys. Rev. D} {\bf 108} (2023) 086009}, [\href{https://arxiv.org/abs/2308.03833}{{\tt 2308.03833}}].

\bibitem{Eckner:2024ggx}
C.~Eckner, F.~Figueroa and P.~Tourkine, \emph{{Regge bootstrap: From linear to nonlinear trajectories}}, \href{http://dx.doi.org/10.1103/PhysRevD.111.126005}{\emph{Phys. Rev. D} {\bf 111} (2025) 126005}, [\href{https://arxiv.org/abs/2401.08736}{{\tt 2401.08736}}].

\bibitem{Haring:2023zwu}
K.~H{\"a}ring and A.~Zhiboedov, \emph{{The stringy S-matrix bootstrap: maximal spin and superpolynomial softness}}, \href{http://dx.doi.org/10.1007/JHEP10(2024)075}{\emph{JHEP} {\bf 10} (2024) 075}, [\href{https://arxiv.org/abs/2311.13631}{{\tt 2311.13631}}].

\bibitem{Gross:1969db}
D.~J. Gross, \emph{{Factorization and the generalized veneziano model with satellites}}, \href{http://dx.doi.org/10.1016/0550-3213(69)90248-X}{\emph{Nucl. Phys. B} {\bf 13} (1969) 467--476}.

\bibitem{Arkani-Hamed:2023jwn}
N.~Arkani-Hamed, C.~Cheung, C.~Figueiredo and G.~N. Remmen, \emph{{Multiparticle Factorization and the Rigidity of String Theory}}, \href{http://dx.doi.org/10.1103/PhysRevLett.132.091601}{\emph{Phys. Rev. Lett.} {\bf 132} (2024) 091601}, [\href{https://arxiv.org/abs/2312.07652}{{\tt 2312.07652}}].

\bibitem{Geiser:2023qqq}
N.~Geiser, \emph{{The Baker-Coon-Romans N-point amplitude and an exact field theory limit of the Coon amplitude}}, \href{http://dx.doi.org/10.1007/JHEP10(2024)010}{\emph{JHEP} {\bf 10} (2024) 010}, [\href{https://arxiv.org/abs/2311.04130}{{\tt 2311.04130}}].

\bibitem{Cheung:2025krg}
C.~Cheung and G.~N. Remmen, \emph{{Multipositivity Bounds}},  \href{https://arxiv.org/abs/2505.05553}{{\tt 2505.05553}}.

\bibitem{Drummond:2008vq}
J.~M. Drummond, J.~Henn, G.~P. Korchemsky and E.~Sokatchev, \emph{{Dual superconformal symmetry of scattering amplitudes in N=4 super-Yang-Mills theory}}, \href{http://dx.doi.org/10.1016/j.nuclphysb.2009.11.022}{\emph{Nucl. Phys. B} {\bf 828} (2010) 317--374}, [\href{https://arxiv.org/abs/0807.1095}{{\tt 0807.1095}}].

\bibitem{Berkovits:2008ic}
N.~Berkovits and J.~Maldacena, \emph{{Fermionic T-Duality, Dual Superconformal Symmetry, and the Amplitude/Wilson Loop Connection}}, \href{http://dx.doi.org/10.1088/1126-6708/2008/09/062}{\emph{JHEP} {\bf 09} (2008) 062}, [\href{https://arxiv.org/abs/0807.3196}{{\tt 0807.3196}}].

\bibitem{Alday:2007hr}
L.~F. Alday and J.~M. Maldacena, \emph{{Gluon scattering amplitudes at strong coupling}}, \href{http://dx.doi.org/10.1088/1126-6708/2007/06/064}{\emph{JHEP} {\bf 06} (2007) 064}, [\href{https://arxiv.org/abs/0705.0303}{{\tt 0705.0303}}].

\bibitem{Bruser:2018jnc}
R.~Br\"user, S.~Caron-Huot and J.~M. Henn, \emph{{Subleading Regge limit from a soft anomalous dimension}}, \href{http://dx.doi.org/10.1007/JHEP04(2018)047}{\emph{JHEP} {\bf 04} (2018) 047}, [\href{https://arxiv.org/abs/1802.02524}{{\tt 1802.02524}}].

\bibitem{Fock:1935vv}
V.~Fock, \emph{{On the Theory of the hydrogen atoms}}, \href{http://dx.doi.org/10.1007/BF01336904}{\emph{Z. Phys.} {\bf 98} (1935) 145--154}.

\bibitem{Henn:2010bk}
J.~M. Henn, S.~G. Naculich, H.~J. Schnitzer and M.~Spradlin, \emph{{Higgs-regularized three-loop four-gluon amplitude in N=4 SYM: exponentiation and Regge limits}}, \href{http://dx.doi.org/10.1007/JHEP04(2010)038}{\emph{JHEP} {\bf 04} (2010) 038}, [\href{https://arxiv.org/abs/1001.1358}{{\tt 1001.1358}}].

\bibitem{Correa:2012nk}
D.~Correa, J.~Henn, J.~Maldacena and A.~Sever, \emph{{The cusp anomalous dimension at three loops and beyond}}, \href{http://dx.doi.org/10.1007/JHEP05(2012)098}{\emph{JHEP} {\bf 05} (2012) 098}, [\href{https://arxiv.org/abs/1203.1019}{{\tt 1203.1019}}].

\bibitem{Polyakov:1980ca}
A.~M. Polyakov, \emph{{Gauge Fields as Rings of Glue}}, \href{http://dx.doi.org/10.1016/0550-3213(80)90507-6}{\emph{Nucl. Phys. B} {\bf 164} (1980) 171--188}.

\bibitem{Korchemsky:1987wg}
G.~P. Korchemsky and A.~V. Radyushkin, \emph{{Renormalization of the Wilson Loops Beyond the Leading Order}}, \href{http://dx.doi.org/10.1016/0550-3213(87)90277-X}{\emph{Nucl. Phys. B} {\bf 283} (1987) 342--364}.

\bibitem{Flieger:2025ekn}
W.~Flieger, J.~Henn, A.~Schreiber and J.~Trnka, \emph{{Two-loop four-point amplitudes on the Coulomb branch of ${\mathcal{N}}=4$ super Yang-Mills}},  \href{https://arxiv.org/abs/2501.09454}{{\tt 2501.09454}}.

\bibitem{Gromov:2015dfa}
N.~Gromov and F.~Levkovich-Maslyuk, \emph{{Quantum Spectral Curve for a cusped Wilson line in $ \mathcal{N}=4 $ SYM}}, \href{http://dx.doi.org/10.1007/JHEP04(2016)134}{\emph{JHEP} {\bf 04} (2016) 134}, [\href{https://arxiv.org/abs/1510.02098}{{\tt 1510.02098}}].

\bibitem{Grabner:2020nis}
D.~Grabner, N.~Gromov and J.~Julius, \emph{{Excited States of One-Dimensional Defect CFTs from the Quantum Spectral Curve}}, \href{http://dx.doi.org/10.1007/JHEP07(2020)042}{\emph{JHEP} {\bf 07} (2020) 042}, [\href{https://arxiv.org/abs/2001.11039}{{\tt 2001.11039}}].

\bibitem{Alday:2023jdk}
L.~F. Alday, T.~Hansen and J.~A. Silva, \emph{{Emergent Worldsheet for the AdS Virasoro-Shapiro Amplitude}}, \href{http://dx.doi.org/10.1103/PhysRevLett.131.161603}{\emph{Phys. Rev. Lett.} {\bf 131} (2023) 161603}, [\href{https://arxiv.org/abs/2305.03593}{{\tt 2305.03593}}].

\bibitem{Alday:2023mvu}
L.~F. Alday and T.~Hansen, \emph{{The AdS Virasoro-Shapiro amplitude}}, \href{http://dx.doi.org/10.1007/JHEP10(2023)023}{\emph{JHEP} {\bf 10} (2023) 023}, [\href{https://arxiv.org/abs/2306.12786}{{\tt 2306.12786}}].

\bibitem{Alday:2024yax}
L.~F. Alday, S.~M. Chester, T.~Hansen and D.-l. Zhong, \emph{{The AdS Veneziano amplitude at small curvature}}, \href{http://dx.doi.org/10.1007/JHEP05(2024)322}{\emph{JHEP} {\bf 05} (2024) 322}, [\href{https://arxiv.org/abs/2403.13877}{{\tt 2403.13877}}].

\bibitem{Alday:2024ksp}
L.~F. Alday and T.~Hansen, \emph{{Single-valuedness of the AdS Veneziano amplitude}}, \href{http://dx.doi.org/10.1007/JHEP08(2024)108}{\emph{JHEP} {\bf 08} (2024) 108}, [\href{https://arxiv.org/abs/2404.16084}{{\tt 2404.16084}}].

\bibitem{Henn:2024qwe}
J.~Henn and P.~Raman, \emph{{Positivity properties of scattering amplitudes}},  \href{https://arxiv.org/abs/2407.05755}{{\tt 2407.05755}}.

\bibitem{Ivanovskiy:2024vel}
V.~Ivanovskiy, S.~Komatsu, V.~Mishnyakov, N.~Terziev, N.~Zaigraev and K.~Zarembo, \emph{{Vacuum Condensates on the Coulomb Branch}},  \href{https://arxiv.org/abs/2405.19043}{{\tt 2405.19043}}.

\bibitem{Berman:2023jys}
J.~Berman, H.~Elvang and A.~Herderschee, \emph{{Flattening of the EFT-hedron: supersymmetric positivity bounds and the search for string theory}}, \href{http://dx.doi.org/10.1007/JHEP03(2024)021}{\emph{JHEP} {\bf 03} (2024) 021}, [\href{https://arxiv.org/abs/2310.10729}{{\tt 2310.10729}}].

\bibitem{Davydychev:1993ut}
A.~I. Davydychev, \emph{{Standard and hypergeometric representations for loop diagrams and the photon-photon scattering}},  in \emph{{7th International Seminar on High-energy Physics}}, 5, 1993.
\newblock \href{https://arxiv.org/abs/hep-ph/9307323}{{\tt hep-ph/9307323}}.

\bibitem{Caron-Huot:2014lda}
S.~Caron-Huot and J.~M. Henn, \emph{{Iterative structure of finite loop integrals}}, \href{http://dx.doi.org/10.1007/JHEP06(2014)114}{\emph{JHEP} {\bf 06} (2014) 114}, [\href{https://arxiv.org/abs/1404.2922}{{\tt 1404.2922}}].

\bibitem{Bourjaily:2011hi}
J.~L. Bourjaily, A.~DiRe, A.~Shaikh, M.~Spradlin and A.~Volovich, \emph{{The Soft-Collinear Bootstrap: N=4 Yang-Mills Amplitudes at Six and Seven Loops}}, \href{http://dx.doi.org/10.1007/JHEP03(2012)032}{\emph{JHEP} {\bf 03} (2012) 032}, [\href{https://arxiv.org/abs/1112.6432}{{\tt 1112.6432}}].

\bibitem{Caron-Huot:2021usw}
S.~Caron-Huot and F.~Coronado, \emph{{Ten dimensional symmetry of $ \mathcal{N} $ = 4 SYM correlators}}, \href{http://dx.doi.org/10.1007/JHEP03(2022)151}{\emph{JHEP} {\bf 03} (2022) 151}, [\href{https://arxiv.org/abs/2106.03892}{{\tt 2106.03892}}].

\bibitem{Dine:1997nq}
M.~Dine and N.~Seiberg, \emph{{Comments on higher derivative operators in some SUSY field theories}}, \href{http://dx.doi.org/10.1016/S0370-2693(97)00899-X}{\emph{Phys. Lett. B} {\bf 409} (1997) 239--244}, [\href{https://arxiv.org/abs/hep-th/9705057}{{\tt hep-th/9705057}}].

\bibitem{Basso:2007wd}
B.~Basso, G.~P. Korchemsky and J.~Kotanski, \emph{{Cusp anomalous dimension in maximally supersymmetric Yang-Mills theory at strong coupling}}, \href{http://dx.doi.org/10.1103/PhysRevLett.100.091601}{\emph{Phys. Rev. Lett.} {\bf 100} (2008) 091601}, [\href{https://arxiv.org/abs/0708.3933}{{\tt 0708.3933}}].

\bibitem{Gross:1987kza}
D.~J. Gross and P.~F. Mende, \emph{{The High-Energy Behavior of String Scattering Amplitudes}}, \href{http://dx.doi.org/10.1016/0370-2693(87)90355-8}{\emph{Phys. Lett. B} {\bf 197} (1987) 129--134}.

\bibitem{Eckner:2024pqt}
C.~Eckner, F.~Figueroa and P.~Tourkine, \emph{{On the number of Regge trajectories for dual amplitudes}},  \href{https://arxiv.org/abs/2405.21057}{{\tt 2405.21057}}.

\bibitem{Maldacena:1998im}
J.~M. Maldacena, \emph{{Wilson loops in large N field theories}}, \href{http://dx.doi.org/10.1103/PhysRevLett.80.4859}{\emph{Phys. Rev. Lett.} {\bf 80} (1998) 4859--4862}, [\href{https://arxiv.org/abs/hep-th/9803002}{{\tt hep-th/9803002}}].

\bibitem{Drukker:2012de}
N.~Drukker, \emph{{Integrable Wilson loops}}, \href{http://dx.doi.org/10.1007/JHEP10(2013)135}{\emph{JHEP} {\bf 10} (2013) 135}, [\href{https://arxiv.org/abs/1203.1617}{{\tt 1203.1617}}].

\bibitem{Correa:2012hh}
D.~Correa, J.~Maldacena and A.~Sever, \emph{{The quark anti-quark potential and the cusp anomalous dimension from a TBA equation}}, \href{http://dx.doi.org/10.1007/JHEP08(2012)134}{\emph{JHEP} {\bf 08} (2012) 134}, [\href{https://arxiv.org/abs/1203.1913}{{\tt 1203.1913}}].

\bibitem{Erickson:1999qv}
J.~K. Erickson, G.~W. Semenoff, R.~J. Szabo and K.~Zarembo, \emph{{Static potential in N=4 supersymmetric Yang-Mills theory}}, \href{http://dx.doi.org/10.1103/PhysRevD.61.105006}{\emph{Phys. Rev. D} {\bf 61} (2000) 105006}, [\href{https://arxiv.org/abs/hep-th/9911088}{{\tt hep-th/9911088}}].

\bibitem{Gromov:2016rrp}
N.~Gromov and F.~Levkovich-Maslyuk, \emph{{Quark-anti-quark potential in $ \mathcal{N} =$ 4 SYM}}, \href{http://dx.doi.org/10.1007/JHEP12(2016)122}{\emph{JHEP} {\bf 12} (2016) 122}, [\href{https://arxiv.org/abs/1601.05679}{{\tt 1601.05679}}].

\bibitem{Correa:2012at}
D.~Correa, J.~Henn, J.~Maldacena and A.~Sever, \emph{{An exact formula for the radiation of a moving quark in N=4 super Yang Mills}}, \href{http://dx.doi.org/10.1007/JHEP06(2012)048}{\emph{JHEP} {\bf 06} (2012) 048}, [\href{https://arxiv.org/abs/1202.4455}{{\tt 1202.4455}}].

\bibitem{Fiol:2012sg}
B.~Fiol, B.~Garolera and A.~Lewkowycz, \emph{{Exact results for static and radiative fields of a quark in N=4 super Yang-Mills}}, \href{http://dx.doi.org/10.1007/JHEP05(2012)093}{\emph{JHEP} {\bf 05} (2012) 093}, [\href{https://arxiv.org/abs/1202.5292}{{\tt 1202.5292}}].

\bibitem{Sakurai:2011zz}
J.~J. Sakurai and J.~Napolitano, \emph{{Modern Quantum Mechanics}}.
\newblock Quantum physics, quantum information and quantum computation. Cambridge University Press, 3~ed., 10, 2020, \href{http://dx.doi.org/10.1017/9781108587280}{10.1017/9781108587280}.

\bibitem{Mandelstam:1974fq}
S.~Mandelstam, \emph{{Dual - Resonance Models}}, \href{http://dx.doi.org/10.1016/0370-1573(74)90034-9}{\emph{Phys. Rept.} {\bf 13} (1974) 259}.

\bibitem{Giombi:2017cqn}
S.~Giombi, R.~Roiban and A.~A. Tseytlin, \emph{{Half-BPS Wilson loop and AdS$_2$/CFT$_1$}}, \href{http://dx.doi.org/10.1016/j.nuclphysb.2017.07.004}{\emph{Nucl. Phys. B} {\bf 922} (2017) 499--527}, [\href{https://arxiv.org/abs/1706.00756}{{\tt 1706.00756}}].

\bibitem{Ferrero:2023znz}
P.~Ferrero and C.~Meneghelli, \emph{{Unmixing the Wilson line defect CFT. Part I. Spectrum and kinematics}}, \href{http://dx.doi.org/10.1007/JHEP05(2024)090}{\emph{JHEP} {\bf 05} (2024) 090}, [\href{https://arxiv.org/abs/2312.12550}{{\tt 2312.12550}}].

\bibitem{Ferrero:2023gnu}
P.~Ferrero and C.~Meneghelli, \emph{{Unmixing the Wilson line defect CFT. Part II. Analytic bootstrap}}, \href{http://dx.doi.org/10.1007/JHEP06(2024)010}{\emph{JHEP} {\bf 06} (2024) 010}, [\href{https://arxiv.org/abs/2312.12551}{{\tt 2312.12551}}].

\bibitem{Cavaglia:2021bnz}
A.~Cavagli\`a, N.~Gromov, J.~Julius and M.~Preti, \emph{{Integrability and conformal bootstrap: One dimensional defect conformal field theory}}, \href{http://dx.doi.org/10.1103/PhysRevD.105.L021902}{\emph{Phys. Rev. D} {\bf 105} (2022) L021902}, [\href{https://arxiv.org/abs/2107.08510}{{\tt 2107.08510}}].

\bibitem{Cavaglia:2023mmu}
A.~Cavagli\`a, N.~Gromov and M.~Preti, \emph{{Computing four-point functions with integrability, bootstrap and parity symmetry}}, \href{http://dx.doi.org/10.1007/JHEP02(2025)026}{\emph{JHEP} {\bf 02} (2025) 026}, [\href{https://arxiv.org/abs/2312.11604}{{\tt 2312.11604}}].

\bibitem{Gromov:2013pga}
N.~Gromov, V.~Kazakov, S.~Leurent and D.~Volin, \emph{{Quantum Spectral Curve for Planar $\mathcal{N} = 4$ Super-Yang-Mills Theory}}, \href{http://dx.doi.org/10.1103/PhysRevLett.112.011602}{\emph{Phys. Rev. Lett.} {\bf 112} (2014) 011602}, [\href{https://arxiv.org/abs/1305.1939}{{\tt 1305.1939}}].

\bibitem{Gromov:2017blm}
N.~Gromov, \emph{{Introduction to the Spectrum of $N=4$ SYM and the Quantum Spectral Curve}},  \href{https://arxiv.org/abs/1708.03648}{{\tt 1708.03648}}.

\bibitem{Cavaglia:2022qpg}
A.~Cavagli\`a, N.~Gromov, J.~Julius and M.~Preti, \emph{{Bootstrability in defect CFT: integrated correlators and sharper bounds}}, \href{http://dx.doi.org/10.1007/JHEP05(2022)164}{\emph{JHEP} {\bf 05} (2022) 164}, [\href{https://arxiv.org/abs/2203.09556}{{\tt 2203.09556}}].

\bibitem{Gromov_QSC}
N.~Gromov. {\textit{Private communication}}.

\bibitem{Chu:2009qt}
S.-x. Chu, D.~Hou and H.-c. Ren, \emph{{The Subleading Term of the Strong Coupling Expansion of the Heavy-Quark Potential in a N=4 Super Yang-Mills Vacuum}}, \href{http://dx.doi.org/10.1088/1126-6708/2009/08/004}{\emph{JHEP} {\bf 08} (2009) 004}, [\href{https://arxiv.org/abs/0905.1874}{{\tt 0905.1874}}].

\bibitem{Forini:2010ek}
V.~Forini, \emph{{Quark-antiquark potential in AdS at one loop}}, \href{http://dx.doi.org/10.1007/JHEP11(2010)079}{\emph{JHEP} {\bf 11} (2010) 079}, [\href{https://arxiv.org/abs/1009.3939}{{\tt 1009.3939}}].

\bibitem{Korchemskaya:1992je}
I.~A. Korchemskaya and G.~P. Korchemsky, \emph{{On lightlike Wilson loops}}, \href{http://dx.doi.org/10.1016/0370-2693(92)91895-G}{\emph{Phys. Lett. B} {\bf 287} (1992) 169--175}.

\bibitem{Bern:2006ew}
Z.~Bern, M.~Czakon, L.~J. Dixon, D.~A. Kosower and V.~A. Smirnov, \emph{{The Four-Loop Planar Amplitude and Cusp Anomalous Dimension in Maximally Supersymmetric Yang-Mills Theory}}, \href{http://dx.doi.org/10.1103/PhysRevD.75.085010}{\emph{Phys. Rev. D} {\bf 75} (2007) 085010}, [\href{https://arxiv.org/abs/hep-th/0610248}{{\tt hep-th/0610248}}].

\bibitem{Cachazo:2006az}
F.~Cachazo, M.~Spradlin and A.~Volovich, \emph{{Four-loop cusp anomalous dimension from obstructions}}, \href{http://dx.doi.org/10.1103/PhysRevD.75.105011}{\emph{Phys. Rev. D} {\bf 75} (2007) 105011}, [\href{https://arxiv.org/abs/hep-th/0612309}{{\tt hep-th/0612309}}].

\bibitem{Beisert:2006ez}
N.~Beisert, B.~Eden and M.~Staudacher, \emph{{Transcendentality and Crossing}}, \href{http://dx.doi.org/10.1088/1742-5468/2007/01/P01021}{\emph{J. Stat. Mech.} {\bf 0701} (2007) P01021}, [\href{https://arxiv.org/abs/hep-th/0610251}{{\tt hep-th/0610251}}].

\bibitem{Benna:2006nd}
M.~K. Benna, S.~Benvenuti, I.~R. Klebanov and A.~Scardicchio, \emph{{A Test of the AdS/CFT correspondence using high-spin operators}}, \href{http://dx.doi.org/10.1103/PhysRevLett.98.131603}{\emph{Phys. Rev. Lett.} {\bf 98} (2007) 131603}, [\href{https://arxiv.org/abs/hep-th/0611135}{{\tt hep-th/0611135}}].

\bibitem{Remiddi:1999ew}
E.~Remiddi and J.~A.~M. Vermaseren, \emph{{Harmonic polylogarithms}}, \href{http://dx.doi.org/10.1142/S0217751X00000367}{\emph{Int. J. Mod. Phys. A} {\bf 15} (2000) 725--754}, [\href{https://arxiv.org/abs/hep-ph/9905237}{{\tt hep-ph/9905237}}].

\bibitem{Makeenko:2006ds}
Y.~Makeenko, P.~Olesen and G.~W. Semenoff, \emph{{Cusped SYM Wilson loop at two loops and beyond}}, \href{http://dx.doi.org/10.1016/j.nuclphysb.2006.05.002}{\emph{Nucl. Phys. B} {\bf 748} (2006) 170--199}, [\href{https://arxiv.org/abs/hep-th/0602100}{{\tt hep-th/0602100}}].

\bibitem{Henn:2019swt}
J.~M. Henn, G.~P. Korchemsky and B.~Mistlberger, \emph{{The full four-loop cusp anomalous dimension in $\mathcal{N}=4$ super Yang-Mills and QCD}}, \href{http://dx.doi.org/10.1007/JHEP04(2020)018}{\emph{JHEP} {\bf 04} (2020) 018}, [\href{https://arxiv.org/abs/1911.10174}{{\tt 1911.10174}}].

\bibitem{Drukker:1999zq}
N.~Drukker, D.~J. Gross and H.~Ooguri, \emph{{Wilson loops and minimal surfaces}}, \href{http://dx.doi.org/10.1103/PhysRevD.60.125006}{\emph{Phys. Rev. D} {\bf 60} (1999) 125006}, [\href{https://arxiv.org/abs/hep-th/9904191}{{\tt hep-th/9904191}}].

\bibitem{Espindola:2016afe}
R.~Esp\'\i{}ndola and J.~A. Garc\'\i{}a, \emph{{Cusp Anomalous dimension and rotating open strings in AdS/CFT}}, \href{http://dx.doi.org/10.1007/JHEP03(2018)116}{\emph{JHEP} {\bf 03} (2018) 116}, [\href{https://arxiv.org/abs/1607.05305}{{\tt 1607.05305}}].

\bibitem{Dunne:2025wbq}
G.~V. Dunne, \emph{{Resurgence of the tilted cusp anomalous dimension$^{*}$}}, \href{http://dx.doi.org/10.1088/1751-8121/add2af}{\emph{J. Phys. A} {\bf 58} (2025) 205401}, [\href{https://arxiv.org/abs/2501.03105}{{\tt 2501.03105}}].

\bibitem{Kravchuk:2024qoh}
P.~Kravchuk, A.~Radcliffe and R.~Sinha, \emph{{Effective theory for fusion of conformal defects}},  \href{https://arxiv.org/abs/2406.04561}{{\tt 2406.04561}}.

\bibitem{Cuomo:2024psk}
G.~Cuomo, Y.-C. He and Z.~Komargodski, \emph{{Impurities with a cusp: general theory and 3d Ising}}, \href{http://dx.doi.org/10.1007/JHEP11(2024)061}{\emph{JHEP} {\bf 11} (2024) 061}, [\href{https://arxiv.org/abs/2406.10186}{{\tt 2406.10186}}].

\bibitem{Drukker:2006xg}
N.~Drukker and S.~Kawamoto, \emph{{Small deformations of supersymmetric Wilson loops and open spin-chains}}, \href{http://dx.doi.org/10.1088/1126-6708/2006/07/024}{\emph{JHEP} {\bf 07} (2006) 024}, [\href{https://arxiv.org/abs/hep-th/0604124}{{\tt hep-th/0604124}}].

\bibitem{Alday:2007he}
L.~F. Alday and J.~Maldacena, \emph{{Comments on gluon scattering amplitudes via AdS/CFT}}, \href{http://dx.doi.org/10.1088/1126-6708/2007/11/068}{\emph{JHEP} {\bf 11} (2007) 068}, [\href{https://arxiv.org/abs/0710.1060}{{\tt 0710.1060}}].

\bibitem{Ferrero:2021bsb}
P.~Ferrero and C.~Meneghelli, \emph{{Bootstrapping the half-BPS line defect CFT in N=4 supersymmetric Yang-Mills theory at strong coupling}}, \href{http://dx.doi.org/10.1103/PhysRevD.104.L081703}{\emph{Phys. Rev. D} {\bf 104} (2021) L081703}, [\href{https://arxiv.org/abs/2103.10440}{{\tt 2103.10440}}].

\bibitem{Becher:2009cu}
T.~Becher and M.~Neubert, \emph{Infrared singularities of scattering amplitudes in perturbative {QCD}}, \href{http://dx.doi.org/10.1103/PhysRevLett.102.162001}{\emph{Phys. Rev. Lett.} {\bf 102} (2009) 162001}, [\href{https://arxiv.org/abs/0901.0722}{{\tt 0901.0722}}].

\bibitem{Gardi:2009qi}
E.~Gardi and L.~Magnea, \emph{Infrared singularities in {QCD} amplitudes}, \href{http://dx.doi.org/10.1088/1126-6708/2009/03/079}{\emph{JHEP} {\bf 03} (2009) 079}, [\href{https://arxiv.org/abs/0901.1091}{{\tt 0901.1091}}].

\bibitem{Catani:1998bh}
S.~Catani, \emph{The singular behaviour of {QCD} amplitudes at two-loop order}, \href{http://dx.doi.org/10.1016/S0370-2693(98)00332-3}{\emph{Phys. Lett. B} {\bf 427} (1998) 161--171}, [\href{https://arxiv.org/abs/hep-ph/9802439}{{\tt hep-ph/9802439}}].

\bibitem{Korchemsky:1992xv}
G.~P. Korchemsky and A.~V. Radyushkin, \emph{Infrared factorization, wilson lines and the heavy quark limit}, \href{http://dx.doi.org/10.1016/0370-2693(92)90385-R}{\emph{Phys. Lett. B} {\bf 279} (1992) 359--366}.

\bibitem{Eichten:1989zv}
E.~Eichten and B.~R. Hill, \emph{{An Effective Field Theory for the Calculation of Matrix Elements Involving Heavy Quarks}}, \href{http://dx.doi.org/10.1016/0370-2693(90)92049-O}{\emph{Phys. Lett. B} {\bf 234} (1990) 511--516}.

\bibitem{Georgi:1990um}
H.~Georgi, \emph{{An Effective Field Theory for Heavy Quarks at Low-energies}}, \href{http://dx.doi.org/10.1016/0370-2693(90)91128-X}{\emph{Phys. Lett. B} {\bf 240} (1990) 447--450}.

\bibitem{Falk:1990pz}
A.~F. Falk, B.~Grinstein and M.~E. Luke, \emph{{Leading mass corrections to the heavy quark effective theory}}, \href{http://dx.doi.org/10.1016/0550-3213(91)90464-9}{\emph{Nucl. Phys. B} {\bf 357} (1991) 185--207}.

\bibitem{Cooke:2017qgm}
M.~Cooke, A.~Dekel and N.~Drukker, \emph{{The Wilson loop CFT: Insertion dimensions and structure constants from wavy lines}}, \href{http://dx.doi.org/10.1088/1751-8121/aa7db4}{\emph{J. Phys. A} {\bf 50} (2017) 335401}, [\href{https://arxiv.org/abs/1703.03812}{{\tt 1703.03812}}].

\bibitem{Agmon:2020pde}
N.~B. Agmon and Y.~Wang, \emph{{Classifying Superconformal Defects in Diverse Dimensions Part I: Superconformal Lines}},  \href{https://arxiv.org/abs/2009.06650}{{\tt 2009.06650}}.

\bibitem{Julius:2024ewf}
J.~Julius and N.~S. Sokolova, \emph{{Unmixing sub-leading Regge trajectories of $ \mathcal{N} $ = 4 Super-Yang-Mills}}, \href{http://dx.doi.org/10.1007/JHEP04(2025)200}{\emph{JHEP} {\bf 04} (2025) 200}, [\href{https://arxiv.org/abs/2409.07529}{{\tt 2409.07529}}].

\bibitem{Alday:2025bjp}
L.~F. Alday, M.~Nocchi and A.~S. Sangar\'e, \emph{{Stringy KLT Relations on $AdS$}},  \href{https://arxiv.org/abs/2504.19973}{{\tt 2504.19973}}.

\bibitem{Baune:2025hfu}
K.~Baune, \emph{{Associators for AdS string amplitude building blocks}},  \href{https://arxiv.org/abs/2505.23385}{{\tt 2505.23385}}.

\bibitem{osti_4155317}
E.~Plahte, \emph{Symmetry properties of dual tree-graph n-point amplitudes.}, \href{http://dx.doi.org/10.1007/BF02824716}{\emph{Nuovo Cim. 66A: 713-33(21 Apr 1970).} (12, 1969) }.

\bibitem{Alday:2025cxr}
L.~F. Alday, R.~S. Pitombo and A.~Str{\"o}mholm~Sangar{\'e}, \emph{{Monodromy Relations for String Amplitudes on AdS}},  \href{https://arxiv.org/abs/2509.02719}{{\tt 2509.02719}}.

\bibitem{Bellazzini:2020cot}
B.~Bellazzini, J.~Elias~Mir{\'o}, R.~Rattazzi, M.~Riembau and F.~Riva, \emph{{Positive moments for scattering amplitudes}}, \href{http://dx.doi.org/10.1103/PhysRevD.104.036006}{\emph{Phys. Rev. D} {\bf 104} (2021) 036006}, [\href{https://arxiv.org/abs/2011.00037}{{\tt 2011.00037}}].

\bibitem{Tolley:2020gtv}
A.~J. Tolley, Z.-Y. Wang and S.-Y. Zhou, \emph{{New positivity bounds from full crossing symmetry}}, \href{http://dx.doi.org/10.1007/JHEP05(2021)255}{\emph{JHEP} {\bf 05} (2021) 255}, [\href{https://arxiv.org/abs/2011.02400}{{\tt 2011.02400}}].

\bibitem{Caron-Huot:2020cmc}
S.~Caron-Huot and V.~Van~Duong, \emph{{Extremal Effective Field Theories}}, \href{http://dx.doi.org/10.1007/JHEP05(2021)280}{\emph{JHEP} {\bf 05} (2021) 280}, [\href{https://arxiv.org/abs/2011.02957}{{\tt 2011.02957}}].

\bibitem{Arkani-Hamed:2020blm}
N.~Arkani-Hamed, T.-C. Huang and Y.-t. Huang, \emph{{The EFT-Hedron}}, \href{http://dx.doi.org/10.1007/JHEP05(2021)259}{\emph{JHEP} {\bf 05} (2021) 259}, [\href{https://arxiv.org/abs/2012.15849}{{\tt 2012.15849}}].

\bibitem{Caron-Huot:2021rmr}
S.~Caron-Huot, D.~Mazac, L.~Rastelli and D.~Simmons-Duffin, \emph{{Sharp boundaries for the swampland}}, \href{http://dx.doi.org/10.1007/JHEP07(2021)110}{\emph{JHEP} {\bf 07} (2021) 110}, [\href{https://arxiv.org/abs/2102.08951}{{\tt 2102.08951}}].

\bibitem{Chiang:2023quf}
L.-Y. Chiang, Y.-t. Huang and H.-C. Weng, \emph{{Bootstrapping string theory EFT}}, \href{http://dx.doi.org/10.1007/JHEP05(2024)289}{\emph{JHEP} {\bf 05} (2024) 289}, [\href{https://arxiv.org/abs/2310.10710}{{\tt 2310.10710}}].

\bibitem{Albert:2024yap}
J.~Albert, W.~Knop and L.~Rastelli, \emph{{Where is tree-level string theory?}}, \href{http://dx.doi.org/10.1007/JHEP02(2025)157}{\emph{JHEP} {\bf 02} (2025) 157}, [\href{https://arxiv.org/abs/2406.12959}{{\tt 2406.12959}}].

\bibitem{Berman:2024wyt}
J.~Berman and H.~Elvang, \emph{{Corners and islands in the S-matrix bootstrap of the open superstring}}, \href{http://dx.doi.org/10.1007/JHEP09(2024)076}{\emph{JHEP} {\bf 09} (2024) 076}, [\href{https://arxiv.org/abs/2406.03543}{{\tt 2406.03543}}].

\bibitem{Berman:2025owb}
J.~Berman, H.~Elvang and C.~Figueiredo, \emph{{Splitting Regions and Shrinking Islands from Higher Point Constraints}},  \href{https://arxiv.org/abs/2506.22538}{{\tt 2506.22538}}.

\bibitem{Huang:2025icl}
Y.-t. Huang, S.~Ricossa, F.~Riva and J.-D. Tsai, \emph{{The Rise of Linear Trajectories}},  \href{https://arxiv.org/abs/2510.07991}{{\tt 2510.07991}}.

\bibitem{Caron-Huot:2024tzr}
S.~Caron-Huot, F.~Coronado and Z.~Zahraee, \emph{{Bootstrapping $ \mathcal{N} $ = 4 sYM correlators using integrability and localization}}, \href{http://dx.doi.org/10.1007/JHEP05(2025)220}{\emph{JHEP} {\bf 05} (2025) 220}, [\href{https://arxiv.org/abs/2412.00249}{{\tt 2412.00249}}].

\bibitem{Li:2023qzs}
Y.-Z. Li, \emph{{Effective field theory bootstrap, large-N {\ensuremath{\chi}}PT and holographic QCD}}, \href{http://dx.doi.org/10.1007/JHEP01(2024)072}{\emph{JHEP} {\bf 01} (2024) 072}, [\href{https://arxiv.org/abs/2310.09698}{{\tt 2310.09698}}].

\bibitem{Guerrieri:2021tak}
A.~Guerrieri and A.~Sever, \emph{{Rigorous Bounds on the Analytic S Matrix}}, \href{http://dx.doi.org/10.1103/PhysRevLett.127.251601}{\emph{Phys. Rev. Lett.} {\bf 127} (2021) 251601}, [\href{https://arxiv.org/abs/2106.10257}{{\tt 2106.10257}}].

\bibitem{Caracciolo:2009bx}
F.~Caracciolo and V.~S. Rychkov, \emph{{Rigorous Limits on the Interaction Strength in Quantum Field Theory}}, \href{http://dx.doi.org/10.1103/PhysRevD.81.085037}{\emph{Phys. Rev. D} {\bf 81} (2010) 085037}, [\href{https://arxiv.org/abs/0912.2726}{{\tt 0912.2726}}].

\bibitem{Simmons-Duffin:2015qma}
D.~Simmons-Duffin, \emph{{A Semidefinite Program Solver for the Conformal Bootstrap}}, \href{http://dx.doi.org/10.1007/JHEP06(2015)174}{\emph{JHEP} {\bf 06} (2015) 174}, [\href{https://arxiv.org/abs/1502.02033}{{\tt 1502.02033}}].

\bibitem{Landry:2019qug}
W.~Landry and D.~Simmons-Duffin, \emph{{Scaling the semidefinite program solver SDPB}},  \href{https://arxiv.org/abs/1909.09745}{{\tt 1909.09745}}.

\bibitem{Kravchuk:2021akc}
P.~Kravchuk, D.~Mazac and S.~Pal, \emph{{Automorphic spectra and the conformal bootstrap}}, \href{http://dx.doi.org/10.1090/cams/26}{\emph{Commun. Am. Math. Soc.} {\bf 4} (2024) 1--63}, [\href{https://arxiv.org/abs/2111.12716}{{\tt 2111.12716}}].

\bibitem{Banks:2013nga}
T.~Banks and T.~J. Torres, \emph{{Two Point Pade Approximants and Duality}},  \href{https://arxiv.org/abs/1307.3689}{{\tt 1307.3689}}.

\bibitem{Costin:2020hwg}
O.~Costin and G.~V. Dunne, \emph{{Physical Resurgent Extrapolation}}, \href{http://dx.doi.org/10.1016/j.physletb.2020.135627}{\emph{Phys. Lett. B} {\bf 808} (2020) 135627}, [\href{https://arxiv.org/abs/2003.07451}{{\tt 2003.07451}}].

\bibitem{Costin:2020pcj}
O.~Costin and G.~V. Dunne, \emph{{Uniformization and Constructive Analytic Continuation of Taylor Series}}, \href{http://dx.doi.org/10.1007/s00220-022-04361-6}{\emph{Commun. Math. Phys.} {\bf 392} (2022) 863--906}, [\href{https://arxiv.org/abs/2009.01962}{{\tt 2009.01962}}].

\bibitem{Baker_Graves-Morris_1996}
G.~A. Baker and P.~Graves-Morris, \emph{Padé Approximants}.
\newblock Encyclopedia of Mathematics and its Applications. Cambridge University Press, 2~ed., 1996.

\bibitem{stahl1997convergence}
H.~Stahl, \emph{The convergence of pad{\'e} approximants to functions with branch points}, {\emph{Journal of Approximation Theory} {\bf 91} (1997) 139--204}.

\bibitem{buslaev2013convergence}
V.~I. Buslaev, \emph{Convergence of multipoint pad{\'e} approximants of piecewise analytic functions}, {\emph{Sbornik: Mathematics} {\bf 204} (2013) 190}.

\bibitem{yattselev2021convergence}
M.~L. Yattselev, \emph{Convergence of two-point pad$\backslash$'e approximants to piecewise holomorphic functions}, {\emph{arXiv preprint arXiv:2104.13549} (2021) }.

\bibitem{froissart1973approximation}
M.~Froissart, \emph{Approximation de pade application {\`a} la physique des particules {\'e}l{\'e}mentaires}, {\emph{Les rencontres physiciens-math{\'e}maticiens de Strasbourg-RCP25} {\bf 16} (1973) 1--13}.

\bibitem{Mazzucchelli:2025gyg}
E.~Mazzucchelli and P.~Raman, \emph{{Canonical Forms as Dual Volumes}},  \href{https://arxiv.org/abs/2509.02239}{{\tt 2509.02239}}.

\bibitem{Hodges:2009hk}
A.~Hodges, \emph{{Eliminating spurious poles from gauge-theoretic amplitudes}}, \href{http://dx.doi.org/10.1007/JHEP05(2013)135}{\emph{JHEP} {\bf 05} (2013) 135}, [\href{https://arxiv.org/abs/0905.1473}{{\tt 0905.1473}}].

\bibitem{Arkani-Hamed:2013jha}
N.~Arkani-Hamed and J.~Trnka, \emph{{The Amplituhedron}}, \href{http://dx.doi.org/10.1007/JHEP10(2014)030}{\emph{JHEP} {\bf 10} (2014) 030}, [\href{https://arxiv.org/abs/1312.2007}{{\tt 1312.2007}}].

\bibitem{Arkani-Hamed:2014dca}
N.~Arkani-Hamed, A.~Hodges and J.~Trnka, \emph{{Positive Amplitudes In The Amplituhedron}}, \href{http://dx.doi.org/10.1007/JHEP08(2015)030}{\emph{JHEP} {\bf 08} (2015) 030}, [\href{https://arxiv.org/abs/1412.8478}{{\tt 1412.8478}}].

\bibitem{Arkani-Hamed:2017mur}
N.~Arkani-Hamed, Y.~Bai, S.~He and G.~Yan, \emph{{Scattering Forms and the Positive Geometry of Kinematics, Color and the Worldsheet}}, \href{http://dx.doi.org/10.1007/JHEP05(2018)096}{\emph{JHEP} {\bf 05} (2018) 096}, [\href{https://arxiv.org/abs/1711.09102}{{\tt 1711.09102}}].

\bibitem{Dixon:2016apl}
L.~J. Dixon, M.~von Hippel, A.~J. McLeod and J.~Trnka, \emph{{Multi-loop positivity of the planar $ \mathcal{N} $ = 4 SYM six-point amplitude}}, \href{http://dx.doi.org/10.1007/JHEP02(2017)112}{\emph{JHEP} {\bf 02} (2017) 112}, [\href{https://arxiv.org/abs/1611.08325}{{\tt 1611.08325}}].

\bibitem{Albert:2022oes}
J.~Albert and L.~Rastelli, \emph{{Bootstrapping pions at large N}}, \href{http://dx.doi.org/10.1007/JHEP08(2022)151}{\emph{JHEP} {\bf 08} (2022) 151}, [\href{https://arxiv.org/abs/2203.11950}{{\tt 2203.11950}}].

\bibitem{Arkani-Hamed:2023epq}
N.~Arkani-Hamed, W.~Flieger, J.~M. Henn, A.~Schreiber and J.~Trnka, \emph{{Coulomb Branch Amplitudes from a Deformed Amplituhedron Geometry}}, \href{http://dx.doi.org/10.1103/PhysRevLett.132.211601}{\emph{Phys. Rev. Lett.} {\bf 132} (2024) 211601}, [\href{https://arxiv.org/abs/2311.10814}{{\tt 2311.10814}}].

\bibitem{Demjaha:2025axy}
R.~Demjaha and K.~Zarembo, \emph{{String integrability on the Coulomb branch}}, \href{http://dx.doi.org/10.1007/JHEP09(2025)154}{\emph{JHEP} {\bf 09} (2025) 154}, [\href{https://arxiv.org/abs/2506.17955}{{\tt 2506.17955}}].

\bibitem{Coronado:2025xwk}
F.~Coronado, S.~Komatsu and K.~Zarembo, \emph{{Coulomb Branch and Integrability}},  \href{https://arxiv.org/abs/2506.07222}{{\tt 2506.07222}}.

\bibitem{Belitsky:2025bgb}
A.~V. Belitsky, L.~V. Bork, R.~N. Lee, A.~I. Onishchenko and V.~A. Smirnov, \emph{{Five W-boson amplitude = near-null decagon}},  \href{https://arxiv.org/abs/2510.16471}{{\tt 2510.16471}}.

\bibitem{Craig:2011ws}
N.~Craig, H.~Elvang, M.~Kiermaier and T.~Slatyer, \emph{{Massive amplitudes on the Coulomb branch of N=4 SYM}}, \href{http://dx.doi.org/10.1007/JHEP12(2011)097}{\emph{JHEP} {\bf 12} (2011) 097}, [\href{https://arxiv.org/abs/1104.2050}{{\tt 1104.2050}}].

\bibitem{Kiermaier:2011cr}
M.~Kiermaier, \emph{{The Coulomb-branch S-matrix from massless amplitudes}},  \href{https://arxiv.org/abs/1105.5385}{{\tt 1105.5385}}.

\bibitem{Herderschee:2019dmc}
A.~Herderschee, S.~Koren and T.~Trott, \emph{{Constructing $ \mathcal{N} $ = 4 Coulomb branch superamplitudes}}, \href{http://dx.doi.org/10.1007/JHEP08(2019)107}{\emph{JHEP} {\bf 08} (2019) 107}, [\href{https://arxiv.org/abs/1902.07205}{{\tt 1902.07205}}].

\bibitem{Bork:2022vat}
L.~V. Bork, N.~B. Muzhichkov and E.~S. Sozinov, \emph{{Infrared properties of five-point massive amplitudes in $ \mathcal{N} $ = 4 SYM on the Coulomb branch}}, \href{http://dx.doi.org/10.1007/JHEP08(2022)173}{\emph{JHEP} {\bf 08} (2022) 173}, [\href{https://arxiv.org/abs/2201.08762}{{\tt 2201.08762}}].

\bibitem{MdAbhishek:2023nvg}
M.~Abhishek, S.~Hegde, D.~P. Jatkar, A.~P. Saha and A.~Suthar, \emph{{On-shell functions on the Coulomb branch of $\mathcal{N}$ = 4 SYM}}, \href{http://dx.doi.org/10.1007/JHEP05(2024)157}{\emph{JHEP} {\bf 05} (2024) 157}, [\href{https://arxiv.org/abs/2311.17763}{{\tt 2311.17763}}].

\bibitem{Bianchi:2015tba}
M.~S. Bianchi, \emph{{A note on scattering amplitudes on the moduli space of ABJM}}, \href{http://dx.doi.org/10.1007/JHEP06(2015)194}{\emph{JHEP} {\bf 06} (2015) 194}, [\href{https://arxiv.org/abs/1502.05352}{{\tt 1502.05352}}].

\bibitem{Drummond:2007aua}
J.~M. Drummond, G.~P. Korchemsky and E.~Sokatchev, \emph{{Conformal properties of four-gluon planar amplitudes and Wilson loops}}, \href{http://dx.doi.org/10.1016/j.nuclphysb.2007.11.041}{\emph{Nucl. Phys. B} {\bf 795} (2008) 385--408}, [\href{https://arxiv.org/abs/0707.0243}{{\tt 0707.0243}}].

\bibitem{Brandhuber:2007yx}
A.~Brandhuber, P.~Heslop and G.~Travaglini, \emph{{MHV amplitudes in N=4 super Yang-Mills and Wilson loops}}, \href{http://dx.doi.org/10.1016/j.nuclphysb.2007.11.002}{\emph{Nucl. Phys. B} {\bf 794} (2008) 231--243}, [\href{https://arxiv.org/abs/0707.1153}{{\tt 0707.1153}}].

\bibitem{Belitsky:2021huz}
A.~V. Belitsky and V.~A. Smirnov, \emph{{An off-shell Wilson loop}}, \href{http://dx.doi.org/10.1007/JHEP04(2023)071}{\emph{JHEP} {\bf 04} (2023) 071}, [\href{https://arxiv.org/abs/2110.13206}{{\tt 2110.13206}}].

\bibitem{Chen:2025ffl}
H.~Chen, P.~F. Monni, Z.~Pang, G.~Vita and H.~X. Zhu, \emph{{Correlation Function/Wilson Loop Duality in Gauge Theory from EFT}},  \href{https://arxiv.org/abs/2510.07377}{{\tt 2510.07377}}.

\bibitem{Chester:2021aun}
S.~M. Chester, R.~Dempsey and S.~S. Pufu, \emph{{Bootstrapping $ \mathcal{N} $ = 4 super-Yang-Mills on the conformal manifold}}, \href{http://dx.doi.org/10.1007/JHEP01(2023)038}{\emph{JHEP} {\bf 01} (2023) 038}, [\href{https://arxiv.org/abs/2111.07989}{{\tt 2111.07989}}].

\bibitem{Chester:2023ehi}
S.~M. Chester, R.~Dempsey and S.~S. Pufu, \emph{{Level repulsion in $ \mathcal{N} $ = 4 super-Yang-Mills via integrability, holography, and the bootstrap}}, \href{http://dx.doi.org/10.1007/JHEP07(2024)059}{\emph{JHEP} {\bf 07} (2024) 059}, [\href{https://arxiv.org/abs/2312.12576}{{\tt 2312.12576}}].

\bibitem{Tseytlin:2010jv}
A.~A. Tseytlin, \emph{{Review of AdS/CFT Integrability, Chapter II.1: Classical AdS5xS5 string solutions}}, \href{http://dx.doi.org/10.1007/s11005-011-0466-0}{\emph{Lett. Math. Phys.} {\bf 99} (2012) 103--125}, [\href{https://arxiv.org/abs/1012.3986}{{\tt 1012.3986}}].

\bibitem{Penedones:2007ns}
J.~Penedones, \emph{{High Energy Scattering in the AdS/CFT Correspondence}},  other thesis, 12, 2007.

\bibitem{Gabai:2025hwf}
B.~Gabai, V.~Gorbenko and J.~Qiao, \emph{{Yang-Mills Flux Tube in AdS}},  \href{https://arxiv.org/abs/2508.08250}{{\tt 2508.08250}}.

\bibitem{Hanany:2010da}
A.~Hanany, D.~Forcella and J.~Troost, \emph{{The Covariant perturbative string spectrum}}, \href{http://dx.doi.org/10.1016/j.nuclphysb.2011.01.002}{\emph{Nucl. Phys. B} {\bf 846} (2011) 212--225}, [\href{https://arxiv.org/abs/1007.2622}{{\tt 1007.2622}}].

\bibitem{Todorov:1970gr}
I.~T. Todorov, \emph{{Quasipotential equation corresponding to the relativistic Eikonal approximation}}, \href{http://dx.doi.org/10.1103/PhysRevD.3.2351}{\emph{Phys. Rev. D} {\bf 3} (1971) 2351--2356}.

\bibitem{Correia:2024jgr}
M.~Correia and G.~Isabella, \emph{{The Born regime of gravitational amplitudes}}, \href{http://dx.doi.org/10.1007/JHEP03(2025)144}{\emph{JHEP} {\bf 03} (2025) 144}, [\href{https://arxiv.org/abs/2406.13737}{{\tt 2406.13737}}].

\bibitem{Schwinger:1964zzb}
J.~Schwinger, \emph{{Coulomb Green's Function}}, \href{http://dx.doi.org/10.1063/1.1931195}{\emph{J. Math. Phys.} {\bf 5} (1964) 1606--1608}.

\bibitem{Feinberg:1970zz}
G.~Feinberg and J.~Sucher, \emph{{General Theory of the van der Waals Interaction: A Model-Independent Approach}}, \href{http://dx.doi.org/10.1103/PhysRevA.2.2395}{\emph{Phys. Rev. A} {\bf 2} (1970) 2395--2415}.

\bibitem{Kos:2013tga}
F.~Kos, D.~Poland and D.~Simmons-Duffin, \emph{{Bootstrapping the $O(N)$ vector models}}, \href{http://dx.doi.org/10.1007/JHEP06(2014)091}{\emph{JHEP} {\bf 06} (2014) 091}, [\href{https://arxiv.org/abs/1307.6856}{{\tt 1307.6856}}].

\bibitem{Chang:2025mwt}
C.-H. Chang, V.~Dommes, P.~Kravchuk, D.~Poland and D.~Simmons-Duffin, \emph{{Accurate bootstrap bounds from optimal interpolation}},  \href{https://arxiv.org/abs/2509.14307}{{\tt 2509.14307}}.

\bibitem{Boyd_Vandenberghe_2004}
S.~Boyd and L.~Vandenberghe, \emph{Convex Optimization}.
\newblock Cambridge University Press, 2004.

\end{thebibliography}\endgroup

\end{document}